\documentclass[a4paper,11pt]{article}
\pdfoutput=1 
             
\usepackage{jheppub} 

\usepackage[utf8]{inputenc}
\usepackage[T1]{fontenc} 

\usepackage{amsmath}
\usepackage{amssymb}
\usepackage{graphicx}
\usepackage{bm}
\usepackage{epsfig}
\usepackage{amsfonts}
\usepackage{color}
\usepackage{mathtools}
\usepackage[section]{placeins}

\newcommand{\be}{\begin{equation}}
\newcommand{\ee}{\end{equation}}

\newcommand{\bea}{\begin{eqnarray}}
\newcommand{\eea}{\end{eqnarray}}
\newcommand{\bean}{\begin{eqnarray*}}
\newcommand{\eean}{\end{eqnarray*}}

\newcommand{\xv}{{\mathbf x}}

\newcommand{\nn}{\nonumber}

\newcommand{\bB}{{\bf B}}

\newcommand{\bDl}{{\bf\Delta}}
\newcommand{\bE}{{\bf E}}
\newcommand{\bsigma}{{\boldsymbol\sigma}}

\graphicspath{{./img/}}

\title{\boldmath Quarkonium in-medium properties from realistic lattice NRQCD}

\author[a]{Seyong Kim,}
\author[b]{Peter Petreczky,}
\author[c,d,1]{Alexander Rothkopf \note{Corresponding author.}}

\affiliation[a]{Department of Physics, Sejong University, Seoul 143-747, Korea}
\affiliation[b]{Physics Department, Brookhaven National Laboratory,
  Upton, NY 11973, USA}
\affiliation[c]{Institut f\"ur Theoretische Physik, Universit\"at Heidelberg, 
Philosophenweg 12, 69120 Heidelberg, Germany}
\affiliation[d]{Faculty of Science and Technology, University of Stavanger, NO-4036 Stavanger, Norway}

\emailAdd{skim@sejong.ac.kr}
\emailAdd{petreczk@quark.phy.bnl.gov}
\emailAdd{alexander.rothkopf@uis.no}

\abstract{We present the final results of our high statistics study on the properties of bottomonium and charmonium at finite temperature. We focus on the temperature range around the crossover transition $150\leq T\leq 410$MeV, relevant for current heavy ion collision experiments. The QCD medium degrees of freedom which consist of dynamical u,d, and s quarks and gluons are captured by realistic state-of-the art ($m_\pi\approx 161$MeV) lattice QCD simulations of the HotQCD collaboration. For the heavy quarks we deploy the non-relativistic effective field theory of QCD, NRQCD. The in-medium properties of quarkonium are deduced from their spectral functions, which are reconstructed using improved and novel Bayesian approaches. Through a systematic analysis we shed light on the origin of the discrepancies in melting temperatures previously reported in the literature, showing that they are owed to underestimated methods uncertainties of the deployed spectral reconstructions. Our simulations corroborate a picture of sequential in-medium modification, ordered according to the vacuum binding energy of the states. As a central quantitative result, our study reveals how the mass of the heavy quarkonium ground state reduces as temperature increases. The observed spectral modifications are interpreted in the light of, and compared to previous studies based on the complex lattice potential for heavy quarkonium. Thus for the first time we provide a robust picture of in-medium heavy quarkonium modification in the quark-gluon plasma consistent among different non-relativistic methods. We also critically discuss the perspectives for improving on these results.}

\begin{document} 
\maketitle
\flushbottom


\section{Introduction}
 \label{sec:intro}

Resolving different temporal stages of a relativistic heavy-ion collision is a crucial requirement to truly understand the genesis and the evolution of the Quark-Gluon Plasma (QGP) \cite{Jacak:2012dx,Muller:2013dea}. The early time stage is characterized by strongly interacting matter far away from even local thermal equilibrium with large momentum anisotropy. In the intermediate to late time stage, local thermal equilibrium may have been achieved and different physical processes may become important. The experimentalists toolkit contains several classes of observables that hold the promise of eventually granting a time resolved window into the collision center. Among them are electromagnetic and hard probes. A representative of the former are photons. The observation of a Boltzmann type distribution in photon yields at low momenta at RHIC \cite{Adare:2008ab} and LHC \cite{Wilde:2012wc} was initially interpreted as a representative signal for the temperature of QGP. Later more detailed studies of photon production in the context of hydrodynamic models \cite{Shen:2013vja} and of photon production at early time \cite{Berges:2017eom} have shown that non-thermal physics can also lead to similar spectral shapes. Thus, distinguishing the photons produced at each stages of the collision remains a challenging subject to date. In contrast, heavy quarkonia, bound states of a heavy quark and antiquark, play a unique role among hard probes \cite {Brambilla:2010cs,Andronic:2015wma}. The reason lies in the fact that a separation of scales exists between the heavy quark masses ($m^{\overline{MS}}_c=1.28(3)$GeV, $m^{\overline{MS}}_b=4.18(4)$GeV) \cite{PDG2018}, the characteristic scale of quantum fluctuations in QCD ($\Lambda_{\rm QCD}\approx 0.2-0.5$ GeV) and the characteristic temperatures of a relativistic heavy-ion collision ($T\lesssim 0.6$GeV). This scale separation allows us to quantitatively understand hadronic processes involving heavy quarks by use of various factorization theorems. According to this understanding, creating heavy quarks and antiquarks is mainly possible in the early stage of a collision, where partons from the incoming nuclei can deposit their kinetic energy into exciting a pair from the vacuum. Subsequently, it is those quark pairs from the early stage that traverse the collision region, interacting with the rest of the bulk matter created therein and thus sampling the full evolution of the collision. In this work, we concentrate on quarkonia.

\subsection{Phenomenological consideration}

Originally, quarkonium, considered in a purely thermal context, had been proposed as gold plated signal for the creation of a QGP \cite{Matsui:1986dk}, since the force binding two quarks together weakens due to the presence of deconfined color charges. Thus a relative depletion of their yields in a heavy-ion collision compared to that in a $p+p$ collision was proposed and indeed subsequently measured at RHIC \cite{Adare:2006ns,Adare:2008sh,Adare:2011yf,Adamczyk:2013tvk,Adamczyk:2013poh,Adare:2014hje} and LHC \cite{Abelev:2013ila,Adam:2016rdg,Chatrchyan:2011pe,Chatrchyan:2012lxa,Abelev:2014nua,Khachatryan:2016xxp,Sirunyan:2017lzi}. Taking a more quantitative angle, it was argued that the suppression would be ordered according to the binding energy of the vacuum quarkonium states \cite{Karsch:2005nk}: more weakly bound states dissociate at lower temperatures than the more strongly bound ones. In turn a detailed measurement of different quarkonium states would allow them to take on the role of a strongly interacting thermometer for heavy-ion collisions \cite{Mocsy:2013syh}.

At the comparatively low RHIC energies, around ${\cal O}(10)$ $c\bar{c}$ and ${\cal O}(1)$ $b\bar{b}$ pairs are expected to be created in each collisions (one has to bear in mind that the nuclear parton distribution functions can have an impact of around $10\%$ for bottomonium and $20\%$ for charmonium on the suppression besides medium effects. This is known as shadowing effect. \cite{Paakkinen:2018zbs,Andronic:2015wma}). However, both charmonium and bottomonium yields show a significant suppression and this experimental observation is commonly expressed in terms of the nuclear modification factor $R_{\rm AA}<1$. This quantity decreases monotonously as more nucleons participate in the collisions. Also, it is interesting to note that at least for charmonium suppression no significant dependence on the beam energy is observed in the RHIC beam energy scan program \cite{Andronic:2015wma}. 

At the higher LHC energies the number of heavy quark pairs increases to ${\cal O}(100)$ for $c\bar{c}$ and ${\cal O}(10)$ for $b\bar{b}$ changing the physics significantly for the charm quarks. The abundance of charm quarks (possibly deconfined) in the collision center allows them to meet at the time of freezeout and to form a quarkonium bound state in significant numbers even if they have become completely decorrelated from the original partner with which they had been produced. This regeneration or recombination production channel will lead to a replenishment of yields for charmonium. This phenomenon had been first predicted in the context of the statistical model of hadronization \cite{BraunMunzinger:2000px,Andronic:2010dt} and is clearly evidenced by measurements of the ALICE collaboration \cite{Abelev:2013ila}. 

At first the success of the statistical model for charmonium may come as a surprise, had one originally thought of as quarkonium as test particle which samples the QGP as it plows through the collision center. However current heavy-ion collisions produce a QGP with an estimated lifetime around $\tau_{\rm QGP}\approx {\cal O}(10)$fm, which leaves enough time for the charm quarks to interact often enough with their surrounding so that partial kinetic equilibration may be achieved. Strong evidence for such a scenario at the LHC is found in the recent measurements of a finite elliptic flow for the $J/\Psi$ particle \cite{ALICE:2013xna,Acharya:2017tgv}, the ground state of the vector channel $c\bar{c}$. This indicates that indeed the individual charm quarks partake in the collective motion of the surrounding bulk. Such an at least partial equilibration in turn entails a loss of memory about the initial conditions of the charm quarks which forms charmonium system. In short, at LHC charmonium will become a messenger of the late stages of the collision. 

Bottomonium at the LHC behaves however very similar to charmonium at RHIC. Di-muon spectra obtained by the CMS collaboration show a clear pattern consistent with a sequential in-medium modification, by which we simply mean that the ratio of excited states to ground states $\Upsilon(2S)/\Upsilon(1S)$ and $\Upsilon(3S)/\Upsilon(1S)$ are consecutively smaller (see e.g. \cite{Khachatryan:2016xxp,Sirunyan:2017lzi}). Between Run 1 and Run 2, the beam energy for heavy ion increased from $\sqrt{s_{NN}}=2.76$TeV to $5.02$TeV which imprinted itself differently on the $R_{AA}$ of the two flavors. While it increased the value for charmonium (consistent with even higher recombination), it slightly lowered that of bottomonium. This again supports the picture of bottomonium suppression being dominated by the melting of initially bound bottom anti-bottom pairs.

The fact that bottomonium interacts with the bulk matter over its full lifetime and that there is of yet no direct evidence for an equilibration of the bottom quarks is promising. It entails that we may extract from its yields properties of the QGP at earlier times than those charmonium informs us about. The combination of charmonium and bottomonium thus provides us an access to different regimes of the heavy-ion collision. One should note however that with increasing beam energy also bottom quarks will eventually becomes kinetically equilibrated. While at Run 1 with $\sqrt{s_{NN}}=2.76$TeV there were indeed no direct hints at a possible equilibration of bottom quarks it is still unclear whether this is still the case at $\sqrt{s_{NN}}=5.02$TeV (see e.g. discussion in \cite{Krouppa:2017jlg}).

A quantitative description of charmonium $R_{\rm AA}$ has been achieved by different phenomenological approaches. Continuous dissolution and regeneration occurring in the QGP is implemented by transport models based on the Boltzmann \cite{Yao:2017fuc} or kinetic rate equations \cite{Rapp:2008tf,Zhao:2010nk,Zhao:2011cv,Zhou:2014kka,Song:2011nu,Emerick:2011xu,Zhou:2014hwa}. For bottomonium, solving a Schr\"odinger equation with a complex potential, embedded in a dynamical hydrodynamic background provides another viable alternative \cite{Strickland:2011mw,Strickland:2011aa,Nendzig:2012cu,Krouppa:2015yoa,Krouppa:2016jcl,Hoelck:2016tqf,Krouppa:2017jlg}. A first principle understanding of both the inputs to, as well as the range of applicability of such phenomenological models is needed to further insight of underlying QCD.

\subsection{Considerations in thermal equilibrium}

The modest goal of this work is to provide quantitative insight into the properties of heavy quarkonium in thermal equilibrium. From the above discussion, our result will be most relevant for an understanding of the late stages of a heavy ion collision. We will in the following consider both charm and bottom quarks immersed in a static medium at fixed temperature. The most basic questions to ask and which we attempt to answer are: at which temperature do heavy quarkonium states dissolve, relating to their role as thermometer, and how do their properties change as they are heated up. Answers to these questions will give us hints as to how their production is modified. If the in-medium mass increases one may expect that at hadronization the number of vacuum states produced from the in-medium states will decrease. Vice versa, a lowering of the in-medium mass may lead to a more abundant production at hadronization \cite{Burnier:2015tda}.

In the language of quantum field theory the properties of bound states can be extracted from so called spectral functions. These objects are defined from heavy quarkonium two-point correlation functions and their functional form offers an intuitive picture of the physics involved. Stable bound states available to the system correspond to delta-peak like structures situated at a certain frequency, denoting the mass of the corresponding particle. In case of a resonance or thermally unstable particle the peak broadens and acquires a finite width according to its inverse lifetime. At finite temperature this width consists of both a loss channel contribution and a thermal excitation contribution, i.e. the width does not only encode the annihilation decay of the quarkonium into other light particles (light quarks and gluons) but also that it may be excited by the medium into another more weakly bound state. In the presence of medium of light quark flavors, it can become energetically favorable for the heavy quarks to pair instead with one of the light quarks forming a D or B meson. Due to the further particles involved, the system may now carry a continuous spectrum of energy leading to a threshold above which the spectral function goes into a extended continuum structure. 

In thermal equilibrium there exists a direct relation between the peak structures in the spectral function and the dilepton decay rates of the corresponding particles \cite{McLerran:1984ay}, allowing a first principles connection from theory to measured yields. This fact is exploited in the study of in-medium modification of light meson spectra. It is important to understand, however, that in the case of a heavy quarkonium in heavy ion collisions such a scenario is never met. The yields measured in experiment are those of $T=0$ states because decays of quarkonium into di-muon mostly happen long after the QGP has ceased to exist. Were the detectors at LHC capable of resolving quarkonium states as well as dedicated quarkonium factories, we would observe that all peaks are those of the vacuum particles. Thus, we need to connect the in-medium properties via the process of hadronization to the yields of vacuum particles, but no consistent theory framework exist yet. A recent proposal to translate in-medium spectral functions to vacuum particle yields has been discussed in \cite{Burnier:2015tda,Burnier:2016kqm} where the area of the in-medium spectral structure was expressed in units of the vacuum spectral function area of the same state, defining the number of produced vacuum states at freezeout in that way. More work on a first principles understanding of hadronization however is called for.

Let us also remark on a concept often discussed in phenomenology, the melting temperature of heavy quarkonium. It is important to note that the popularity of this concept historically stems from the use of real-valued model potentials (see e.g. \cite{Satz:2008zc}) with which a Schr\"odinger equation was solved. In such a setting a more and more screened potential will eventually be unable to support a bound state and the temperature where this happens in uniquely defined \cite{Mocsy:2007jz}. In the modern view of heavy quarkonium as open quantum system \cite{Young:2010jq, Akamatsu:2011se, Borghini:2011ms, Akamatsu:2014qsa,Blaizot:2015hya, Akamatsu:2015kaa,Katz:2015qja,Brambilla:2016wgg,Brambilla:2017zei,Blaizot:2017ypk,DeBoni:2017ocl,Blaizot:2018oev} where ``open'' means a contact with a thermal bath, it is known that the potential between heavy quarks actually takes on complex values \cite{Laine:2006ns,Beraudo:2007ky,Brambilla:2008cx,Rothkopf:2011db,Burnier:2014ssa,Burnier:2016mxc,Petreczky:2017aiz}, related to kicks of the medium onto the color string binding the pair. In the language of spectral functions, a complex potential induces a thermal width in the in-medium state, which then merges smoothly with the continuum \cite{Burnier:2007qm,Petreczky:2010tk,Burnier:2015tda,Burnier:2016kqm}. This makes a definition of the melting temperature rather ambiguous. 

Currently, the most commonly used definition of the melting temperature is that this temperature is the point at which the in-medium binding energy equals the thermal width. Let us reiterate that the thermal width is not just a measure of the heavy quark pair annihilating into gluons but instead also quantifies the probability of the bound state being excited to higher lying states by interactions with the medium. The development of a fully dynamical description of heavy quarkonium as open quantum system is an active and fast moving field of research. Our study of the equilibrium properties of quarkonium can provide an important input to this field as the benchmark toward which any dynamical evolution in a static medium should thermalize.

In order to compute heavy quarkonium spectral functions at temperatures relevant in heavy ion collisions, we need to resort to genuinely non-perturbative methods. The temperature range of the current generation of heavy ion collisions spans $120<T<600$MeV. As experimentally intended, it includes the crossover transition region from hadronic matter to the QGP \cite{Bazavov:2011nk, Borsanyi:2013bia,Bazavov:2014pvz,Borsanyi:2016ksw,Bazavov:2017dsy}. There the QGP is strongly correlated, 
as indicated by the large value of the trace anomaly (also known as the interaction measure) \cite{Borsanyi:2013bia,Bazavov:2014pvz}
and the effective coupling defined in terms of static quark anti-quark free energy \cite{Bazavov:2018wmo}.
Even at $T=600$MeV, however, it is not clear a priori, how well perturbative methods can describe various aspects of 
quark gluon plasma. Some quantities likes the pressure \cite{Borsanyi:2016ksw,Bazavov:2017dsy} and quark number susceptibilities
\cite{Ding:2015fca,Bellwied:2015lba,Bazavov:2013uja} are reasonably well described by perturbative methods, other quantities that are sensitive
to chromo-electric screening, e.g. the free energy of a static quark, are described by perturbative calculations only at much higher
temperatures \cite{Berwein:2015ayt,Bazavov:2016uvm,Bazavov:2018wmo}.
In order to capture QCD in a non-perturbative setting, 
we thus deploy lattice QCD, a proven first principles approach to the strong interactions. 

While lattice QCD has made possible vital insight into the static properties of the QCD medium it remains challenging to extract dynamical properties, such as spectral functions from these simulations because lattice simulations are carried out in Euclidean time. Contrary to the Minkowski time domain, computing spectral functions in Euclidean time requires us to tackle an ill-posed inverse problem. In this study we use methods from Bayesian inference to provide a regularization of the unfolding task. Besides deploying novel methods, such as the BR method \cite{Burnier:2013nla} and a smooth extension thereof \cite{Fischer:2017kbq}, we also compare to established approaches, 
such as the Maximum Entropy Method (MEM) \cite{Asakawa:2000tr,Wetzorke:2001dk}.

A further challenge is the inclusion of heavy quark degrees of freedom in the lattice simulation, since in the standard relativistic formulation they require very fine lattice spacings compared to the typical scales resolved in the QCD medium. This separation of scales however will be turned into an advantage in this study by deploying instead a non-relativistic effective field theory (EFT) on the lattice called NRQCD \cite{Thacker:1990bm,Lepage:1992tx,Brambilla:2004jw}. NRQCD has been established as precision tool for vacuum spectroscopy (see e.g. \cite{Colquhoun:2015fuw}) and is straight forwardly applicable to heavy quarks also at finite temperature in the nonperturbative regime close to $T_c$. Another technical benefit of NRQCD is that its correlation functions are not periodic in Euclidean time and thus the full temporal extent remains accessible for use in spectral reconstructions, and is free from the zero mode problem in the two-point correlation function associated with susceptibility \cite{Umeda:2007hy,
Aarts:2002cc,Petreczky:2008px}.

The study of quarkonium in-medium spectral properties has a long history using both relativistic 
and non-relativistic heavy quark formulations. The former has been deployed mostly in the study of charmonium 
\cite{Karsch:2002wv,Asakawa:2003re,Datta:2003ww,Jakovac:2006sf,Iida:2006mv,Ohno:2011zc,Ding:2012sp,Aarts:2007pk,Borsanyi:2014vka,Ohno:2014uga,Ikeda:2016czj,Ding:2017std,Burnier:2017bod,Kelly:2018hsi}, where also full QCD simulations can be realized. 
Bottomonium on the other hand requires so small lattice spacings 
that currently only quenched QCD results exist \cite{Jakovac:2006sf,Ohno:2014uga,Jin:2017jhq}. 
The FASTSUM collaboration on the other hand has studied bottomonium in lattice NRQCD on anisotropic lattices 
over the past years \cite{Aarts:2011sm,Aarts:2012ka,Aarts:2013kaa,Aarts:2014cda}. 
Besides the use of anisotropy their approach further differs from ours in that they deploy a fixed scale approach, 
i.e. temperature is varied by changes in the number of Euclidean lattice sites with a fixed lattice spacing. In addition the pion mass on their second generation ensembles currently in use remains higher than that available on the latest generation isotropic lattices used here.

An important part of this study is to understand how different Bayesian methods may have lead to disagreement on e.g. melting temperatures, in particular for the P-wave quarkonium states. We propose a reconciliation by systematically investigating the uncertainty of the results using different methods, placing particular care on the influence of the role of smoothing.

Preliminary results of this study have been presented at various conferences and workshops (e.g. \cite{Rothkopf:2016vsn,Kim:2017aio}).

\subsection{The main results}

For the reader foremost interested in the physics content of this study we here summarize the four main results presented in the manuscript:

\begin{itemize}
\item We have extended our previous analysis of in-medium NRQCD correlation functions from bottomonium to include also charmonium and significantly reduced the statistical uncertainties (Fig. \ref{Fig:FiniteTCorrelatorRatios}). The inspection of these correlation functions corroborates with high significance a picture of sequential in-medium modification\footnote{Note that this does not automatically imply a sequential suppression to be observed in a heavy-ion collision}. I.e. the effect of the medium manifests itself more strongly in states, which have a lower vacuum binding energy, or correspondingly which have a larger spatial vacuum extent. 

\item We provide improved estimates of the melting temperatures from an inspection of the spectral functions from three different reconstruction prescriptions, the BR method, a smooth variant thereof and the Maximum Entropy Method (Fig.\ref{Fig:FiniteTModelCmpBottom} and Fig.\ref{Fig:FiniteTModelCmpCharm}).
\begin{align}
T_{\Upsilon} > 407 {\rm MeV}, \quad T_{\chi_{b1}} \in [185,223]{\rm MeV}, \quad T_{J/\Psi} \in  [200,210]{\rm MeV}, \quad T_{\chi_{c1}} \lessapprox 185 {\rm MeV}
\end{align}
As a crosscheck, using the same reconstruction as other studies in the literature (e.g. MEM) we obtain very similar results. However the use of different reconstructions allows us to shed light on the systematic uncertainties of the obtained temperatures, leaving us with a melting region instead of a single temperature.
 
\item This study provides an improved determination of in-medium mass shifts (Fig.\ref{Fig:FiniteTMassShifts}) of bottomonium and charmonium, shown below are the value at an intermediate temperature:
 \begin{align}
 \nonumber&\Delta m_{\Upsilon}(T=185{\rm MeV})=-13(4){\rm MeV}, \quad \Delta m_{\chi_{b1}}(T=185{\rm MeV})=-55(12){\rm MeV}\\
& \Delta m_{J/\Psi}(T=185{\rm MeV})=-50(6){\rm MeV},\quad \Delta m_{\chi_{c1}}(T=185{\rm MeV})=-115(35){\rm MeV}
 \end{align}
The obtained negative values are again ordered in magnitude with the vacuum binding energy of the vacuum state as suggested by sequential in-medium modification. They furthermore are fully consistent with the non-perturbative implementation of the effective field theory pNRQCD. On the other hand they tell us that weakly coupled pNRQCD, which predicts a positive mass shift is not applicable in the temperature range considered here. In order to obtain these results the selection of a the correct vacuum reference was crucial (gray squares vs. blue crosses in Fig.\ref{Fig:FiniteTMassShifts}). I.e. while our raw results for the in-medium masses are very similar to those obtained in previous studies (e.g. by the FASTSUM collaboration) our final result and conclusion differs significantly. 

\item Mock data tests reveal that in order to improve on the results presented in this study, significant efforts are needed in the simulation of quarkonium in lattice QCD. We find that simply going towards the continuum limit does not significantly improve the reconstruction of the bound state peaks but will at least provide more reliable insight into the continuum structure of the spectral function. As a first route, focus should be placed on bringing simulations on anisotropic lattices to the same level of realism as those on isotropic lattices (in terms of pion masses) but ultimately genuinely novel ideas, such as an extension of the multilevel algorithm to full QCD will be needed.
\end{itemize}
 
A more detailed discussion of the methods and results will be presented in the following chapters, starting in sec.\ref{sec:nummethods} with a summary of the lattice QCD simulation setup (sec. \ref{sec:latdetails}) followed by a detailed description of the implementation of lattice NRQCD (sec. \ref{Num:LatNRQCD}). We introduce the Bayesian approach to spectral function reconstruction in sec. \ref{sec:bayesrec} and discuss our strategy how to reconcile results from different Bayesian methods in sec.\ref{sec:RecBayesRes}. The study of correlators and spectral functions at zero temperatures in sec.\ref{sec:T0res} provides a benchmark for the reliability of the NRQCD approximation (sec.\ref{sec:T0groundstate}) as well as the Bayesian reconstruction (sec.\ref{sec:prepFiniteT}). We present our finite temperature results in sec.\ref{sec:physres} starting with the raw correlators (sec\ref{sec:secmod}) and a subsequent investigation of their spectral functions (sec.\ref{sec:FiniteTspecfunc}). Our main quantitative result of the in-medium mass shifts is discussed in sec.\ref{sec:FiniteTMassShifts}. We close in sec.\ref{sec:disc} with a summary and an outlook, which critically discusses possible routes towards improving on our results.

\section{Numerical methods}
\label{sec:nummethods}

In this section we outline the individual methods we combine in our study to non-perturbatively determine the in-medium properties of heavy quarkonium at finite temperature. We first introduce the lattice QCD simulations used to capture the gluons and light dynamical quarks. Next, we provide a detailed description of our implementation of the lattice effective field theory NRQCD for the heavy quarks. Finally we discuss the improved and novel Bayesian methods, which are deployed in the reconstruction of spectral functions from lattice NRQCD correlation functions.

\begin{table}[t]
\begin{center}
\vspace*{0.2cm}
\begin{tabular}{|c|c|c|c|r|r|r|r|r|r|}
\hline
$\beta$	& Volume      &$L$[fm]  & $a$[fm]& $u_0$   &$M_b a$&$M_c a$ \\
\hline
6.664 &$32^3\times 32$&3.74& 0.117 &$0.87025$  & 2.76 & 0.757 \\
6.740 &$48^3\times 48$&5.22& 0.109 &$0.87288$  & 2.57 & 0.704 \\
6.800 &$32^3\times 32$&3.28& 0.103 &$0.87485$  & 2.42 & 0.664 \\
6.880 &$48^3\times 48$&4.56& 0.095 &$0.87736$  & 2.24 & 0.615\\
6.950 &$32^3\times 32$&2.85& 0.089&$0.87945$  & 2.10& 0.576 \\
7.030 &$48^3\times 48$&3.96& 0.082&$0.88173$  & 1.95 & 0.534 \\
7.150 &$48^3\times 64$&3.54& 0.074&$0.88493$  & 1.74 & 0.477 \\
7.280 &$48^3\times 64$&3.14& 0.066&$0.88817$  & 1.55 & 0.424 \\
7.373 &$48^3\times 64$&2.89& 0.060&$0.89035$  & 1.42 & -\\
7.596 &$64^3\times 64$&3.15& 0.049&$0.89517$  & 1.16 & -\\
7.825 &$64^3\times 64$&2.58& 0.040&$0.89962$  & 0.95 & - \\
\hline
\end{tabular}
\vspace*{0.2cm}
\caption{Parameters of the HISQ lattice configurations at $T=0$,
  which are used to calibrate the NRQCD energy shift and in cross-checks of
  the Bayesian reconstruction methods. For each lattice spacing there are
  400 configurations available.}
\label{tab:parameterT0}
\end{center}
\end{table}
\begin{table}[t]
\begin{center}
\vspace*{0.2cm}
\begin{tabular}{|c|c|c|r|r|r|r|r|r|}
\hline
$\beta$	& T    & $T/T_c$& $a$(fm)& $u_0$   &$M_b a$&$M_c a$& $N_{\rm conf}^{b\bar{b}}$\\
\hline
6.664	&140& 0.882  & 0.117& 0.87025  & 2.76 & 0.76 & 420 \\
6.700	&145& 0.914  & 0.113&  0.87151   & 2.67& 0.73  & 690 \\
6.740	&151& 0.950  & 0.109&  0.87288   & 2.57 & 0.70 &  2940 \\
6.770	&155& 0.978  & 0.106&  0.87388   & 2.49& 0.68 & 680  \\
6.800	&160& 1.01  & 0.103&  0.87485   & 2.42& 0.66  & 3690 \\
6.840	&166& 1.05  & 0.0987&  0.87612  & 2.33& 0.64  &  830 \\
6.880	&173& 1.09  & 0.0950& 0.87736  & 2.24& 0.62  & 1000 \\
6.910	&178& 1.12  & 0.0924&  0.87827  & 2.18& 0.60  & 460 \\
6.950	&185& 1.16  & 0.0889&  0.87945  & 2.10& 0.58  & 4110\\
6.990	&192& 1.21  & 0.0856&  0.88060  & 2.02& 0.55  &520 \\
7.030	&199& 1.25  & 0.0825&  0.88173  & 1.95& 0.53 & 4070 \\
7.100	&212& 1.34  & 0.0773&  0.88363  & 1.82& 0.50 &  990 \\
7.150	&223& 1.40  & 0.0738&  0.88493  & 1.74 & 0.48 & 3740 \\
7.280	&251& 1.58  & 0.0655&  0.88817  & 1.55& 0.42  & 3630 \\
7.373	&273& 1.72  & 0.0602&  $0.89035$  & 1.42& - &4030 \\
7.596	&333& 2.09  & 0.0493&  $0.89517$  & 1.16& -  & 4000 \\
7.825	&407& 2.56  & 0.0403&  $0.89962$  & 0.95& -  & 4060\\
\hline
\end{tabular}
\vspace*{0.2cm}
\caption{Parameters of the HISQ lattice configurations at $T>0$ with size $48^3\times 12$, which underlie
  the extraction of the in-medium spectral functions. For charmonium we have a evaluated the correlators on 
  $N_{\rm conf}^{c\bar{c}}=400$ configurations each.
  }
\label{tab:parameters}
\end{center}
\end{table}

\subsection{Lattice simulations for the QCD medium}
\label{sec:latdetails}

In order to obtain a realistic description of the QCD medium in which the heavy quark anti-quark pair will be immersed in, we take advantage of high statistics state-of-the-art lattice QCD simulations of the HotQCD collaboration, including fully dynamical u,d and s quarks. Being interested in the temperature range of $150<T<410$MeV, relevant for the late stages of current heavy-ion collisions at RHIC and LHC, simulations with $(u,d,s)$ quarks are sufficient. The contribution of charm and bottom quarks to the thermodynamic properties of the medium at these temperatures is indeed small \cite{Borsanyi:2016ksw}.

We combine finite temperature lattices from two studies \cite{Bazavov:2014pvz, Bazavov:2011nk} originally designed for the study of the transition temperature and the QCD equation of state. The number of grid points on the lattices is fixed at $48^3 \times 12$ and temperature is changed by a variation of the lattice spacing.  Dynamical quarks, based on the Highly Improved Staggered Quark (HISQ) action for $N_f=2+1$ flavors are included, where the strange quark mass $m_s$ is tuned to take on physical values. The light quark masses are set to $m_l=m_s/20$, which leads to an almost physical pion mass in the continuum limit of $m_\pi=161$ MeV.

The continuum limit value of the chiral transition temperature determined from these lattices lies at $T_{\rm pc}=154(9)$ MeV. Its value at finite lattice spacing and Euclidean extent $N_\tau=12$ on the other hand is $T_{\rm pc}^{\rm lat}=159(3)$ MeV, which is however compatible with the physical result within its uncertainties. The updated results for the transition temperature from HotQCD collaboration, $T_c=156.5(1.5)$ MeV
also agrees with the above number within errors \cite{Steinbrecher:2018phh}. We therefore refer in the following simply to $T_{\rm pc}$ and in some cases quote temperatures in units thereof. 

For calibration purposes and methods cross-checks, configurations at zero temperatures are also used. They are computed on $32^3$ and $64^3$ spatial lattices at the same values of the lattice parameter $\beta$ as the corresponding finite temperature configurations. 

Compared to our previous study on in-medium heavy quarkonium, we not only enlarge the temperature range surveyed from $T_{\rm max}^{\rm old}=251$MeV $=1.58T_c$ to $T_{\rm max}^{\rm new}=407$MeV $=2.56T_c$ for bottomonium but at the same time also increase the number of available lattice configurations significantly. It amounts to a four-fold increase at zero temperature to 400 per lattice spacing, while at finite temperature we have an eight to ten-fold increase to 3000 or 4000 configurations at those temperatures, for which also zero temperature calibration lattices are available. Statistics at the other lattice spacings can reach up to three times the previous value. Details about the physical box size, temperature, lattice spacing and statistics can be found in Tab.\ref{tab:parameterT0} for the zero temperature ensembles and in Tab.\ref{tab:parameters} for the high temperature ensembles. Note that we follow the improved determination of lattice spacings in \cite{Bazavov:2014pvz}, which leads to percent-level differences in the values of the lattice spacing and temperature compared to our previous work.

\subsection{Lattice NRQCD for heavy quarks}
\label{Num:LatNRQCD}

As mentioned in the introduction, we choose to deploy the non-relativistic effective field theory NRQCD \cite{Brambilla:2004jw}, discretized on the lattice \cite{Lepage:1992tx}, in order to describe the heavy quark degrees of freedom. This choice is a compromise between accuracy and feasibility, since a fully relativistic description of heavy quarks e.g. with the same HISQ action as used for the u,d, and s quarks would require a much finer lattice spacing if bottom quarks are to be considered. In addition relativistic correlation functions necessarily obey the KMS relation, i.e. they are symmetric around the center of the Euclidean time domain, in essence depriving us of half of the usable information contained in the simulation. 

The EFT approach on the other hand relies on a systematic expansion of the QCD Lagrangian in increasing powers of $v \sim {\mathbf p}/(M_q
a)$ (in contrast to an expansion directly in $(M_qa)^{-1}$, as in the EFT called heavy quark effective theory \cite{HQET1,HQET2}). Radiative corrections to NRQCD on the lattice have been investigated in detail \cite{Hammant:2013sca} for a discretization scheme that is
different from ours.  Therefore, the role of radiative corrections in our calculations is unknown. We take the acceptable reproduction of the vacuum mass shifts as indication that our simulations are reliable at all lattice spacing and quark masses considered. This stability argument also underlies our decision to limit the investigation of charmonium in NRQCD to temperatures up to $T=1.58T_c$ or $a=0.066$fm, since for finer lattices the convergence of the NRQCD expansion was not satisfactory anymore.

Let us discuss the details of the implementation, where we follow the lattice discretization to order ${\cal O} (v^4)$ of the NRQCD Lagrangian in Euclidean time originally developed in \cite{Lepage:1992tx, Davies:1994mp} and applied early on at finite temperature in \cite{Fingberg:1997qd}.

It amounts to an expansion of the QCD heavy-quark Lagrangian in the form 
\begin{equation}
\label{LNRQCD}
{\cal L} = {\cal L}_0 + \delta {\cal L},
\end{equation}
where the leading order contribution reads
\begin{equation}
\label{LNRQCD_1}
{\cal L}_0 = \psi^\dagger \left(D_\tau - \frac{{\bold D^2}}{2M_q} \right) \psi +
\chi^\dagger \left(D_\tau + \frac{{\bold D^2}}{2M_q} \right) \chi,
\end{equation}
and its corrections are summarized in 

\begin{align}
\nonumber \delta {\cal L} = &\hm - \frac{c_1}{8M_q^3} \left[\psi^\dagger ({\bold D^2})^2
  \psi - \chi^\dagger ({\bold D^2})^2 \chi \right]\\ 
  &+ c_2 \frac{ig}{8M_q^2}\left[\psi^\dagger \left({\bold
    D}\cdot{\bold E} - {\bold E}\cdot{\bold D}\right) \psi +
  \chi^\dagger \left({\bold D}\cdot {\bold E} - {\bold E}\cdot{\bold
    D} \right) \chi \right] \nonumber \\   
\nonumber& \hm - c_3 \frac{g}{8M_q^2}\left[\psi^\dagger
  \bsigma\cdot\left({\bold D}\times{\bold E}-{\bold E}\times{\bold
    D}\right)\psi + \chi^\dagger {\bold
    \sigma}\cdot\left({\bold D}\times{\bold E}-{\bold E}\times{\bold
    D}\right)\chi \right]\\
    &- c_4 \frac{g}{2M_q} \left[\psi^\dagger {\bold \sigma}\cdot{\bold B} \psi -
    \chi^\dagger {\bold \sigma} \cdot {\bold B} \chi \right].
    \label{LNRQCD_2}
\end{align}

We denote by $D_\tau$ and ${\bold D}$ the gauge covariant derivatives
in temporal and spatial direction respectively and use $\psi$ and $\chi$
to refer to the the two-component Pauli spinor fields of the heavy quark and
antiquark. We use the leading order for the Wilson coefficients $c_i$ in eq.\eqref{LNRQCD_2}, i.e.
$c_i=1$, since perturbative calculations of these coefficients
for the lattice discretization used by us are not available. 
It is important to note that some of the effects of higher energy scales  $(\geq M_q)$
reflected otherwise in a deviation of $c_i$ from unity will be incorporated
instead by the use of tadpole improvement.

Since the quark and antiquark fields are fully decoupled to this order in the NRQCD
Lagrangian, the forward and backward propagating modes contributing equally to the
relativistic correlator can actually be distinguished. Instead of having to solve a 
boundary value problem for the propagation of the original four-component Dirac spinor,
we need to solve an initial value problem for each of the two two-component Pauli
spinors present. Each quark and antiquark propagator then becomes a non-periodic
function in Euclidean time, which combined appropriately would nevertheless fashion an 
approximation of the symmetric relativistic correlator. On the level of a 
spectral function we may think of each correlator representing either
the positive frequency or negative frequency regime of the antisymmetric relativistic
spectral function we originally started out from in QCD. 

The initial value problem for the NRQCD propagator can be cast in the form
of a diffusion-like equation, which when discretized with an explicit
forward scheme reads
\begin{eqnarray}
G (\xv, \tau_0) = &&\hm S(\xv), \nn\\
G (\xv, \tau_1) = &&\hm  \left(1 - \frac{H_0}{2n}\right)^n
U_4^\dagger(\xv, \tau_0) \left(1 - \frac{H_0}{2n}\right)^n G(\xv,\tau_0), \nn \\
G (\xv, \tau_i ) = &&\hm  \left(1 - \frac{H_0}{2n}\right)^n
U_4^\dagger(\xv, \tau_{i-1}) \left(1 - \frac{H_0}{2n}\right)^n \nonumber\\
&&\times \left(1 -\delta H \right)  G (\xv, \tau_{i-1}).\label{NRQCDEvolEq}
\end{eqnarray}
To handle efficiently the spatial delta function that the propagator has to equal at
initial time $\tau_0=0$, we introduce here a complex valued random point source
$S(\xv)$, diagonal in spin and color. It helps to improve the signal to noise ratio
in the actual computation, where multiple correlators are computed starting 
from random sources placed on different time slices $\tau_{\rm start}$ among
the available Euclidean times, i.e.
\begin{align}
\nonumber S^{\Upsilon}(\xv,\tau_{\rm start})=\eta(\xv,\tau_{\rm start}),\\
 \langle \eta^\dagger(\xv,\tau_{\rm start}) \eta(\xv',\tau_{\rm start}) \rangle=\delta_{\xv\xv'}. \label{Eq:sourcesNRQCD}
\end{align}

The discretized evolution equation for the propagator may now be used to define what
is meant by the lattice NRQCD Hamiltonian, as we have already suggestively used the letter 
$H$ in eq.\eqref{NRQCDEvolEq}. We are lead to the leading order term
\be 
H_0 = - \frac{\Delta^{(2)}}{2M_q}, 
\ee and higher order corrections of the form
\begin{eqnarray}
\delta H = &&\hm - \frac{(\Delta^{(2)})^2}{8 M_q^3} + \frac{ig}{8 M_q^2}
(\bDl^{\pm}\cdot \bE - \bE\cdot \bDl^{\pm})\nonumber \\
&&- \frac{g}{8 M_q^2} \bsigma \cdot
  (\bDl^{\pm} \times \bE - \bE\times \bDl^{\pm})   \nonumber \\
&&\hm
- \frac{g}{2 M_q} \bsigma\cdot\bB
 + \frac{a^2\Delta^{(4)}}{24 M_q} - \frac{a (\Delta^{(2)})^2}{16 n M_q^2}.
\label{eq:deltaH}
\end{eqnarray}

In the above expression the lattice covariant derivatives $\Delta$ are 
defined via 
\begin{eqnarray}
a \Delta_i^{+} \psi (\xv, \tau) &=& U_i (\xv, \tau) \psi (\xv + \hat{i} a ,
\tau) - \psi (\xv, \tau) \nonumber \\
a \Delta_i^{-} \psi (\xv, \tau) &=& \psi (\xv, \tau) - U_i^\dagger (\xv
- \hat{i} a, \tau) \psi (\xv - \hat{i} a , \tau) \nonumber \\
\Delta^{(2)} &=& \sum_{i=1}^{3} \Delta_i^{+} \Delta_i^{-}, \;
\Delta^{(4)} = \sum_{i=1}^{3} (\Delta_i^{+} \Delta_i^{-})^2,
\end{eqnarray}
and the expressions for the chromo-electric $({\bold E})$ 
and the magnetic field $({\bold B})$ are taken to arise from from clover-leaf plaquettes. 
The link variables entering these plaquettes are modified according to the naive tadpole
improvement procedure, i.e. they are divided by the fourth root of the single plaquette
expectation value as listed in Tab.\ref{tab:parameters}. The last two terms of Eq. \eqref{eq:deltaH} 
do not contribute in the continuum and are added to diminish finite lattice spacing errors.

Starting with eq.\eqref{NRQCDEvolEq} we have introduced a parameter $n$, conventionally
christened the Lepage parameter, which allows us to select an effective step size in Euclidean time during the
evolution of the heavy quark propagator on the lattice. Its values have a direct
effect on the UV behavior of the propagator $G$ and need to be chosen
so that the theory is well defined. On the other hand, as long the value of
$n$ is not too large the effect it has on the non-perturbative IR regime, where
bound state signals are expected, is small, as we had shown in our previous 
work \cite{Kim:2014iga}. 

The stability of the NRQCD expansion, i.e. the behavior of the correlator in the UV,
is closely related with the value of $M_q a$. In this study our goal is to study
in-medium effects and not a precision determination of vacuum properties. This
on the one hand means that we will refrain from using overlap-optimized interpolating
operators and on the other hand forego tuning of the heavy quark masses. Instead we simply use the values
$M_b=4.65$GeV and $M_c=1.275$GeV. 

The Lepage parameter then needs to be chosen large enough so that NRQCD remains stable and 
at the same time small enough to 
not affect the low energy regime. Here we use $n=4$ for the bottom quarks
and $n=8$ for charm. Even with this relatively large value for $n$ the dynamics
of charmonium is only reliably captured on our lattices up to $a=0.0656$fm, corresponding
to $\beta=7.280$, beyond which we only consider bottomonium. The 
individual values of $M_b a \in [2.759 \ldots 1.559]$ and $M_c a \in  [0.757 \ldots 0.427]$ are listed in Tab.\ref{tab:parameters}.

To reduce the computational burden in solving Eq.\eqref{NRQCDEvolEq} we furthermore
resort to a well established procedure in which covariant derivatives are replaced
by simple finite differences, while being evaluated on Coulomb gauge fixed
lattices. 

Having computed the heavy quark propagator we can turn to constructing the 
 heavy quarkonium correlation function projected onto appropriate quantum numbers
\begin{align}
\nonumber D({\mathbf x},\tau)&=\sum_{{\mathbf x}_0}\langle O({\mathbf x},\tau)
G({\mathbf x},\tau) O^\dagger({\mathbf x}_0,\tau_0) G^\dagger({\mathbf
  x},\tau) \rangle_{\rm med} .
\end{align}
As we are interested in a purely static scenario, we will consider only the spatially
averaged correlator $D$, which corresponds to heavy quarkonium at vanishing 
momentum ${\mathbf p}=0$. Note that this correlator can be considered as the
NRQCD counterpart of the unequal time correlation function of static wave-functions 
considered in the effective theory of pNRQCD \cite{Brambilla:2004jw}.

Different channels are distinguished through the vertex operator inserted in between the
heavy quark propagators. Their explicit form for the channels considered in this study reads
\begin{align}
\nonumber &O(^3S_1;{\mathbf x},\tau)=\sigma_i, \quad O(^1S_0;{\mathbf x},\tau)=1\\
\nonumber&O(^1P_1;{\mathbf
  x},\tau)=\overset{\leftrightarrow_s}{\Delta}_i, \quad   O(^3P_0;{\mathbf
  x},\tau)=\sum_i\overset{\leftrightarrow_s}{\Delta}_i\sigma_i\\
  \nonumber&  O(^3P_1;{\mathbf
  x},\tau)=\overset{\leftrightarrow_s}{\Delta}_i\sigma_j-\overset{\leftrightarrow_s}{\Delta}_j\sigma_i,\\
  &O(^3P_2;{\mathbf
  x},\tau)=\overset{\leftrightarrow_s}{\Delta}_i\sigma_i-\overset{\leftrightarrow_s}{\Delta}_j\sigma_j
  \end{align}
Here we use $\chi^\dagger\overset{\leftrightarrow_s}{\Delta}_i\psi=-\Big[\frac{1}{4}\big(\Delta^+_i+\Delta^-_i\big)\chi\Big]^\dagger\psi +\chi^\dagger\Big[\frac{1}{4}\big(\Delta^+_i+\Delta^-_i\big)\psi\Big]$
as defined originally in \cite{Thacker:1990bm}.

To describe either the $\Upsilon$ or $J/\Psi$ vector channel $(^3S_1)$, 
the correlator takes on the following explicit form

\begin{eqnarray}
\langle ({\chi^\dagger}_a  (\sigma_x)_{ab} \psi_b (x^\prime))^\dagger
\chi^\dagger_c (\sigma_x)_{cd} \psi_d (x) \rangle &=& 2 \langle G_{+ +}
(x^\prime;x)^\dagger G_{- -} (x^\prime;x) + G_{+ -} (x^\prime;x)^\dagger G_{- +}
(x^\prime;x) \rangle \nonumber \\
\langle (\chi^\dagger_a  (\sigma_y)_{ab} \psi_b (x^\prime))^\dagger
\chi^\dagger_c (\sigma_y)_{cd} \psi_d (x) \rangle &=& 2 \langle G_{+ +}
(x^\prime;x)^\dagger G_{- -} (x^\prime;x) - G_{+ -} (x^\prime;x)^\dagger G_{- +}
(x^\prime;x) \rangle \nonumber \\
\langle (\chi^\dagger_a  (\sigma_z)_{ab} \psi_b (x^\prime))^\dagger
\chi^\dagger_c (\sigma_z)_{cd} \psi_d (x) \rangle &=& \langle G_{+ +}
(x^\prime;x)^\dagger G_{+ +} (x^\prime;x) + G_{-
  -}(x^\prime;x)^\dagger G_{- -} (x^\prime;x) \nonumber \\
& & - G_{+ -}(x^\prime;x)^\dagger G_{+ -} (x^\prime;x) - G_{-
  +}(x^\prime;x)^\dagger G_{- +} (x^\prime;x) \rangle, \nonumber \\
\end{eqnarray}

In the above expression $+(-)$ refers to the spin-up (spin-down) component
and $\langle \cdots \rangle$ to the thermal average over the lattice QCD ensembles.

The P-wave $\chi_{b1}$ or $\chi_{c1}$ correlator $(^3P_1)$ including finite angular momentum
on the other hand reads

\begin{eqnarray}
\langle \left[\chi^\dagger_a
  \left(\stackrel{\leftrightarrow}{\Delta}_i (\sigma_j)_{ab} -
  \stackrel{\leftrightarrow}{\Delta}_j (\sigma_i)_{ab} \right)\psi_b
(x^\prime)\right]^\dagger \chi^\dagger_c \left(\stackrel{\leftrightarrow}{\Delta}_i
(\sigma_j)_{cd} - \stackrel{\leftrightarrow}{\Delta}_j (\sigma_i)_{cd}
\right) \psi_d
(x) \rangle \nonumber \\
= \langle {\rm tr} \left[G^\dagger (x^\prime;x) \sigma_i
  \Delta^+_j G_V (x^\prime;x,j) \sigma_i \right] \rangle
- \langle {\rm tr} \left[{G_V}^\dagger (x^\prime;x,j) \sigma_i
  \Delta^+_j G (x^\prime;x) \sigma_i \right] \rangle
\nonumber \\
+  \langle {\rm tr} \left[G^\dagger (x^\prime;x) \sigma_j
  \Delta^+_i G_V (x^\prime;x,i) \sigma_j \right] \rangle
- \langle {\rm tr} \left[{G_V}^\dagger (x^\prime;x,i) \sigma_j
  \Delta^+_i G (x^\prime;x) \sigma_j \right] \rangle
\nonumber \\
- \langle {\rm tr} \left[G^\dagger (x^\prime;x) \sigma_i
  \Delta^+_j G_V (x^\prime;x,i) \sigma_j \right] \rangle
+ \langle {\rm tr} \left[{G_V}^\dagger (x^\prime;x,j) \sigma_i
  \Delta^+_i G (x^\prime;x) \sigma_j \right] \rangle
\nonumber \\
- \langle {\rm tr} \left[G^\dagger (x^\prime;x) \sigma_j
  \Delta^+_j G_V (x^\prime;x,i) \sigma_i \right] \rangle
+ \langle {\rm tr} \left[{G_V}^\dagger (x^\prime;x,j) \sigma_j
  \Delta^+_i G (x^\prime;x) \sigma_i \right] \rangle
\nonumber
\end{eqnarray}

where  $\chi^\dagger \stackrel{\leftrightarrow}{\Delta}_i \psi \equiv
\chi^\dagger \Delta^+_i \psi - \Delta^+_i \chi^\dagger
\psi$. By deploying point split sources along the $i$-th direction
we achieve a further reduction in computational cost when 
evolving $G_V(x^\prime ;x, i)$ and ``${\rm tr}$'' denotes the color and spin trace.

\subsection{Bayesian spectral reconstruction}
\label{sec:bayesrec}

While the correlation functions of NRQCD already allow us to study the overall effects of in-medium modification as will be discussed in sec.\ref{sec:secmod}, our interest lies in understanding the modification in a more differential manner, i.e. for each of the original vacuum bound states separately. The pertinent information of particle mass and lifetime may be conveniently read-off from the position and width of peak structures in the spectral function underlying the Euclidean correlator.

Extracting spectral functions from numerical simulation data however presents a formidable challenge
that we tackle in this study through the use of Bayesian inference \cite{skilling1991,Jarrell:1996rrw,bishop:2006:PRML} . We work predominantly with the recent
BR method \cite{Burnier:2013nla} and an extension thereof \cite{Fischer:2017kbq} but also compare our results to well known approaches,
such as the Maximum Entropy Method (MEM)\cite{Asakawa:2000tr}. We will learn how to reconcile apparently different results that were obtained through these methods in the literature leading to consistent estimates for in-medium quarkonium properties.

The starting point for the reconstruction is the integral relation between Euclidean correlators $D$ and their spectral function $\rho(\omega)$. Conventionally in lattice QCD we consider the correlator $D(\tau)$ in Euclidean time, which in NRQCD fulfills
\begin{align}
 D(\tau)=D({\bf p}=0,\tau)=\sum_{\bf x}D({\bf x},\tau)=\int_{-2M_q}^\infty
 d\omega \; e^{-\omega\tau}\;\rho(\omega).\label{Eq:SpecConv}
\end{align}
with a temperature independent but exponentially suppressed kernel $K(\tau,\omega)={\rm
exp}[-\omega\tau]$. 

Note that by construction of NRQCD, the origin of the frequency axis here is shifted to a value close to $2 M_q$. On the lattice the value of this shift depends on the lattice spacing. In turn the spectrum may actually reach into the negative frequency domain, which requires us in practice to choose a priori a minimal frequency $\omega_{\rm min}$ when setting up the spectral reconstruction. The freedom to choose $\omega_{\rm min}$, as will be discussed in more detail below, also reveals that some reconstruction methods, such as the MEM carry an intrinsic dependence on the choice of reconstruction interval, which is usually hidden in a relativistic scenario, where only positive frequencies are considered.

Another benefit of using NRQCD is that we do not have to deal with the periodic kernel of the relativistic theory here. This removes the ``constant contribution'' issue discussed in \cite{Umeda:2007hy,
Aarts:2002cc,Petreczky:2008px}.

On the other hand the Euclidean correlator may equally well be represented in imaginary 
frequencies, where it is known as the Matsubara correlator. A Fourier transform (FT) connects
the two, which also allows to translate the kernel from one scenario the the other, still harboring
the same spectral function
\begin{align}
 D(\omega_n)&=D({\bf p}=0,\omega_n)=\int_{-2M_q}^\infty
 d\omega \; \frac{1-{\rm exp}[-\beta \omega + i\beta\omega_n]}{\omega-i\omega_n} \rho(\omega).
 \label{Eq:SpecConvKL}
 \end{align}
 This is the non-relativistic lattice counterpart of the K{\"a}ll{\'e}n-Lehmann kernel. Since the
 NRQCD correlator in Euclidean time is not periodic $D(\omega_n)$ takes on
 complex values. Note that the kernel here for large real-time frequencies is much less damped
 than in the standard case. One may thus be worried about the weak decay in the 
 denominator, however for any finite lattice spacing the expression
 is well defined no matter the UV scaling of the operator.
 
 Since the correlators of eq.\eqref{Eq:SpecConv} or eq.\eqref{Eq:SpecConvKL}
 are sensitive to different parts of the spectrum, the former more to the low frequency
 regime, while the latter more to the large frequency regime, we will combine 
 both data sets in the analysis. While one may argue that a FT is simply a linear transformation
 and thus nothing can be gained from switching between eq.\eqref{Eq:SpecConv} or eq.\eqref{Eq:SpecConvKL},
 we find that since the inversion process is highly non-linear the results do benefit
 from increased stability if both correlators are considered.
 
Inverting either eq.\eqref{Eq:SpecConv} or eq.\eqref{Eq:SpecConvKL} to
obtain $\rho$ is an ill-posed problem since our input data is limited in two ways.
Due to the space-time discretization in lattice QCD the correlator is evaluated only at a finite
number of points $(N_\tau=N_{\omega_n})$ and since the simulation prescription relies on Monte-Carlo
sampling, each correlator is known only to a finite precision. Anticipating the
intricate structure encoded in the spectral function one discretizes the real-time
frequency range with a number of bins $N_\omega\gg N_\tau, N_{\omega_n}$.
Determining from the available simulation data the $N_\omega$ parameters
representing the spectrum, is at first sight a severely under-determined problem.
 
Up to this point we have however not made use of a powerful resource at our
disposal, i.e. analytic knowledge about the spectral properties in QCD, be it the
positive definiteness of hadronic spectral functions or even their pole structure. 
Such prior knowledge should be incorporated systematically in the inversion task,
and Bayesian inference provides a mathematically sound framework to do so.
 
Bayesian reasoning assigns a probability distribution to the following question:
How probable is it that a test function $\rho$ is the correct spectral function, given
the simulated correlator $D$ and the prior information $I$ at hand. This so called
posterior probability $P[\rho|D,I]$ enter Bayes theorem
\begin{align}
 P[\rho|D,I]= P[D|\rho,I]P[\rho|I]/P[D|I]\label{eq:BayesTheorem}
\end{align}
on the LHS. The right hand side consists of three terms, the likelihood probability
$P[D|\rho,I]$, the prior probability $P[\rho|I]$ and what is colloquially termed the
evidence $P[D|I]$.

For statistically sampled data, due to the central limit theorem, the likelihood 
probability $P[D|\rho,I]={\rm exp}[-L]$ may be written using the $\chi^2$ functional 
\begin{align}
 L[\rho]=\frac{1}{2}\sum_{ij}(D_i-D^\rho_i)C_{ij}(D_j-D^\rho_j),
\end{align}
where $C_{ij}$ refers to the covariance matrix of the simulation data and
$D^\rho_i$ refers to the Euclidean correlator 
\begin{align}
 D^\rho_i=\sum_{l=1}^{N_\omega}\;K_{il}\;\rho_l\;\Delta\omega_l\label{eq:discrcorr}
\end{align}
corresponding to the test function $\rho$. In case of combining Euclidean time 
and Matsubara correlators we keep a block diagonal form of $C_{ij}$ counting 
only correlations among the points of each correlator species.
Maximizing the likelihood alone corresponds to a naive $\chi^2$ fit, which would
not converge due to the many flat directions present in $L$.

Prior knowledge may already enter this probability distribution (see eq.\eqref{eq:BayesTheorem}) , 
and we use this freedom to encourage the selection of spectra that fulfill the additional constraint $L=N_\tau$.
The correct spectral function inserted into eq.\eqref{eq:discrcorr} and sampled
with Gaussian noise would lead on average to this value. In effect we are
preventing the over-fitting of noise.

The majority of prior information is encoded in the prior probability, which
quantifies how compatible the test function rho is to that information. A particular choice for the functional form of the prior probability 
$P[\rho|I]={\rm exp}[S]$ selects among the many 
degenerate maximum likelihood solutions the one which adheres most
closely to the constraints given by prior information. 

It is important to note that in this construction we always guarantee that
the reconstructed spectrum also reproduces the input data, which is not the
case in other competing approaches, such as the Backus Gilbert \cite{Press:1992:NRC:148286} and
Pad\'e methods \cite{Tripolt:2018xeo}.

We will deploy two different prior probabilities, i.e. regulator functionals,
in this study. The first is the one of the standard BR method, which has been
designed specifically with spectral reconstruction in field theories in mind.
It enforces (1) the positivity of the reconstructed spectrum, (2) guarantees
that the end result remains unaffected by the units of $\rho(\omega)$ and
(3) produces a smooth function whenever simulation data has not imprinted
sharp peak structures on the spectrum. Its functional form reads
\begin{align}
 S_{\rm BR}[\rho]=\alpha\sum_l\;\Big( 1-\frac{\rho_l}{m_l}+{\rm
   log}\Big[\frac{\rho_l}{m_l}\Big]\Big)\Delta \omega_l.
\end{align}
which differs from the Shannon-Jaynes entropy used
in the Maximum Entropy Method 
\begin{align}
 S_{\rm SJ}[\rho]=\alpha\sum_l\;\Big( \rho_l - m_l - \rho_l {\rm
   log}\Big[\frac{\rho_l}{m_l}\Big]\Big)\Delta \omega_l.
\end{align}
In particular, the absence of the factor $\rho$ in front of the logarithm
means $S_{BR}$ is not an entropy and the prior probability is instead related to 
the gamma distribution. 

The functional form of $S$ is one piece of the prior information entering
the Bayesian approach. There is furthermore the function $m(\omega_l)=m_l$,
the so called default model, which by definition corresponds to the correct spectrum in
the absence of data. It represents the unique extremum of $S$.

We will use a constant $m(\omega)={\rm const.}$ in the following as we do not
wish to assume any additional knowledge beyond positive definiteness and smoothness.
From the Euclidean time correlator at $\tau=0$ we know the area under the
spectrum, to which we normalize the default model. Representative examples of 
the default model dependence of our spectral reconstructions are provided in 
Appendix \ref{sec:dmdep}. All quantitative results presented in the following
contain in their error bars estimates of such default model dependence.

There is one more ingredient in the standard BR method we need to take care of,
the hyperparameter $\alpha$ that weighs prior information versus data. Due to the
particular form of $S_{\rm BR}$ it is possible to integrate it out a priorly using a flat
overall hyperprior.
\begin{align}
 P[\rho|D,I]=P[D|I]\int_0^\infty d\alpha P[\rho|m,\alpha]
\end{align}
This procedure liberates us from using the usually poor Gaussian approximation
entering the $\alpha$ probability estimation of the MEM (for details
see ref.\cite{Burnier:2013nla}).

The most probable spectrum in the Bayesian sense is then obtained 
as point estimate of the maximum of the posterior probability
\begin{align}
 \left.\frac{\delta}{\delta \rho}P[\rho|D,I]\right|_{\rho=\rho^{\rm
     BR}}=0
\label{BayesStationary}
\end{align}
It is important to note that in contrast to the MEM the search space
from which $\rho^{\rm BR}$ is selected is not restricted but we carry out
a numerical optimization procedure in the full $N_\omega$ degrees of freedom.
One may wonder, whether a unique solution can be found if the search
space is not restricted. However as was shown in \cite{Asakawa:2000tr},
supplying the regulator functional and the $N_\omega$ pieces of prior
information in form of the default model are sufficient to guarantee the
uniqueness of the solution of eq.\eqref{BayesStationary} if a solution exists.

There are two sources of uncertainty entering our determination of $\rho^{\rm BR}$.
One is related to the errors present in the input correlator, the other 
related to the choice of default model. Bayesian methods, such as the BR method,
as well as the MEM allow for a internal measure of robustness \cite{Asakawa:2000tr}, which is related to
the depth of the global extremum found in the posterior probability. Usually
this entails a Gaussian approximation of the posterior which we have found often
lead to an underestimation of the true uncertainties.

Therefore in our study we explicitly check the dependence of the end result on
the statistics by carrying out a binned Jackknife procedure: we repeat the reconstruction
using an averaged correlator based on only $90\%$ of the available configurations,
each time removing a different but consecutive block of $10\%$. The variance among
these results is the basis for the Jackknife errorbars attached to our results below.

The default model dependence is estimated by varying both the amplitude and functional form of 
$m(\omega)$. We use $m(\omega)=m_0 (\omega-\omega_{\rm min}+1)^{\gamma}, 
\omega>\omega_{\rm min}$ with $m_0\in[0.1,10]$, as well as $\gamma=\{0,1,2,-1,-2\}$.

\subsection{Reconciling results from different Bayesian methods}
\label{sec:RecBayesRes}

In our previous study \cite{Kim:2014iga} we found that the spectral structures obtained with different Bayesian methods, such as the MEM and the BR method using the same correlator data sets were significantly different. This then translated into methods differences in the determination of e.g. the disappearance of the ground state peak, i.e. its melting temperature. In this section we discuss how this disagreement arises and introduce a novel regulator that interpolates between the two different methods in a genuinely Bayesian fashion. In turn we will be able to reconcile the apparent differences.

We would like to note that the apparent differences between our previous results \cite{Kim:2014iga} and other NRQCD based studies from e.g. the FASTSUM collaboration \cite{Aarts:2014cda} cannot be solely attributed to different Bayesian methods. In addition to using the MEM instead of the BR, their lattice QCD setup was also significantly different from ours, in particular featuring a more than twice as large pion mass.

In this section we focus solely on differences between the MEM and the BR methods. We argue below that Bayesian methods with local regulators such as BR and MEM, share a common disadvantage, i.e. ringing may contaminate the reconstruction. This artifact, closely related to the Gibbs phenomenon in the inverse Fourier transform, is e.g. clearly observed in the BR method when reconstructing from a small number of correlator points, i.e. $N_\tau<16-20$, as shown in our previous study.  The MEM on the other hand is usually considered safe from ringing artifacts, an impression, which however is unfortunately incorrect on the level of the regulator as we illustrate here. It is instead due to the ad-hoc restricted search space with dimension equal to the number of data points, that the MEM contains an additional smoothing prescription, which in turn suppresses ringing for a small number of data points. This however does not mean that the MEM produces a more faithful reconstruction of the correct spectral function, even though its result is more visually pleasing. For an explicit counter example to the proof underlying the search space restriction in the MEM see e.g. \cite{Rothkopf:2011ef}, where it is also shown that the smoothing of the MEM depends crucially on the choice of $\omega_{\rm min}$. 

On the other hand we have shown in our previous study \cite{Kim:2014iga} that the BR method results only show a very weak dependence on the choice of frequency interval. In particular its results are robust against extending the frequency interval, contrary to the MEM.

Let us show that there exist situations in which both the BR regulator and the Shannon-Jaynes entropy of the MEM give preference to wiggly spectra, even though they were designed with smooth functions in mind. In particular we give an example that counters the intuitive notion that a regulator based on the concept of entropy will always prefer curves with the minimal amount of structure. Let us consider a constant default model $m=1$ and compare the prior probability assigned to a function $\rho_{\rm broad}$ that contains a single broad peak, as well as to a function $\rho_{\rm wiggle}$ which contains the same broad peak but modified by a single added wiggle. The area under both curves is chosen to be the same. 

\begin{figure}[t]
\centering
\includegraphics[scale=0.59]{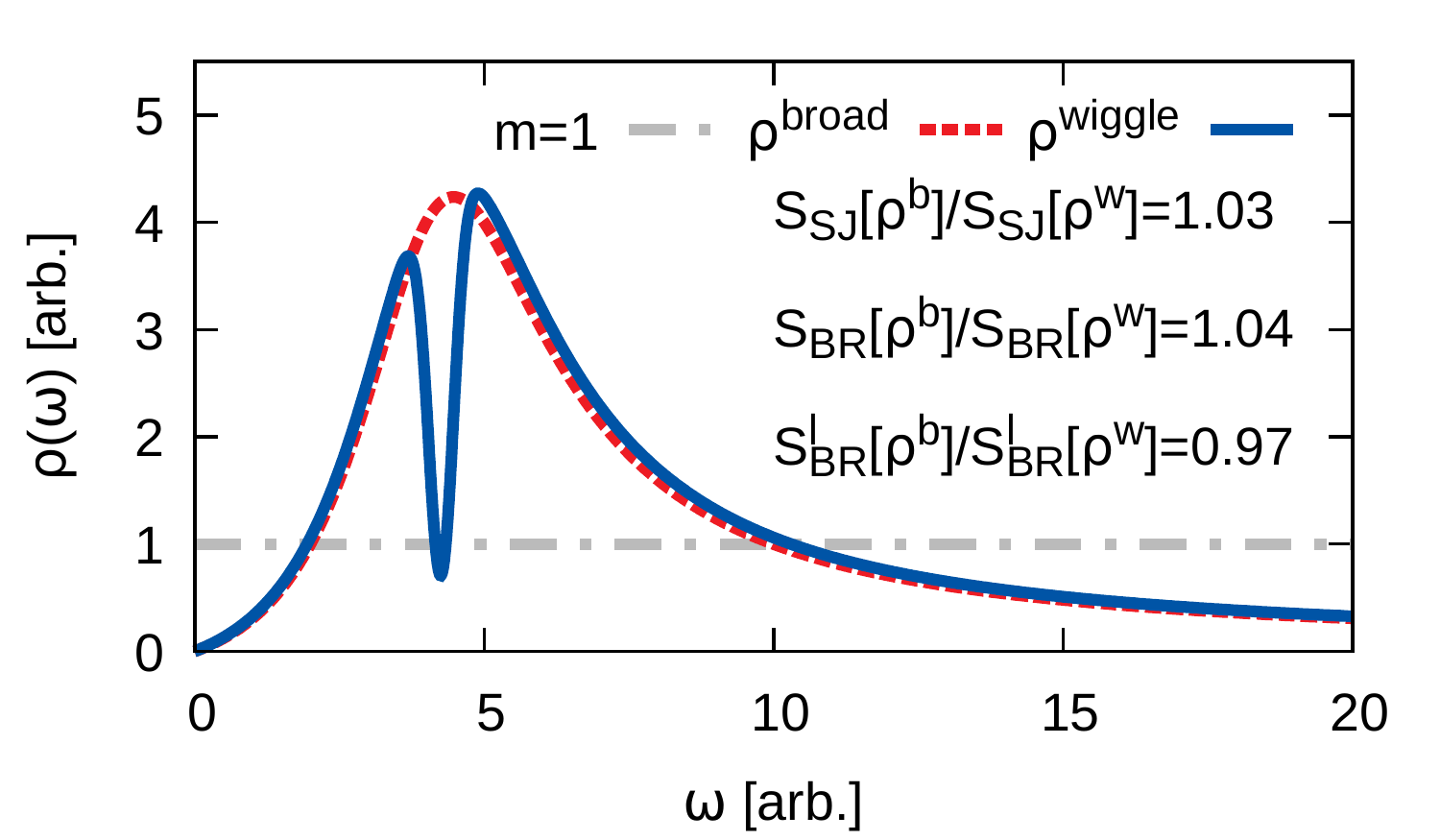}
\caption{Illustration of the possibility for local regulators, such as $S_{\rm BR}$ and the Shannon-Jaynes Entropy $S_{\rm SJ}$ to favor a wiggly function instead of a smooth one, where both curves cover the same area. Since the values for $S$ are by construction negative, a ratio $>1$ corresponds to a higher probability for the wiggly function. Only the non-local BR regulator $S^\ell_{\rm BR}$ manages to assign a higher penalty to the wiggly function here.}
\label{fig:ringingcheck}
\end{figure}

The entropy by definition is negative, so that ${\rm exp}[S]$ remains normalizable. As is shown in Fig.\ref{fig:ringingcheck} both local regulators $S_{\rm BR}$ and $S_{\rm SJ}$ show a ratio greater than unity when dividing the result for the smoother curve by that of the wiggly curve. In turn they assign a higher probability to the wiggly curve, which in a Bayesian analysis may lead to the occurrence of ringing.

\begin{figure*}[t]
\includegraphics[scale=0.5]{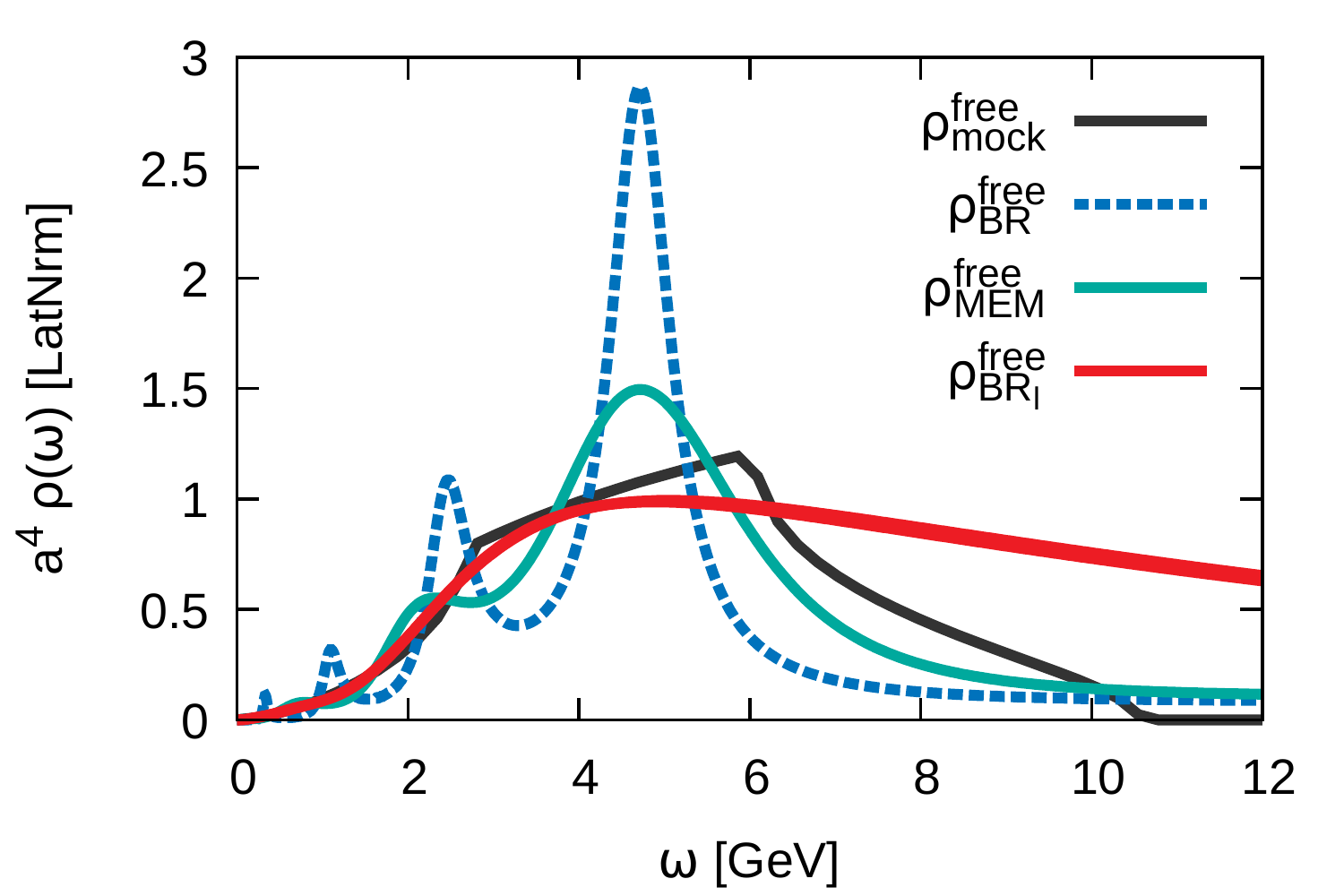}
\includegraphics[scale=0.5]{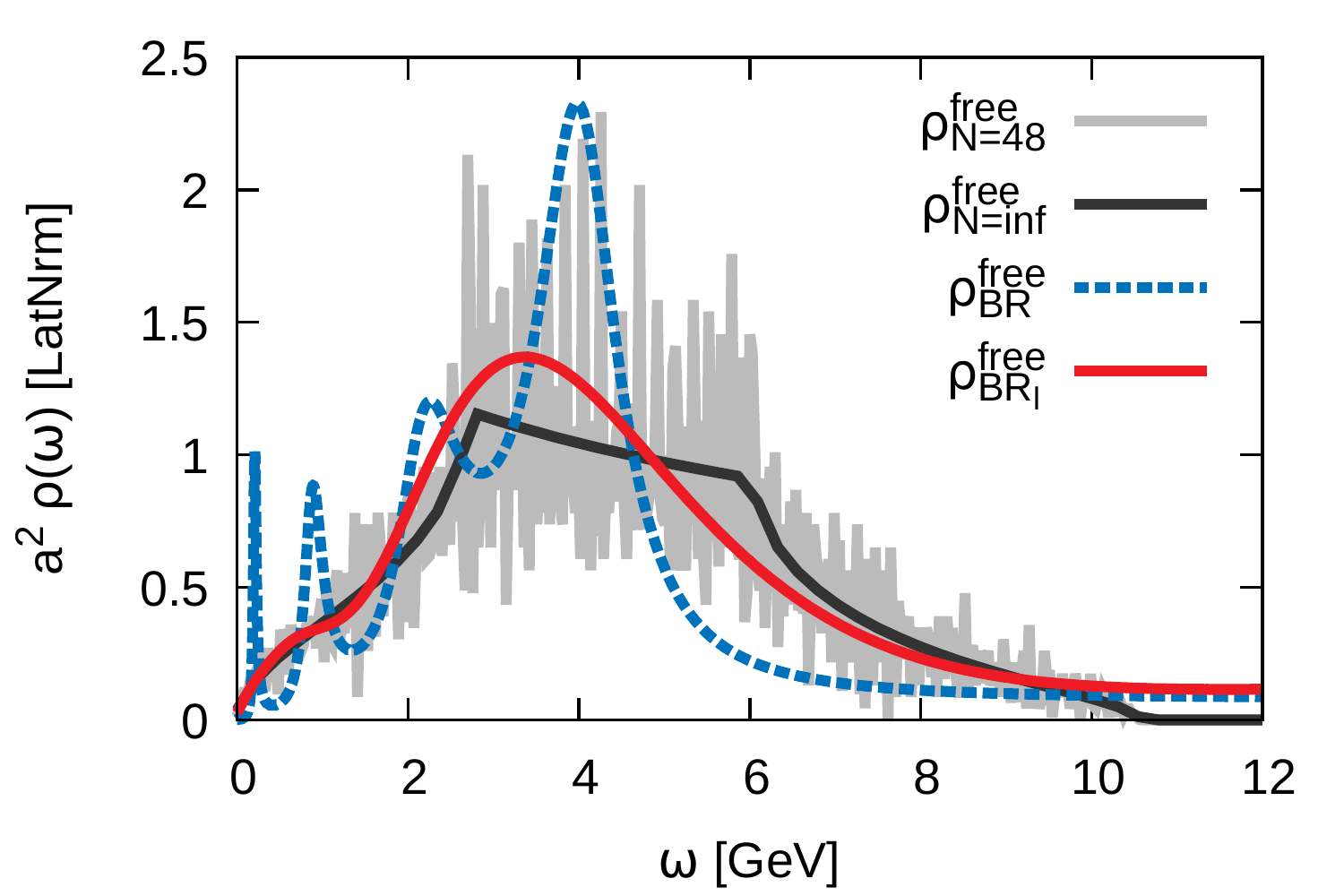}
\caption{(left) Representative examples of mock data reconstruction based on the infinite volume free P-wave spectral function (black solid) encoded in twelve correlator points along Euclidean time ($\Delta D/D=10^{-3}$). Three different Bayesian methods are shown: the standard BR method (blue dashed), the MEM (green) and the modified BR method with $\kappa=1$. In the phenomenologically relevant regime up to the first kink, our new method is accurate and does not show any ringing. (right) Reconstruction of the S-wave free spectral function from actual $N_\tau=12$ lattice data ($N_{\rm conf}=400$). The black solid line denotes the infinite volume spectral functions, while the light gray line show the spectrum actually encoded on the $48^3\times12$ lattice when taking the same frequency resolution as in the Bayesian methods. Here we show the outcome of the standard BR method (blue dashed) and the new modified method with $\kappa=1$ which again up to $\omega=2$GeV shows good accuracy and no ringing.}\label{Fig:TuningHyp}
\end{figure*}

The mathematical criterion which distinguishes more and less wiggly spectra, even if they cover the same area, is the arc length of the curves. Hence we introduce here a modification to the BR regulator which explicitly penalizes the arc length  $\ell=\int d\omega\sqrt{1+(d\rho/d\omega)^2}$. Since we wish to remain as close as possible to the form of the original BR method we thus add  $\ell^2$ and subtract unity, leaving us with
\begin{align}\nonumber 
  &S_{\rm BR}^{\ell}\\[2ex]
  =&\int d\omega \left( - \kappa \left(\frac{d\rho(\omega)}{d
             \omega}\right)^2  + 1- \frac{\rho(\omega)}{m(\omega)} + 
             {\rm log}\big[ \frac{\rho(\omega)}{m(\omega)} \big]\right). \label{Eq:BRellS}
\end{align}

Note that we have introduced an additional hyperparameter $\kappa$, which characterizes how strongly arc length should contribute to the overall penalty produced by $S_{\rm BR}^\ell$. It has to be treated on the same footing as the hyperparameter $\alpha$. In the presence of the derivative term, it is not possible to a priorily marginalize the hyperparameter $\alpha$ and we resort to the "historic MEM" prescription, in which one adjusts $\alpha$ such that $(L-N_{\rm data})<10^{-1}$ when deploying the regulator $S^\ell_{\rm BR}$.

How does the new modified BR method relate to the original BR\footnote{The newly added term actually violates the scale invariance axiom of the BR method, which also is absent from the MEM. Similar to the construction of the BR method for non-positive definite spectra \cite{Rothkopf:2016luz} it can be restored by introducing an additional default model related function $h(\omega)$ which encodes our confidence in the knowledge of the derivative of the spectrum.} and MEM? It essentially represents an interpolation between the two. Setting $\kappa=0$ we end up with the latter, setting $\kappa>0$ and large, we can reproduce the washed out features characteristic for MEM reconstructions with small search spaces. What is missing now is a prescription to select the hyperparameter $\kappa$ self consistently.

To this end we follow the supervised learning paradigm often deployed in the context of machine learning. We take a set of realistic (mock) input data, whose spectrum is known to us, e.g. the non-interacting lattice theory, and tune the value of $\kappa$ such that no ringing is present in the end result. While the strength of smoothing is now unrelated to the number of data points, we will carry our this tuning on lattices with the same extent as those at $T>0$ from HotQCD.

The best parameter that allows us to reproduce the nontrivial structure in the mock spectra while removing any ringing was $\kappa\approx 1$, which we will use in the subsequent parts of this work. We arrive at this choice by reconstructing two different test cases, each from a twelve data-point correlator. The first is a genuine mock data test, where the input data is computed from the analytically determined infinite volume spectrum of P-wave bottomonium in the non-interacting lattice theory and subsequently distorted by Gaussian noise with strength $\Delta D/D=10^{-3}$. 

In order for the mock spectrum to resemble as closely as possible the free lattice NRQCD spectrum in our numerical setup we have to take into account the particular normalization arising from our choice of point sources. I.e. we normalize the mock correlator at $\tau=0$ to have the same value as the one measured on a unit link lattice.

The resulting free spectrum is shown as black curve in the left panel of Fig.\ref{Fig:TuningHyp} and details on the free theory computation can be found in App.\ref{app:freetheory}. In previous studies, BR reconstructions from this input data showed very strong ringing. And indeed the standard BR method, i.e. for $\kappa=0$ (blue dashed) also here shows sizable ringing at low frequencies. The MEM produces a result that is much less contaminated by ringing, while still exhibiting weak wiggles close to the ringing peaks of the BR method. The strength of smoothing in the MEM crucially depends on the choice of $\omega_{\rm min}$ and the used number of basis functions. On the other hand the modified BR method with $\kappa=1$ manages to reproduce faithfully the region below $\omega=2$GeV without introducing further ringing. Changing $\kappa$ from zero to unity one manages to obtain results that are very close to those of the MEM for a particular choice of  $\omega_{\rm min}$.  At higher frequencies all Bayesian methods eventually go towards the default model $m=1$. 

The second test case shown in the right panel of Fig.\ref{Fig:TuningHyp} is a reconstruction based on actual non-interacting lattice data from grids of extend $48^3\times 12$. Here we use $N_{\rm conf}=400$ S-wave correlators, which are computed using different random sources on a unit-link lattice. We set the only parameter $M_qa=2.76$ to take on the value obtained for bottom quarks on the $\beta=6.664$ lattices. 

Often intuition is derived from the infinite volume free spectral functions shown as black solid line, while the actual spectral function present on a $48^3\times 12$ lattice, resolved at the same frequencies as the Bayesian reconstructions is given by the light gray lines. Comparing the outcome of the standard BR method (blue dashed) and the new modified method with $\kappa=1$ we find that the former again up to $\omega=2$GeV shows very good accuracy and no ringing.

These tests give us confidence that the modified BR method can be used to robustly determine the presence of peak structures in the reconstructed spectra, liberating us from the complicated tests on ringing that previously needed to be performed for the standard BR method \cite{Kim:2014iga}. 

In essence the standard $\kappa=0$ and modified $\kappa=1$ BR methods can be understood as two different theory detectors for heavy quarkonium, one with high noise and high gain and one with low noise and low gain. While in the presence of a well pronounced peak signal, the former has been shown to be very efficient in extracting the peak position and width, as we discussed above, it may introduce ringing. On the other hand the latter method will not show ringing but as a consequence is not able to resolve pronounced peaks equally sharply. In the following we will thus combine the two approaches, first determining with $\kappa=1$ whether a genuine peak structure is present and if so will use $\kappa=0$ to extract the peak position. The efficiency of this strategy is tested in the subsequent section, first at $T\approx0$ before being applied to $T>0$ lattice data in sec.\ref{sec:physres}.

\section{Lattice NRQCD results at $T=0$}
\label{sec:T0res}

The study of correlation functions and spectral functions at zero temperatures in this section provides the necessary baseline upon which we build our study of quarkonium in-medium modification. 

\subsection{Ground state properties}
\label{sec:T0groundstate}

\begin{figure}[t]
\includegraphics[scale=0.5]{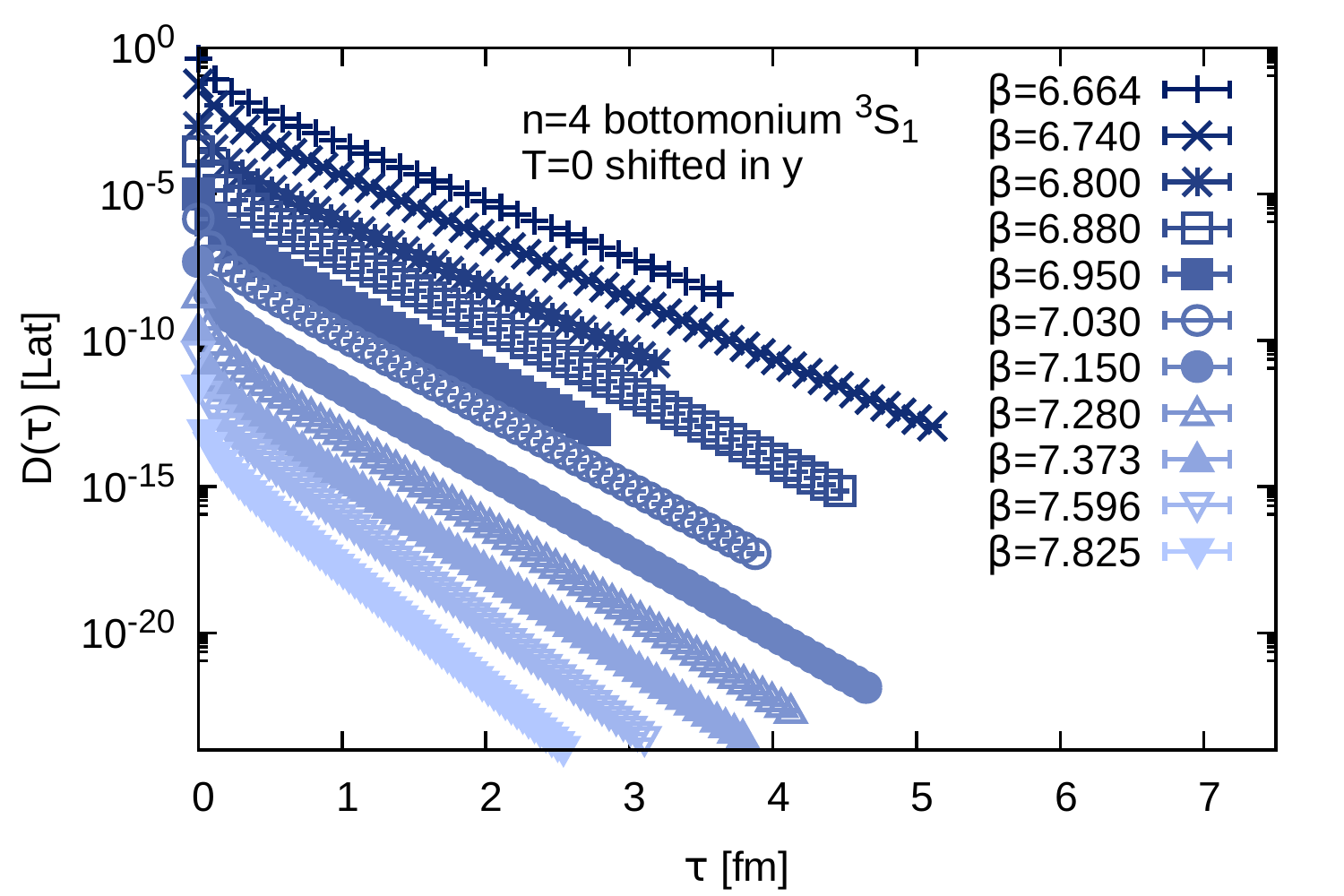}
\includegraphics[scale=0.5]{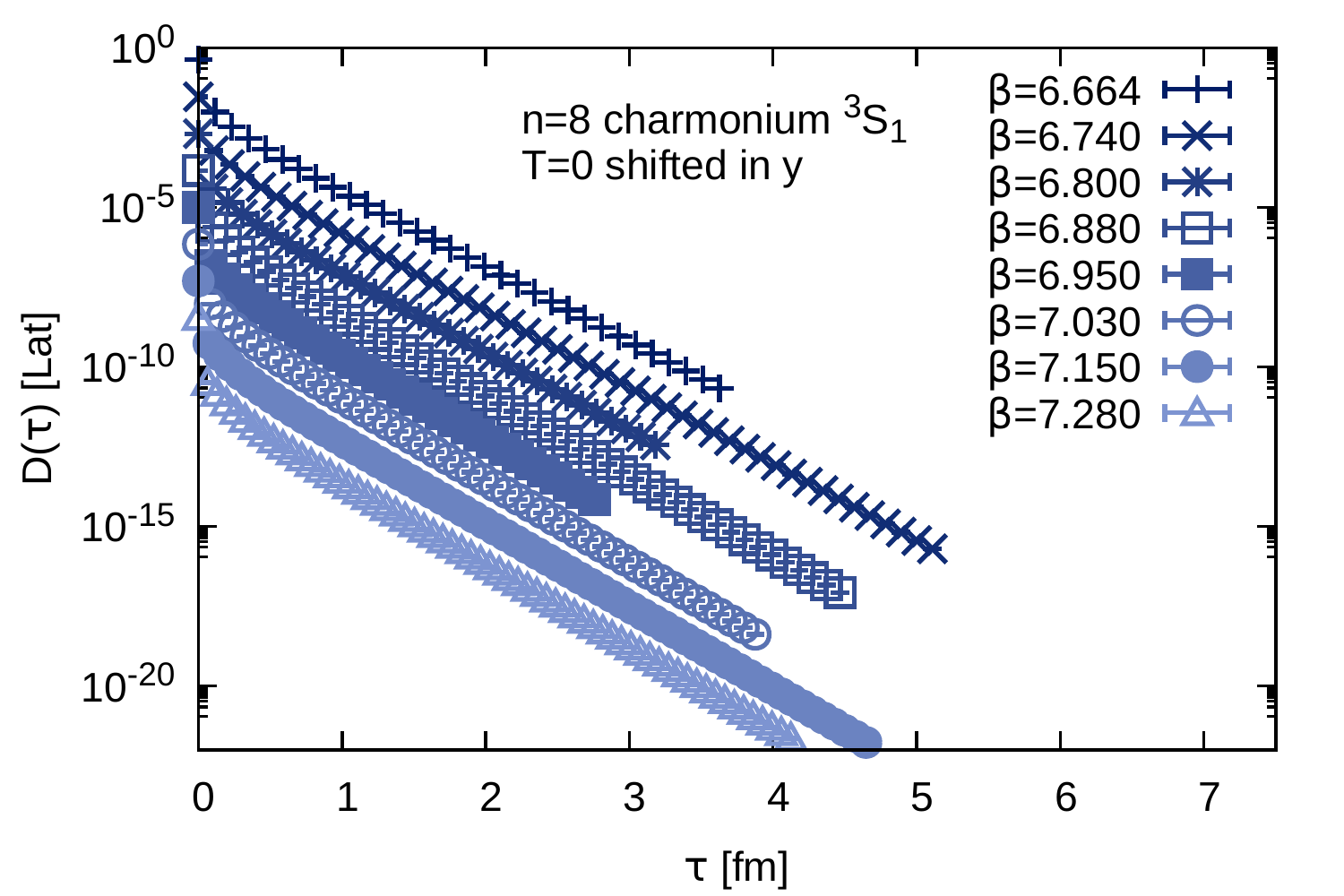}
\includegraphics[scale=0.5]{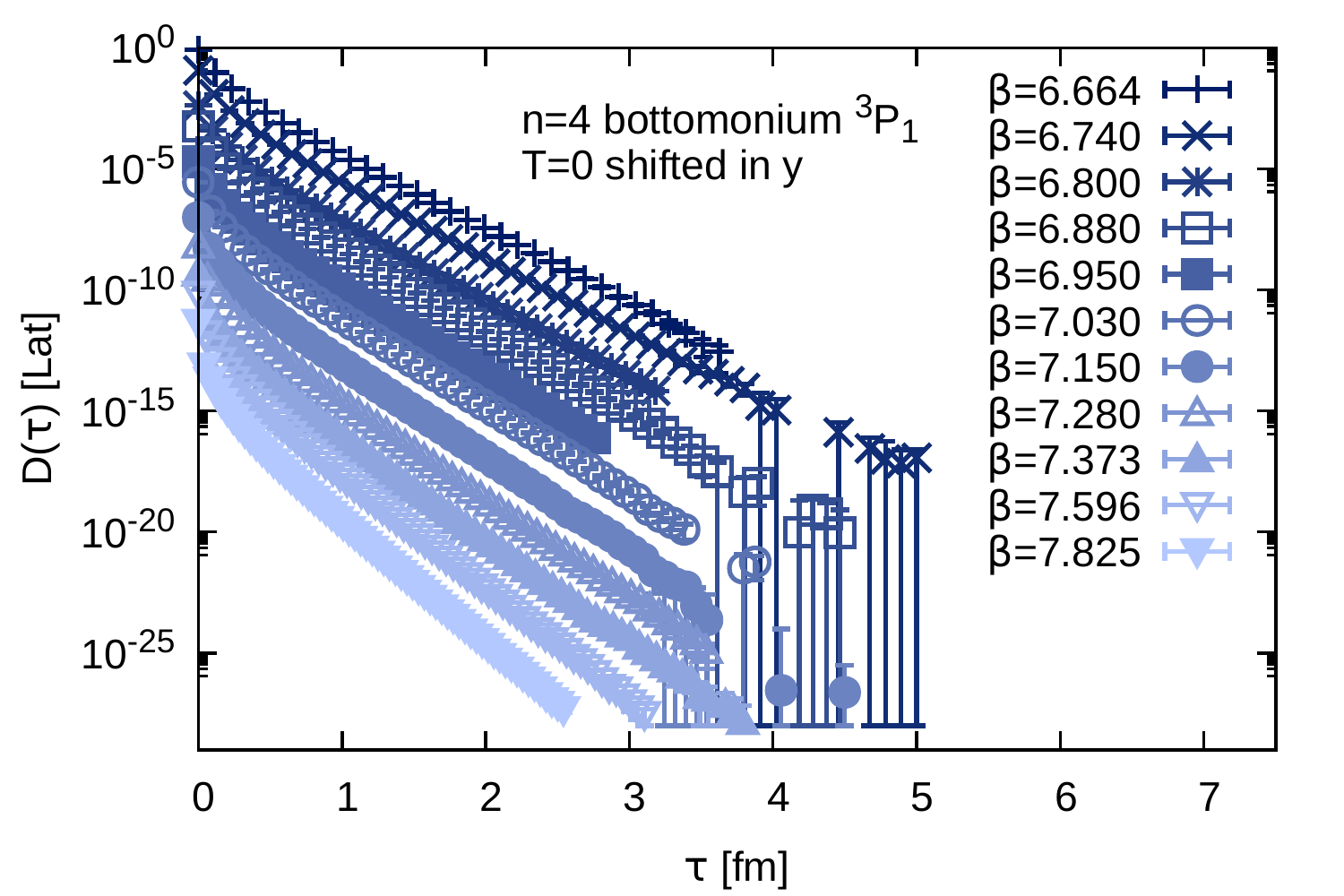}
\includegraphics[scale=0.5]{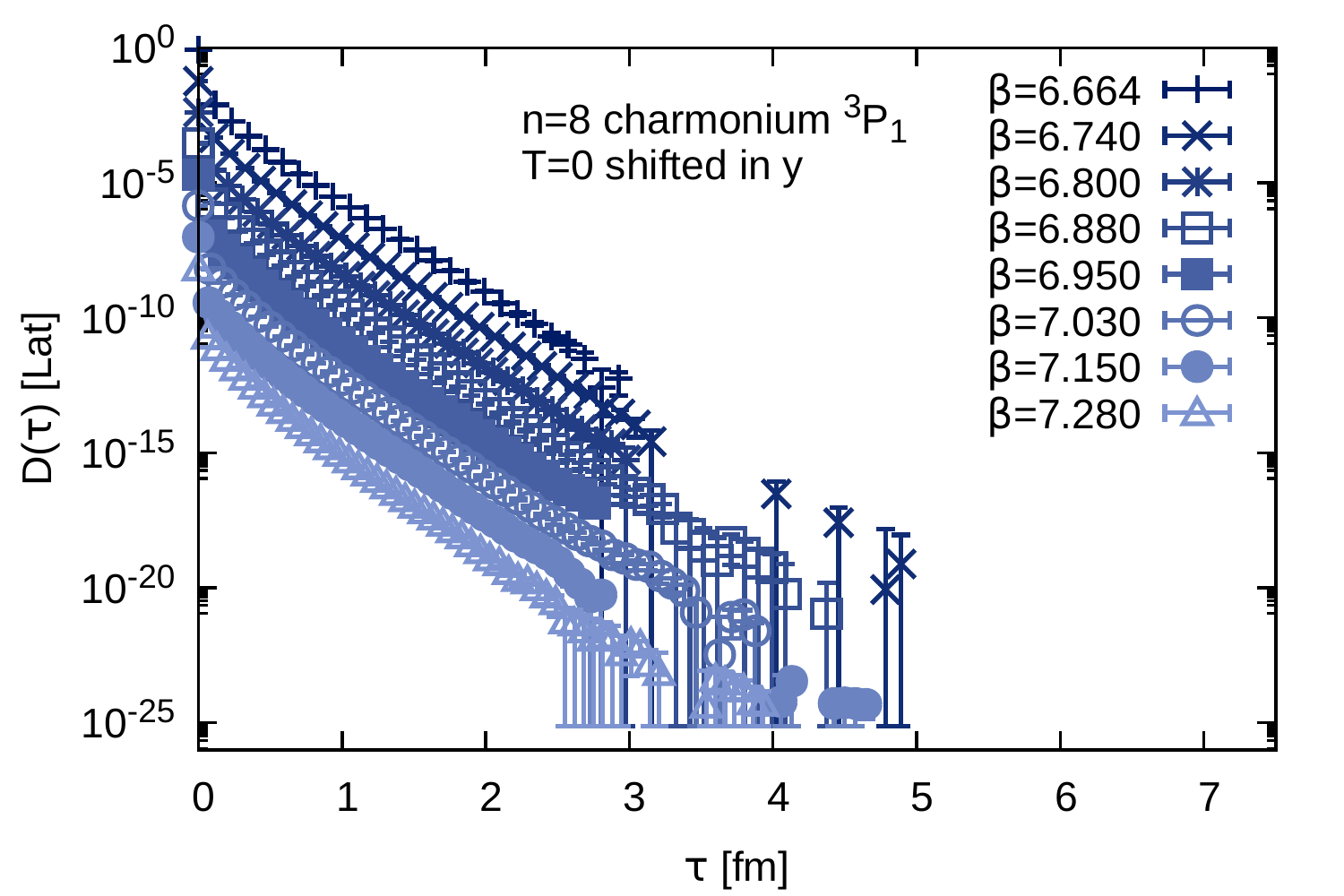}
\caption{Zero temperature correlators with bottomonium results on the left and charmonium on the right. (top) The correlation functions of the $^3S_1$ channel shifted for better visibility. The well pronounced exponential falloff at late times indicated the presence of a well defined ground state signal. (bottom) The correlation functions of the $^3P_1$ channel shifted for better visibility. The well pronounced exponential falloff at late times indicated the presence of a well defined ground state signal.}\label{Fig:T0Correlators}
\end{figure}

Let us start with an inspection of the ground state properties, which can be straight forwardly extracted from the simulated correlation functions without the need for a spectral reconstruction. To this end, at each of the eleven (eight) values of $\beta$ for which we consider bottomonium (charmonium), 400 realizations of the corresponding $^1S_0$, $^3S_1$, $^1P_1$ and $^3P_{0,1,2}$ channel correlators are computed. Since the behavior of the two S-wave and the four P-wave correlators is very similar, we will show in the following raw data only for the $^3S_1$ and $^3P_1$ cases.
\begin{figure}[t]
\includegraphics[scale=0.5]{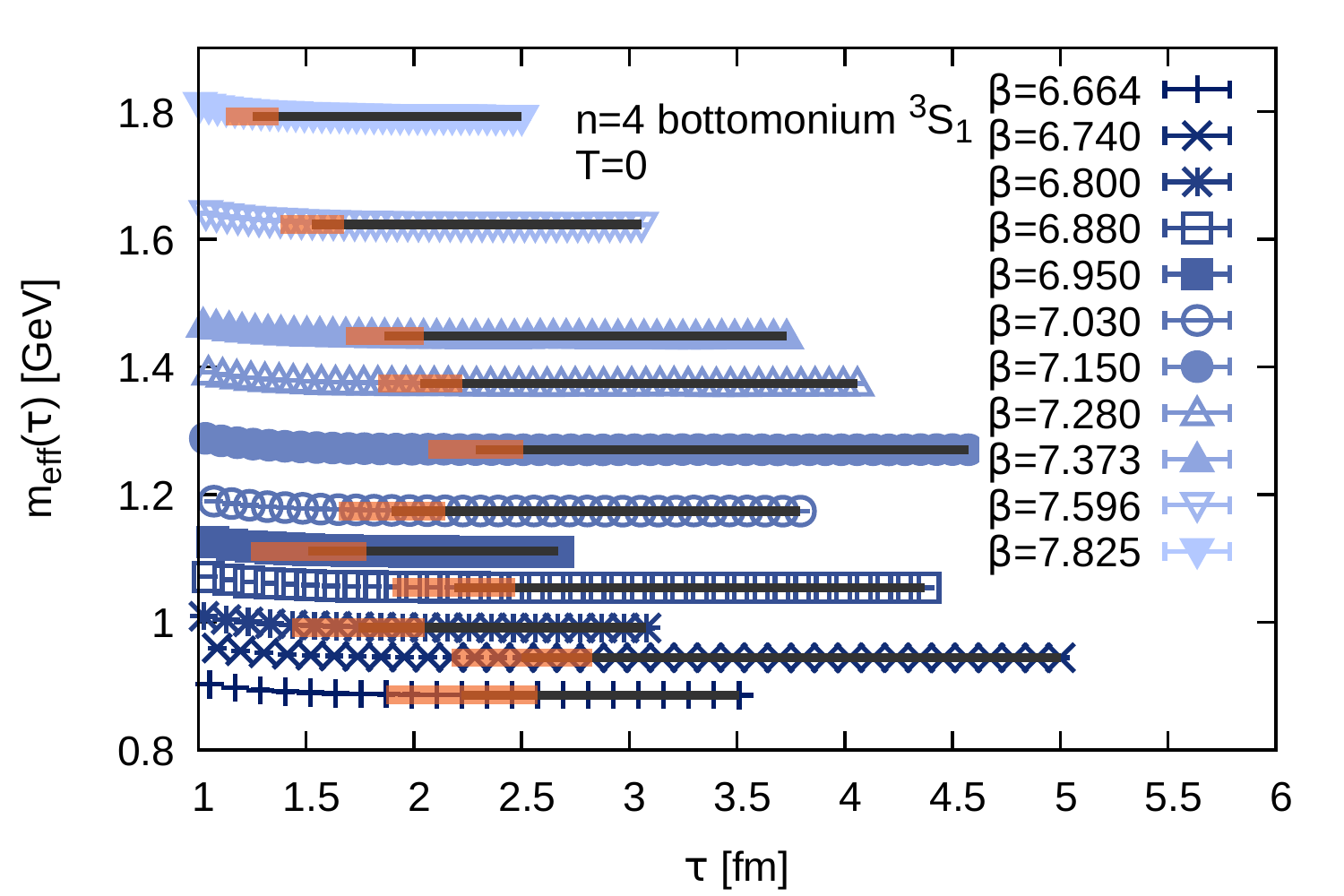}
\includegraphics[scale=0.5]{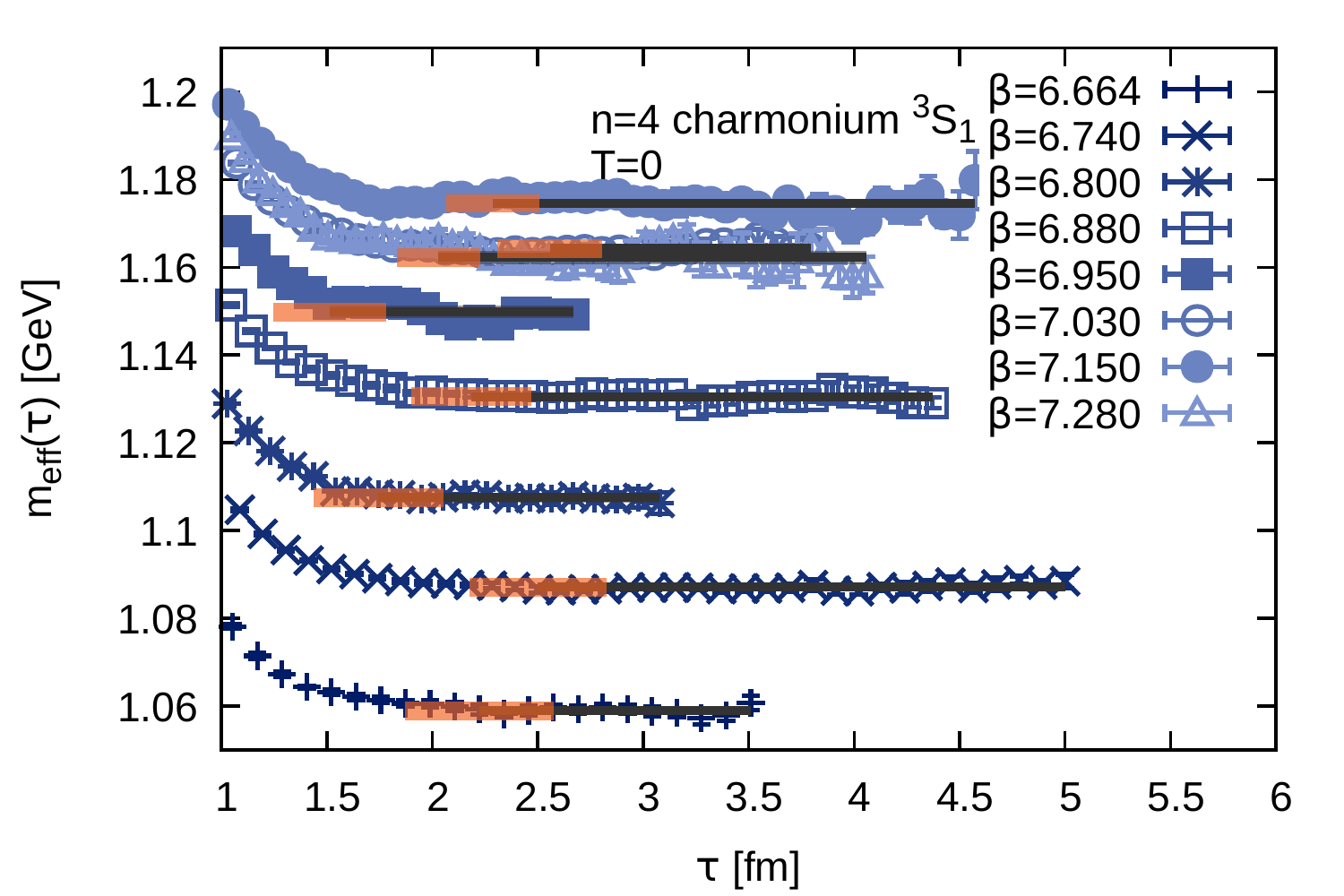}
\caption{Zero temperature effective masses of the S-wave channel with bottomonium results on the left and charmonium on the right. We also show the average fit interval (black solid) and its variation (orange)}\label{Fig:T0EffMassesSwave}
\end{figure}
\begin{figure}[t]
\includegraphics[scale=0.5]{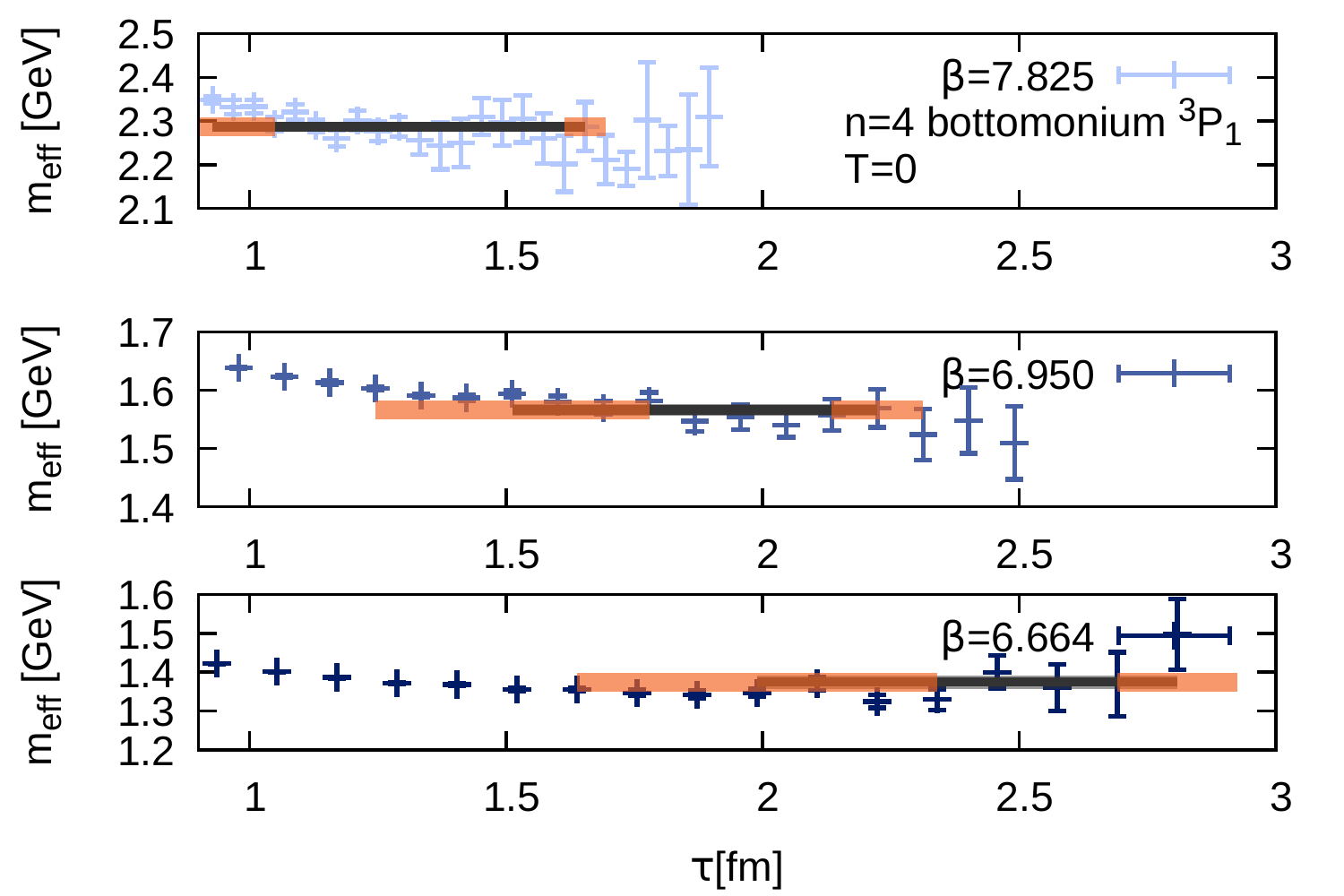}
\includegraphics[scale=0.5]{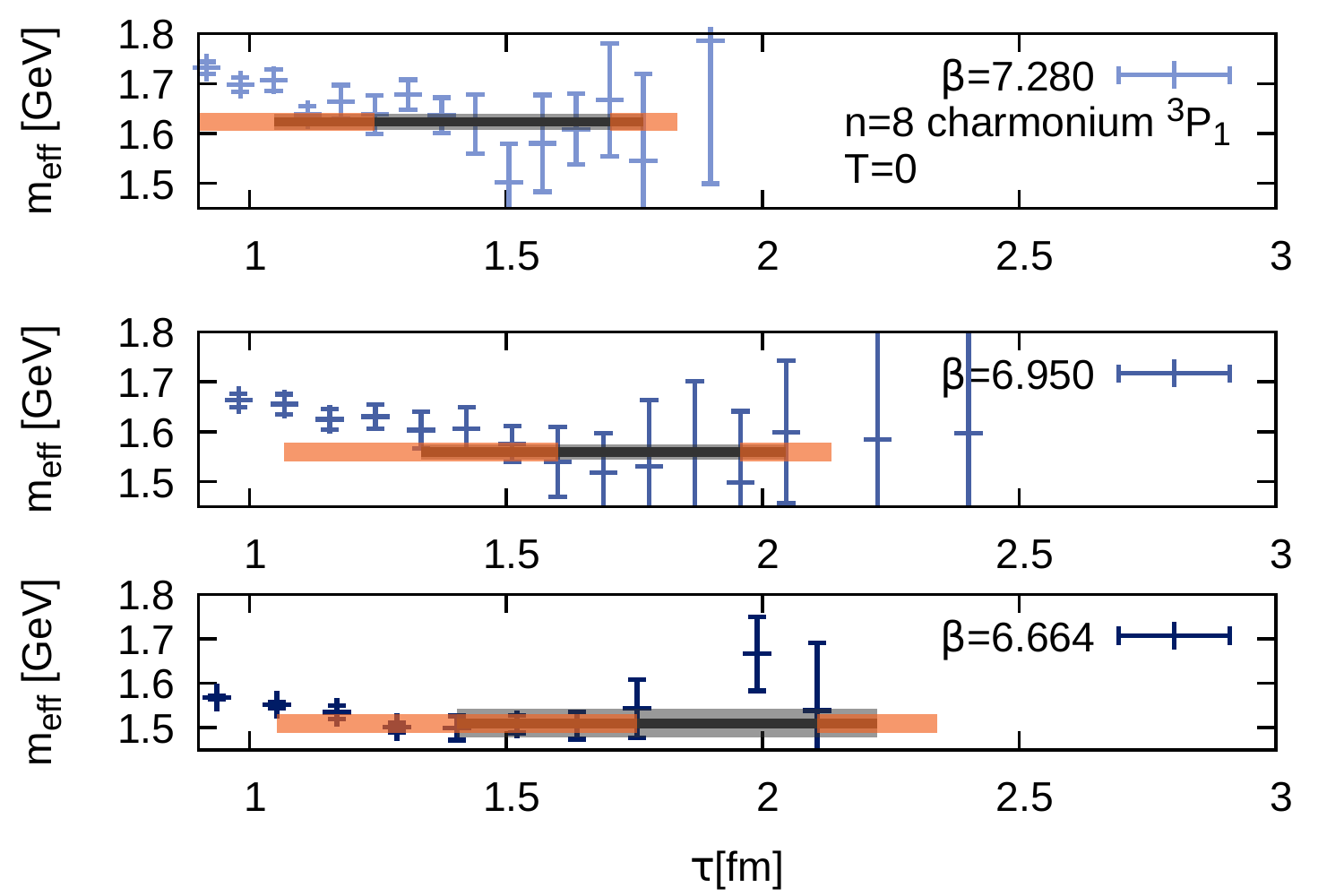}
\caption{Selected zero temperature effective masses of the P-wave channel with bottomonium results on the left and charmonium on the right. We also show the average fit interval (black solid) and its variation (orange)}\label{Fig:T0EffMassesPwave}
\end{figure}

In Fig.\ref{Fig:T0Correlators} the raw correlators of bottomonium at different lattice spacing are plotted on the left, their charmonium counterparts on the right. In order to better distinguish the different data sets they are shifted by hand in the logarithmic plot. Note that the correlators are not symmetric around $\tau=1/2T$ and thus, contrary to the relativistic formulation, the information along the full length of the Euclidean domain is accessible to us.

The S-wave correlators all show a clear single exponential fall-off at late Euclidean times, indicating the presence of a well defined and well separated ground state spectral feature. The P-wave correlators, due to the large mass of their ground states, exhibit a stronger falloff and hence a lower signal to noise ratio at late Euclidean times. A naive inspection by eye
still espies a pronounced single exponential decay while the wiggly
behaviors at large $\tau$ may indicate loss of signal.

Another aspect of the NRQCD discretization is the existence of an additional lattice spacing dependent (additive) shift of the overall energy scale. It arises from the fact that in NRQCD the hard scale of the heavy quark mass has been integrated out by dropping the term $2M_q$ from the Lagrangian. This bare shift becomes renormalized due to the quantum fluctuations resolved in the simulation and is already visible in the slope of the correlators. Their exponential falloff steepens with increasing beta, even though they all describe the same physical state.

The ground state masses may be extracted from the correlators by use of the effective mass
\begin{align}
a m_{\rm eff}(\tau)={\rm log}\big[ \frac{G(\tau+a)}{G(\tau)} \big]\label{eq:effmass}
\end{align}
which for a genuine single exponential decay in $G(\tau)$ simply reads out its exponent. In the presence of excited state contributions one needs to identify a constant plateau region in \eqref{eq:effmass} through which the ground state mass is extracted. As expected from the behavior of the raw correlators and as explicitly shown in Fig.\ref{Fig:T0EffMassesSwave} this is easily achieved in the S-wave channel but already is a challenge in the P-wave case (Fig.\ref{Fig:T0EffMassesPwave}) due to the lower signal to noise ratio there. Because of the large statistical errors in the latter, a meaningful comparison in a single plot is not possible, which is why we only provide here a representative selection of results for the P-wave.

In addition the spectral weight of the ground state in the two channels is different. In the S-wave it is given by $|R_{^3S_1}(0)|^2$, while for the P-wave it is $|R^\prime_{^3P_1}(0)|^2/M_q^2$ and thus suppressed by the mass. The third complication in the P-wave channel is related to the scaling of the spectral function in NRQCD. For the S-wave channel, it proceeds as $\omega^{1/2}$ and for the P-wave channel, as $\omega^{3/2}$. In turn the continuum part overshadows the ground state signal more strongly in the latter channel.

It is known that the use of extended sources in eq.\eqref{Eq:sourcesNRQCD} may provide improved overlap in particular for the more extended P-wave states. On the other hand such non-local sources also alter spectral properties, such as thermal widths. Using a Gaussian source extent of $\sigma = 2.5 a$ we did not yet find significant improvements. Since with a larger extent we expect to start interfering with the in-medium structure of the spectra, we refrained from further pursuing extended source correlators.

We identify the plateau region from an inspection by eye and repeat the fit to its value several times, changing the starting $\tau$ point, and in the case of the P-wave channel also the end-point of the fitting interval by different combinations of up to four steps in $a$. The fitting intervals are indicated in Fig.\ref{Fig:T0EffMassesSwave} and Fig.\ref{Fig:T0EffMassesPwave} by the black solid lines, the variation in the fit range via the orange bands. 

\begin{figure}[t]
\includegraphics[scale=0.5, trim=0 0.9cm 0 0, clip=true]{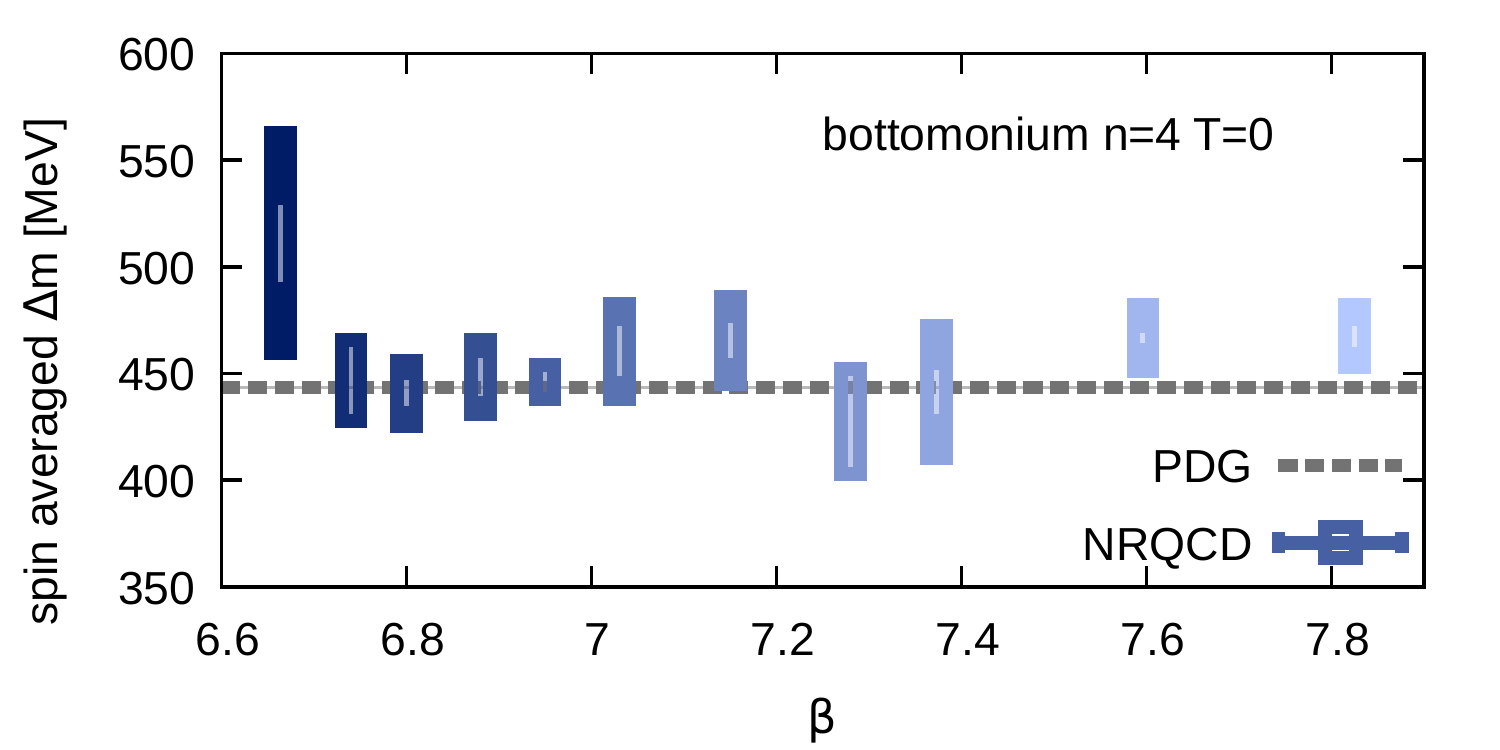}
\includegraphics[scale=0.5, trim=0 0.9cm 0 0, clip=true]{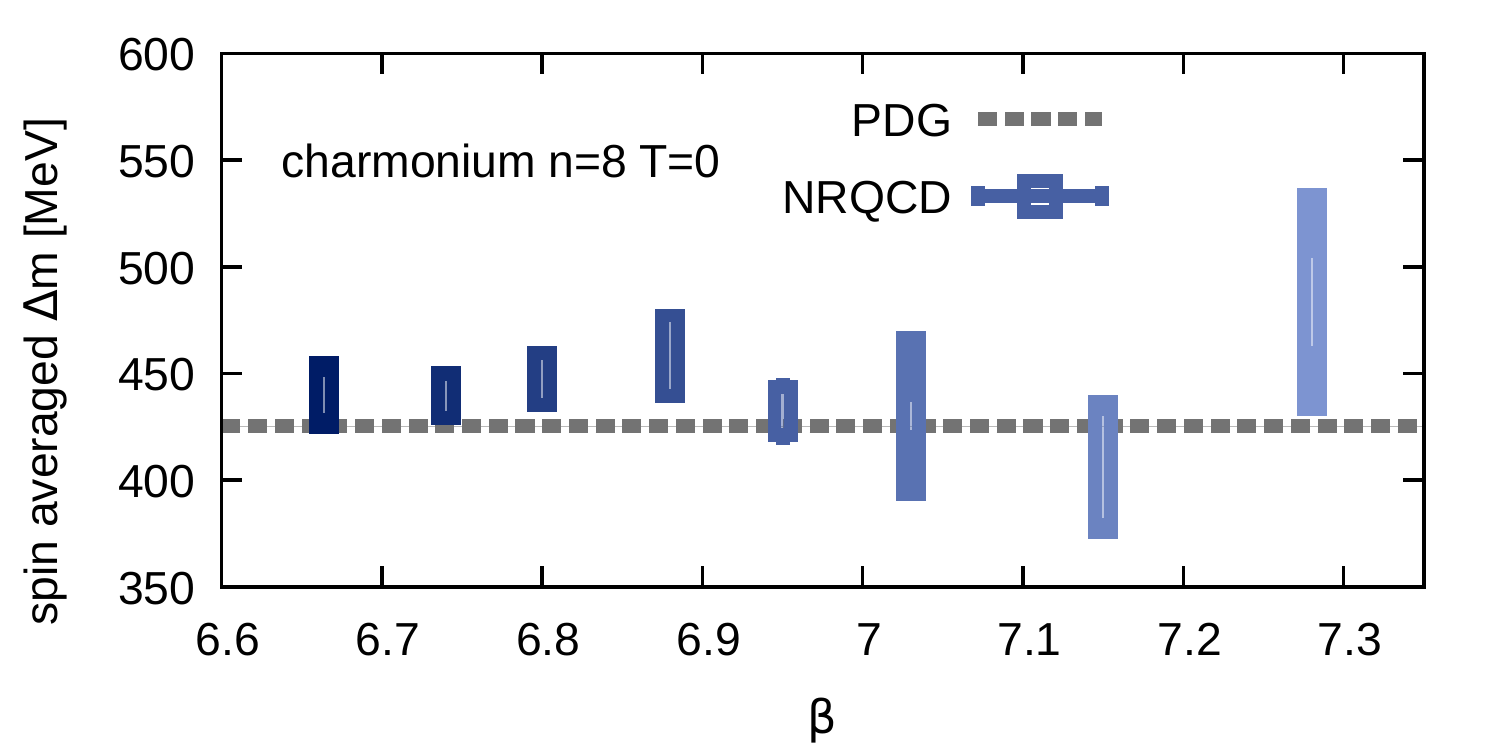}
\includegraphics[scale=0.5, trim=0 0.9cm 0 0, clip=true]{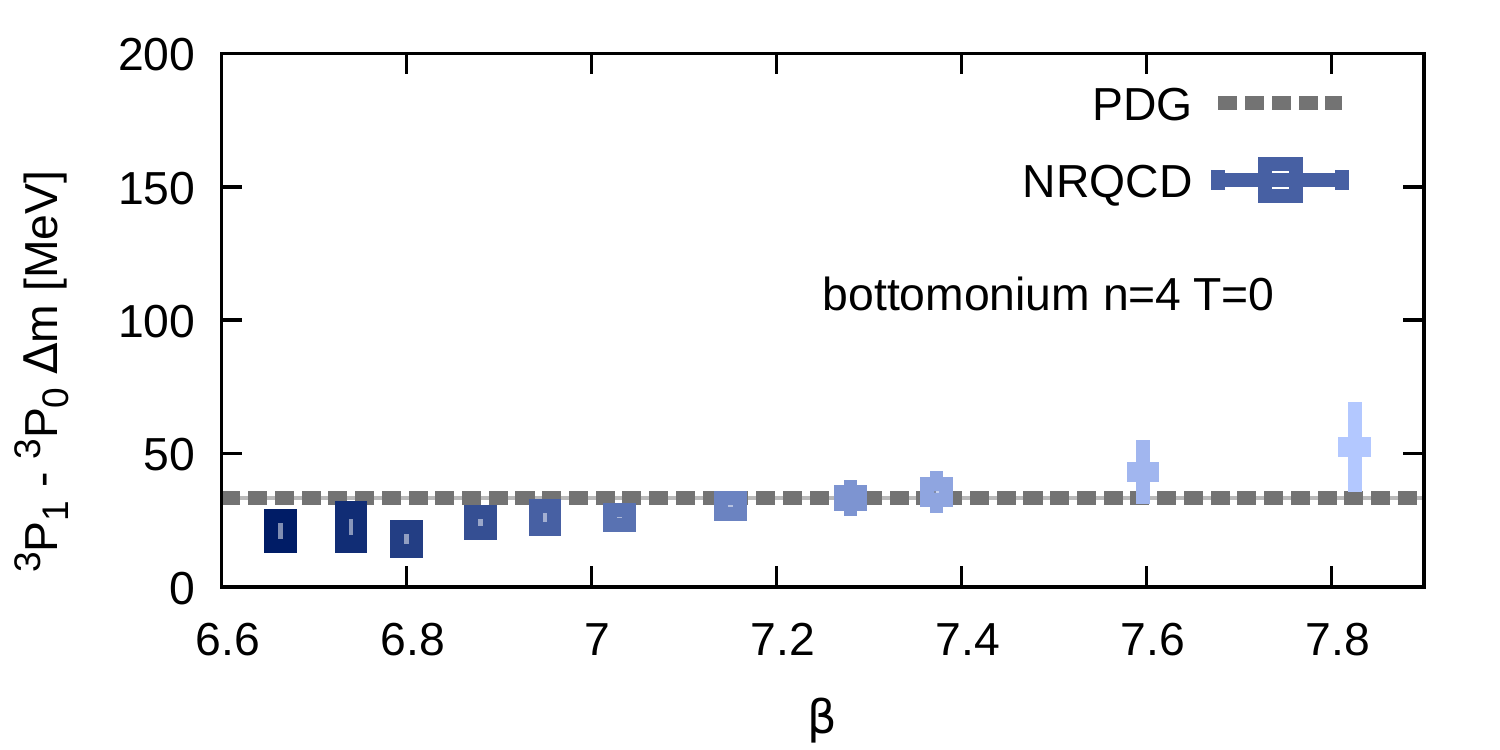}
\includegraphics[scale=0.5, trim=0 0.9cm 0 0, clip=true]{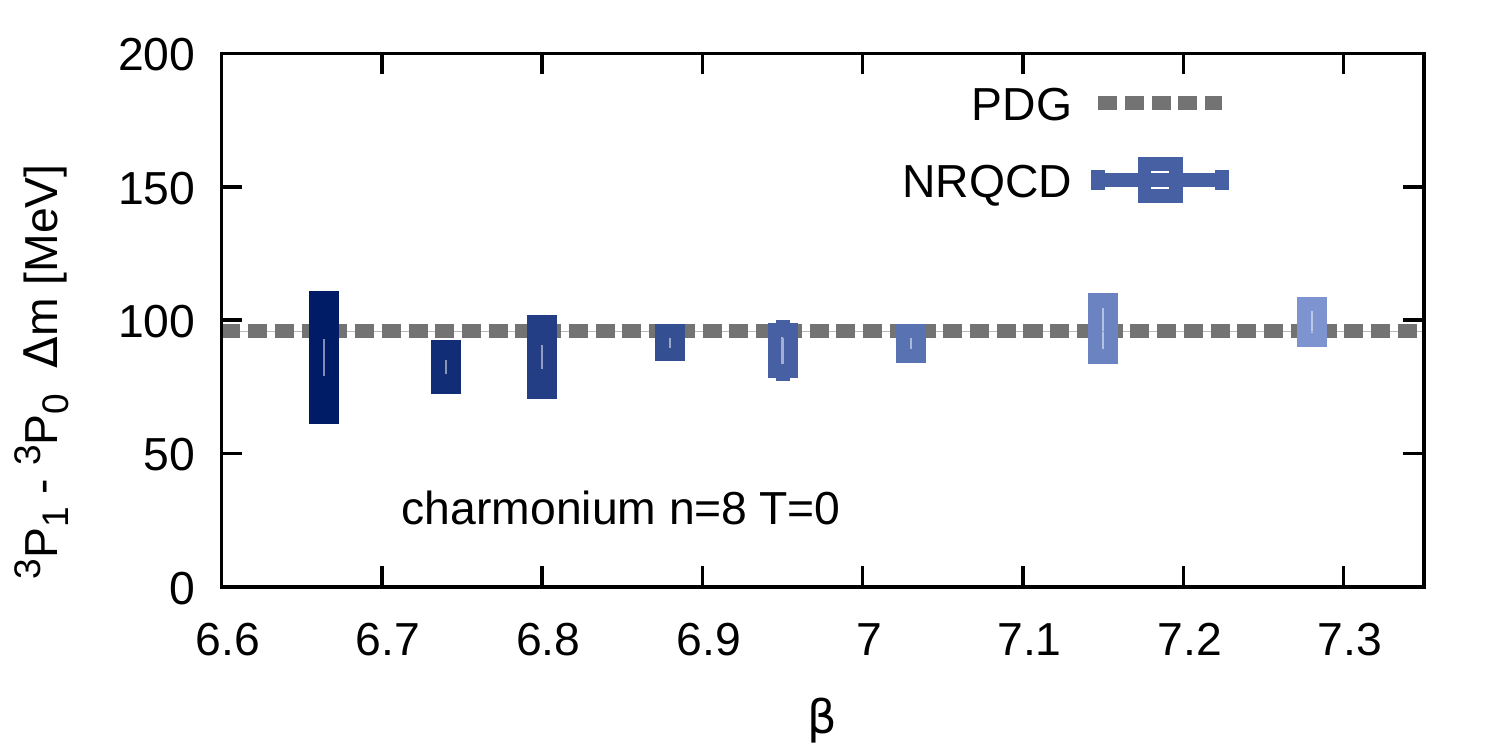}
\includegraphics[scale=0.5, trim=0 0.9cm 0 0, clip=true]{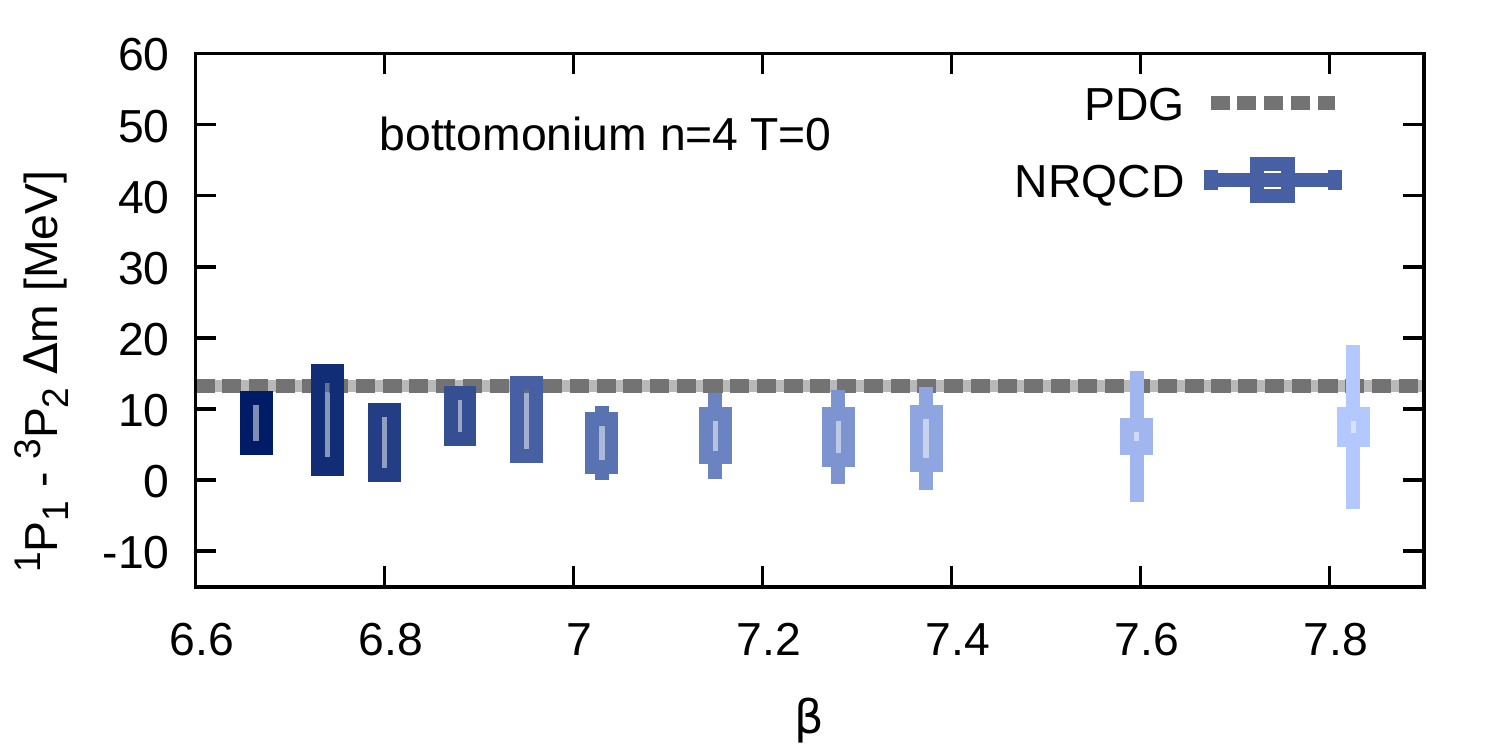}
\includegraphics[scale=0.5, trim=0 0.9cm 0 0, clip=true]{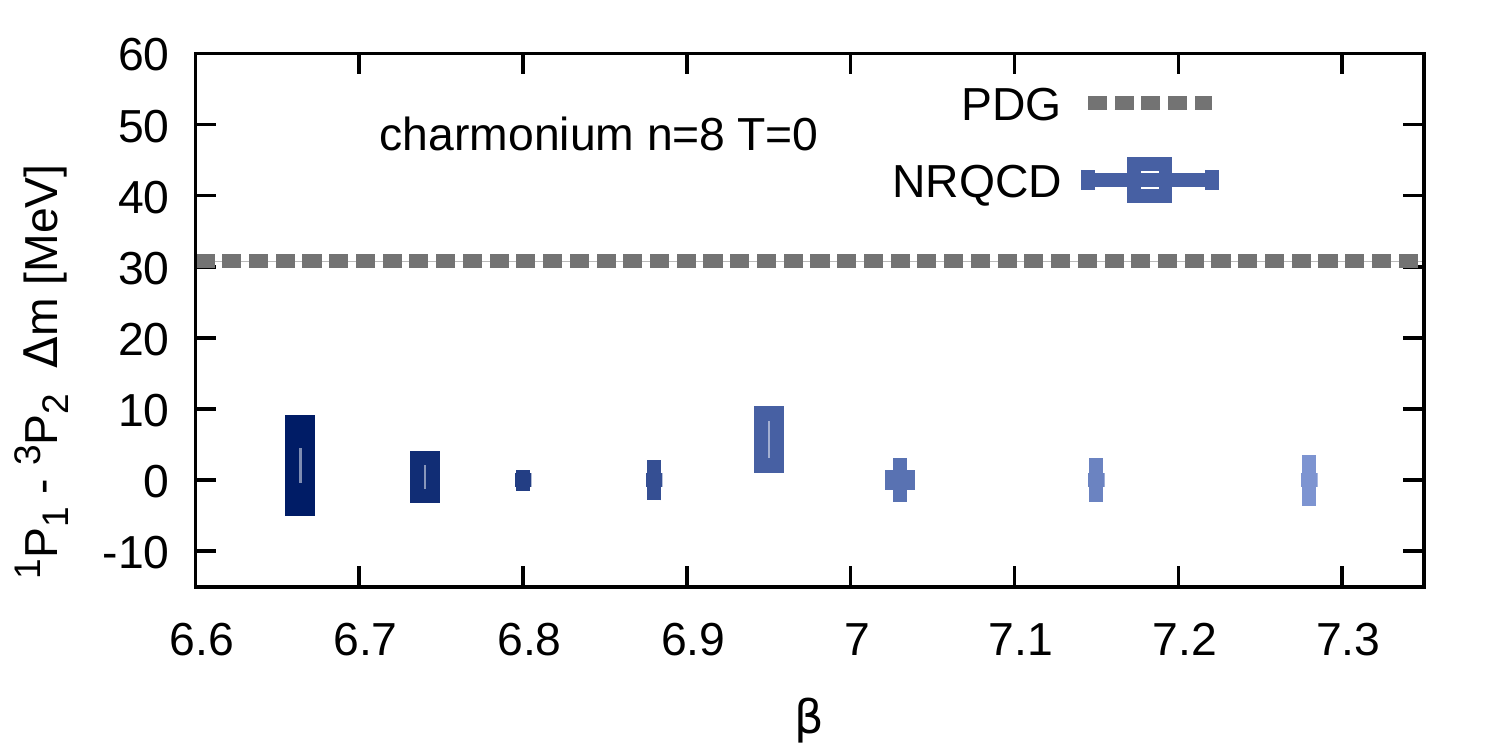}
\includegraphics[scale=0.5, trim=0 0.9cm 0 0, clip=true]{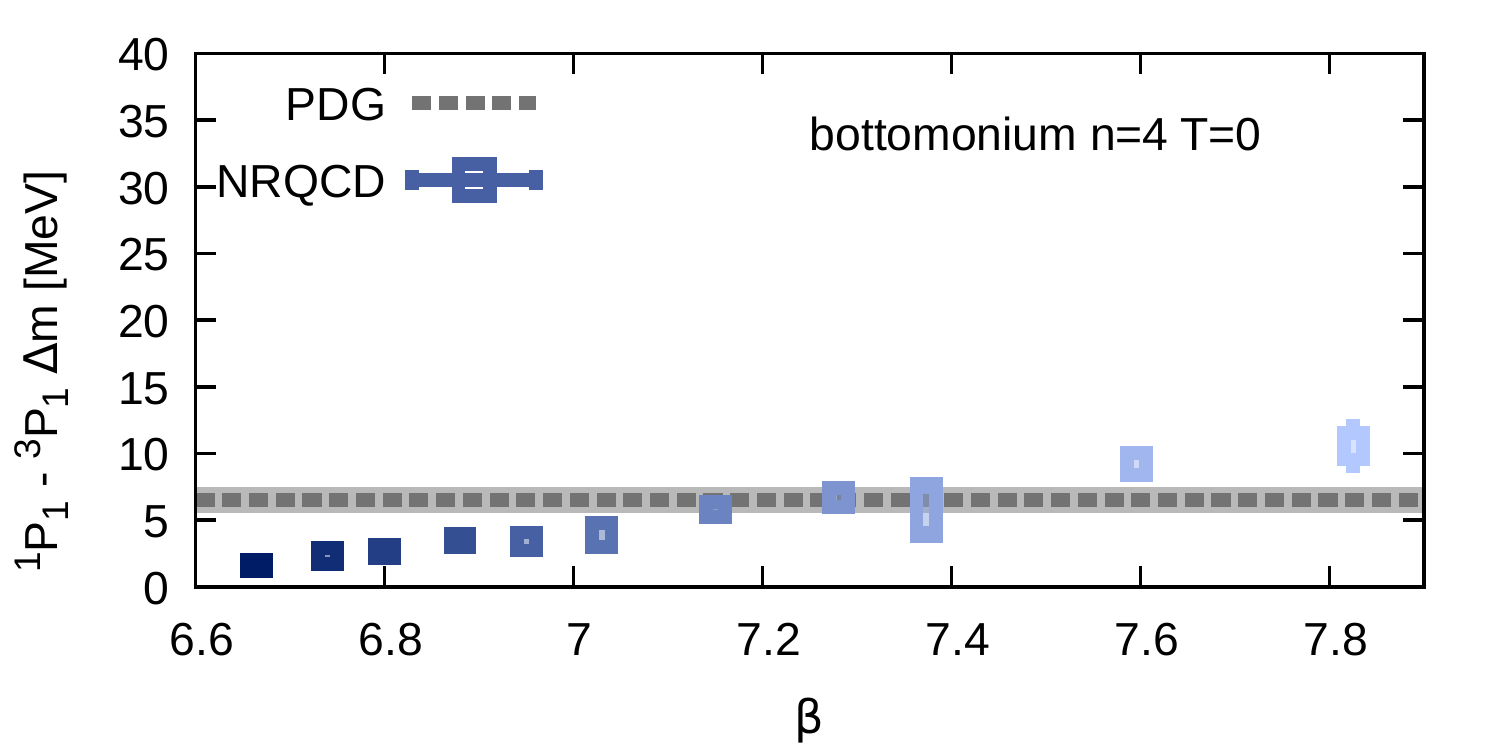}
\includegraphics[scale=0.5, trim=0 0.9cm 0 0, clip=true]{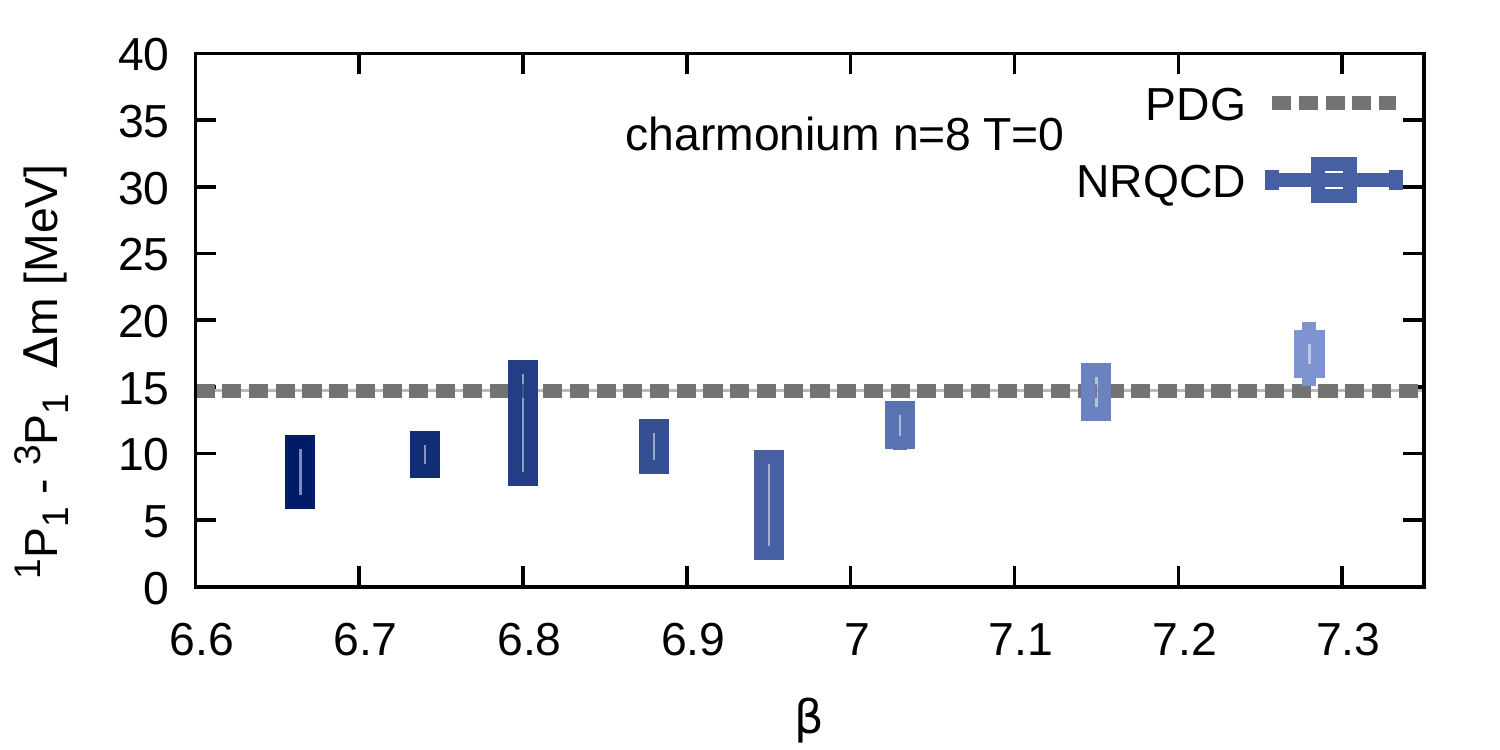}
\includegraphics[scale=0.5]{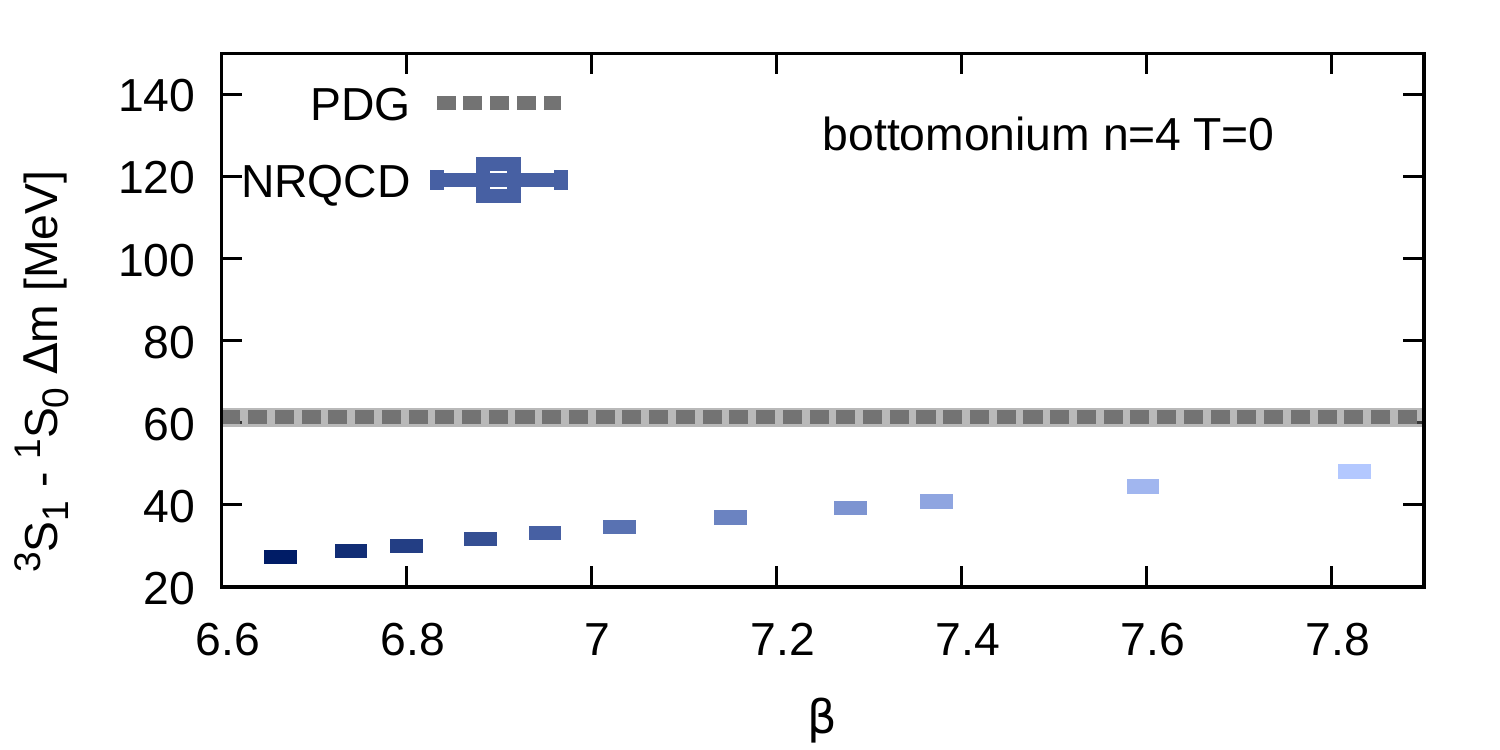}
\includegraphics[scale=0.5]{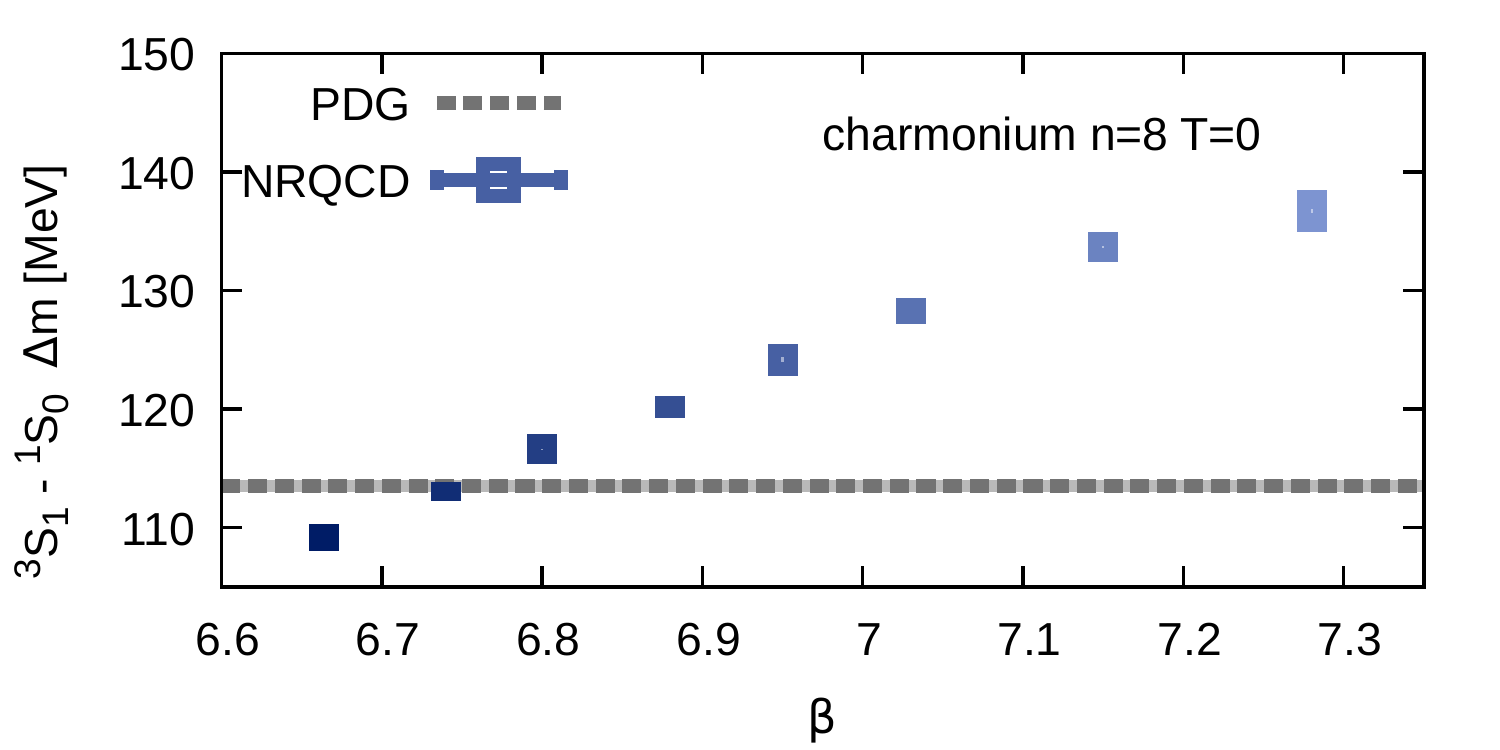}
\caption{Quantitative test of our NRQCD simulation setting with bottomonium results on the left and charmonium on the right.(top) The mass splitting between the three spin weighted $^3P_{0,1,2}$ states and the spin weighted $^1S_0$ and $^3S_1$ ground states. (second from top) The inter-triplet spacing between the scalar and vector states $^3P_{0}$ and $^3P_{1}$. (center) The spacing between the $^1P_{1}$ and the tensor states $^3P_{2}$. (second from bottom) The hyperfine structure between the P-wave  $^1P_1$ and $^3P_1$ ground states. (bottom)The S-wave hyperfine splitting between the ground states of the $^3S_1$ and $^1S_0$ channel.}\label{Fig:T0Splittings}
\end{figure}

At this point we can proceed to the main test of the robustness and accuracy of the NRQCD approximation by considering the mass splitting between the different states. The additive energy shift simply drops out from those quantities. We have selected five different mass splittings with different expected levels of difficulty for a successful reproduction in NRQCD. The values we obtain from the effective mass fits are plotted in Fig.\ref{Fig:T0Splittings} with bottomonium in the left column and charmonium in the right column.

In all figures, in which particle properties are estimated, we quote two sources of uncertainty, statistical and systematic. The former will be indicated by a bold error box, while the latter via a thin error bar. We plot both errors separately, so that the dominant source of uncertainty can be visually distinguished. In Fig.\ref{Fig:T0Splittings} the statistical errors are obtained from a ten-bin Jackknife and the systematic errors arise from the variation due to changing of the fit interval. 

Let us have a look at the individual mass splittings in Fig.\ref{Fig:T0Splittings}, where our NRQCD results are given in shades of blue and the PDF value of the splitting is indicated by a gray dashed line \cite{PDG2018}. 

We start out with the spin averaged splitting, where we compare the spin degeneracy weighted average of the $^1S_0$ and $^3S_1$ ground state mass to that of the three $^3P_{0,1,2}$ P-wave states. Deriving intuition from an analogy with a potential based computation, this splitting is expected to depend only on the central potential, i.e. be most easily reproduced within NRQCD. And indeed we find that our simulation reproduces this splitting excellently. 
Here we note that in the standard potential picture spin-spin interactions are proportional to a Dirac delta function and
therefore, they do no affect the P-waves ($R_P(0)=0$). This means that the center of mass of $^3P_J$ quarkonium states should agree
with the mass of $^1 P_1$ quarkonium ($h_c$ or $h_b$). We find that indeed within the errors the center of mass 
of $^3P_J$ quarkonium states agrees with $^1P_1$ mass.

Next we consider the mass splittings between different $P$ states. These are shown in the second, third and fourth rows (from the top) 
of Fig.\ref{Fig:T0Splittings}. These splittings come from the spin orbit interactions. Taking again intuition from a potential based picture, 
these splittings are expected to be significantly smaller than the spin averaged one, as these are suppressed 
with one additional power of the mass.  The $^3P_1-^3P_0$ splitting is well reproduced by our NRQCD calculations
both for bottomonium and charmonium. The $^3P_2-^1P_1$ splitting is described in a satisfactory manner for bottomonium
but not for charmonium. Finally, we have a satisfactory description for $^1P_1-^3P_1$ splitting, except for bottomonium
at the three smallest lattice spacings.

In high precision studies of quarkonium in vacuum it has been found that the most challenging splitting is the S-wave hyperfine splitting, which in our naive setup is missed by at most $35$MeV in bottomonium and $25$MeV in charmonium in our calculations (bottom row). 
It is known that for bottomonium, NRQCD to order ${\cal O}(v^4)$ is prone to underestimating this splitting, while ${\cal O}(v^6)$ has been found to overestimate it. 
As mentioned above, the spin-spin interactions responsible for the hyperfine splittings are ultra-local and thus are not easy to be captured on the lattice, unless the lattice spacing is very small.
The effects of the UV sector, i.e. energy scales much higher than the soft scale $M_qv$ are encoded in NRQCD via Wilson coefficients of the NRQCD Lagrangian. In our setup only the trivial tree level values $c_i=1$ are deployed (see eq.\eqref{eq:deltaH}) and thus we do not
expect to capture the physics, relevant for the S-wave hyperfine splitting, adequately. And indeed including corrections of the order ${\cal O}(\alpha_Sv^4)$ may have the same order of magnitude as ${\cal O}(v^6)$ corrections but appear to contribute with opposite sign, so that only the combination of one loop order radiative corrections in concert with an otherwise ${\cal O}(v^6)$ NRQCD Lagrangian are capable of reproducing the S-wave hyperfine splitting.

The level of agreement between the computed mass splittings in our particular NRQCD setup and experimental measurements shown here is satisfactory. From the hyperfine splitting results we deduce that an overall systematic uncertainty of around $35$MeV is present, related to our choice of the NRQCD Lagrangian. By using advanced zero temperature techniques for lattice NRQCD such as incorporation of radiative corrections, operator engineering, mass tuning and smearing, the accuracy of the vacuum masses certainly can be further enhanced. Many of these techniques however are not applicable at $T>0$. Since our focus lies on uncovering the in-medium modification of states in the precision range of $\lesssim 10\%$, we are confident that the present NRQCD setup provides adequate quantitative insight into the in-medium properties of the $^3S_1$ and $^3P_1$ states.

\subsection{NRQCD energy calibration}
\label{sec:nrqcdcalib}

\begin{figure}[t]
\centering
\includegraphics[scale=0.5]{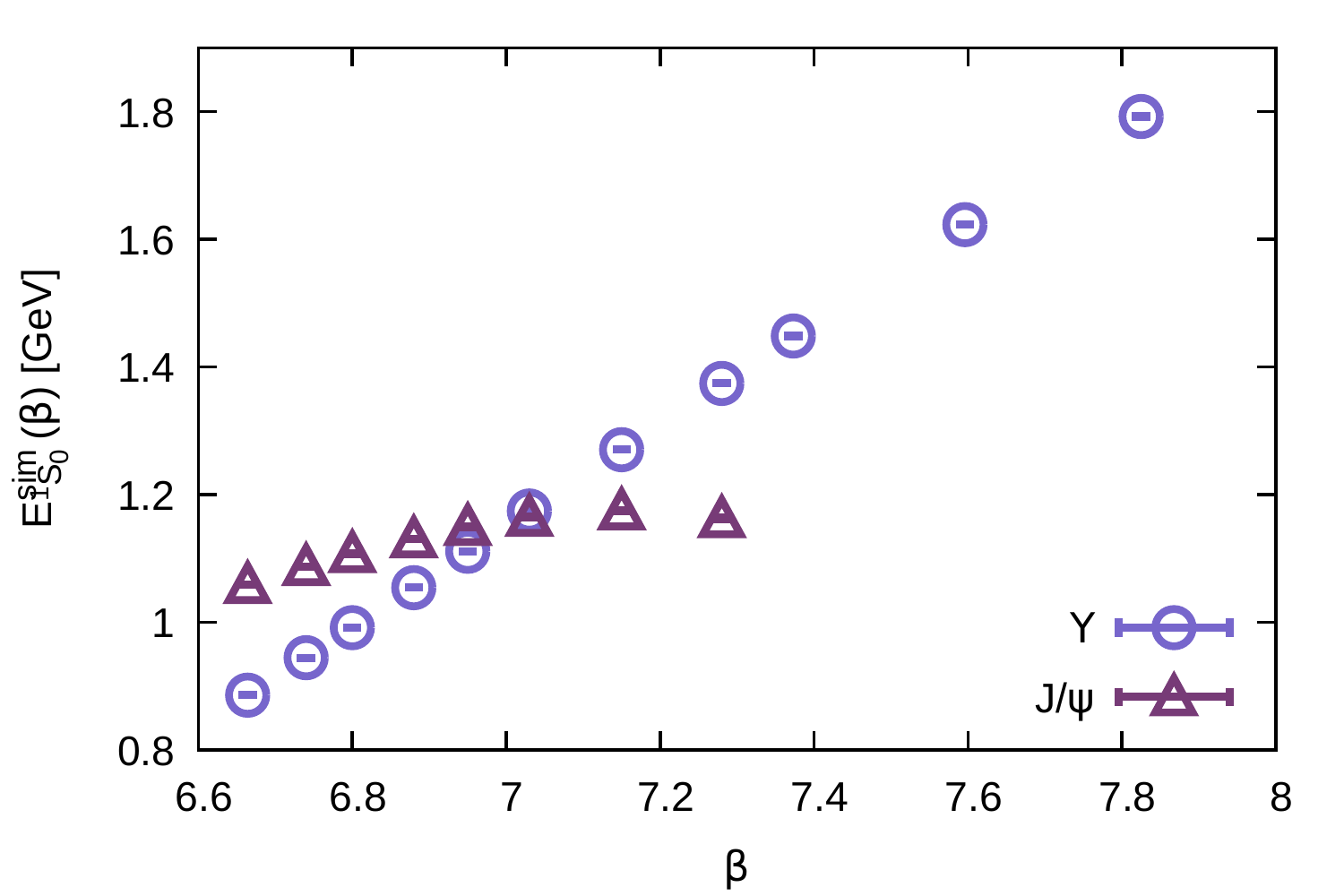}
\caption{The lattice spacing dependent energy $E^{\rm sim}_{^1S_0}$ of the $^1S_0$ ground state for bottomonium (circles) and charmonium (triangles), as obtained from raw lattice NRQCD correlators. Together with the PDG masses it enters the shift function $C(\beta)=M^{\rm exp}_{^1S_0}-E^{\rm sim}_{^1S_0}$.}\label{Fig:T0Calibration}
\end{figure}

Before we can investigate the quarkonium spectrum in terms of absolute values of particle masses, we need to take care of the NRQCD energy shift, absent in relativistic quark formulations. Its value needs to be determined from a comparison with experimental input. The physical mass of a quarkonium state may be written as
\begin{equation}
M_{\rm ^sX_J}^{\rm exp} = E^{\rm sim}_{\rm ^sX_J} + 2 (Z_{M_q} M_q - E_0) ,
\end{equation}
where $E^{\rm sim}_{\rm ^sX_J}$ denotes the energy of a particular quarkonium state computed in the raw NRQCD simulation. $Z_{M_q}$ encodes the multiplicative renormalization of the heavy quark mass and $E_0$ denotes an additional energy shift \cite{Davies:1994mp}. The physical quarkonium mass $M_{\Upsilon}^{\rm exp}$ is obtained by adding the lattice spacing dependent energy shift parameter  
\begin{equation}
C_{\rm shift}(\beta)=2 (Z_{M_q} M_q - E_0),
\end{equation}
to the NRQCD energy value. 

The calibration has to be carried out for each lattice spacing and for each of the two bare heavy quark mass parameters. We may however choose any channel for the comparison to experiment and select in the following the $^3S_1$ ground states $\Upsilon$ and $J/\Psi$ for this task. Not only is their mass known very precisely in experiment but also numerically a good signal to noise ratio is present in their correlators.

In Fig.\ref{Fig:T0Calibration} we show the raw energies of the $^3S_1$ ground state for bottomonium (circles) and charmonium (triangles), which translates into the NRQCD energy shift when subtracted from the PDG mass. For bottom quarks the mass shift remains linear over the whole range of lattice spacings considered, while for charmonium it exhibits a deviation from linearity at around $\beta=7$, flattening off above. The fact that the latter starts off with a higher value and that its $\beta$ dependence is not linear is already an indication for the stronger influence of radiative correction for the lighter charm quark in NRQCD. In order to provide calibration also to the additional $\beta$ values considered at finite temperature, we fit the bottomonium result with a simple linear ansatz, while for the charmonium results a spline interpolation of second order is used.

With the energy scale set, we can proceed to quote our results for the quarkonium ground state masses in vacuum. Since the NRQCD calibration is carried out via the $^3S_1$ states, that channel agrees by construction with the PDG value, while all other masses are a genuine ab-initio results of the simulation. The lattice spacing averaged values of the masses, as well as the PDG reference are listed in Tab.\ref{Tab:zeroTmass}.

\begin{table}[t]
\centering
\begin{tabular}{|c|c|c|c|}\hline
Particle 			& Channel  & $m$ [GeV] & PDG [GeV] \\ \hline
$\eta_b$ 			& $^1S_0(1)$ 	& $9.428(1)$ & 9.399(2) \\ 
$\Upsilon$ 		& $^3S_1(1)$ 	& $9.4603*$ & 9.4603(3) \\ 
$h_{b}$ 			& $^1P_1(1)$ 	& $9.935(7)$ & 9.8993(1)  \\ 
$\chi_{b0}$ 		& $^3P_0(1)$ 	& $9.899(4)$ & 9.8594(6) \\
$\chi_{b1}$ 		& $^3P_1(1)$ 	& $9.927(7)$ & 9.8928(4) \\
$\chi_{b2}$ 		& $^3P_2(1)$ 	& $9.941(6)$ & 9.9122(4) \\ \hline \hline
$\eta_c$ 			& $^1S_0(1)$ 	& $2.979(2)$ & 2.9839(5) \\ 
$J/\Psi$ 			& $^3S_1(1)$ 	& $3.09692*$ & 3.096900(6) \\ 
$h_{c}$ 			& $^1P_1(1)$ 	& $3.563(6)$ & 3.5254(1) \\ 
$\chi_{c0}$ 		& $^3P_0(1)$ 	& $3.453(3)$ & 3.4147(3) \\ 
$\chi_{c1}$ 		& $^3P_1(1)$ 	& $3.538(6)$ & 3.51067(5) \\ 
$\chi_{c2}$ 		& $^3P_2(1)$ 	& $3.529(5)$ & 3.55617(7) \\ \hline \hline
\end{tabular}

\caption{Mass estimates for the ground states of bottomonium and charmonium at zero temperature from the effective mass fits, compared to the experimental PDG values.}
\label{Tab:zeroTmass}
\end{table}  

We continue to the first genuine Bayesian reconstructions of this study, carried out on the zero temperature correlators discussed above.

\subsection{Spectral functions}
\label {sec:T0specfunc}

\begin{figure}[t!]
\includegraphics[scale=0.5]{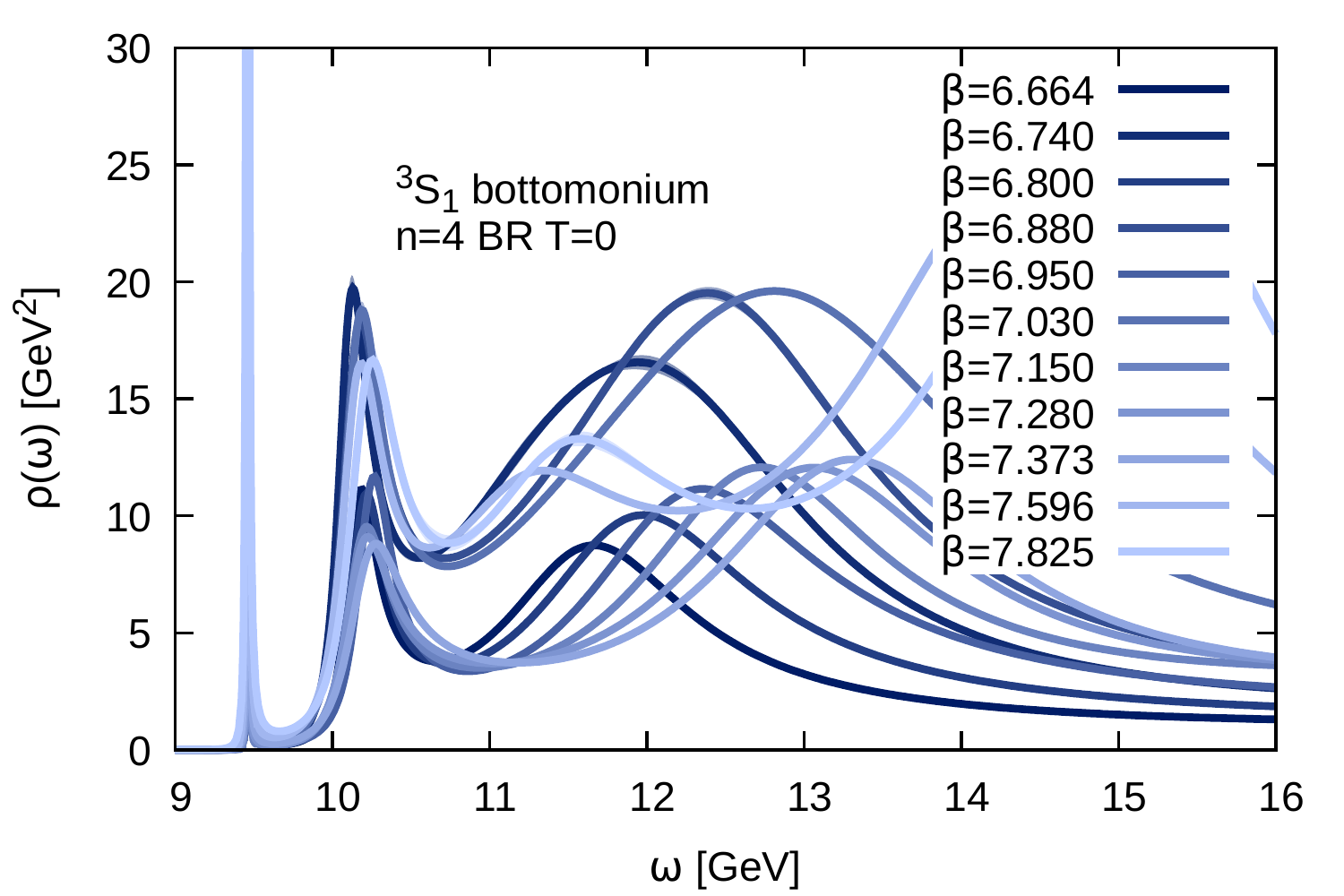}
\includegraphics[scale=0.5]{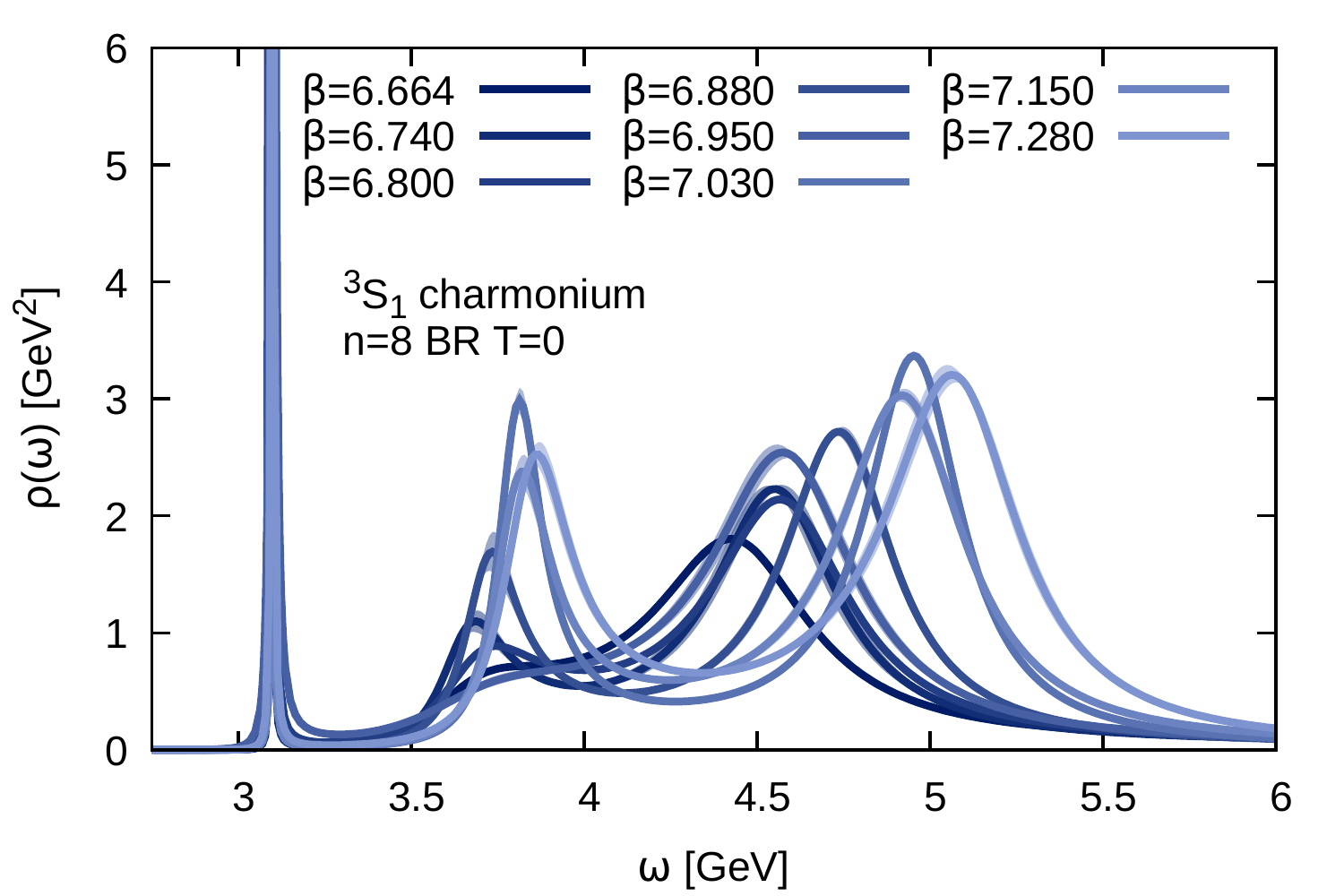}
\includegraphics[scale=0.5]{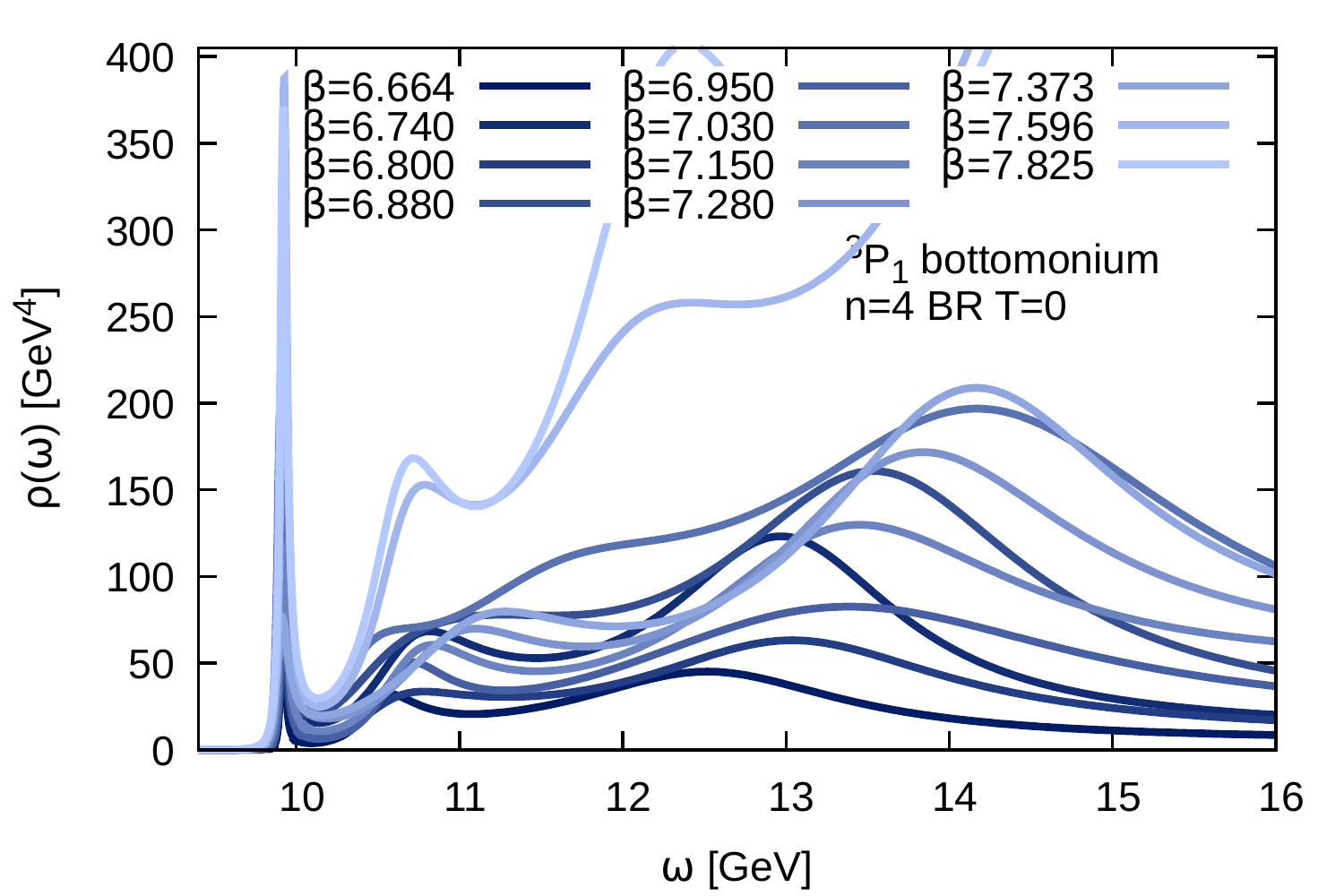}
\includegraphics[scale=0.5]{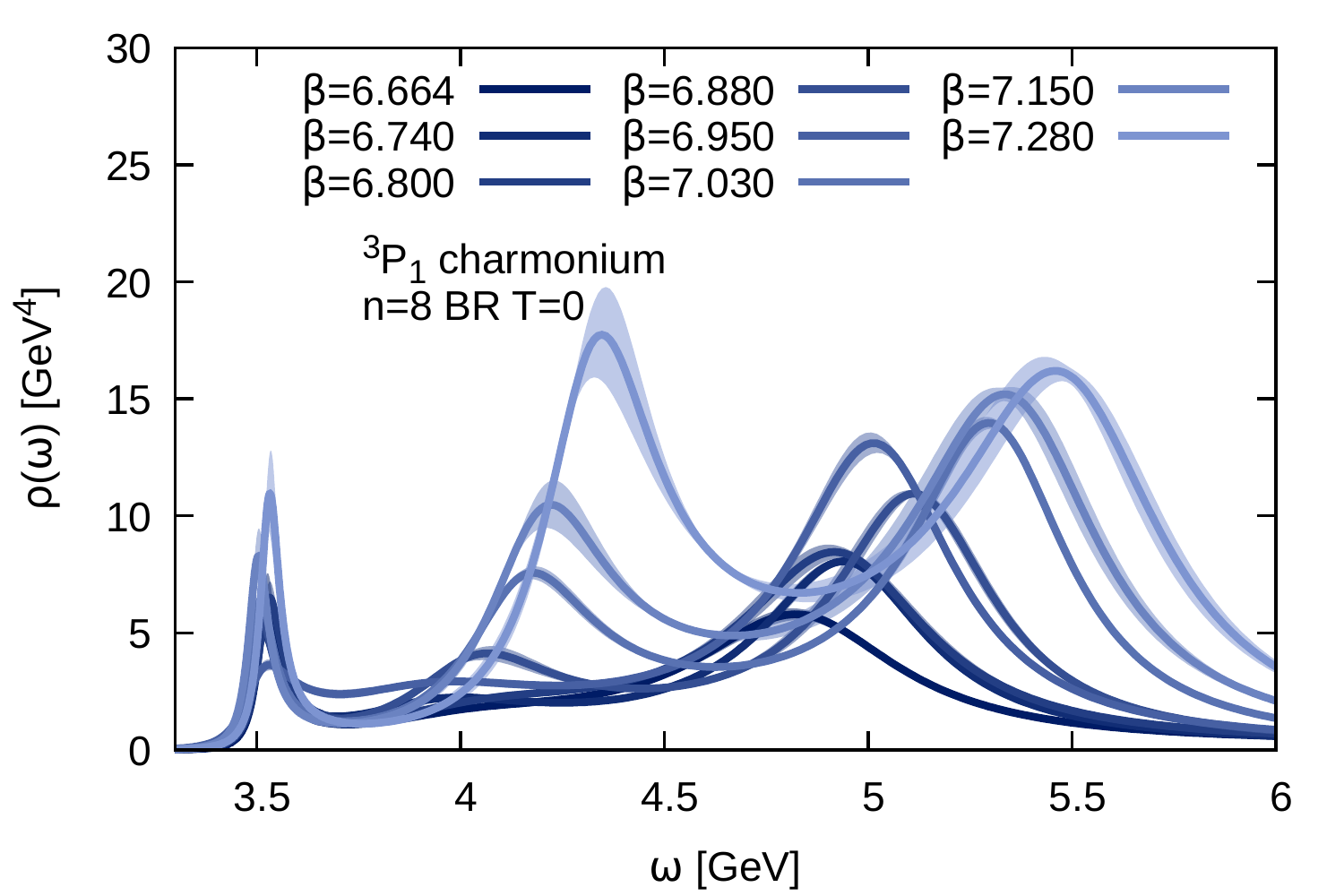}
\caption{Zero temperature reconstruction of bottomonium (left) and charmonium (right) spectral functions for different lattice spacings. The $^3S_1$ S-wave channel (top) shows a very sharp ground state peak and at least for bottomonium also clear signs of the first excited state, which in charmonium is less pronounced. The P-wave spectra in the $^3P_1$ channel due to a lower signal to noise ratio show weaker but still clearly defined signs of a ground state. The excited state structure remains unresolved. Errorbands shown here only denote the statistical uncertainty from the ten-bin Jackknife.}\label{Fig:T0BRSpectra}
\end{figure}

Bayesian spectral reconstructions allow the comprehensive study of spectral structures encoded in hadronic correlators. Their flexibility of resolving not just masses, i.e. the positions, but also the functional form of spectral features however comes at the price of requiring high quality simulation data to keep artifacts, such as default model dependence and numerical ringing under control. As discussed in sec.\ref{sec:bayesrec} we will deploy three different Bayesian approaches in this study, the standard {\rm BR} method, its smoothed cousin ${\rm BR}_\ell$ (i.e. with $S^\ell_{BR}$ given by Eq.\eqref{Eq:BRellS}), as well as the Maximum Entropy Method.

We perform the standard Bayesian reconstruction along $N_\omega=3000$ bins, spanning an interval of $\omega^{\rm num}\in[-5,20]$ in dimensionless real-time frequencies\footnote{In our numerical setup we re-scale the Euclidean
time extent such that for each lattice spacing the maximal time extent is
$\tau^{num}_{max}=\beta^num=20$. The intention behind operating with a fixed relation between $\tau$ and $\beta$ is to remove scale conversion as additional source of systematic uncertainty.}. From among the $N_\omega=3000$ points we use $N_\omega^{\rm hp}=600$ to define a high precision window around the ground state peak, in order to extract its position as precisely as possible. 

In this study we combine the correlator in the Euclidean time and imaginary frequency representation, as the spectral relations \eqref{Eq:SpecConv} and \eqref{Eq:SpecConvKL} indicate that the former will provide better control over the small frequency regime, while the latter may fix more efficiently the high frequency regime. And indeed we find that the combined reconstruction shows less variability at high frequencies compared to the case where only Euclidean data is taken into account.

The reconstructions are carried out with a constant default model normalized via the value of the Euclidean correlator $D(0)$ at zero imaginary time. In order to estimate the systematic uncertainty we repeat the reconstruction both varying the functional form and amplitude of $m(\omega)$. The additional constraint to avoid over-fitting of the data is enforced down to $|L-N_\tau|<10^{-4}$. Statistical errors are estimated with a ten-bin blocked Jackknife procedure. For completeness let us note that our reconstruction code deploys $384$ bits of precision arithmetic throughout the computation to avoid precision loss when an exponentially damped kernel is involved.

The spectral functions of bottomonium (left column) and charmonium (right column) at zero temperature in the S-wave (top row) and P-wave  (bottom row) channel reconstructed with the standard BR method are presented in Fig.\ref{Fig:T0BRSpectra}. We have calibrated the energy axis with the lattice spacing dependent shift function $C(\beta)$ determined from the effective mass fits in the previous section.

As was to be expected from the behavior of the correlators, both S-wave channels show very sharp ground state features. Their amplitude is much larger than the maximum range of the y-axis chosen here, for better visibility of the higher lying structures. In addition we are able to identify clear signs of a first excited state peak for bottomonium at all lattice spacings, while for charmonium the strength of the second peak was not as well pronounced compared to the continuum structures. Compared to our previous study the variability of the reconstructed spectra in the continuum region is markedly reduced both due to the twice higher available statistics and the combination of Euclidean and imaginary frequency data.

The P-wave states, as discussed before, have an intrinsically smaller amplitude of their ground state peak compared to the continuum contribution and we are unable to unambiguously identify a first excited state peak structure. The lower signal to noise ratio of the underlying correlators also leads to a larger (artificial) width of the ground state peak, compared to the S-wave channel.

\begin{figure}[t]
\includegraphics[scale=0.5]{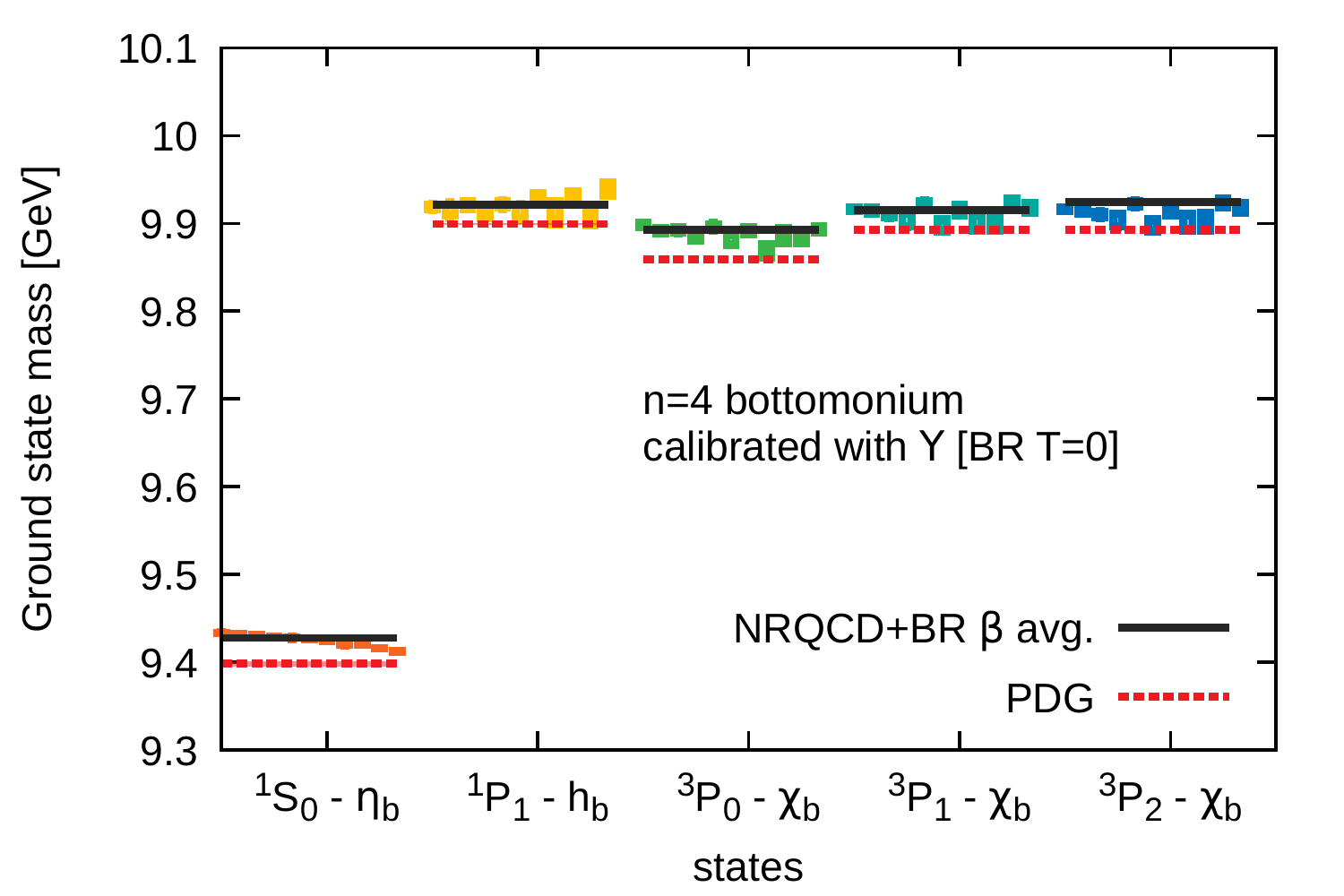}
\includegraphics[scale=0.5]{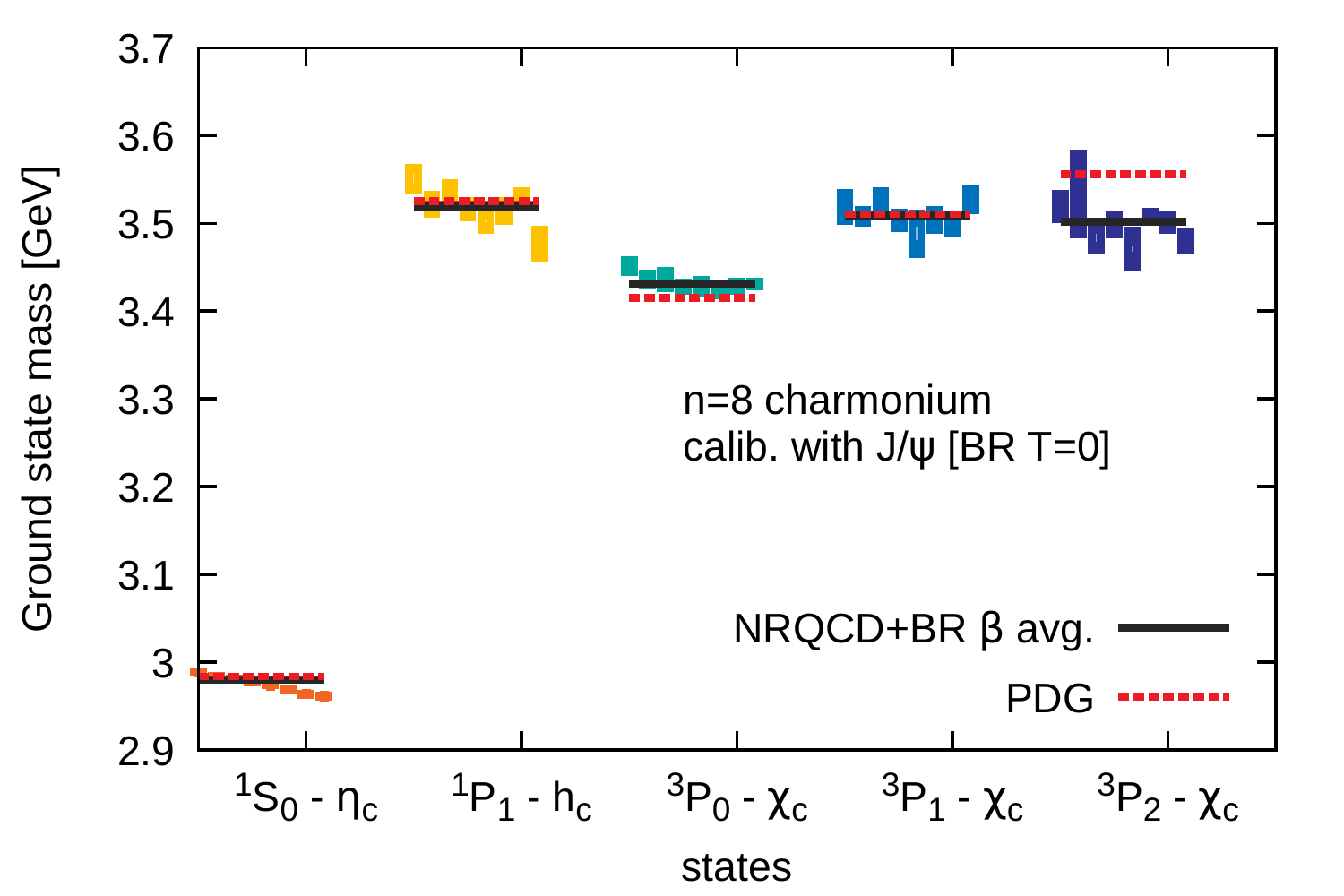}
\caption{Compilation of the ground state masses obtained from the BR method at zero temperatures, the bottomonium results being plotted on the left, the charmonium results on the right. Error boxes denote the statistical uncertainty, while error bars correspond to the default model dependence of the masses.}\label{Fig:T0BRMassesCmp}
\end{figure}

We proceed to obtain the ground and excited state masses from the Bayesian spectral reconstruction. To this end we fit the lowest, and if distinguishable the next higher lying peak structure with a Breit-Wigner in order to extract the position of these features. Besides carrying out a ten-bin Jackknife to estimate the statistical uncertainty, we also repeat the determination of the mass for all different default models considered, which provides an estimate of the systematic errors involved. The values of the lattice spacing averaged masses from the BR method are compiled in Tab.\ref{Tab:zeroTmassBR} and the individual results for the ground states are plotted in the two panels of Fig.\ref{Fig:T0BRMassesCmp}, bottomonium on the left, charmonium on the right. We provide these values for completeness, reminding the reader that due to our focus on in-medium spectral properties we neither use correlators with improved ground state overlap nor specifically tuned quark masses. The reproduction of vacuum ground state properties here is satisfactory but not competitive with dedicated $T=0$ studies.

\begin{table}[t]
\centering
\begin{tabular}{|c|c|c|c|}\hline
Particle 			& Channel  & $m$ [GeV] & PDG [GeV] \\ \hline
$\eta_b$ 			& $^1S_0(1)$ 	& $9.428(1)$ & 9.399(2)  \\ 
$\eta_b^\prime$ 	& $^1S_0(2)$ 	& $10.14(2)$ & 9.999(4)  \\ 
$\Upsilon$ 		& $^3S_1(1)$ 	& $9.46047(4)*$ & 9.4603(3)  \\ 
$\Upsilon^\prime$ 	& $^3S_1(2)$ 	& $10.14(2)$ & 10.0233(3) \\ 
$h_{b}$ 			& $^1P_1(1)$ 	& $9.921(2)$ & 9.8993(1)  \\ 
$h_{b}^\prime$ 		& $^1P_1(2)$ 	& $10.52(5)$ & 10.260(2) \\ 
$\chi_{b0}$ 		& $^3P_0(1)$ 	& $9.893(2)$ & 9.8594(6) \\ 
$\chi_{b0}^\prime$ 	& $^3P_0(2)$ 	& $10.51(3)$ & 10.2325(8)\\ 
$\chi_{b1}$ 		& $^3P_1(1)$ 	& $9.915(2)$ & 9.8928(4) \\ 
$\chi_{b1}^\prime$ 	& $^3P_1(2)$ 	& $10.48(7)$ & 10.2555(5)\\ 
$\chi_{b2}$ 		& $^3P_2(1)$ 	& $9.924(2)$ & 9.9122(4) \\ 
$\chi_{b2}^\prime$ 	& $^3P_2(2)$ 	& $10.57(4)$ & 10.2687(6)\\ \hline \hline
$\eta_c$ 			& $^1S_0(1)$ 	& $2.979(2)$ & 2.9839(5)  \\ 
$\eta_c^\prime$ 	& $^1S_0(2)$ 	& $3.70(2)$ & 3.638(1) \\ 
$J/\Psi$ 			& $^3S_1(1)$ 	& $3.0967(2)*$ & 3.096900(6) \\ 
$\Psi^\prime$ 		& $^3S_1(2)$ 	& $3.77(3)$ & 3.68610(3) \\ 
$h_{c}$ 			& $^1P_1(1)$ 	& $3.520(6)$ & 3.5254(1) \\ 
$\chi_{c0}$ 		& $^3P_0(1)$ 	& $3.432(3)$ & 3.4147(3) \\ 
$\chi_{c1}$ 		& $^3P_1(1)$ 	& $3.509(4)$ & 3.51067(5) \\ 
$\chi_{c2}$ 		& $^3P_2(1)$ 	& $3.499(5)$ & 3.55617(7) \\ \hline \hline
\end{tabular}
\caption{Mass estimates for the ground and first excited states of bottomonium and charmonium at zero temperature from the BR method, compared to the experimental PDG values. While the S-wave ground state values agree with the results from the effective mass fit, the P-wave results from the BR method lie slightly closer to the correct values. A possible reason is the higher flexibility of the BR method to disentangle ground and excited states contributions.}
\label{Tab:zeroTmassBR}
\end{table}  

The first sanity check is the obtained value for the mass of the $^3S_1$-wave ground state. Since we have carried out the calibration using the effective mass fits, the value here does not necessarily agree with the PDG value. Reassuringly we find that the S-wave ground state mass from the BR method takes on very similar values as with the effective mass fit, the two methods agree within errors. In general the S-wave masses are very similar between the two methods. On the other hand we find that for the $^1P_1$ and $^3P_1$ states, on which we will focus in the $T>0$ study, the BR method provides systematically better values, i.e. closer to the PDG benchmark. In case of charmonium this improvement is significant, bringing the absolute value of the ground state mass into agreement with the PDG. One reason for the difference may be the higher flexibility of the BR method in distinguishing ground state and excited state contributions, which may have been intertwined by the selection of the fitting range of the effective masses by eye. The $^3P_0$ channel is reasonable well reproduced in both bottomonium and charmonium, while, as was expected from the inspection of the splittings, the $^3P_2$ channel of charmonium shows sizable deviations from the PDG value. 

When perusing Tab.\ref{Tab:zeroTmassBR}, the reader may wonder why our estimates for excited states show rather sizable deviations from the PDG values. The reason is our use of unimproved correlators in anticipation of the study of in-medium effects and the absence of a heavy quark mass tuning, not a deficiency of NRQCD itself. The information loss affecting the correlators, i.e. from taking the convolution over the spectral function,  makes the reconstruction of the excited state region extremely challenging. As there are several structures present in the UV regime, the spectral reconstruction tends to summarize all of the unresolved structure into a single spectral feature, which then lies at a frequency higher than the lowest contributing actual spectral peak. This artificial phenomenon is well known from mock data tests, see e.g. Fig.\ref{Fig:MockDataTests} in Sec.\ref{Sec:FutureProsp}. It in turn leads to a systematic over-prediction of the excited state mass in Tab.\ref{Tab:zeroTmassBR}, which, when unimproved correlators are used, may only be countered by an exponential reduction of the statistical error on the input data.

As we have seen, at zero temperatures, where the individual spectral structures are known to be sharp delta-like peaks, the standard BR method is able to provide highly accurate reconstructions of the ground state features. Their positions are in quantitative agreement with those obtained from effective mass or multi-exponential fits.

In the following we inspect how different Bayesian methods are able to reproduce the ground state features, taking the standard BR method as benchmark. For the smooth BR method with $\kappa=1$ we use a slightly different setup of $N_\omega=2000$ numerical frequency bins along $\omega^{\rm num}\in[-1.5,10]$ and we have checked that changing to such a frequency interval does not significantly change the outcome of the reconstruction. In addition, since the smooth BR method will not produce equally sharply resolved peak structures as the standard BR method, we refrain from defining a high precision interval around the lowest lying peak.

The results of the standard Maximum Entropy Method in Bryan's formulation on the other hand have a clear dependence on the choice of the frequency interval. The reason is that since the search space is chosen by a singular value decomposition of the integral Kernel, the oscillatory part of its basis functions is located close to $\omega_{\rm min}$. As was shown in a dedicated analysis of Bryan's implementation in \cite{Rothkopf:2011ef}, as well as confirmed in our previous study \cite{Kim:2014iga}, the resolution of the MEM decreases as $\omega_{\rm min}$ takes on larger negative values, which in the NRQCD setup are admissible. Thus via the choice of $\omega_{\rm min}$ we can determine how washed out the MEM end result eventually is. Here we choose $\omega^{\rm num}\in[-0.15,20]$ in order to have as high as possible resolution for the ground state, while still being able to accommodate the extrusion of the reconstructed spectrum into the negative frequency regime.

\begin{figure}[t]
\includegraphics[scale=0.5]{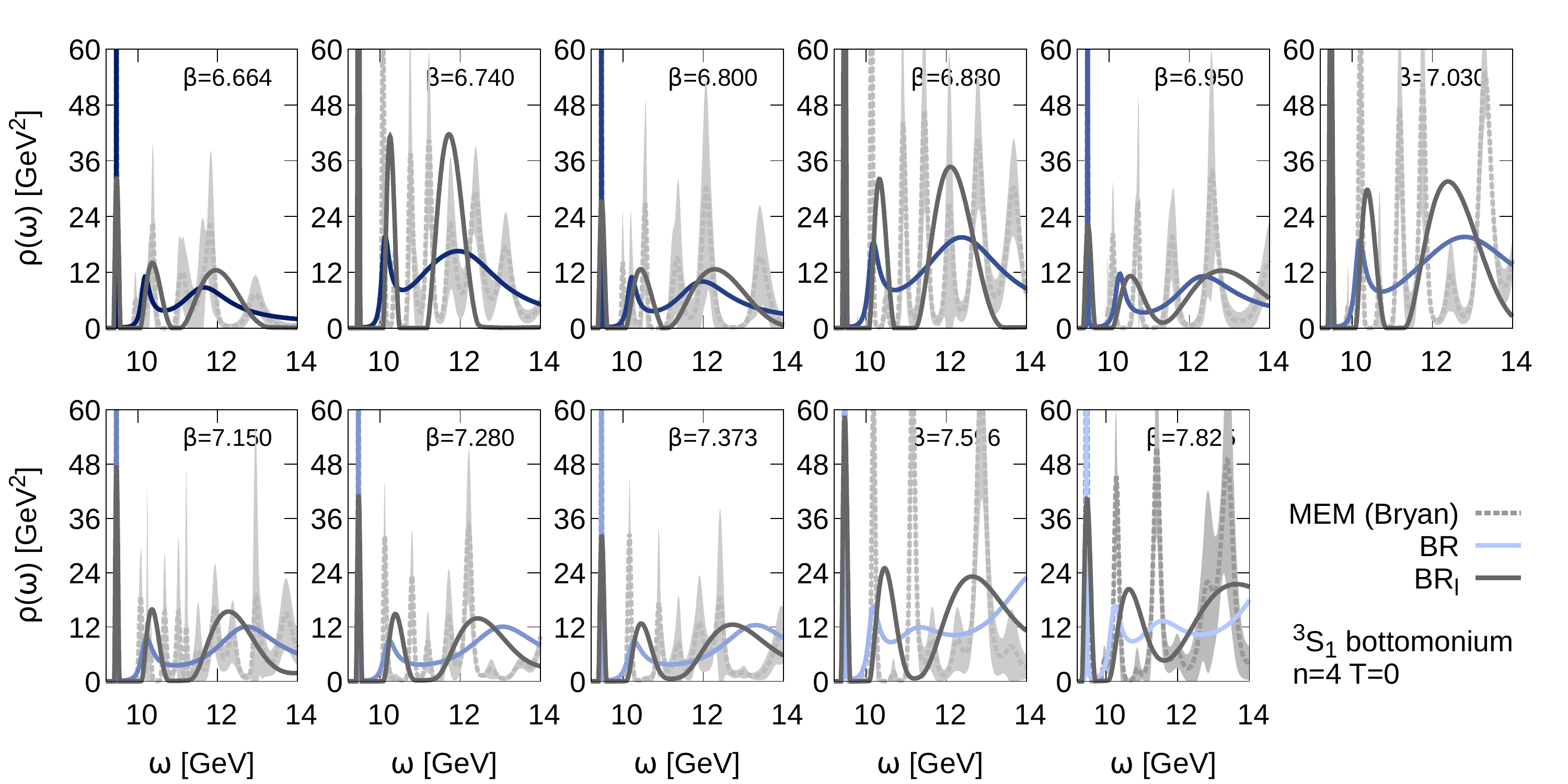}
\includegraphics[scale=0.5]{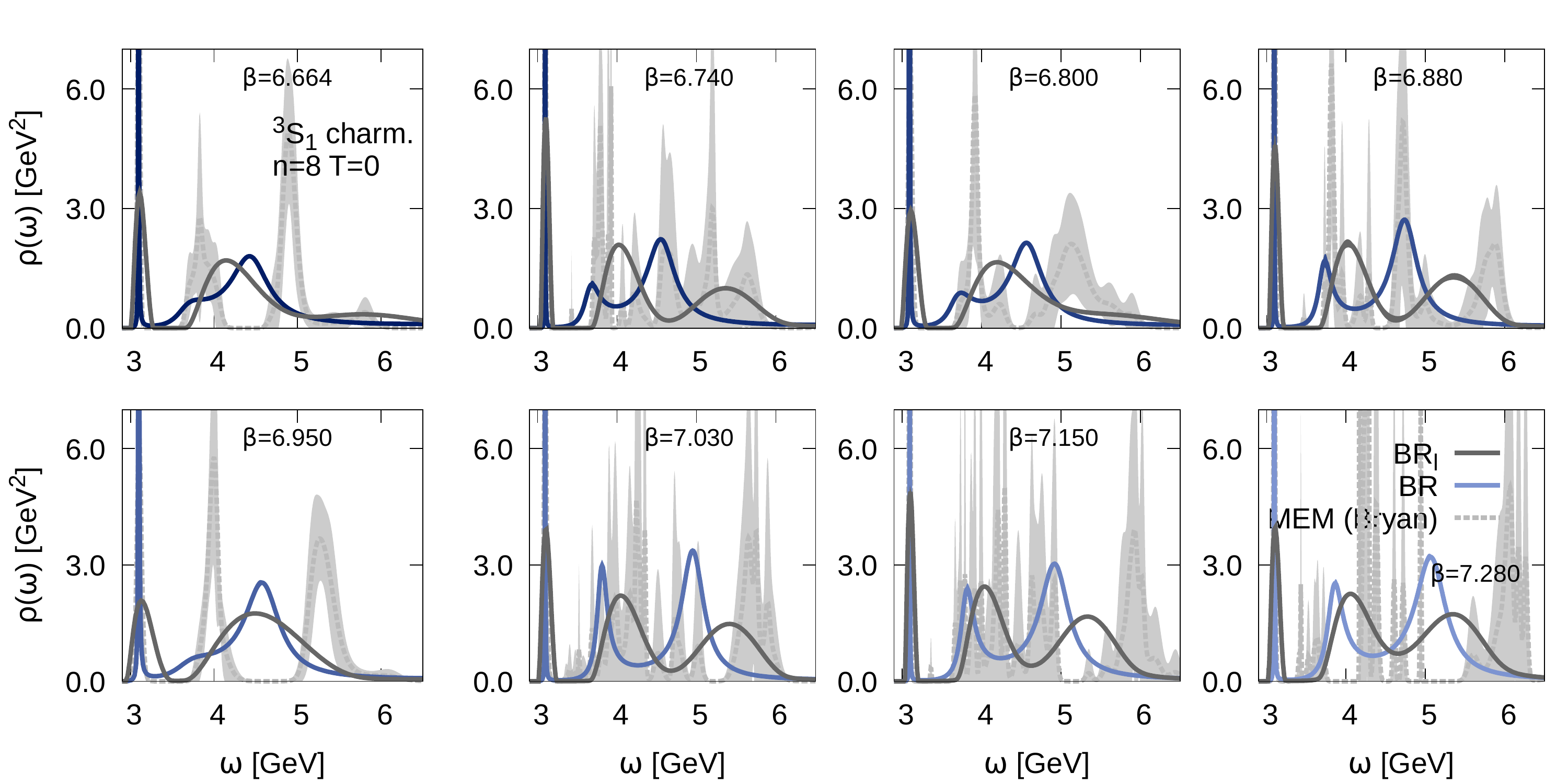}
\caption{Comparison of the spectral reconstruction of the bottomonium (top two rows) and charmonium (bottom two rows) $^3S_1$ S-wave ground state peak among different reconstruction algorithms: $BR$ method (colored solid), Maximum Entropy Method (gray dashed) and the smooth BR$_\ell$ method (gray solid). Note that with the settings $\kappa=1$ the BR$_\ell$ method produces the peaks with lowest amplitude and largest (artificial) width, followed by the MEM and the BR method, as expected produces the sharpest features. Note that the peak apex from the BR$_\ell$ method remains very close to the position located by the standard BR method.}\label{Fig:T0SpectraCmpMethods}
\end{figure}

In Fig.\ref{Fig:T0SpectraCmpMethods} we have compiled individual reconstructions of the S-wave ground state with all three different methods for bottomonium and charmonium respectively. We can clearly see that the standard BR method provides the highest resolution of the ground state, followed by the MEM (based on this one particular choice of $\omega_{\rm min}$). What is important to see is that also the smooth BR method manages to locate the ground state peak reliably at all lattice spacings considered. While we already made sure that the ${\rm BR}_\ell$ manages to avoid numerical ringing via the reconstructions of free spectral functions, we now see that it nevertheless picks up on genuine peak signals if present in the data. Of course the width of the reconstructed peaks in ${\rm BR}_\ell$ is much larger than both in the standard BR method and in the MEM. In the case of charmonium in Fig.\ref{Fig:T0SpectraCmpMethods} the ${\rm BR}_\ell$ results also show a significant peak asymmetry, the peak apex however remains close to the position of the peak resolved by the standard BR method.

This outcome further supports the following strategy for the study of in-medium effects, where peak structures and continuum are much less clearly separated: we will use the ${\rm BR}_\ell$ method first to ascertain whether a genuine peak is present in the data and then use the standard BR method to extract quantitatively its properties.

\subsection{Interpreting the $T= 0$ spectra}

\begin{figure}[t]
\includegraphics[scale=0.5]{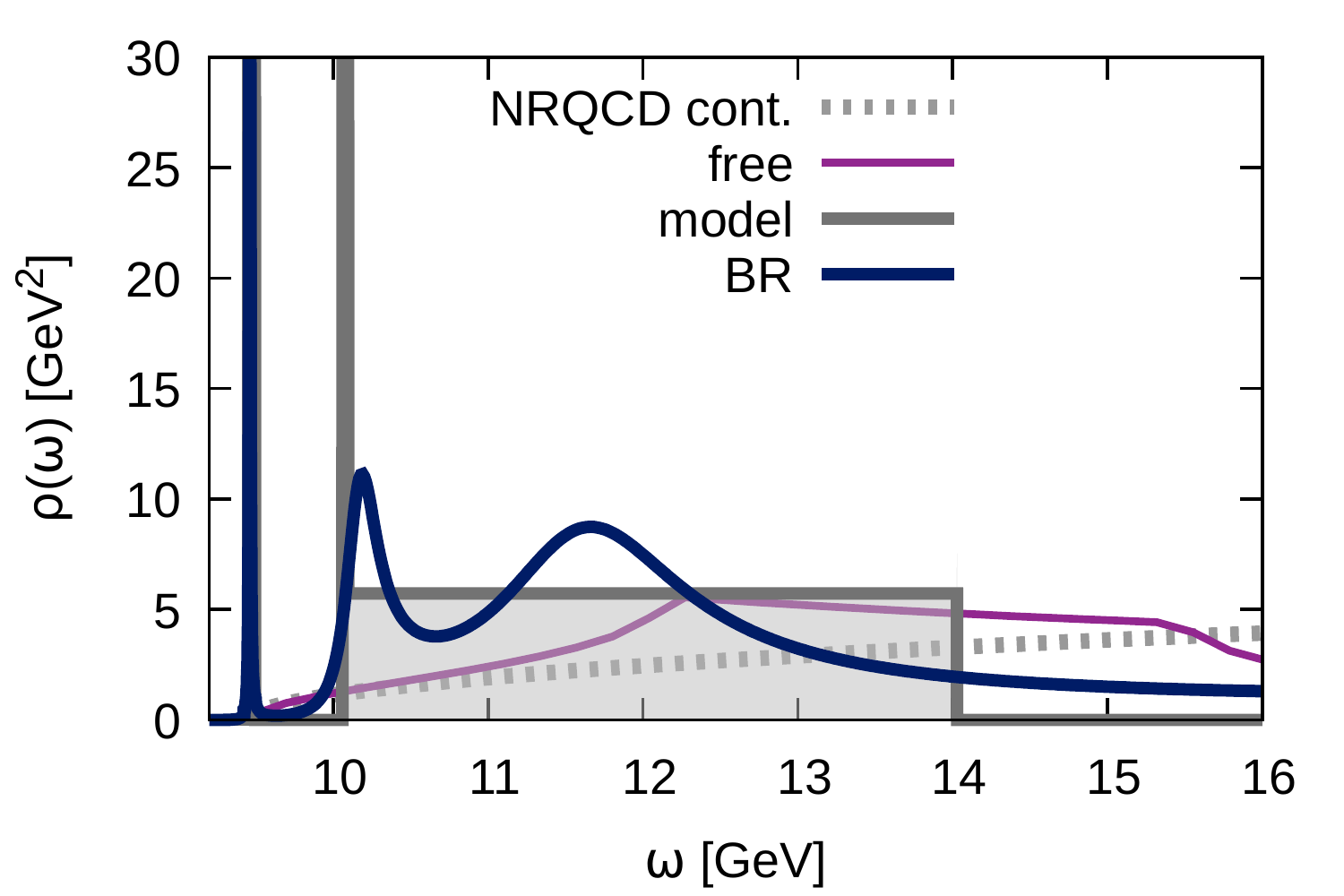}
\includegraphics[scale=0.5]{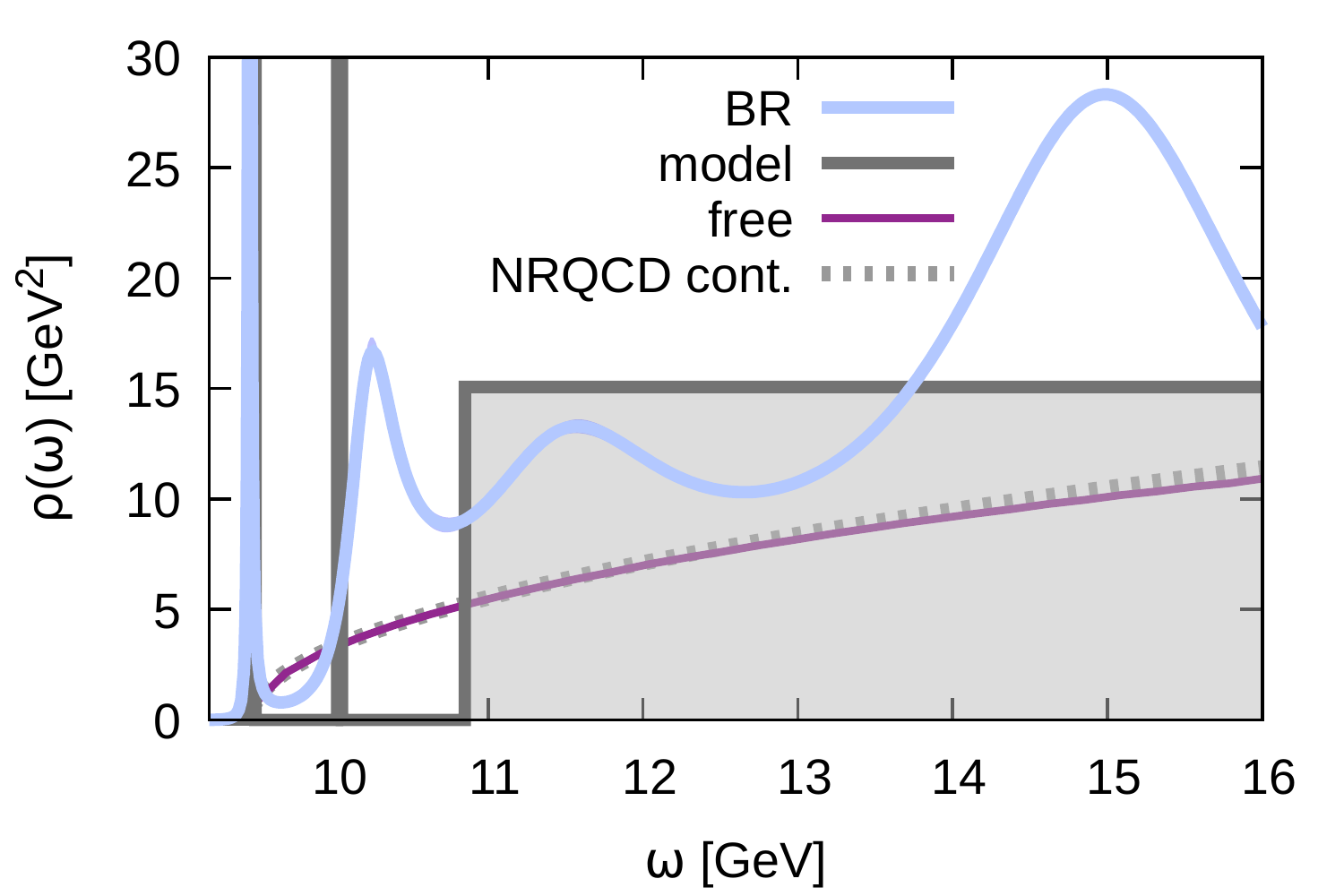}
\includegraphics[scale=0.5]{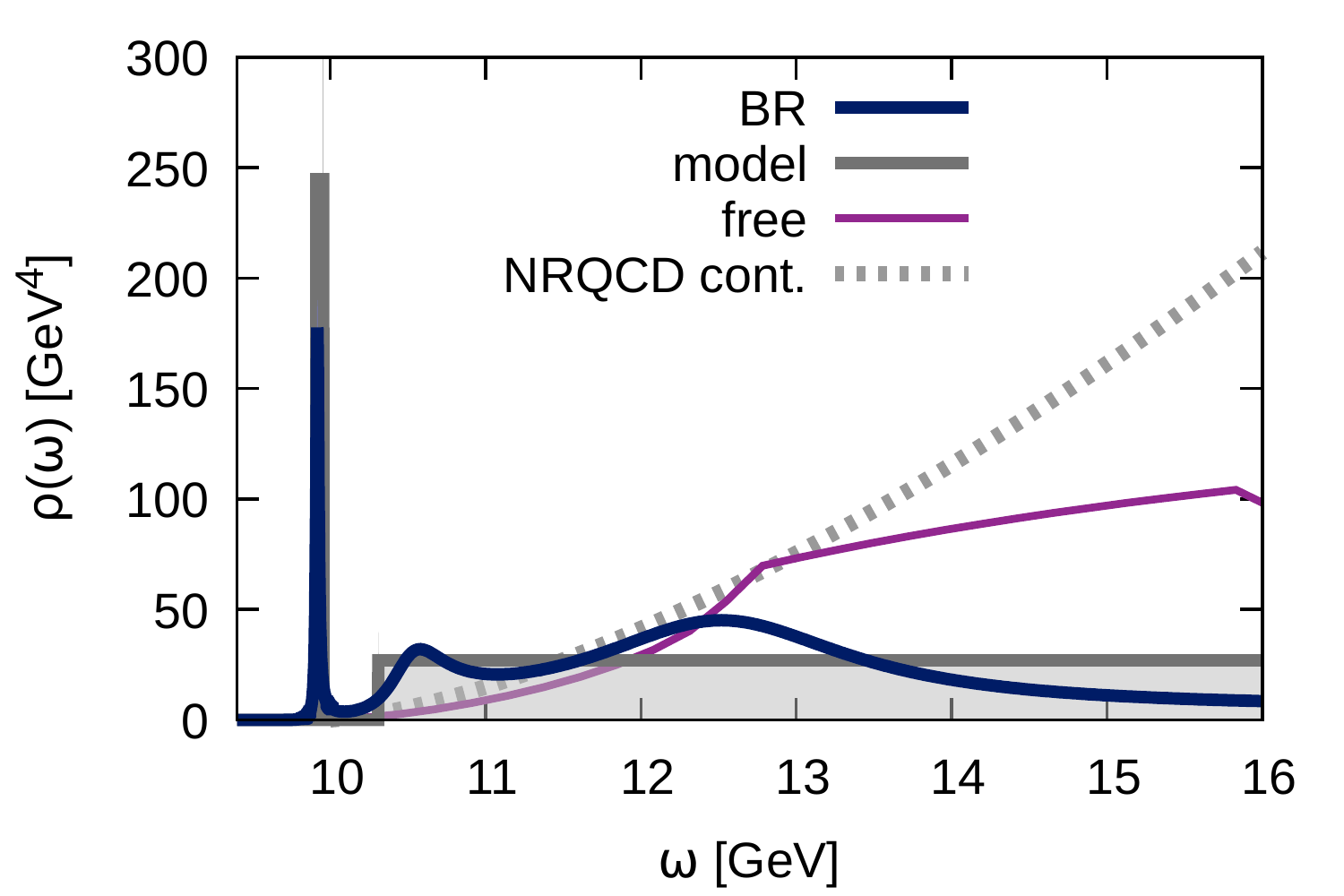}
\includegraphics[scale=0.5]{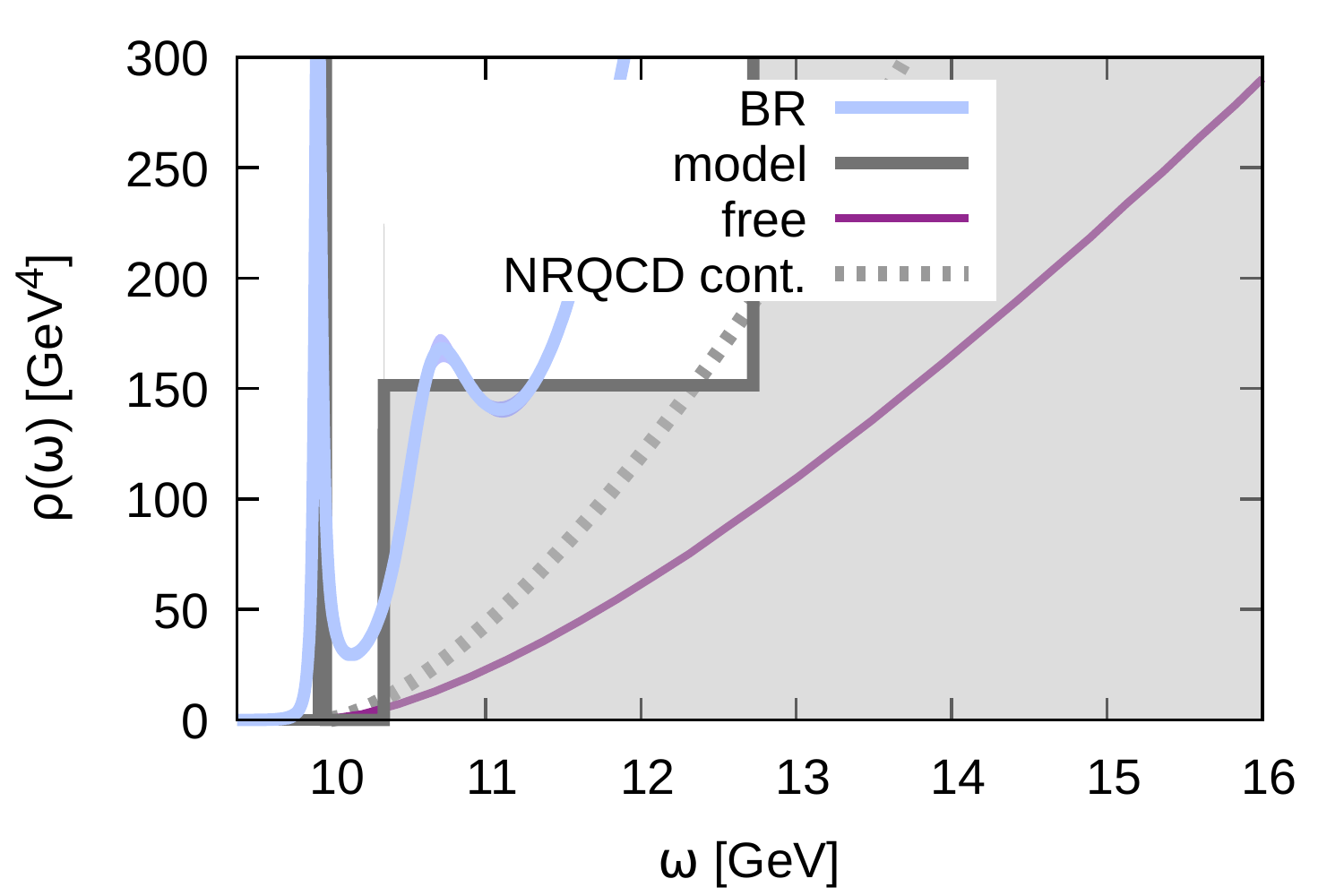}
\caption{Constrained model fits (gray) compared to the spectral reconstruction based on the standard BR method at $T= 0$.The top two panels show S-wave bottomonium on the coarsest $\beta=6.664$ (right) and finest $\beta=7.825$ (left) lattices. The lower two panels on the other hand contain the results for P-wave bottomonium again for both the coarsest $\beta=6.664$ (right) and finest $\beta=7.825$ (left) lattices. }\label{Fig:T0BRModelCmpBottom}
\end{figure}

The Bayesian spectral reconstruction carried out in the previous section revealed robust signs for the ground state and partial signals for possible first excited state contributions encoded in the correlator. However the first excited state did not agree well with the experimental values, being situated at a higher mass. This behavior of the spectral reconstruction (as well as in the exponential fit) is related to the presence of a region densely populated with excited states around the open heavy flavor 
threshold. If the input simulation data quality is not sufficient to resolve individual peaks the spectral reconstruction tends to summarize all structures in one large feature, which then will lie above the lowest of those structures.

Here we wish to find out how well the correlator lends itself to an interpretation that is based on our physical knowledge of the vacuum spectrum. To this end we construct a model spectrum, which incorporates both a well defined ground and excited state as delta-like peaks. In addition we connect to the first excited state peak two boxes of variable width and height. The first of which is added in order to provide an effective description of the higher excited state regime below the threshold, while the latter may provide a description of the continuum. We know the position of the ground state from the effective mass fit and fix it. For the distance to the first excited state we allow a window of around 25MeV for the S-wave and 50MeV for the P-wave around the known PDG value, while leaving the width and height parameters for the two boxes fully free.

\begin{figure}[t]
\includegraphics[scale=0.5]{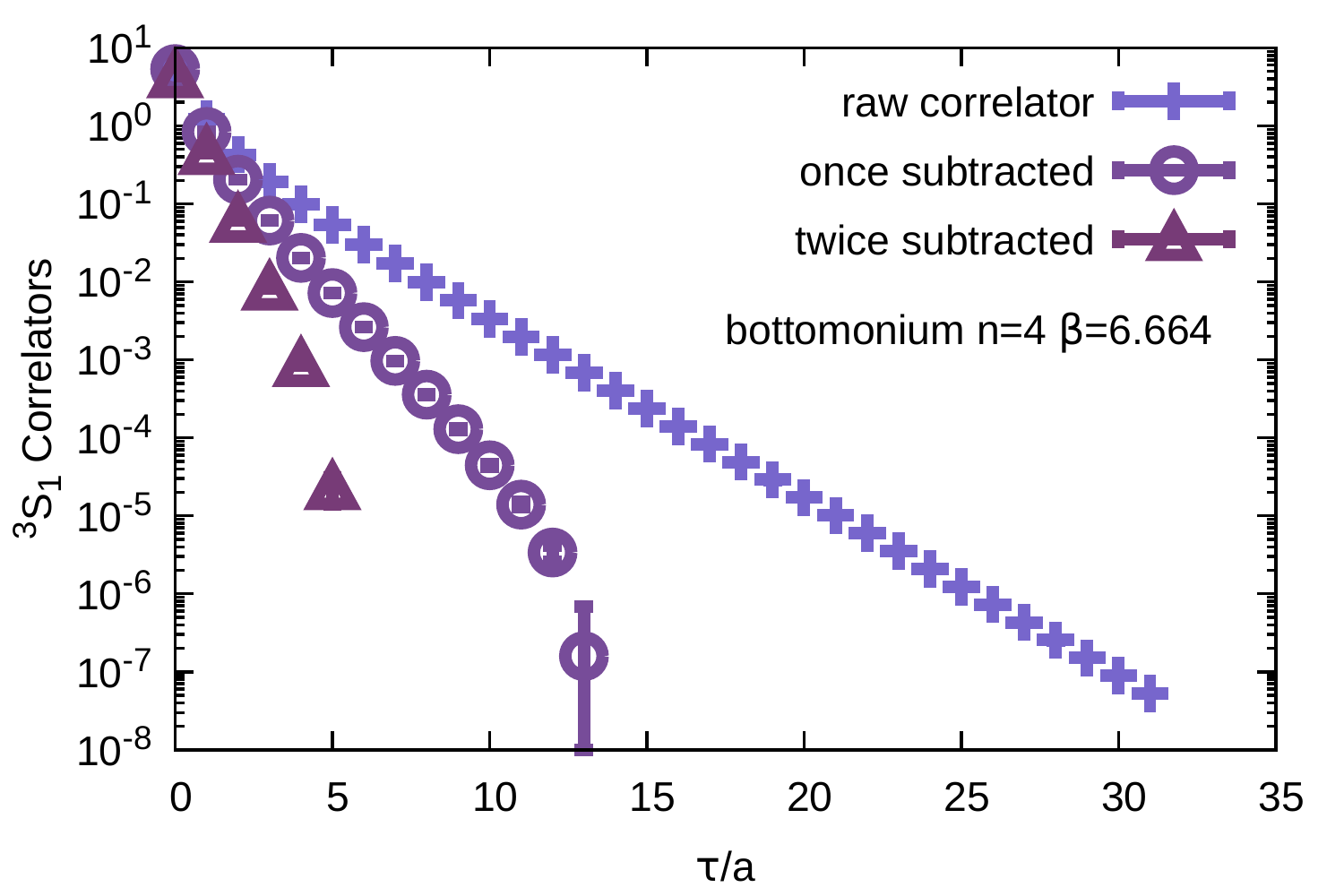}
\includegraphics[scale=0.5]{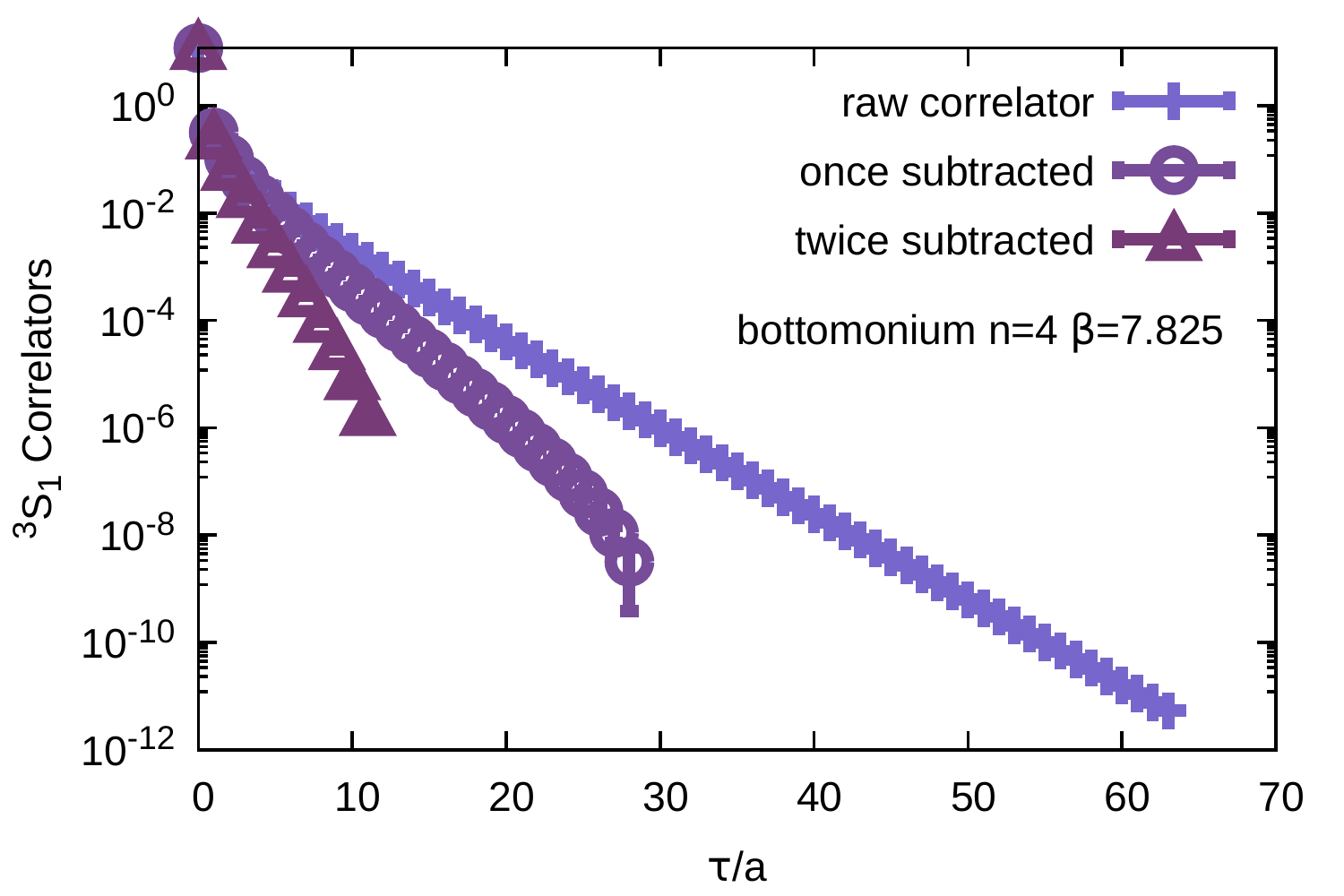}
\caption{Representative examples of the raw S-wave bottomonium correlator (light blue crosses), as well as after subtraction of the ground state contribution (open circle) and after further subtracting the first excited state contribution. Note that both at $\beta=6.664$ with $N_\tau=32$ (left) and even at $\beta=7.825$ with $N_\tau=64$ (right) only a very small number of convex points remain to encode the UV regime including the continuum structures.}\label{Fig:T0SubtrCorr}
\end{figure}

At first we attempted to fix the upper edge of the second box from a simple one-box fit to the small Euclidean time region of the correlators but found that in contrast to relativistic correlators the small and late $\tau$ regime in NRQCD are not separated well enough for an individual determination of the UV properties without taking into account the IR structure present.

The best fit results for bottomonium S-wave (top two panels) and the P-wave (bottom two panels) on the coarsest $\beta=6.664$ (right) and finest $\beta=7.825$ (left) lattices are shown in Fig.\ref{Fig:T0BRModelCmpBottom}. The best fit model spectrum in gray is contrasted to the corresponding standard BR spectral reconstruction as colored solid curve. The shown best fit model spectra allow us to reproduce the correlators with a $\chi^2/d.o.f.\approx 1$, i.e. they provide an admissible explanation of the physics content of the correlator. (The corresponding charmonium results are given in App.\ref{app:ConstrModT0}). As a further guide we also plot the free NRQCD spectral function in continuum (gray dashed) and the corresponding non-interacting lattice NRQCD spectral function. They have been shifted by an offset arbitrarily set to the mass of the ground state peak. 

\begin{figure}[t]
\includegraphics[scale=0.5]{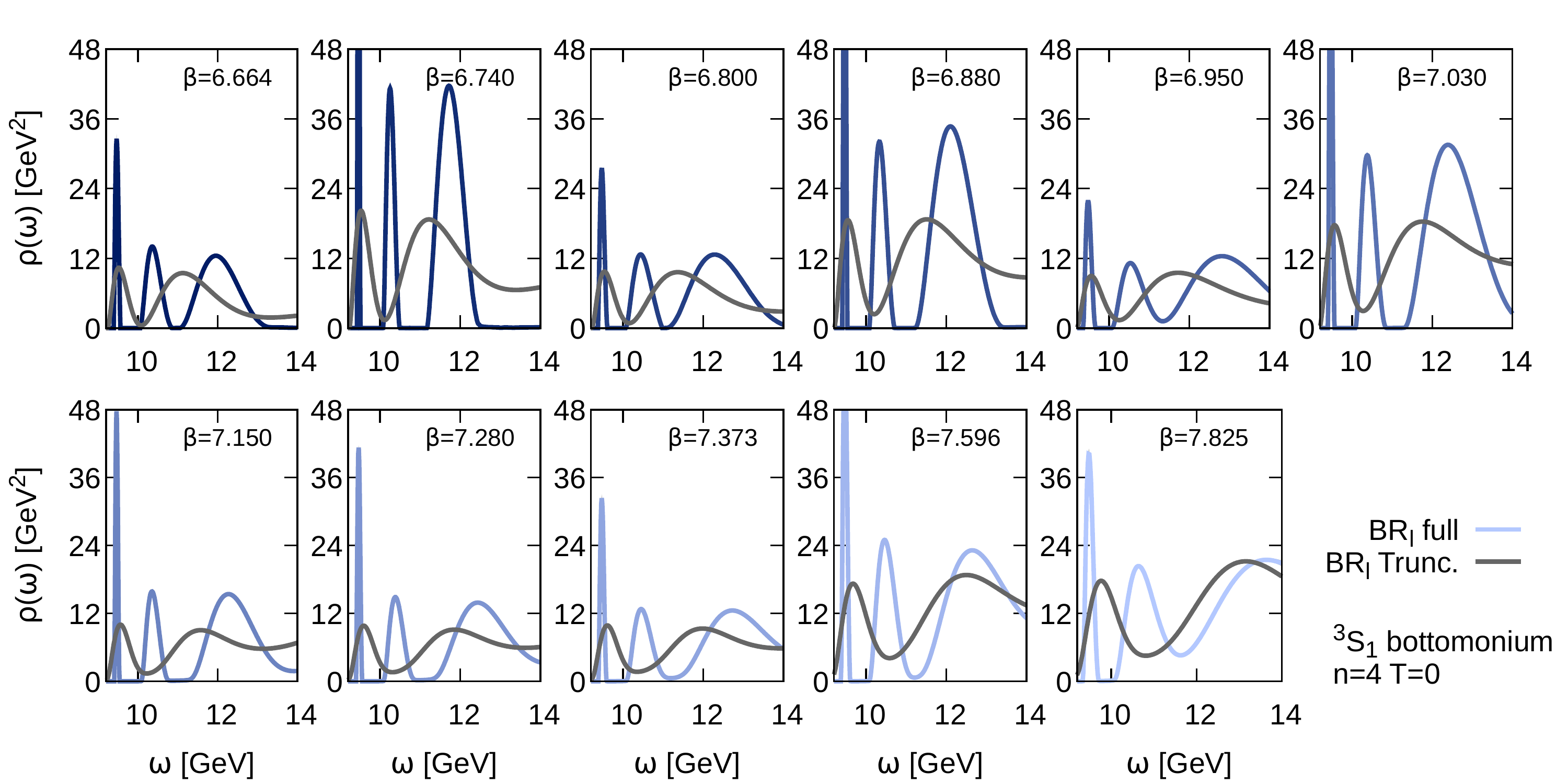}
\includegraphics[scale=0.5]{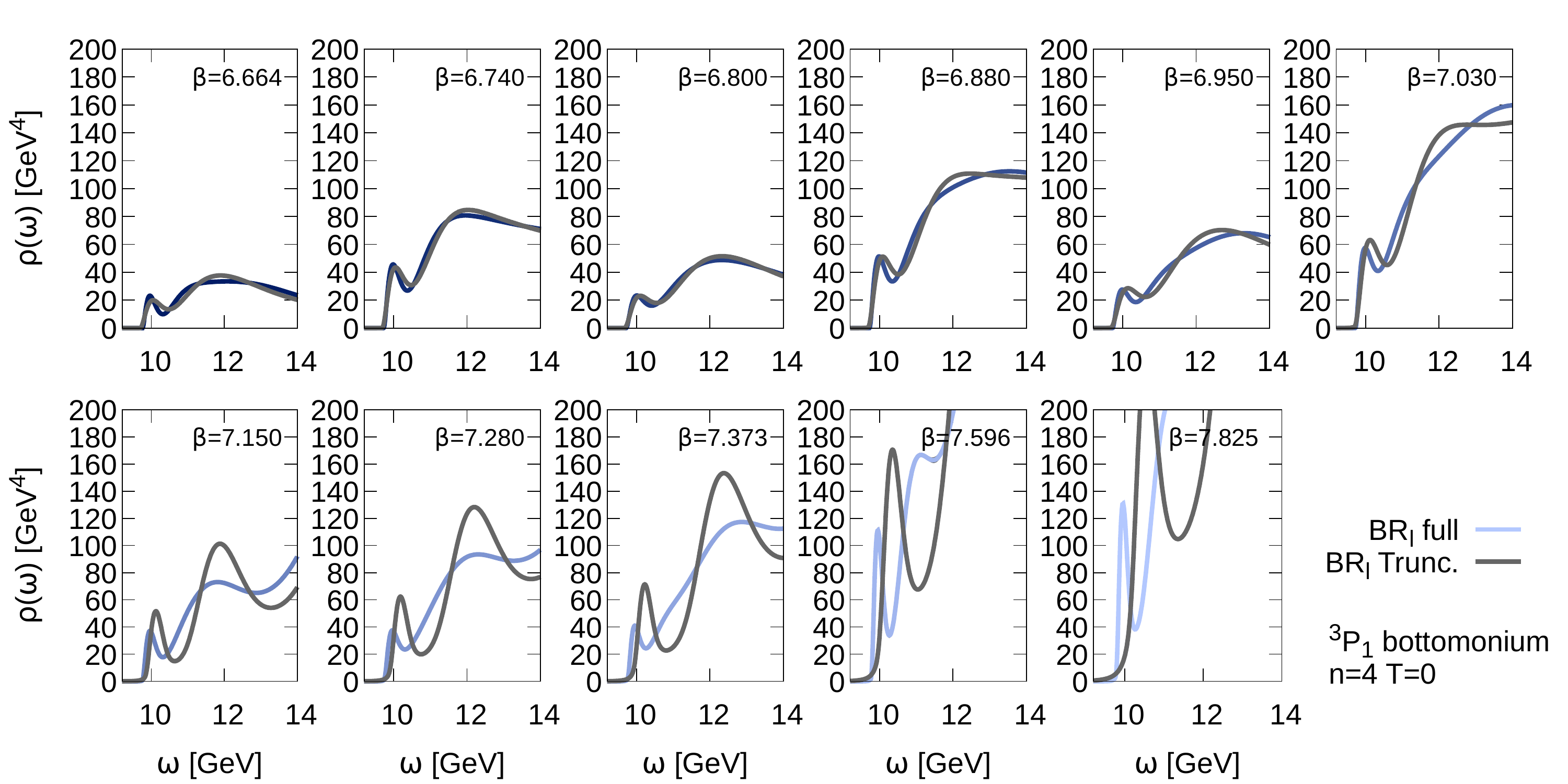}
\caption{Influence on the smooth BR reconstruction from a truncation of the low-temperature simulation data to the same $\tau_{\rm max}/a_\tau=12$ Euclidean time extent available at finite temperature. The spectra shown here correspond to the $^3S_1$ (top) and $^3P_1$ channel of bottomonium with the full reconstruction given by colored solid lines, while the result after truncation is given by the gray solid curves.}\label{Fig:T0BRSmoothSpectraCmpTrunc}
\end{figure}

In addition to being amenable to a spectrum with ground and excited state close to their physical value, we see that the constrained fit produces a continuum structure, which is compatible with what is found directly from the BR method. The finer the lattice, the further the NRQCD spectrum extends into the UV, even beyond the point of $\pi/a$, as is known from the free theory. In the P-wave case, in particular at small lattice spacings the strong rise in the continuum is both manifest in the spectral reconstruction, as well as in the constrained model fit.

On the one hand it is reassuring to see that the constrained model fit provides an intuitive interpretation of the BR spectral functions. It distinguishes four physically relevant structures, ground and first excited state, the densely populated excited state region around threshold and the continuum. All of these are found in a similar fashion also in the Bayesian spectral reconstruction.

\begin{figure}[t]
\includegraphics[scale=0.5]{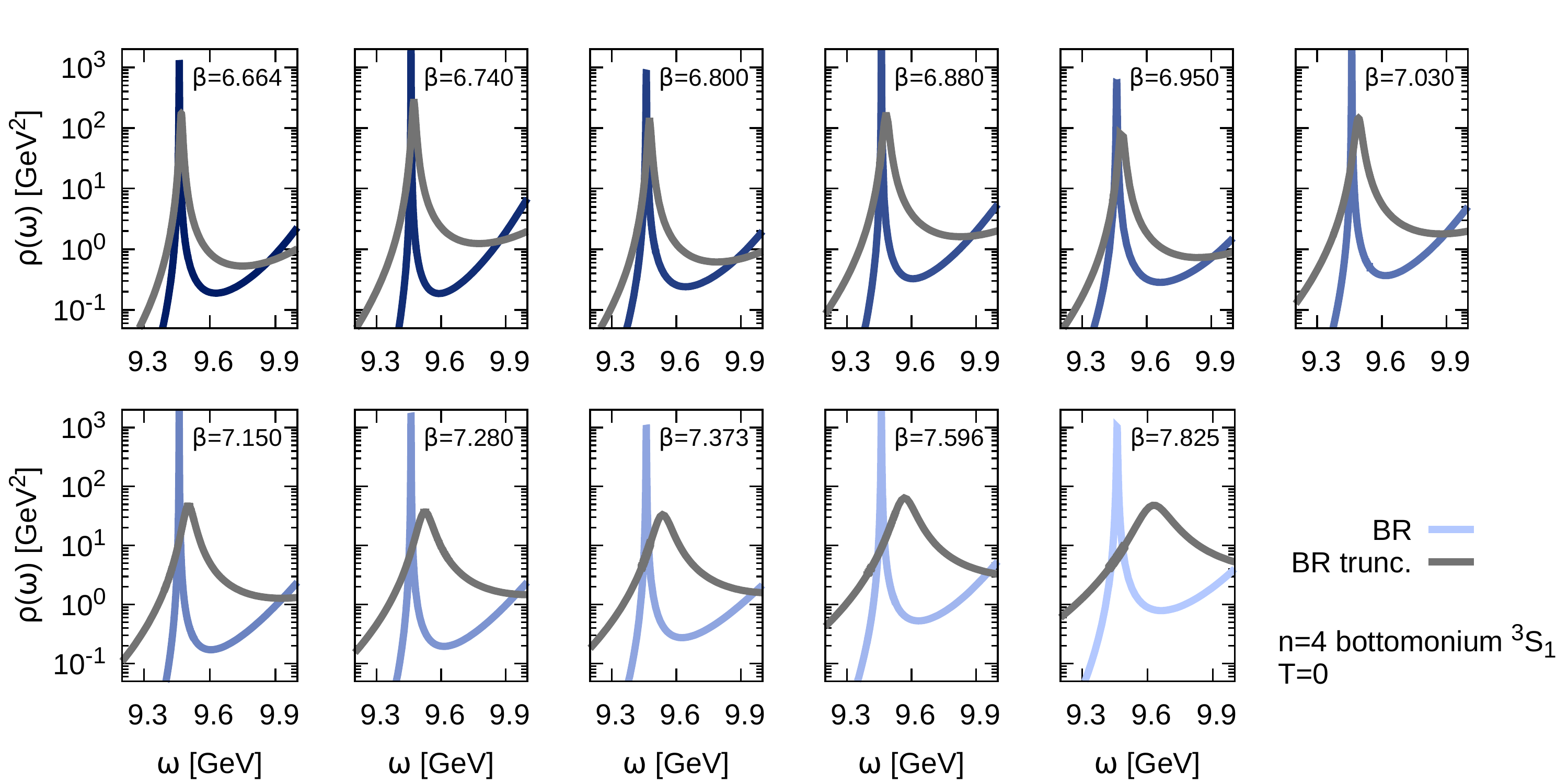}
\caption{Representative example of the influence of truncating the low-temperature simulation data to the same $\tau_{\rm max}/a_\tau=12$ Euclidean time extent available at finite temperature. The spectra shown here correspond to the $^3S_1$ channel of bottomonium with the full reconstruction given by colored solid lines, while the result after truncation is given by the gray solid curves. Clear systematics are visible: a reduction in the input data quality leads to a shift of the reconstructed ground state peak to higher frequencies, as well as to a significant broadening.}\label{Fig:T0BRSpectraCmpTrunc}
\end{figure}

On the other hand the result also tells us that the amount of physical information contained in the correlators does not come close to a detailed description of the region close to the threshold. Our simple two-box model was already capable of reproducing the correlator within its uncertainties, leaving no hope to distinguish between the physically present higher lying states and the continuum. Indeed let us consider what remains of the $T=0$ correlators after subtracting the first and second excited states contribution as shown in Fig.\ref{Fig:T0SubtrCorr}. It is evident that both at $\beta=6.664$ with $N_\tau=32$ (left) and even at $\beta=7.825$ with $N_\tau=64$ (right) only five to eight usable points (i.e. points which exhibit a convex behavior) remain to encode the wealth of structure present above the first excited state. Attempting a detailed extraction of spectral features in that region therefore is highly challenging. One possible option here would be to change to an anisotropic lattice setting, which should provide a significantly larger number of Euclidean time correlator points dominated by the UV regime.

One conclusion of these checks is that we are unable to extract the in-medium binding energy of the quarkonium states, as it is formally defined by the difference between the particle mass and the onset of the continuum. In turn also melting temperatures cannot be defined with a similar rigor, as is possible in potential based computations, where dissociation is defined at the temperature, where the in-medium thermal width equals the  binding energy. Instead in the following will resort to simply identifying at which temperature no signal of a remnant bound state can be identified in the reconstruction.

\subsection{Preparations for $T>0$ studies}
\label{sec:prepFiniteT}

Before we can turn to the investigation of in-medium spectral properties with Bayesian methods, we need to scrutinize a significant source of systematic uncertainty. In order to reach higher temperatures in standard Euclidean simulations one reduces the extent of the Euclidean temporal axis on the lattice. Here two effects combine: on the one hand the diminished Euclidean extent makes the identification even of a well defined ground state more complicated, as the regime where it dominates the falloff of the correlator is not accessible anymore and its signal becomes mixed up with that of higher lying structures. On the other hand if one stays at a fixed lattice spacing the number of data points available for the reconstruction reduces leading to a diminished resolution of the Bayesian method often accompanied with a higher default model dependence.

\begin{figure}[t]
\includegraphics[scale=0.5]{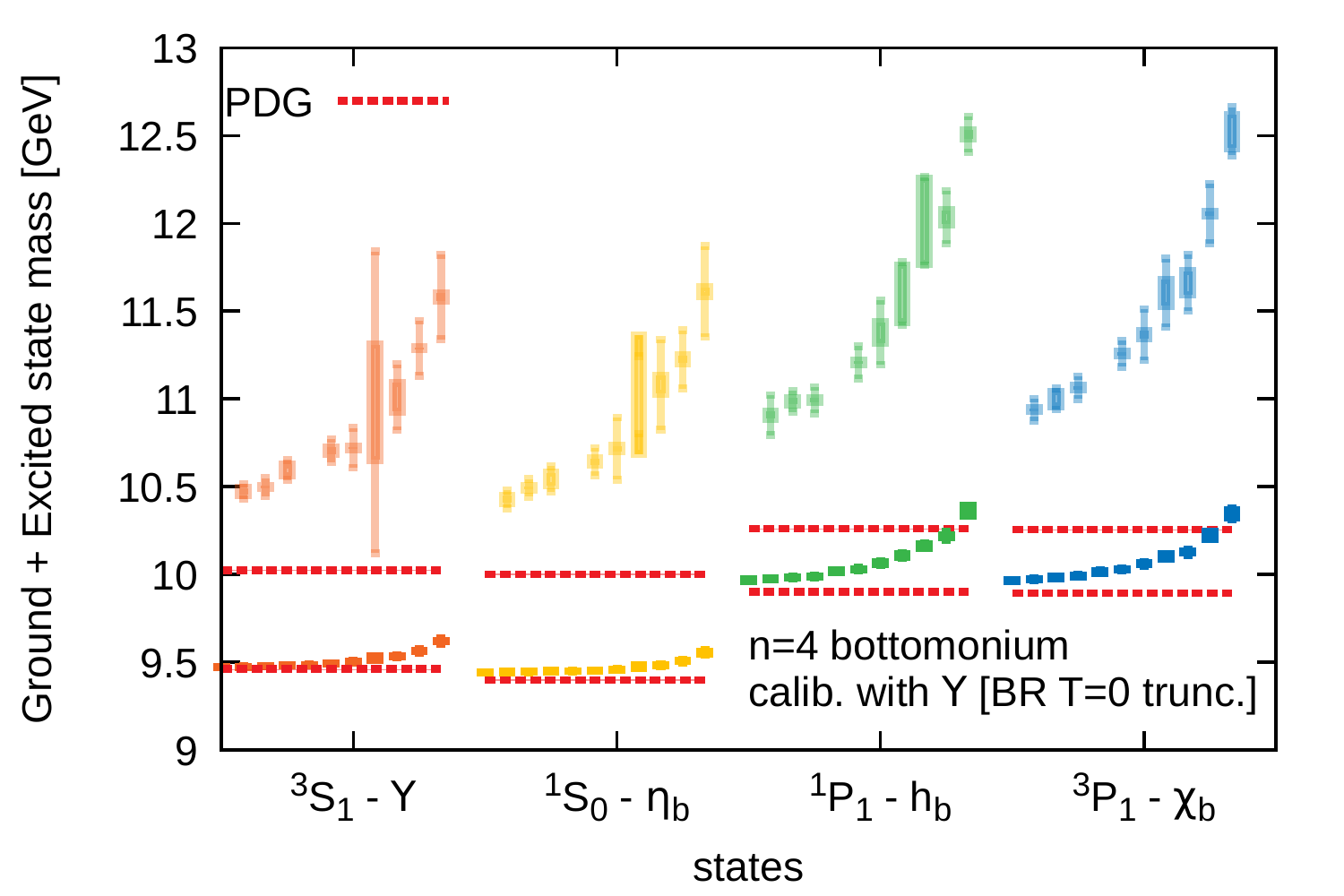}
\includegraphics[scale=0.5]{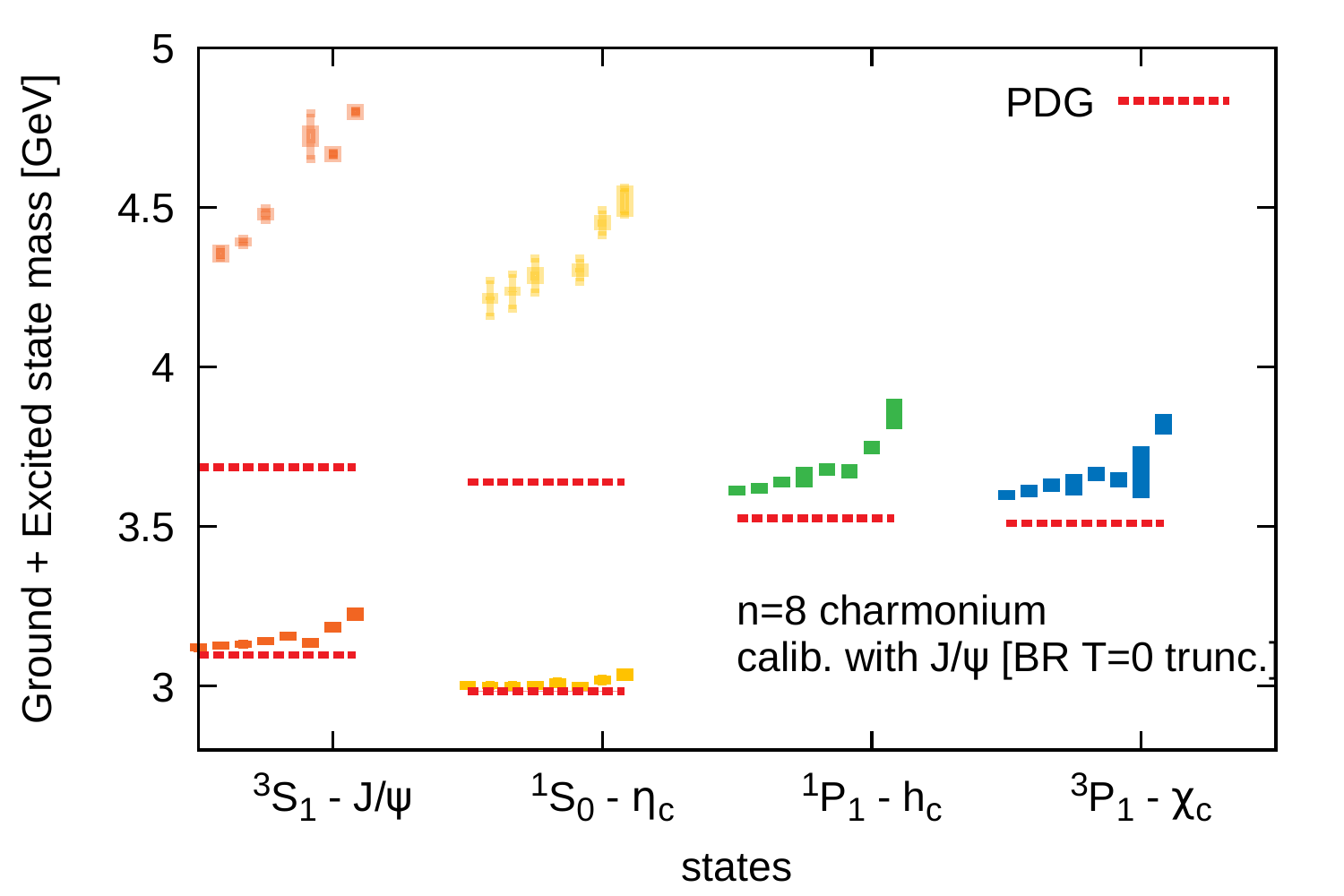}
\caption{Representative examples of the extracted values for the ground and first excited states masses from the BR method based on truncated input. The systematic shift of the obtained value is manifest in every channel, a stark reminder that such an effect needs to be taken into account when interpreting the the results of a genuine finite temperature spectral reconstruction.}\label{Fig:T0BRMaseesCmpTrunc}
\end{figure}

Our intent of using the smooth BR method is to judge, whether a genuine bound state signal is present in the correlator data at finite temperature, before extracting its properties with the standard BR method. Fig.\ref{Fig:T0SpectraCmpMethods} already told us that the smooth BR method is capable at $T=0$ to identify bound state peaks, now we have to make sure that it is able to do so also in the $T>0$ environment. 

We follow thus the following strategy. As the integral kernel in eq.\eqref{Eq:SpecConv} is temperature independent we may simply truncate the Euclidean time correlator data set to a Euclidean time smaller than the original $N_\tau$. This truncated correlator corresponds to a correlator of a simulation at higher temperature, if the same zero temperature spectral function were present at those higher temperatures. In a relativistic quark formulation where the Euclidean kernel does carry a temperature dependence, one would have to compute the {\it reconstructed correlator} instead, see e.g. \cite{Datta:2003ww,Kelly:2018hsi}.

Fig.\ref{Fig:T0BRSmoothSpectraCmpTrunc} shows what happens when the smooth BR method is fed with the truncated $T=0$ data in case of bottomonium S-wave (top two rows) and P-wave (bottom two rows). For the S-wave the results are consistent and congruent with expectations: less data means that the excited states peak structure becomes washed out and merges into the continuum part at high frequencies and at the same time the resolution for the ground state also diminishes. For the P-wave the situation is slightly different, up to $\beta=7.030$ again the loss of data points simply weakens the ground state feature reproduction but at the finest lattices it also starts to bring about ringing. This is not contradictory to how we constructed the method, which only had to make sure that there is no ringing if no actual peaks are present. On the other hand we see here that if peaks are present in the correlator and we only have very few data points available the smooth BR method will announce their presence by exhibiting ringing. 

Once we have established the presence or absence of bound state peaks, we aim at using the standard BR method to extract their properties. Hence we have to also quantify the degradation in the accuracy of such reconstruction result, when facing small Euclidean data sets at $T>0$.

In Fig.\ref{Fig:T0BRSpectraCmpTrunc} we present the outcome of the truncation test using S-wave bottomonium as an example. Note that here we use a logarithmic plot, where the changes in the reconstructed ground state peak are more clearly visible. On the coarsest lattices $\beta=6.664$ where NRQCD works best, the truncation from $\tau_{\rm max}/a_\tau=32$ to $\tau_{\rm max}/a_\tau=12$ only induces a change in the peak position by around $9$MeV. The finer the lattice the more pronounced the effect of truncation becomes, leading to a shift of around $170$MeV at $\beta=7.825$. This is because the first twelve data points cover
a smaller and smaller Euclidean time extent as the lattice spacing is reduced. At the same time also the width of the peak starts to broaden artificially by a significant amount. (The qualitatively very similar results for the other quarkonium channels can be found in App.\ref{app:truncTstT0}).

The effect of this reduction of both Euclidean time extent and number of data points is summarized in Fig.\ref{Fig:T0BRMaseesCmpTrunc}. Compared to the reconstruction based on the full data set in Fig.\ref{Fig:T0BRMassesCmp} we find that each channel shows a characteristic upward trend related to the lattice spacing. This shift characterizes the degradation of the reconstruction method while the underlying spectrum remains the same.

It is this systematic shift, which we will have to take into account carefully, when investigating the spectral properties at finite temperature in the next section, where now in addition also the spectrum itself changes. I.e. the baseline to which we shall compare in-medium masses to is not the results from the $\tau_{\rm max}/a_\tau > 32$ reconstruction but that using the same $\tau_{\rm max}/a_\tau=12$ extent. 

\section{Physics results at $T>0$}
 \label{sec:physres} 

In this section we present and critically discuss the main results of this study, the in-medium properties of both bottomonium and charmonium from lattice NRQCD. We start out first with an investigation of the overall  finite temperature effects based on the raw simulated correlators themselves, before carrying out spectral reconstructions to shed light on the modification of the ground state particles.

Compared to our previous study the results here have been improved in three main aspects. On the one hand we have extended the bottomonium computations to higher temperatures and increased the statistics of the simulations by up to a factor four. In addition through the simulation of charm quark degrees of freedom we now possess a second set of states with different binding properties, which help us to uncover the systematics of in-medium modification.

\subsection{Sequential in-medium modification}
\label{sec:secmod}

The starting point are the $T>0$ NRQCD correlators of the quarkonium $^3S_1$ and $^3P_1$ channels, which we focus on in the subsequent qualitative and quantitative analysis. The main difference to $T=0$ is the significantly reduced Euclidean time extent, leaving even at temperatures below $T_c$ only a small range, where one might identify the presence of a single exponential falloff by eye. In total we consider seventeen temperatures $T=[140-407]$MeV for bottomonium and, as laid out in sec.\ref{Num:LatNRQCD} and fourteen temperatures for charmonium $T=[140-251]$MeV.

To obtain a first glimpse of the strength of the overall in-medium modification we turn next to the ratio of the finite temperature correlator to its $T=0$ counterpart. This ratio is physically meaningful, since in NRQCD the integral Kernel of the spectral representation is temperature independent. In a relativistic formulation we would have to compute the ratio with the corresponding {\it reconstructed} correlator instead.

\begin{figure}[t]
\includegraphics[scale=0.56]{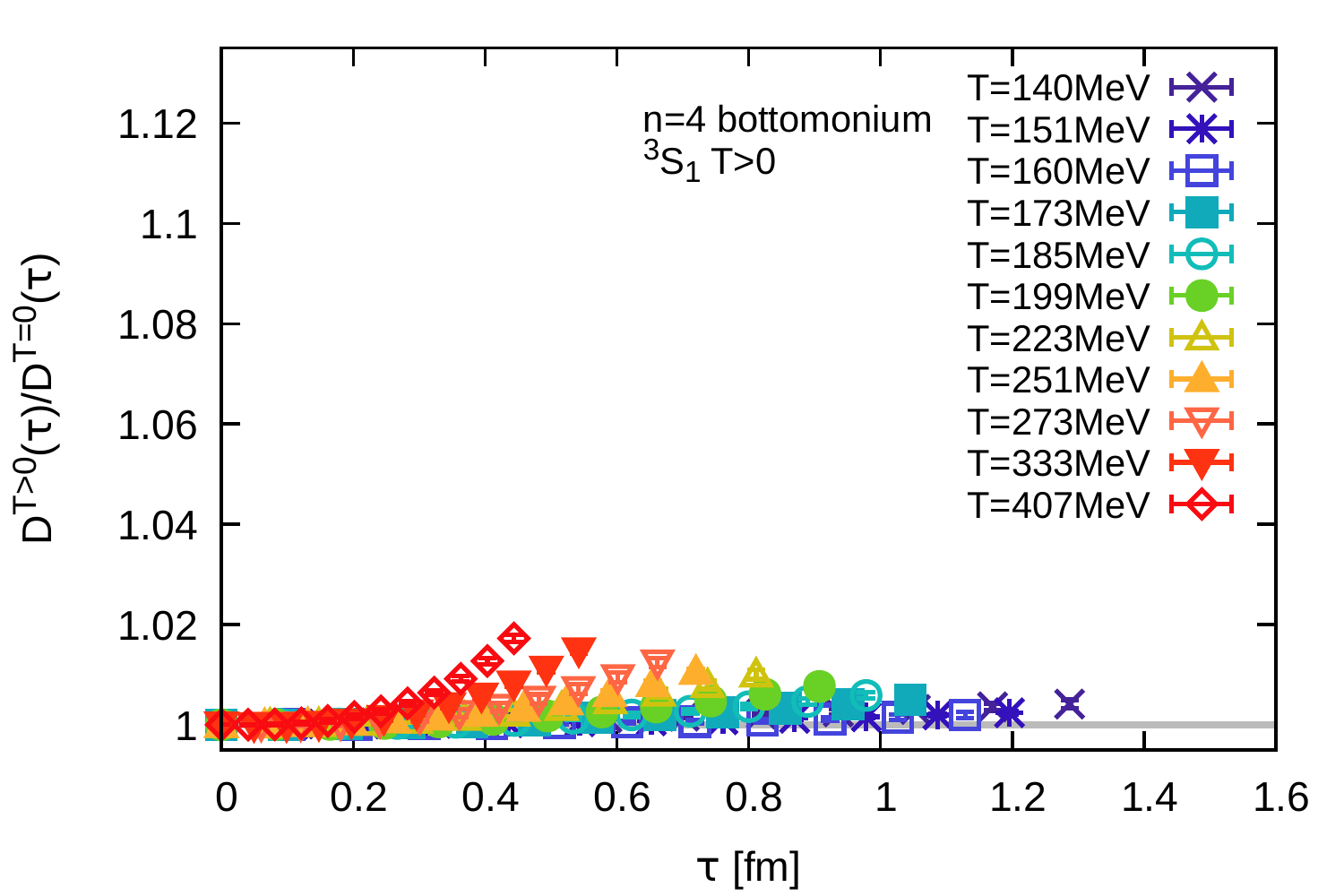}
\includegraphics[scale=0.56]{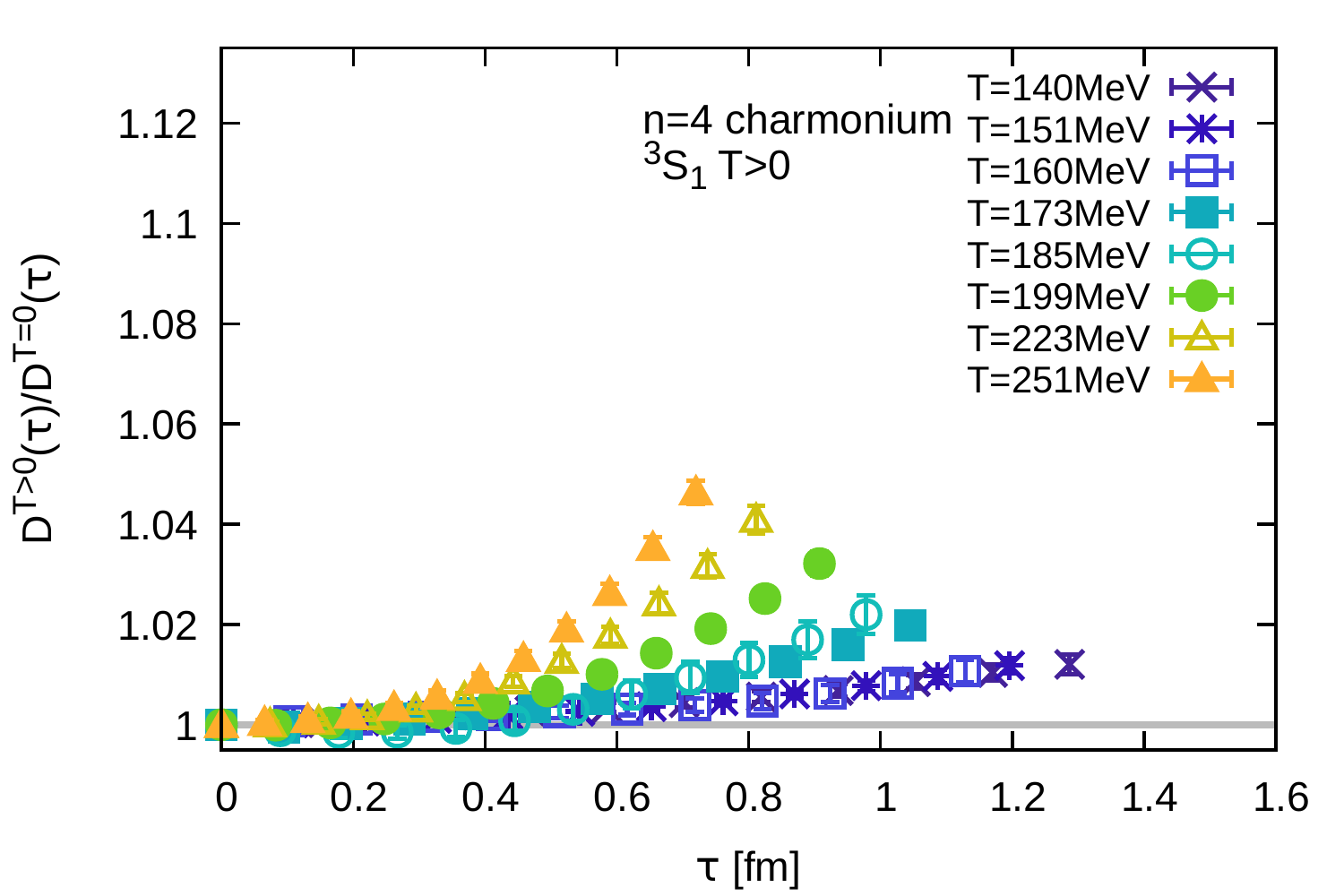}
\includegraphics[scale=0.56]{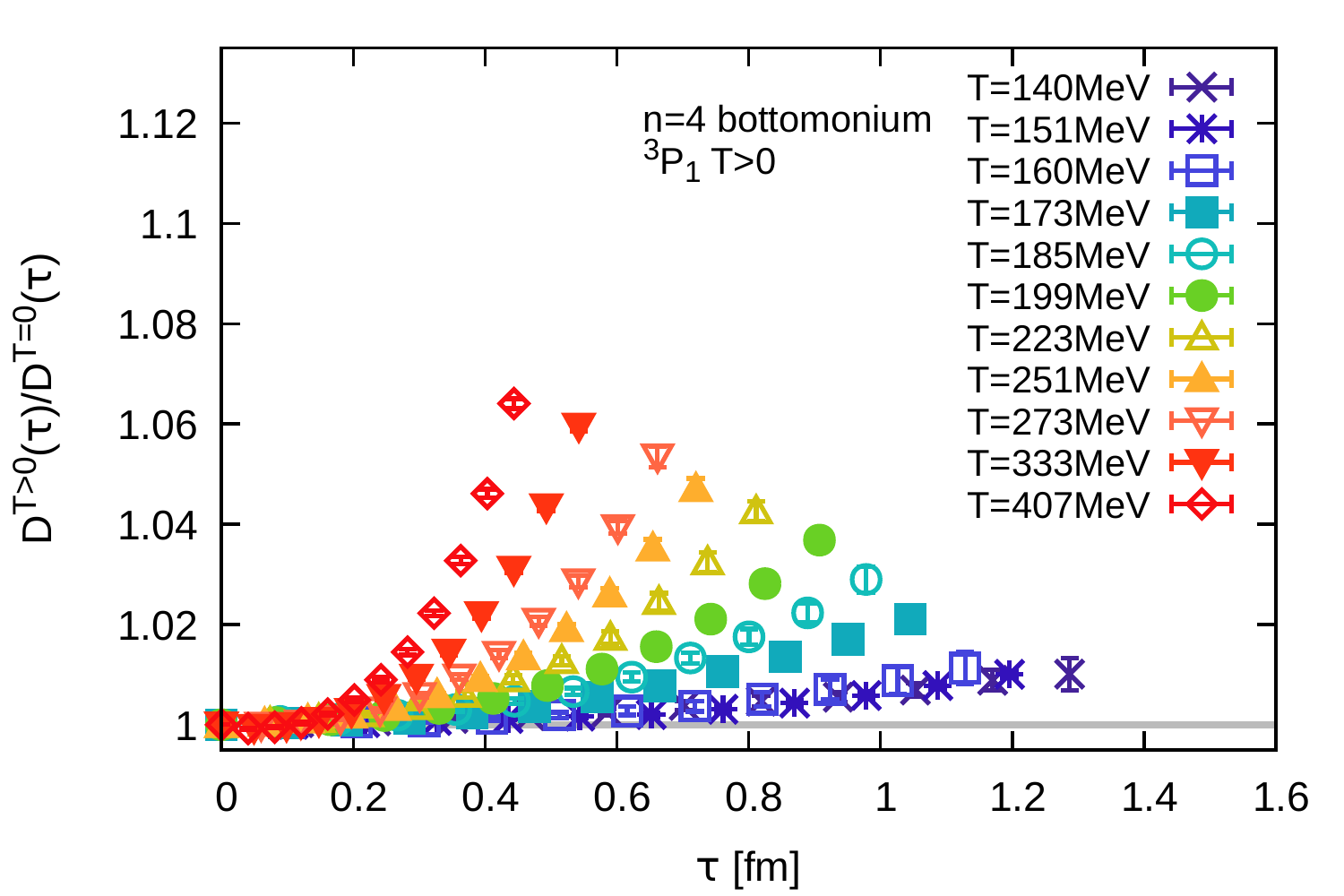}
\includegraphics[scale=0.56]{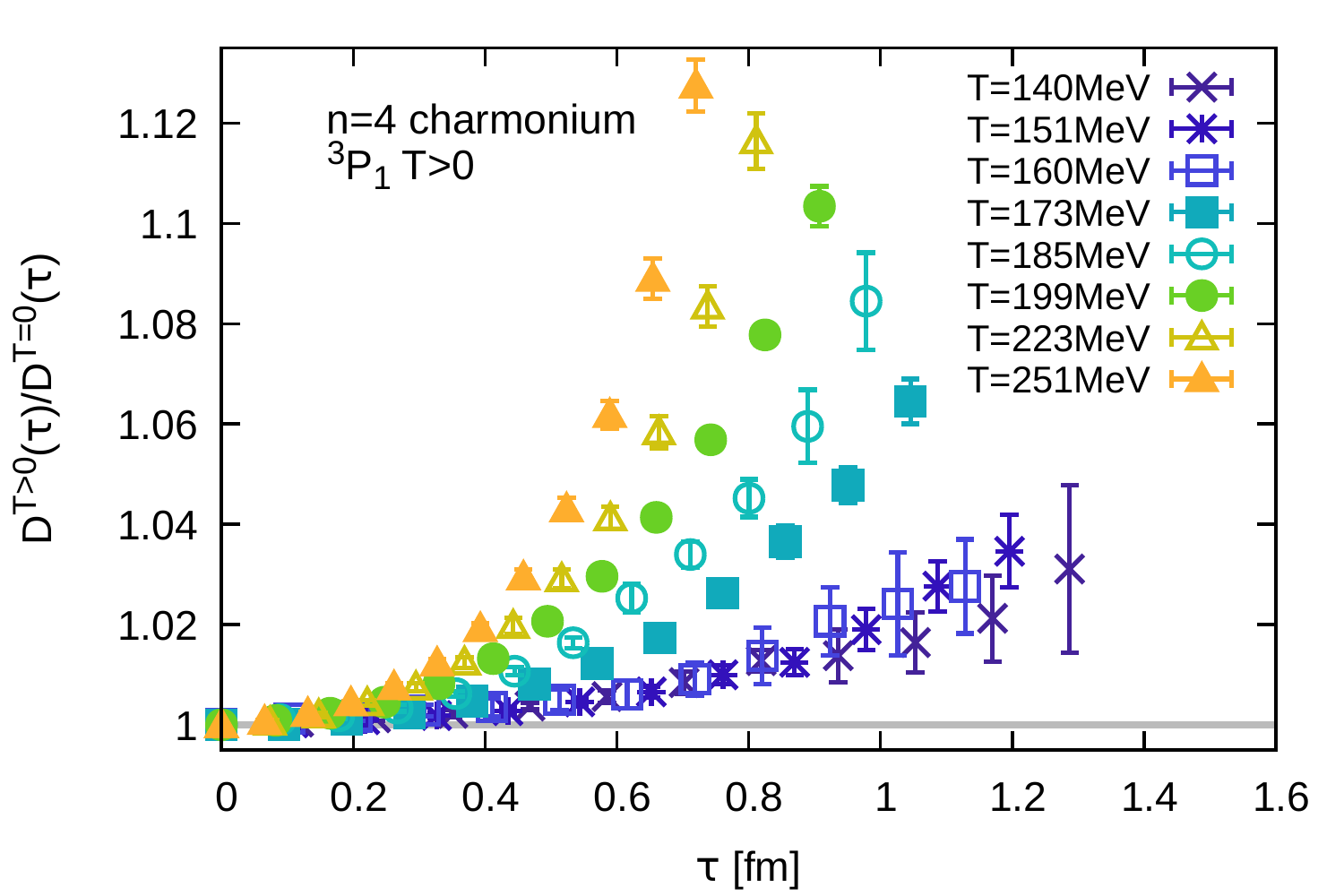}
\caption{Ratio of the raw NRQCD correlators at finite temperature with their corresponding $T=0$ counterparts. We show bottomonium results on the right and charmonium on the left. The top row features the ratios for the $^3S_1$ channel while in the bottom row the $^3P_1$ results can be found. Note that the strength of the upward bend, when compared at the same temperature shows a clear ordering according to the vacuum binding energy of the ground state in that channel $E^{\rm \Upsilon(1S)}_{\rm b}\simeq 1.1{\rm GeV} >  E^{\rm \chi_b(1P)}_{\rm b} \approx E^{\rm J/\psi(1S)}_{\rm b} \simeq 640{\rm MeV} > E^{\rm \chi_c(1P)}_{\rm b} \simeq 200{\rm MeV}$.}\label{Fig:FiniteTCorrelatorRatios}
\end{figure}

Note that the Euclidean correlation functions computed in NRQCD represent a convolution of all spectral structures according to eq.\eqref{Eq:SpecConv}. Due to the exponential damping of the involved integral Kernel even significant changes in the underlying spectral features only appear as small changes in the correlator itself. Furthermore, we have to keep in mind that in-medium modifications in different parts of the spectrum may counterbalance each other after the convolution. With this caveat in mind we present in Fig.\ref{Fig:FiniteTCorrelatorRatios} the raw correlator ratios from lattice NRQCD. The results are ordered as in the previous figures with bottomonium on the left and charmonium on the right. The top row contains the S-wave, the bottom row the P-wave data.

Fig.\ref{Fig:FiniteTCorrelatorRatios} contains valuable insight and corroborates the findings of our previous study on sequential in-medium modification with respect to the vacuum binding energy, now with improved precision. We find, at first sight, that in the hadronic phase the ratios all lie close to each other, showing deviation from
unity by $(0.4-3)\%$ depending on the binding energy of the
corresponding vacuum bound state. These deviations are
smallest for $\Upsilon$ and largest for $\chi_c$.

On the other hand, once temperature rises further, significant temperature dependence is observed, which bends the ratio upwards. The strength of this upward trend is closely correlated with the binding energy of the vacuum ground state, exhibiting a clearly sequential pattern.

For the $\Upsilon$ channel with a vacuum binding energy of $1.1$GeV the maximum deviation from unity at $T=407$MeV is around $1.75\%$, while the $\chi_b$ channel with a $T=0$ binding energy of around $640$MeV already shows a $6.5\%$ change. The specific mass hierarchy of bottom and charm quarks provides us in the charmonium sector with a particle of almost the same binding energy as $\chi_b$, which is $J/\Psi$. Since we only consider charmonium up to $T=251$MeV let us compare the correlator ratios at this temperature. We find that indeed both $J/\Psi$ and $\chi_b$ show a very similar ratio of $1.05$ at that point. Supporting the sequential modification further one finds that the $\chi_c$ state with around $200$MeV vacuum binding energy already features a deviation from unity above $12\%$.

\begin{figure}[t]
\centering
\includegraphics[scale=0.56]{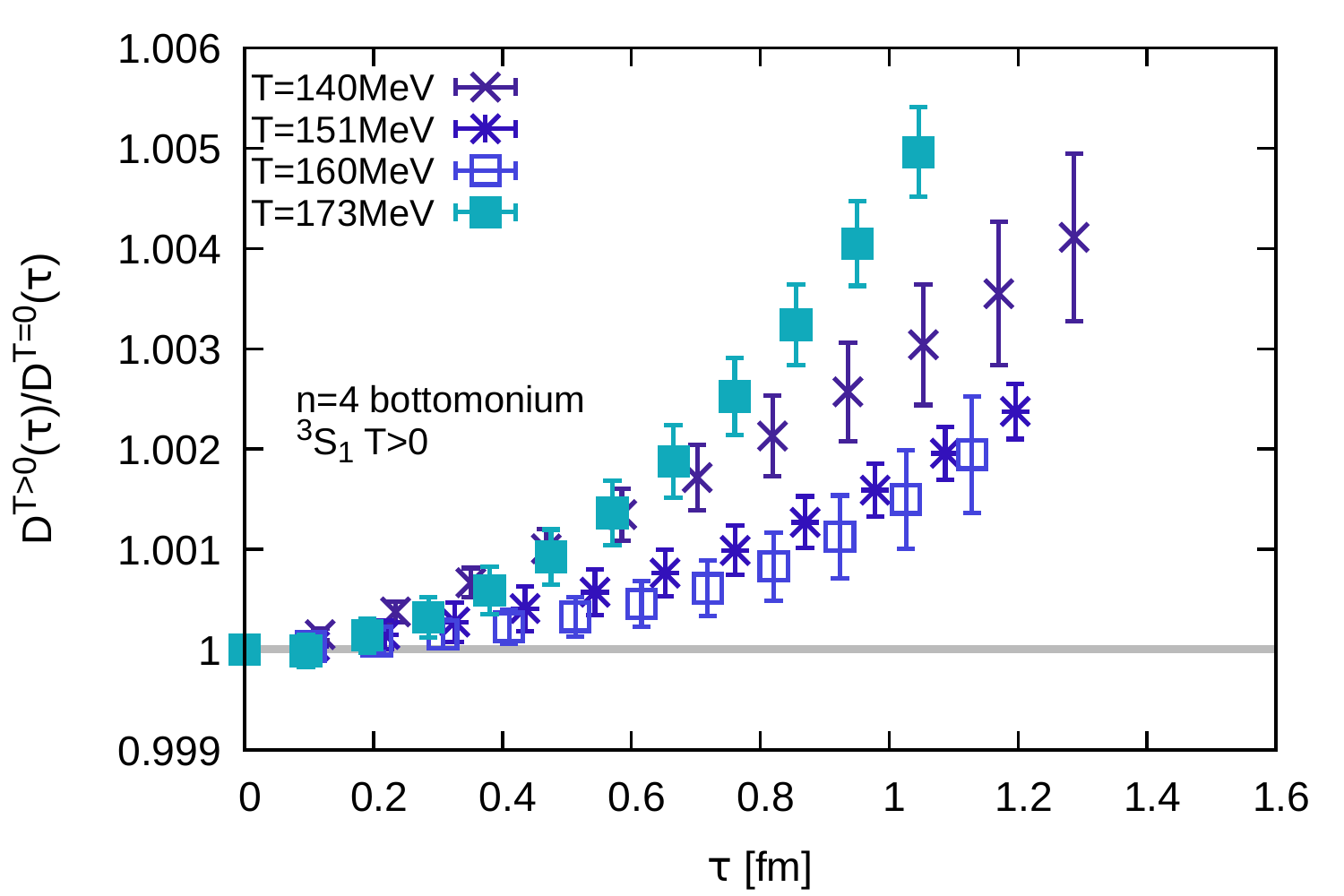}
\caption{Low temperature region of the bottomonium $^3S_1$ channel. Note the non-monotonous behavior around $T_c$ and the fact that already in the hadronic phase the ratio deviates from unity.}\label{Fig:FiniteTCorrelatorRatiosLowT}
\end{figure}

Let us have a closer look at the lower temperatures. In particular  the bottomonium S-wave ratio shows a unique behavior in that the data points at the lowest temperature actually lie above those around $T_c$, as shown in Fig.\ref{Fig:FiniteTCorrelatorRatiosLowT}. While we found first indications for such behavior in our previous study, now with improved statistics it can be confirmed by more than two sigma. There exists a non-monotonic behavior of the ratio as one approaches and passes through the crossover transition. No other channel shows a similar behavior for the lowest three temperatures.

Since in the correlator ratios we only observe the aggregate in-medium modification of all states in a particular channel, we attempt to reverse engineer and interpret the different modification patterns by use of a potential based computation of the in-medium spectra. To this end we utilize the spectral functions computed in a tree-level pNRQCD approach, based on the complex potential between two static quarks obtained recently from lattice QCD \cite{Burnier:2015tda}. The spectra in that study show a clear pattern of sequential in-medium modification with the highest lying peaks being affected first. The changes in the in-medium potential, i.e. a monotonous increase of the screening of its real-part and a concurrent growth of its imaginary part leads to the following changes. The continuum moves to lower and lower values, as ${\rm Re}[V]$ asymptotes at smaller distances. On its way the continuum first pushes the individual in-medium bound state peaks to lower masses before they melt. At the same time the peaks broaden significantly before they dissolve into the continuum structure.

\begin{figure}[t]
\includegraphics[scale=0.5]{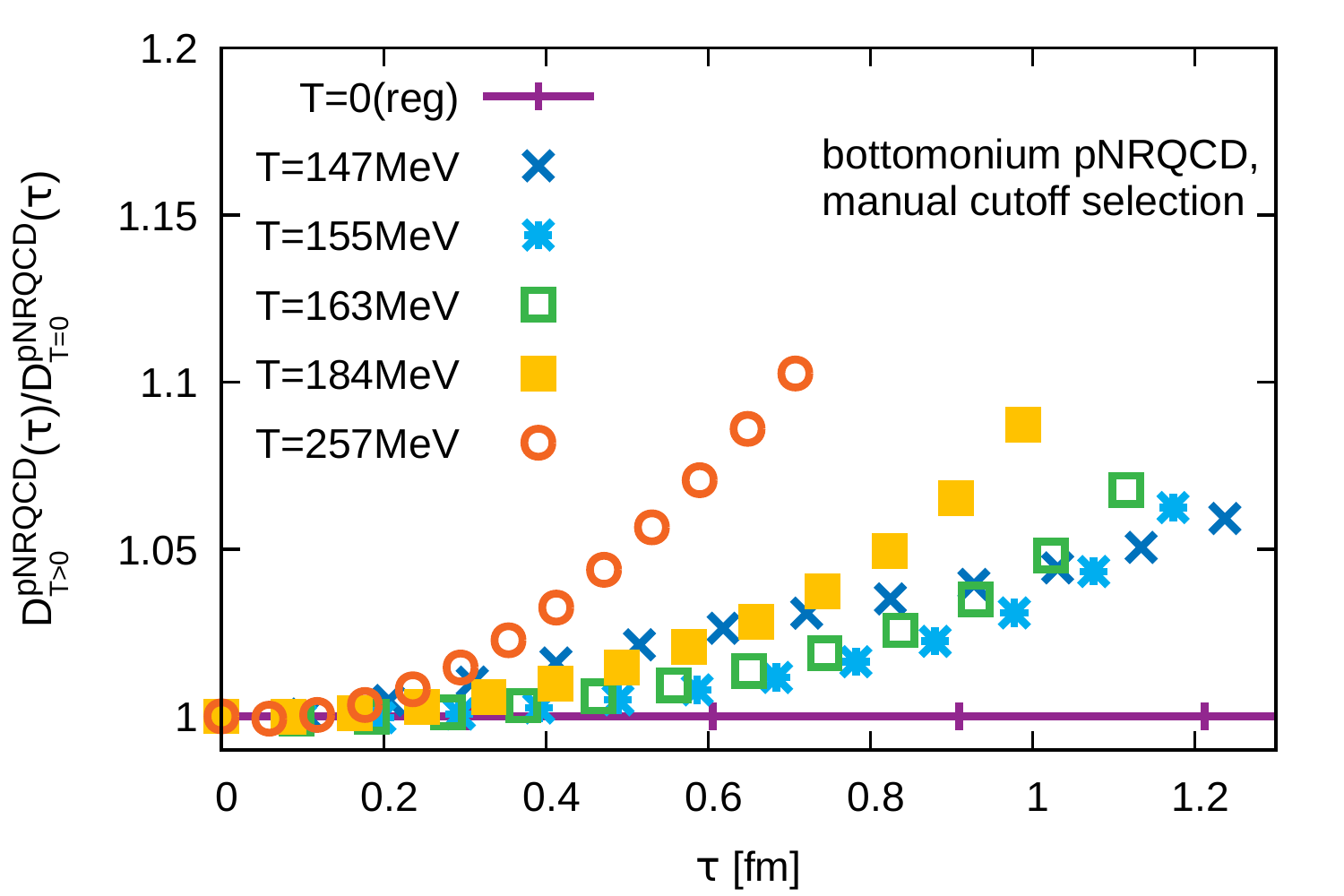}
\includegraphics[scale=0.5]{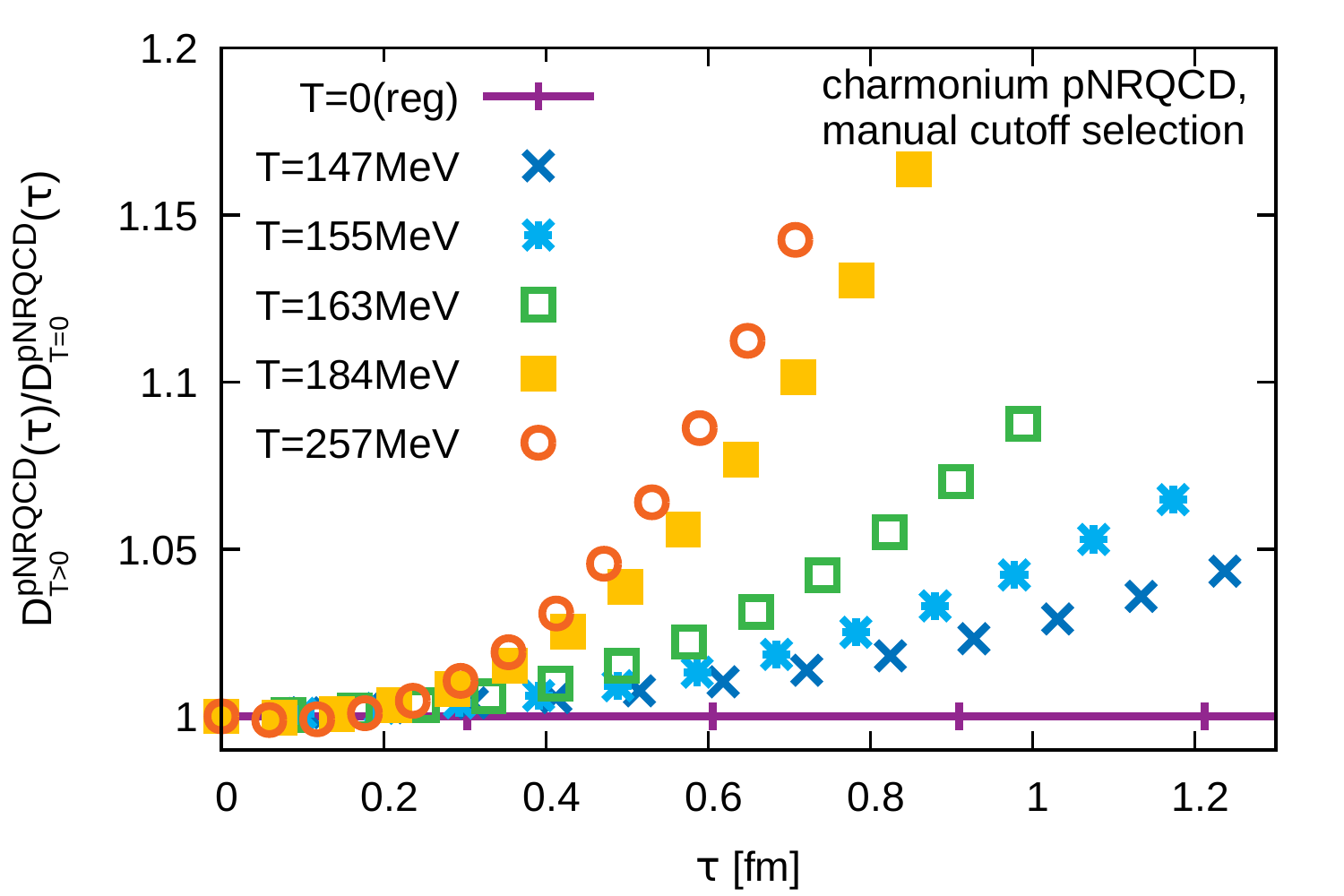}
\caption{Ratio of the in-medium to the zero temperature correlator from a tree-level pNRQCD computation based on a recent determination of the complex heavy quark potential from the lattice. These ratios are taken as a qualitative guide for interpreting the lattice NRQCD results. Note that for bottomonium (left) the early melting of the higher excited states induces a non-monotonicity in the ratio similar to that observed in the NRQCD correlator and which is absent for charmonium (right).}\label{Fig:FiniteTCorrelatorRatioCmpModel}
\end{figure}

Several adjustments are needed to compute from a pNRQCD spectral function the corresponding Euclidean correlator, which may be compared to our raw NRQCD data. On the one hand the ground state peak needs to be shifted to the position it takes on in lattice NRQCD before calibration. On the other hand an IR and UV cutoff need to be selected by hand, in order to take into account deviations of the lattice spectrum from its continuum form. While this may change the specific value of the ratio, we find that the ordering between temperatures and the general trends remain stable even under a variation of these settings. In the following we use the pNRQCD correlator ratios only as qualitative guide for interpretation. Two examples of such ratios for the bottomonium and charmonium S-wave channel are shown in Fig.\ref{Fig:FiniteTCorrelatorRatioCmpModel}

Several characteristics of the in-medium modification can be understood from this exercise. We find that the main upward bend above $T_c$ is compatible with the ground state peak starting to move to lower masses. On the other hand the non-monotonicity around $T_c$ seems to be induced by changes in the excited states, where in the case of bottomonium there are at least three, while charmonium only features one. That is, the early melting of $\Upsilon(3S)$ seems to be counterbalanced by the changes in $\Upsilon(2S)$, since at lower temperatures $\Upsilon(1S)$ is not affected by the medium in a significant fashion.

In our previous study we also saw that at small Euclidean time there were signs of an undershoot of the ratio below unity. This region is dominated by effects in the UV and the less pronounced nature of this dip below one, may be related to our choice of a different Lepage parameter.

We see that the correlator ratios alone provide us with valuable insight into the systematics of quarkonium in-medium modification, in particular when coupled with intuition derived from potential based descriptions. The next step is to investigate the reconstructed spectral functions directly in order to extract the in-medium modification of individual states from the NRQCD simulations.

\subsection{In-medium spectral functions}
\label{sec:FiniteTspecfunc}

For the spectral reconstruction of in-medium quarkonium states we deploy again three different methods, the standard BR method, its smooth variant and the Maximum Entropy Method. We keep the same settings for the reconstructions as in the $T=0$ case and simply use the finite temperature correlators as input. 

The calibration of the absolute energy scale has been performed at zero temperature for a subset of the lattice spacings considered at $T>0$. As mentioned in sec.\ref{sec:nrqcdcalib}, we use a linear fit for bottomonium and a spline interpolation for charmonium to evaluate the mass shift $C(\beta)$ for all other lattice spacings.

\begin{figure}[t]
\includegraphics[scale=0.5]{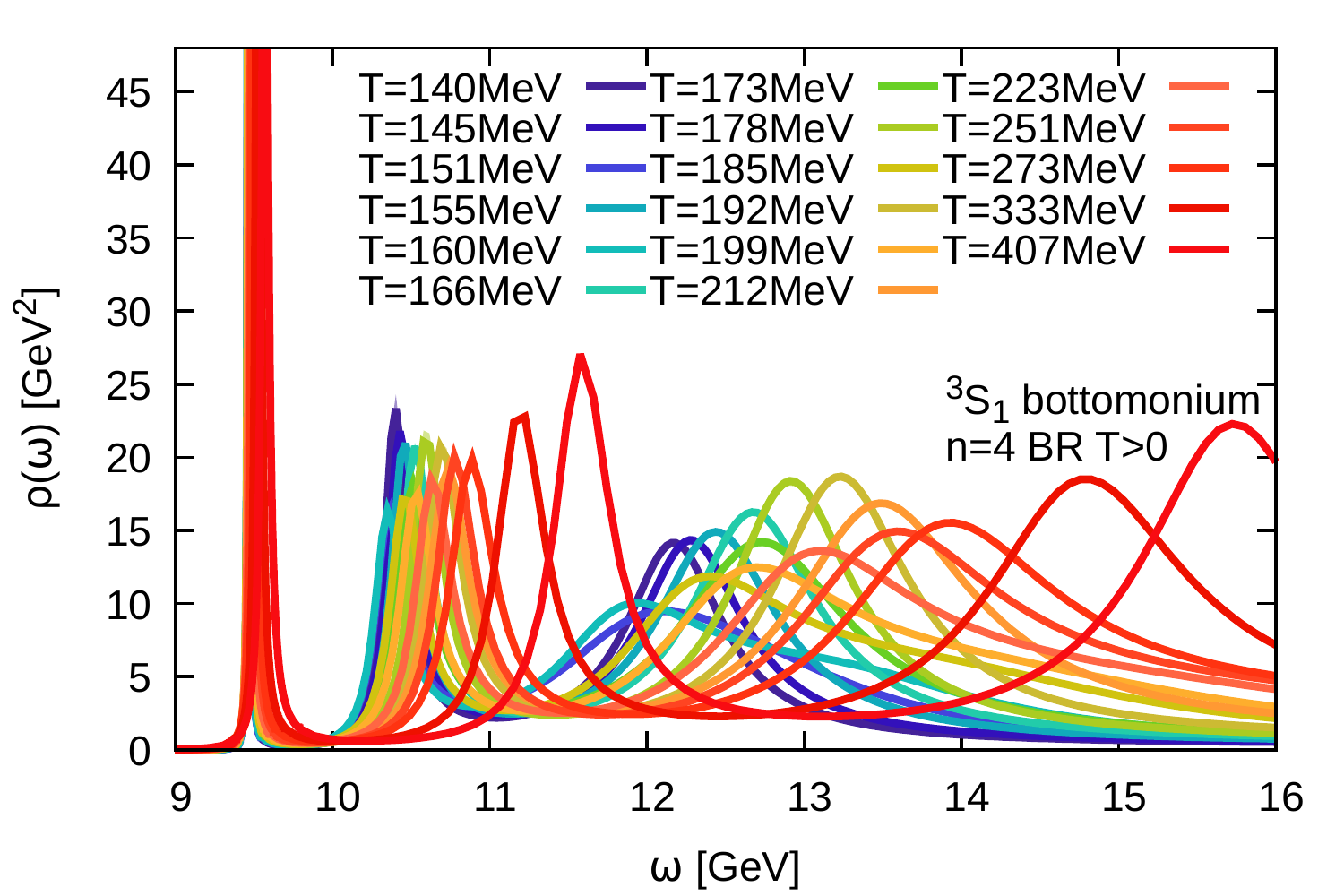}
\includegraphics[scale=0.5]{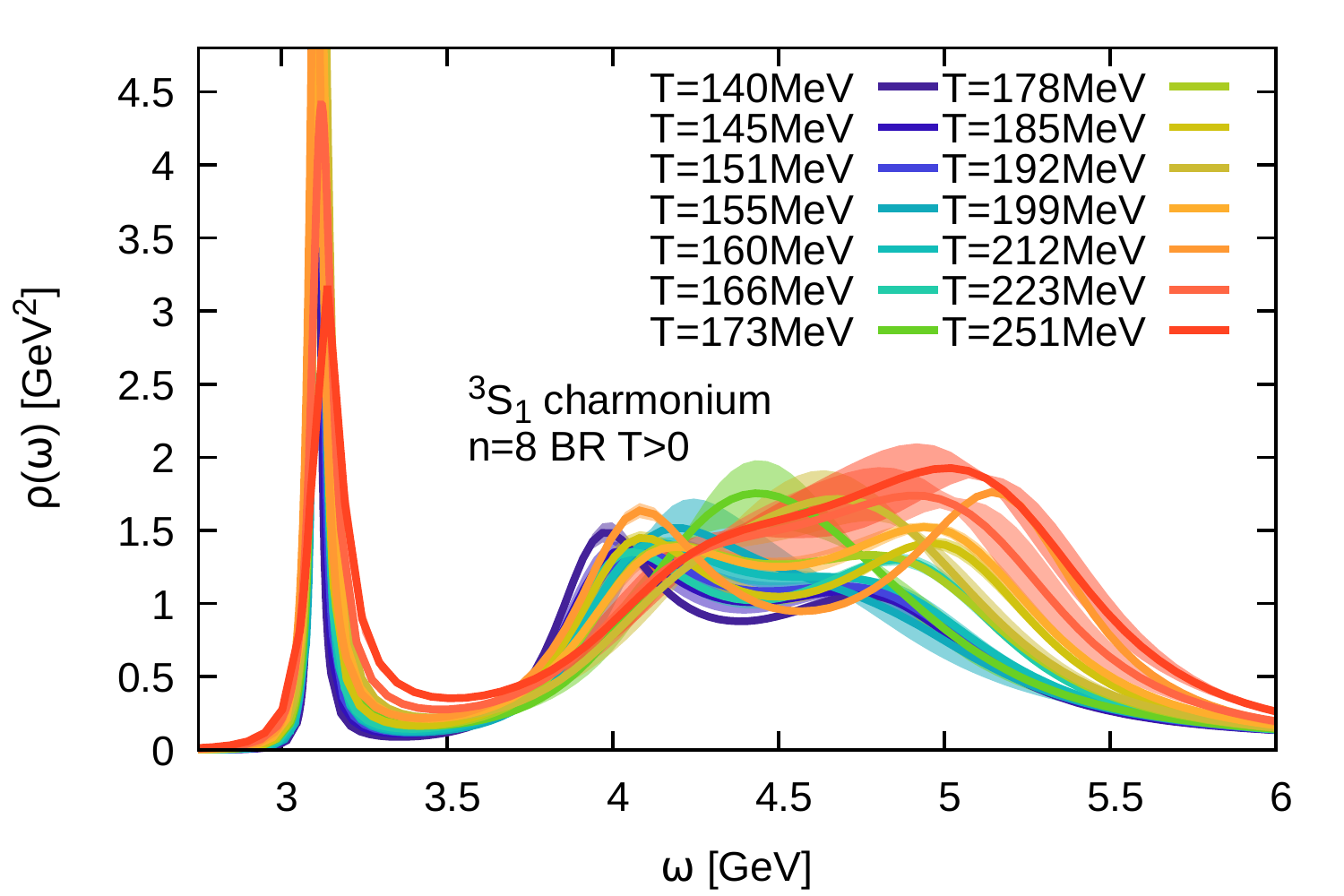}
\includegraphics[scale=0.5]{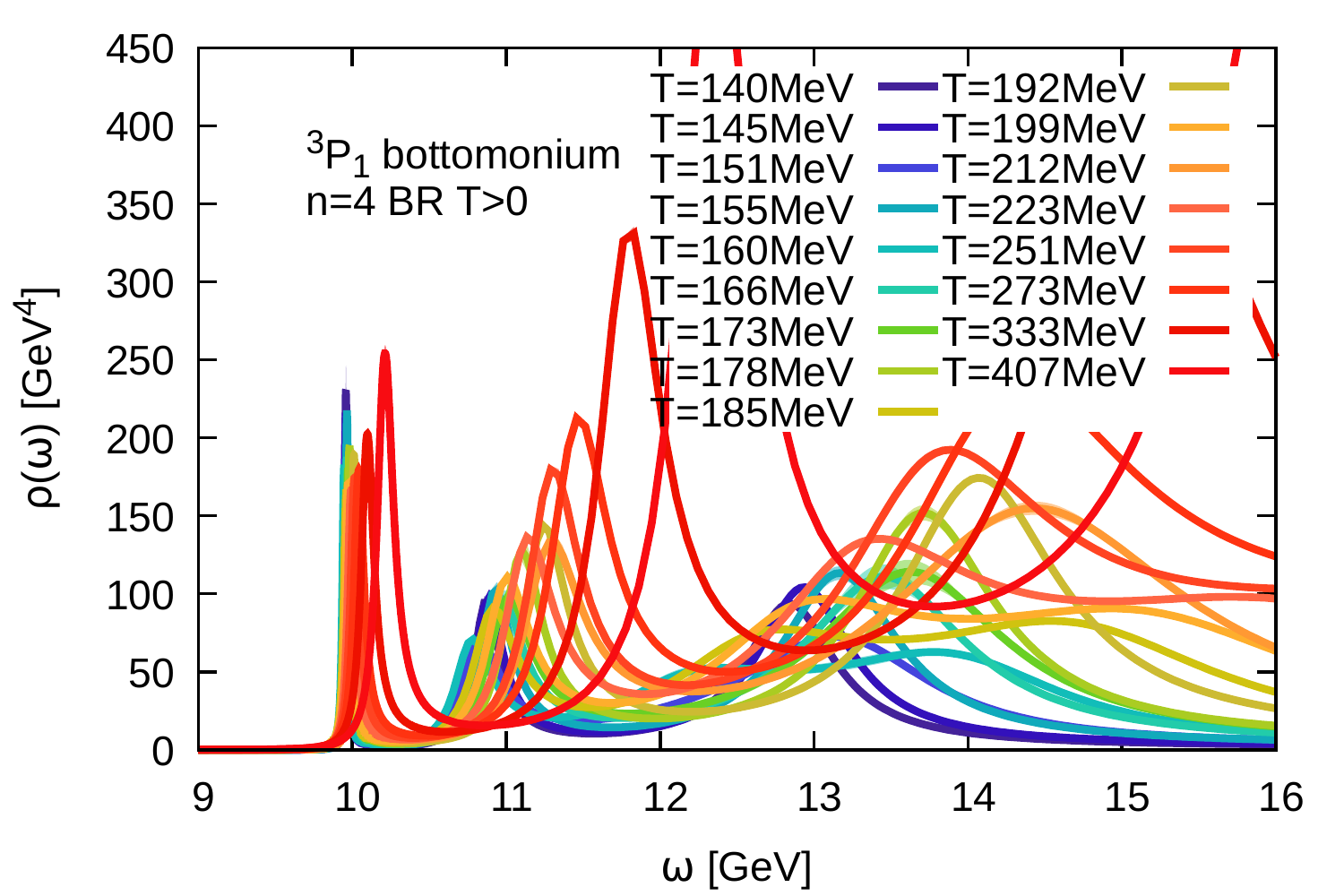}
\includegraphics[scale=0.5]{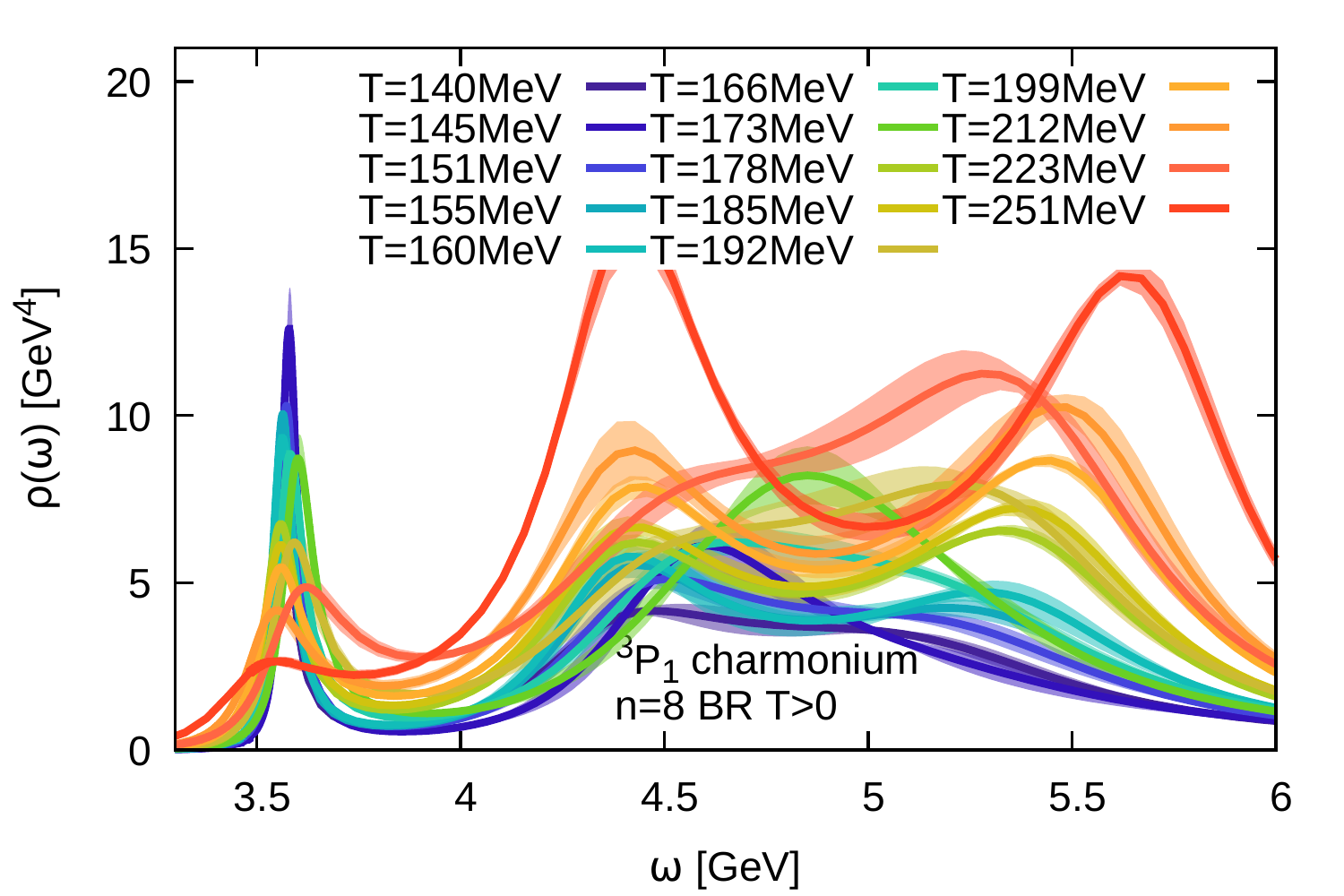}
\caption{Finite temperature bottomonium (left) and charmonium (right) spectral functions reconstructed using the BR method. The top row contains the spectral functions in the $^3S_1$ channel, while the bottom row features the $^3P_1$ channel.}\label{Fig:FiniteTSpectra}
\end{figure}

\begin{figure}[t]
\includegraphics[scale=0.5]{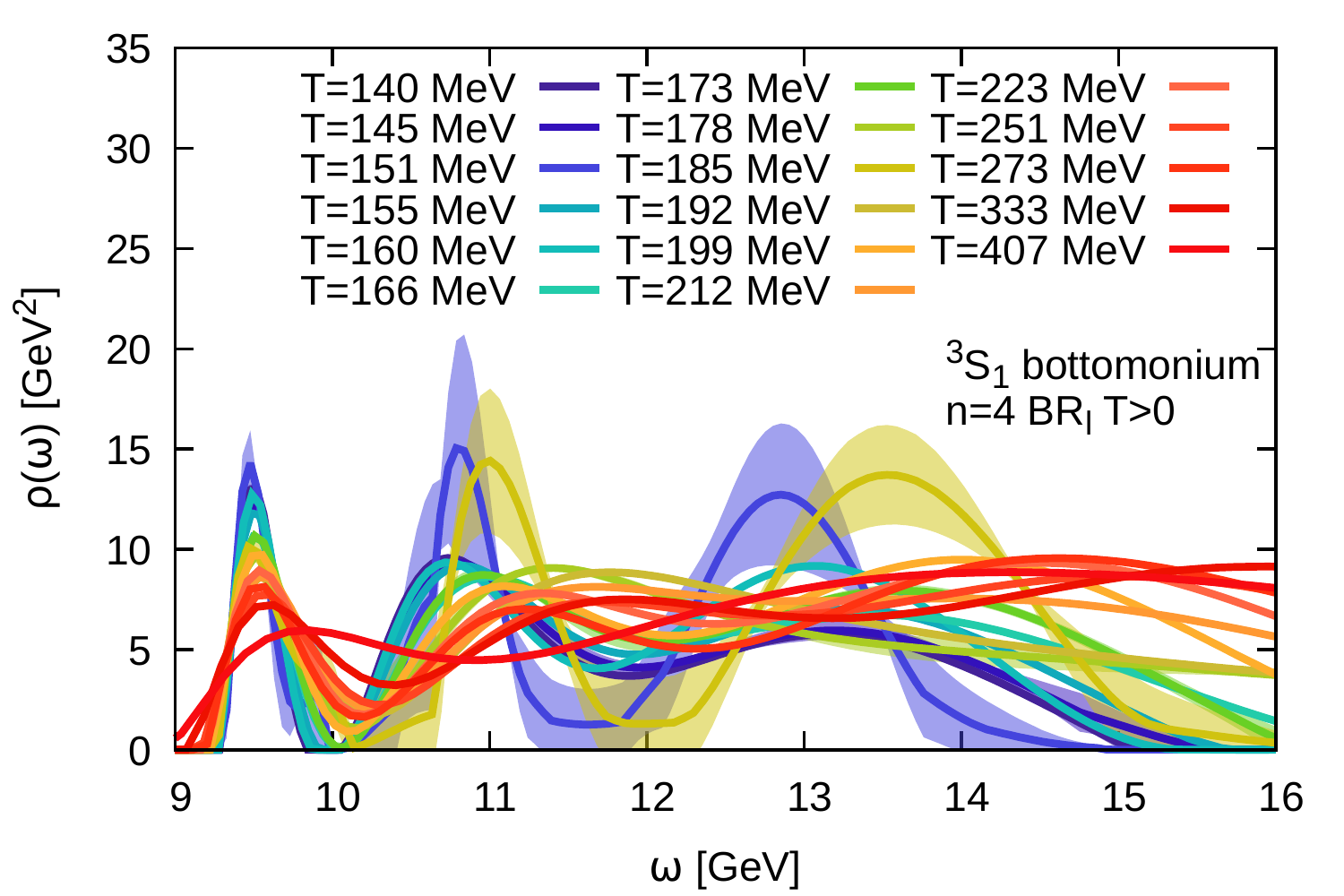}
\includegraphics[scale=0.5]{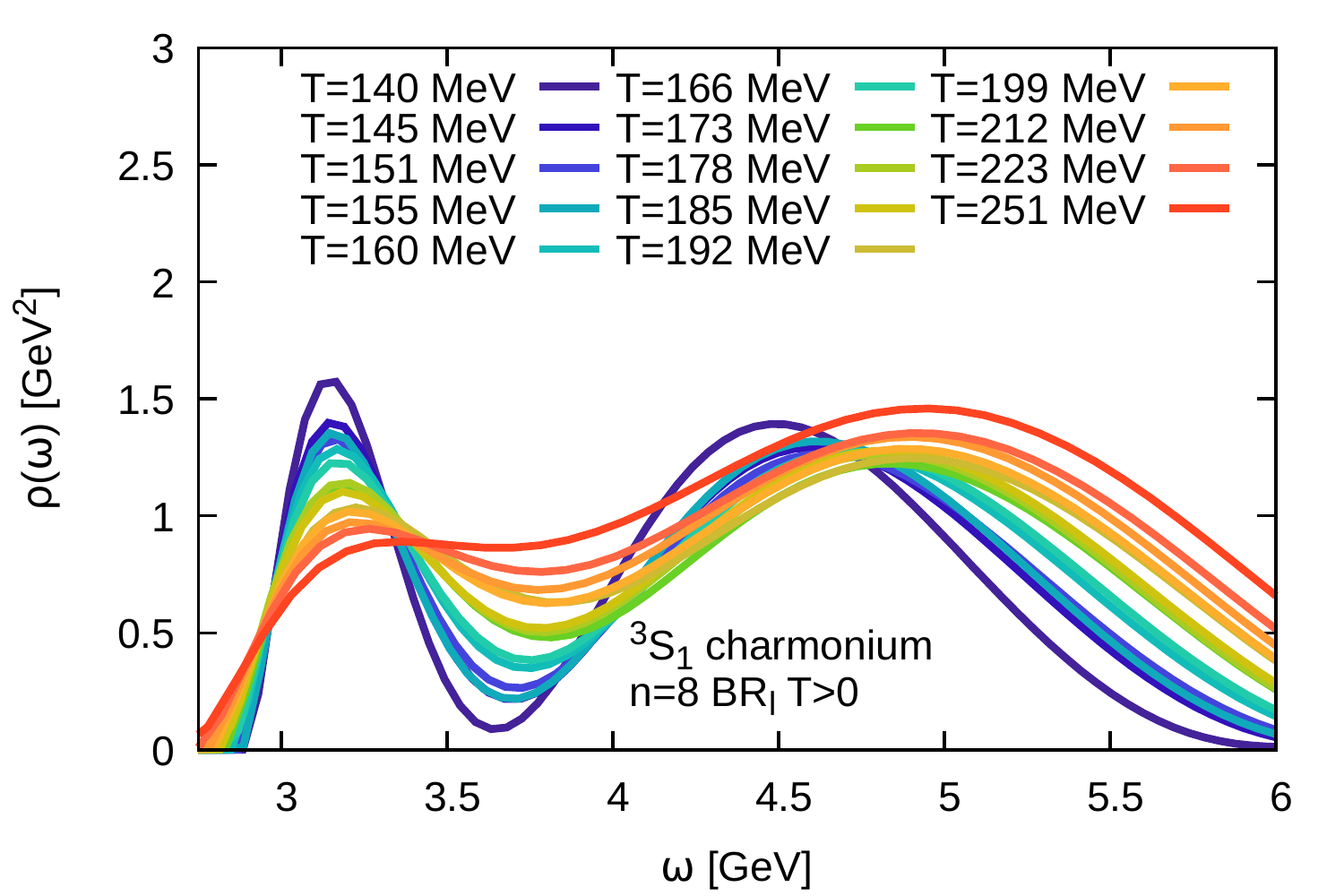}
\includegraphics[scale=0.5]{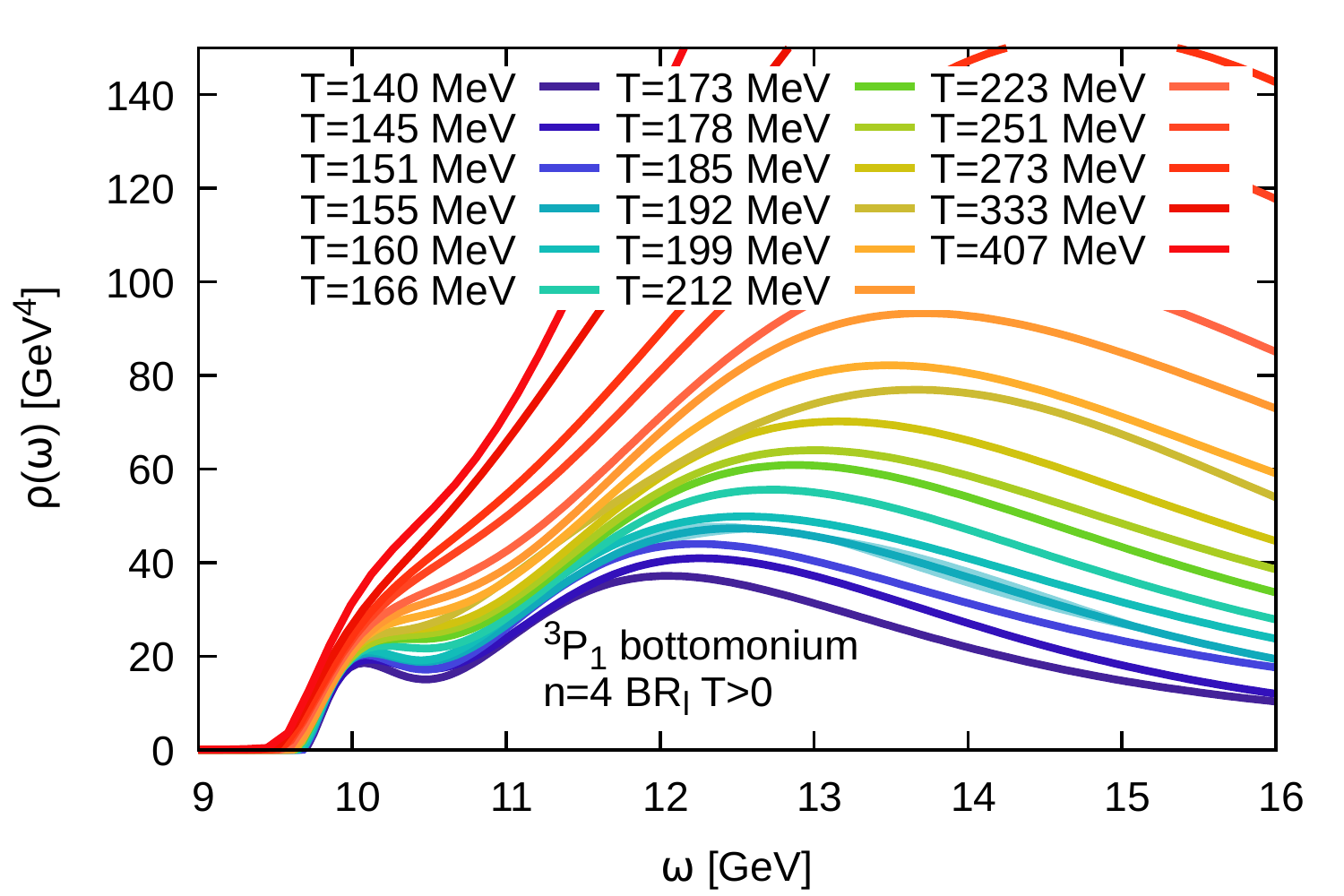}
\includegraphics[scale=0.5]{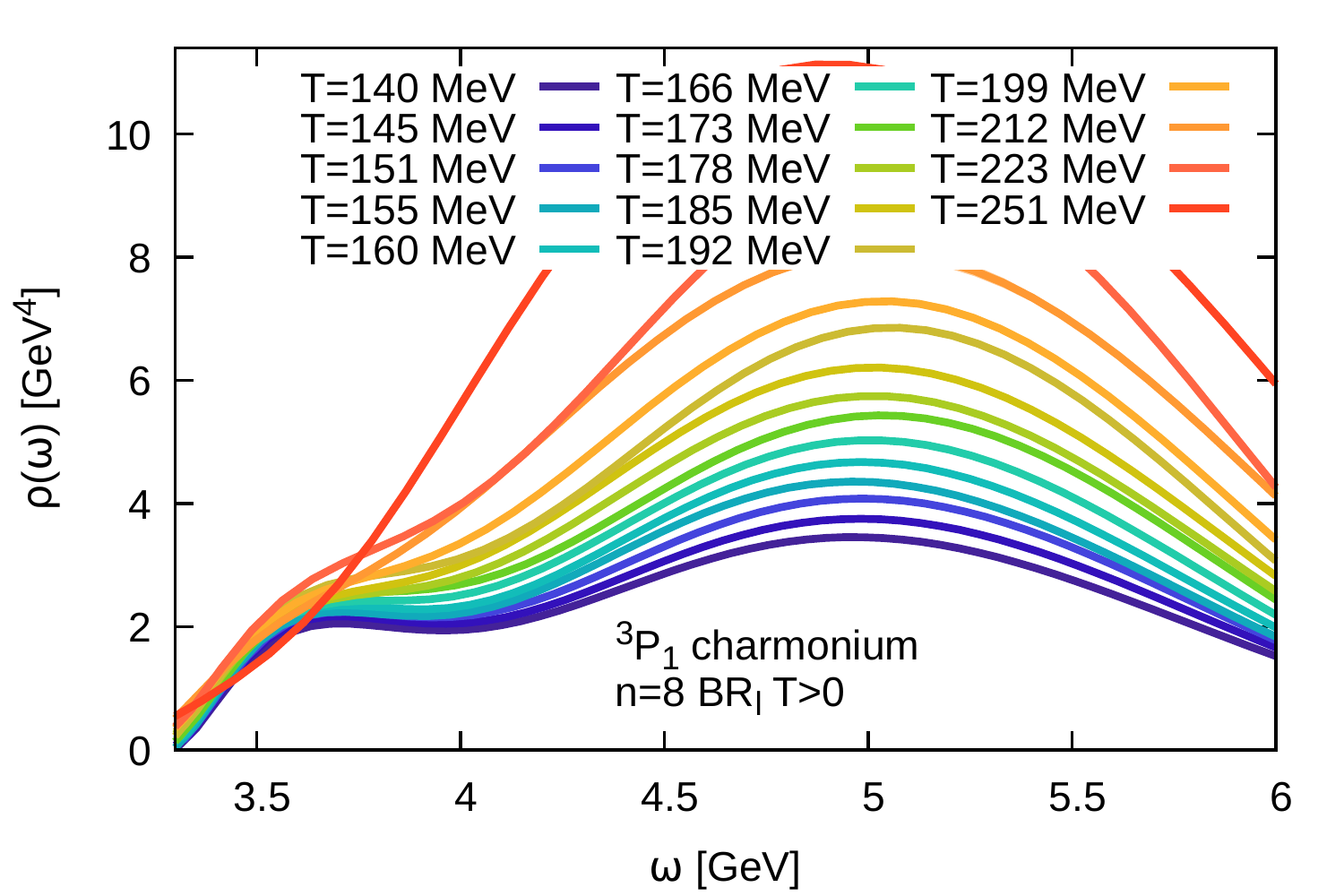}
\caption{Finite temperature bottomonium (left) and charmonium (right) spectral functions reconstructed using the smooth BR method. The top row contains the spectral functions in the $^3S_1$ channel, while the bottom row features the $^3P_1$ channel.}\label{Fig:FiniteTSpectraBRsmooth}
\end{figure}

Fig.\ref{Fig:FiniteTSpectra} and Fig.\ref{Fig:FiniteTSpectraBRsmooth} show a collection of reconstructed spectral functions for bottomonium (left) and charmonium (right) at finite temperature, which were obtained via the standard BR and the smooth BR method respectively. 

In case of the BR method, due to higher statistics and our improved reconstruction, based on both Euclidean and imaginary frequency space data, our results show a smoother structure at high frequencies than in our previous study. Thus all figures can be presented in linear scale. 

As was the case in our previous study we find from the BR method that from a naive inspection by eye a well discernible lowest lying peak is present in all bottomonium spectra considered up to $T=407$MeV. On the other hand in the charmonium case only in the S-wave channel a well defined peak remains up to $T=251$MeV while in the P-wave channel it has disappeared already.

Taking a look at the smooth BR reconstructions the conclusions for the S-wave remain largely unchanged. A well pronounced ground state peak is visible at all temperatures for bottomonium. For charmonium up to $T=212$MeV we also find a discernible peak structure, while above the ground state peak becomes too washed out.

For the P-wave the $BR_\ell$ results are, by construction, much smoother than those of the standard BR method and show a disappearance of a well defined ground state signal at intermediate temperatures, for bottomonium at around $T=199$MeV for charmonium above $T=185$MeV. In the following section we tighten these first impressions by systematic comparison of the different reconstructions at the individual temperatures.

\subsection{Melting temperatures}

The concept of a melting temperature becomes difficult to define, once quarkonium is adequately understood as a dynamical system, in which each state may be temporarily excited to another state, due to medium interactions. In other words, once there is a finite thermal width present in the spectral function the state is not simply swallowed instantaneously by the continuum but smoothly merges with it. One often deployed definition of the melting temperature is the point at which the in-medium binding energy falls below the thermal width of a state.

As we saw that even at zero temperatures with a significantly larger number of data points we were unable to extract the excited states and continuum in a robust fashion this will be even less possible at higher temperatures. Therefore here when we speak about melting we simply mean: what is the highest temperature at which a peak structure for a certain state is observed.

Studies involving different Bayesian approaches have lead to partially conflicting results on the melting temperature of in-medium bottomonium states. In particular it was found that studies based on the MEM would give consistently lower values as compared to the standard BR method. In our previous study we argued that particular care needs to be taken to understand the intrinsic methods artifacts of each Bayesian approach to extract a meaningful value for $T_{\rm melt}$. 

The standard BR method was found to suffer from Gibbs ringing, i.e. in case of a small number ${\cal O}(10)$ of input data points the BR method would imprint wiggly structures into the end result, which are not representative of a signal in the data, in turn mimicking the presence of a remnant bound state. On the other hand we also showed that the MEM in Bryan's implementation with a restricted search space , when faced with an equally small number of ${\cal O}(10)$ of input data points may instead lead to an overly smooth result, indicating the disappearance of an in-medium state, while it still remains encoded in the input data. By extending Bryan's search space we found that we are able to obtain reconstructions that start to resemble those of the standard BR method including the artificial ringing.

From the discussion in sec.\ref{sec:RecBayesRes} we now understand that neither the MEM nor the BR regulator are capable of intrinsically avoiding ringing artifacts. The MEM on the other hand introduces a manual smoothing via the additional choice of a restricted SVD search space, which however is not guaranteed to house the correct solution. In order to implement a form of smoothing that prevents contamination of the reconstruction by ringing in a genuinely Bayesian fashion, i.e. implementing it via the regulator, we have developed the smooth BR method. We showed in this study that using the hyperparameter $\kappa=1$ it both avoids ringing in the free spectrum case and allows us to pick up the ground state signals in the actual interacting $T=0$ case.

\begin{figure}[t]
\includegraphics[scale=0.5]{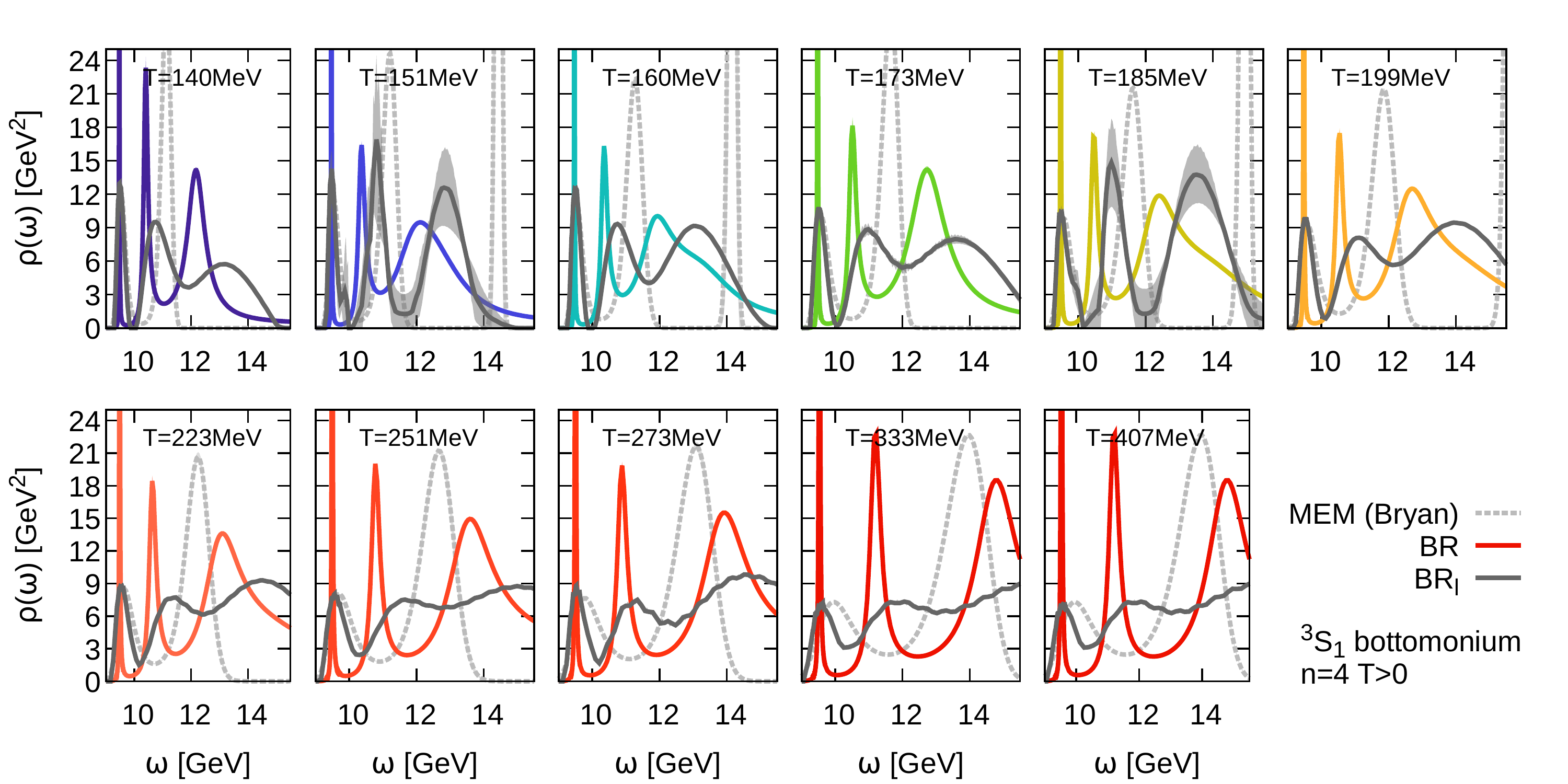}
\includegraphics[scale=0.5]{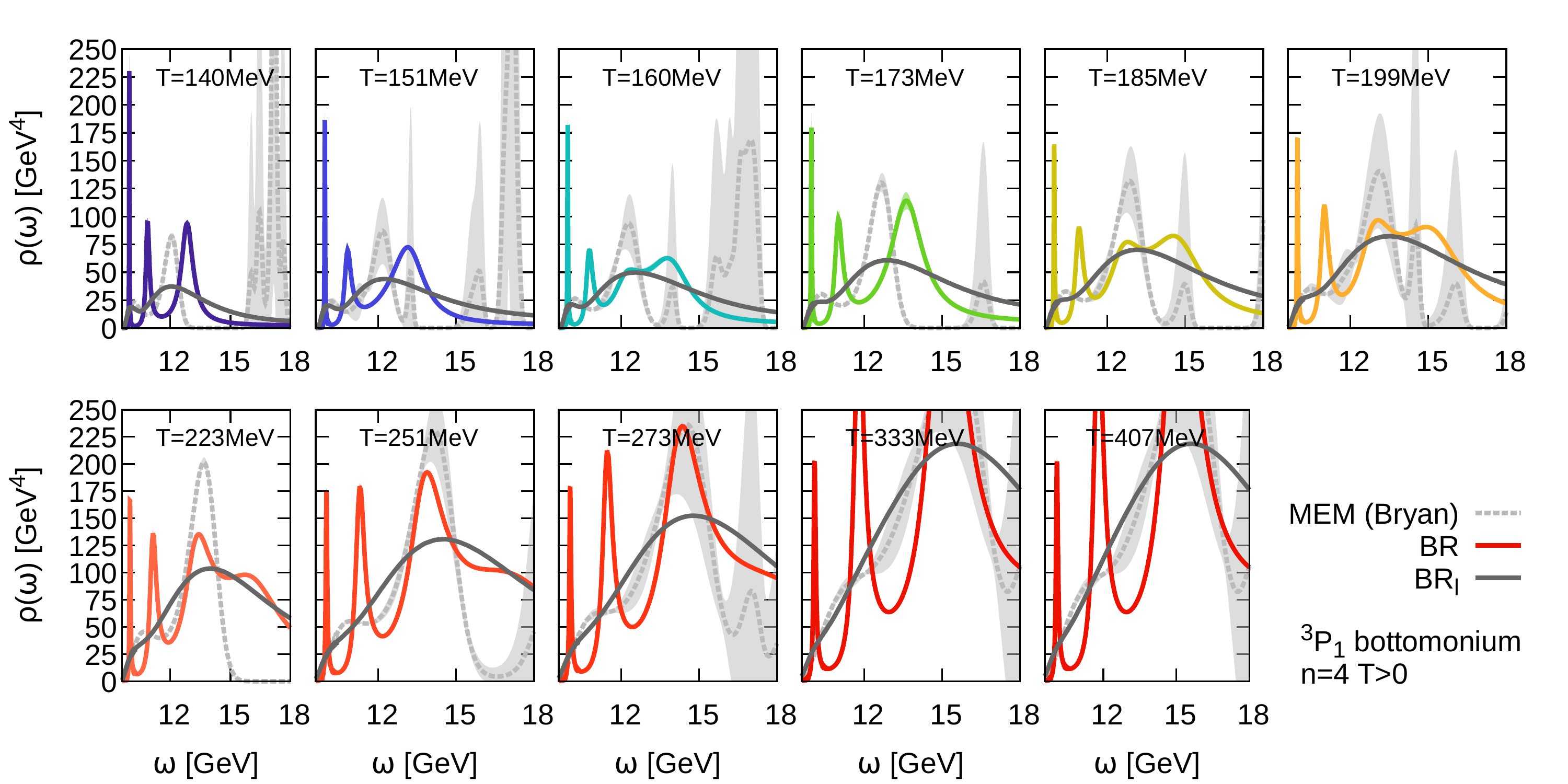}
\caption{Finite temperature spectral functions for bottomonium obtained from the standard BR method (colored solid) the MEM (gray dashed) and the smooth BR method (dark gray solid) The top two rows contain the results for the $^3S_1$ channel, each panel showcasing a different temperature. The lower two rows on the other hand contain the $^3P_1$ spectra.}\label{Fig:FiniteTModelCmpBottom}
\end{figure}

\begin{figure}[t]
\includegraphics[scale=0.5]{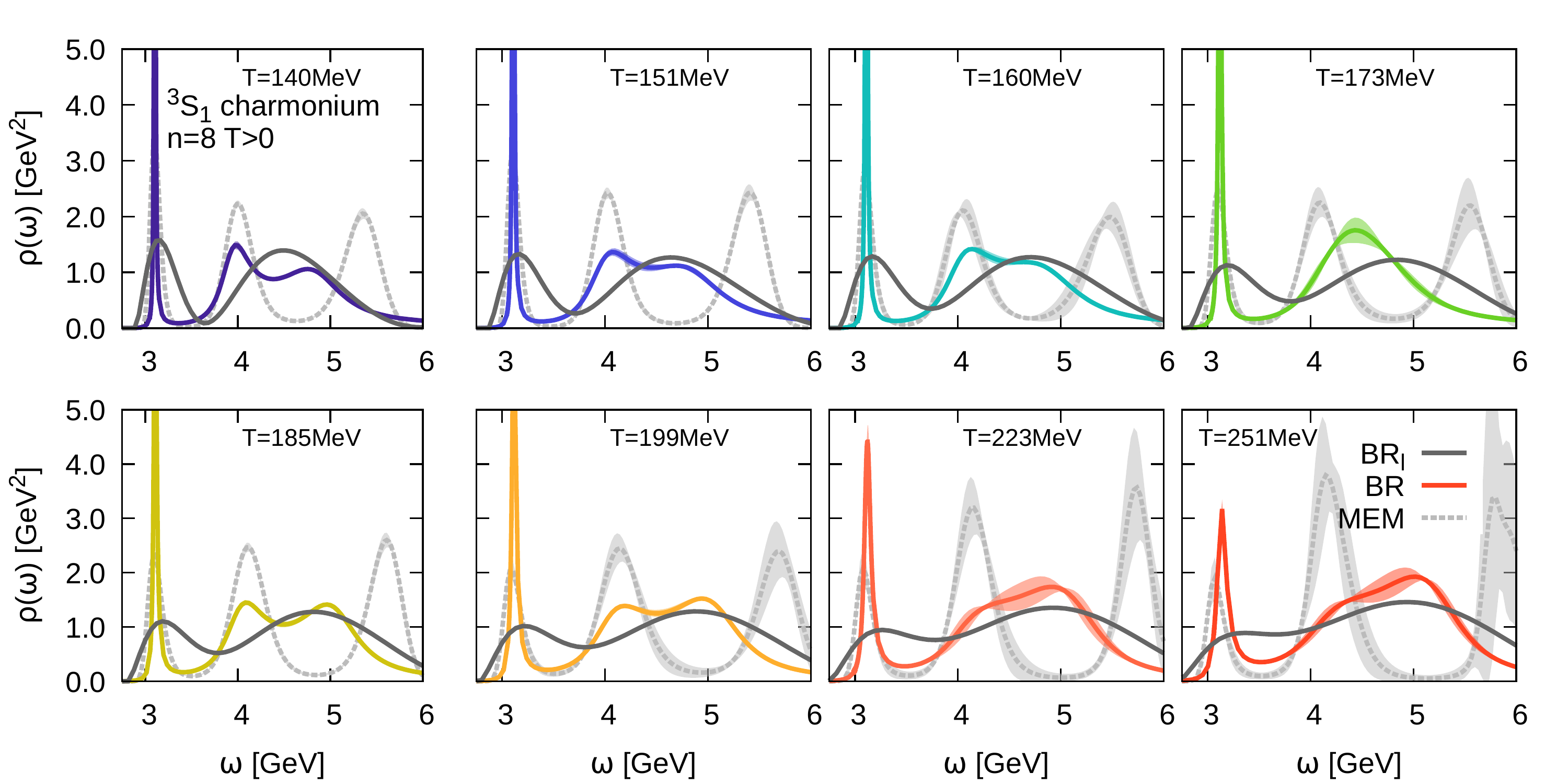}
\includegraphics[scale=0.5]{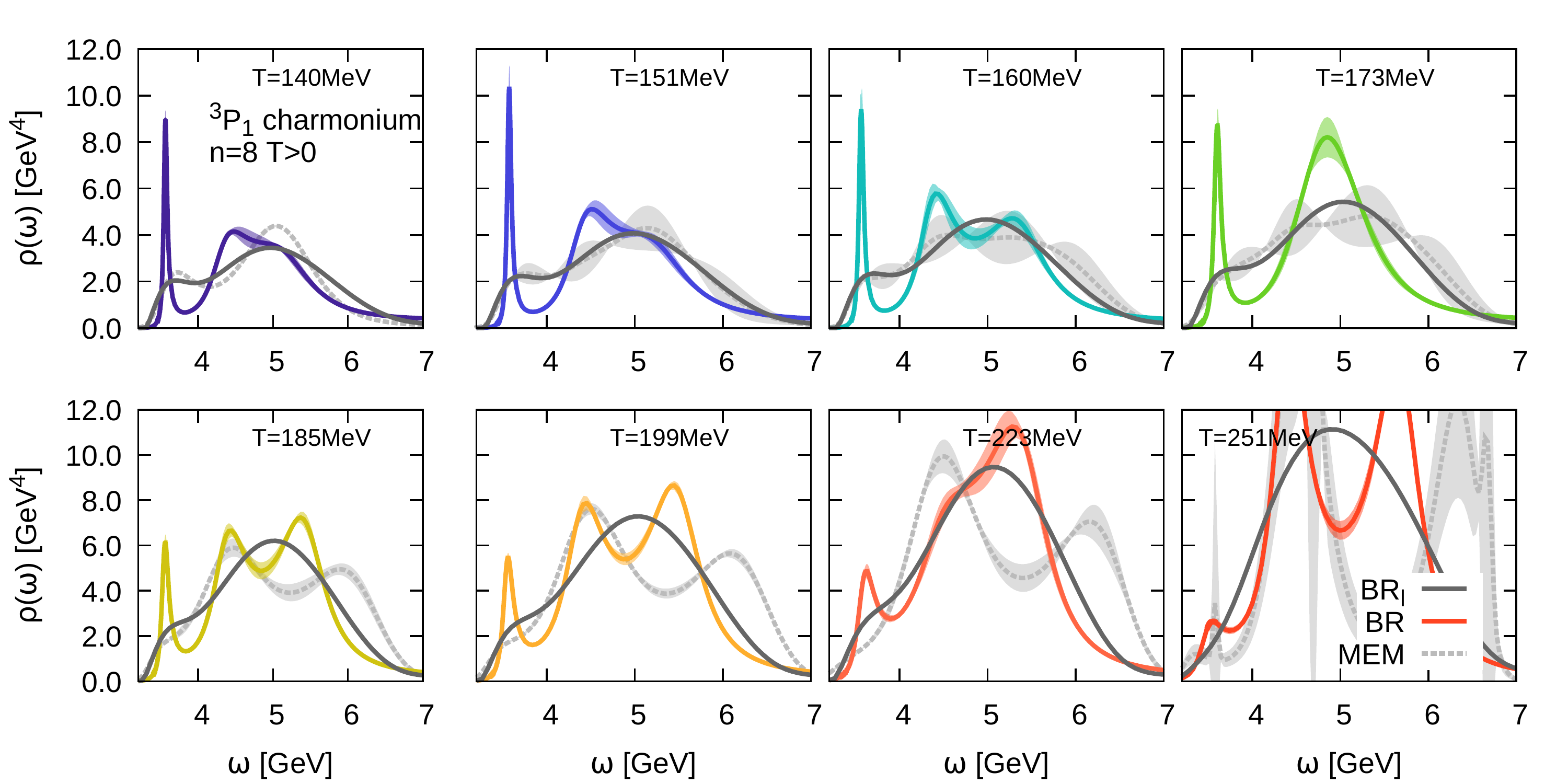}
\caption{Finite temperature spectral functions for charmonium obtained from the standard BR method (colored solid) the MEM (gray dashed) and the smooth BR method (dark gray solid) The top two rows contain the results for the $^3S_1$ channel, each panel showcasing a different temperature. The lower two rows on the other hand contain the $^3P_1$ spectra.}\label{Fig:FiniteTModelCmpCharm}
\end{figure}

Let us have a detailed look at how the three methods fare for the ground state peak in the finite temperature case in Fig.\ref{Fig:FiniteTModelCmpBottom} and Fig.\ref{Fig:FiniteTModelCmpCharm}. We start with bottomonium in the top two rows of Fig.\ref{Fig:FiniteTModelCmpBottom} in the S-wave channel. The standard BR method result is shown as colored solid line, while the MEM result as dashed gray lines. The smooth BR method result is plotted as the dark gray solid curve. 

Several observations can be made: as was expected from the results at $T=0$ the standard BR method produces well defined sharp peaks at zero temperatures but also even at the highest $T=407$MeV. On the other hand the smooth BR method shows much broader reconstructed peaks, whose peak position however is located very close to that of the standard BR method. Interestingly, now that we have a smaller number of data points available, the MEM does not produce sharper peak features than the smooth BR method, even at the lowest temperature $T=140$MeV, where we expect that the ground state signal has not changes significantly from that at $T=0$. Instead it is the much smaller dimensionality of the search space, which now leads to a broader reconstructed peak. Note that the amount of smoothing in the modified BR method is a priorly governed only by $\kappa$ and not the number of available data points.

Moving to higher and higher temperatures, we find that the MEM not only shows slightly lower amplitudes for the ground state peak than the smooth BR method but also moves the position of the lowest lying peak to larger and larger frequencies. This effect is reminiscent of an artifact of the MEM encountered in the study of the in-medium heavy quark potential, where in the presence of UV structure in the spectral function the restricted search space tends to pull the lowest lying spectral feature to too high frequencies \cite{Rothkopf:2011db,Burnier:2013fca}.

With the novel smooth BR method we instead find consistent results for the peak position with the standard BR method and confirm that even up to $T=407$MeV a well defined remnant bump remains in the reconstructed spectrum. Even if we do not trust the position of the MEM quantitatively we have agreement between three methods that a remnant in-medium peak signal survives up to $T=407$MeV, consistent with what was observed in pNRQCD computations based on the complex lattice potential.

Now let us proceed to the more contended P-wave bottomonium states, where the main discrepancy between the different Bayesian methods had been reported. Here the difference between the MEM and the smooth BR method is even more pronounced, in that the former already at $T=140$MeV indicates a significant mass shift, while the smooth BR method again locates the peak close to the position found by the standard BR method. Here we also see that now the MEM (with our choice of parameters, in particular $\omega_{\rm min}$) actually shows a stronger signal for the presence of the P-wave state than the smooth BR method. 

It is interesting to note that in contrast to the truncation tests shown in Fig.\ref{Fig:T0BRSmoothSpectraCmpTrunc} here at finite temperature the smooth BR method indeed gives consistently smoother reconstructions at all temperatures. This is a further indication that in the underlying correlator at $T>185$MeV there is no genuine peak structure remaining.

We conclude that the maximum temperature at which we can identify any form of a peak-remnant in all three methods, be it in the form of a residual threshold enhancement only, is at $T=185$MeV. At higher temperatures only the BR method shows clear peaks. They may be bound state remnants or artifacts due to ringing. One indication for the latter is that at $T=223$MeV the first peak becomes higher than the others peaks. Thus, it is conceivable that below that temperature it still corresponds to a $\chi_b$ state. Therefore, we estimate that the melting temperature of $\chi_b$ lies between $185$MeV and $223$MeV. 

When comparing our results with those from the FASTSUM collaboration, two issues have to be kept in mind. First, the results we obtain with the MEM alone are similar to what has been reported by FASTSUM. Our point here is that using one method alone does not provide an adequate estimate of the uncertainty in the reconstructed spectrum. Second, the light sea quark masses used in the two calculations are different. This differences results in different deconfinement transition temperatures and thus, different screening properties of the medium in the considered temperature region. Therefore, some differences are expected already at the level of the correlation functions in the two calculations. For the completeness let us note that the tree-level pNRQCD analysis predicts a melting at $T\approx220$MeV.

We next turn to charmonium spectral functions shown in Fig.\ref{Fig:FiniteTModelCmpCharm}, again with the S-wave results in the top two rows and the P-wave results in the bottom two rows. It is interesting to see that for the S-wave the MEM shows rather strong spiky features in the UV, while both variants of the BR method show a much smoother continuum regime. Here the ordering of the three methods is similar as in the $T=0$ case, where the MEM resolves peaks with a sharpness in between the standard and smooth BR methods. We find that while the standard BR method and the MEM show a well defined ground state peak up to the highest temperature $T=251$MeV the smooth BR method only find a significant feature up to $T=212$MeV, with the structure essentially washed out at $T=251$MeV.

In the charmonium P-wave channel the MEM at low temperatures $T<185$MeV shows a very similar behavior to the smooth BR method, while starting to produce more wiggly artifacts in the UV at higher temperatures. We thus find a similar result among the two methods for the disappearance of a well defined ground state structure at or above $T=185$MeV.

\begin{table}[t]
\centering
\begin{tabular}{|c|c|c|c|}\hline
Particle 			& $^sX_J(n)$  & $T_{\rm melt}^{\rm NRQCD}$ [MeV]& $T_{\rm melt}^{\rm pNRQCD}$ [MeV]\\ \hline
$\Upsilon$ 		& $^3S_1(1)$ 	& $>407$ &  $412(76)$\\ 
$\chi_{b1}$ 		& $^3P_1(1)$ 	& $ 185-223$  & $220(10)$ \\ \hline \hline
$J/\Psi$ 			& $^3S_1(1)$ & $200-210$ & $212(13)$ \\ 
$\chi_{c1}$ 		& $^3P_1(1)$ 	& $\approx 185$ & $183(13)$\\ \hline \hline
\end{tabular}
\caption{Estimates of the melting temperatures from lattice NRQCD (left column) compared to the predictions from tree-level pNRQCD based on the complex heavy quark potential from $N_f=2+1$ lattice QCD.}
\label{Tab:meltingTemp}
\end{table}  

The values for the melting temperatures found in this study lie close to the values predicted by a tree level pNRQCD computation based on a recently determined complex heavy quark potential from the lattice \cite{Burnier:2015tda,Burnier:2016kqm}. While this is encouraging from the point of view of consistency between non-relativistic approaches, several caveats are present. First of all, as mentioned, the definition of the melting temperature in the two computations is different. Secondly, the discrepancy for the bottomonium P-wave states indicates that the agreement for charmonium P-wave may be simply accidental. And indeed we saw that already at zero temperature the P-wave states were more challenging to capture due to the larger contribution from the continuum and the intrinsically smaller strength of the ground state. 

The above discussion of melting temperatures relied on the determination of the spectral property of peak height or peak area, which is known to be rather challenging for Bayesian reconstruction methods in general. On the other hand the reconstruction of the position of spectral features is much more robust and often accessible from simulation data, for which a determination of peak widths is still out of reach. We will proceed to an investigation of the corresponding in-medium mass shifts in the next section.

\begin{figure}[t]
\includegraphics[scale=0.5]{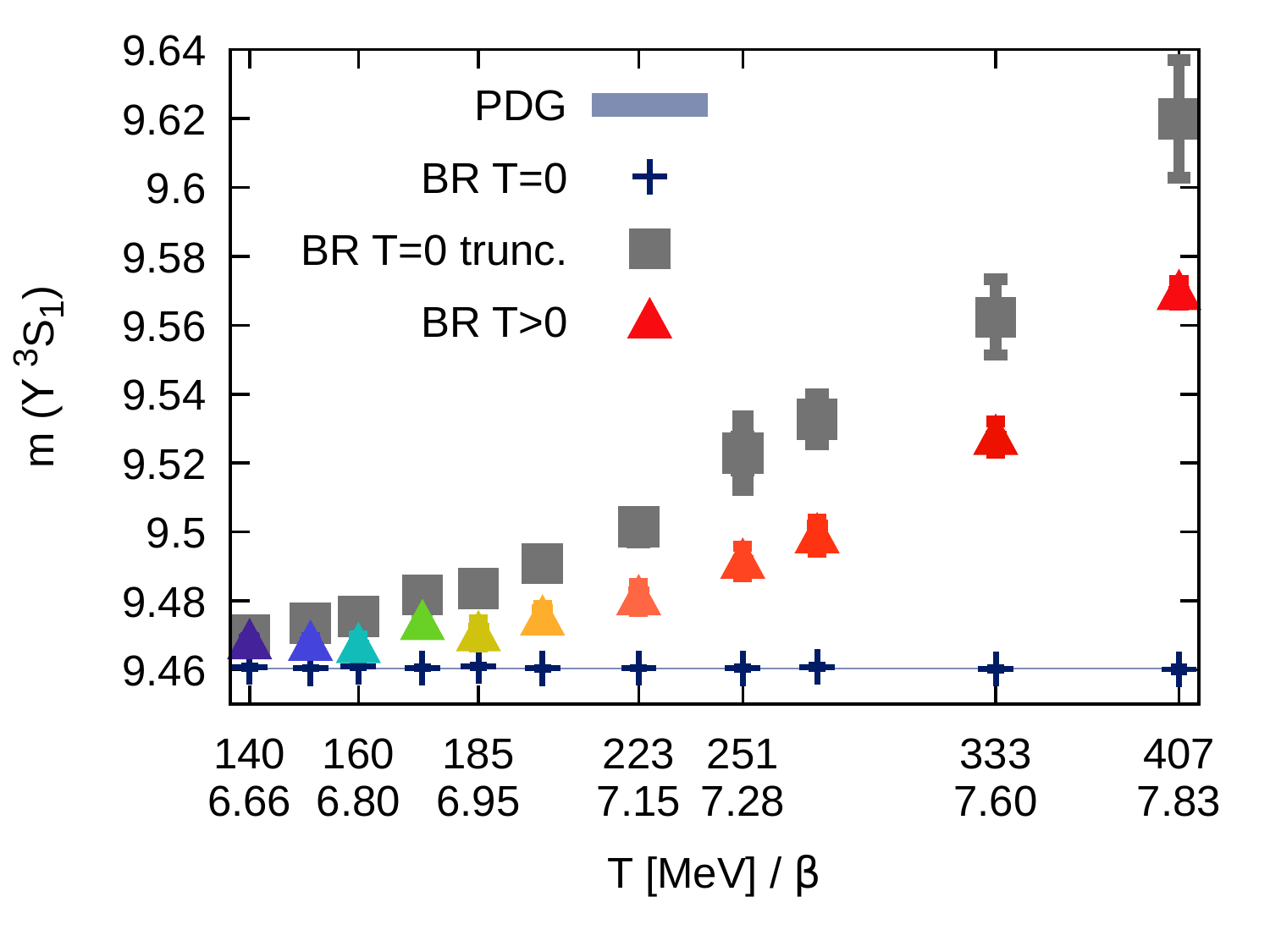}
\includegraphics[scale=0.5]{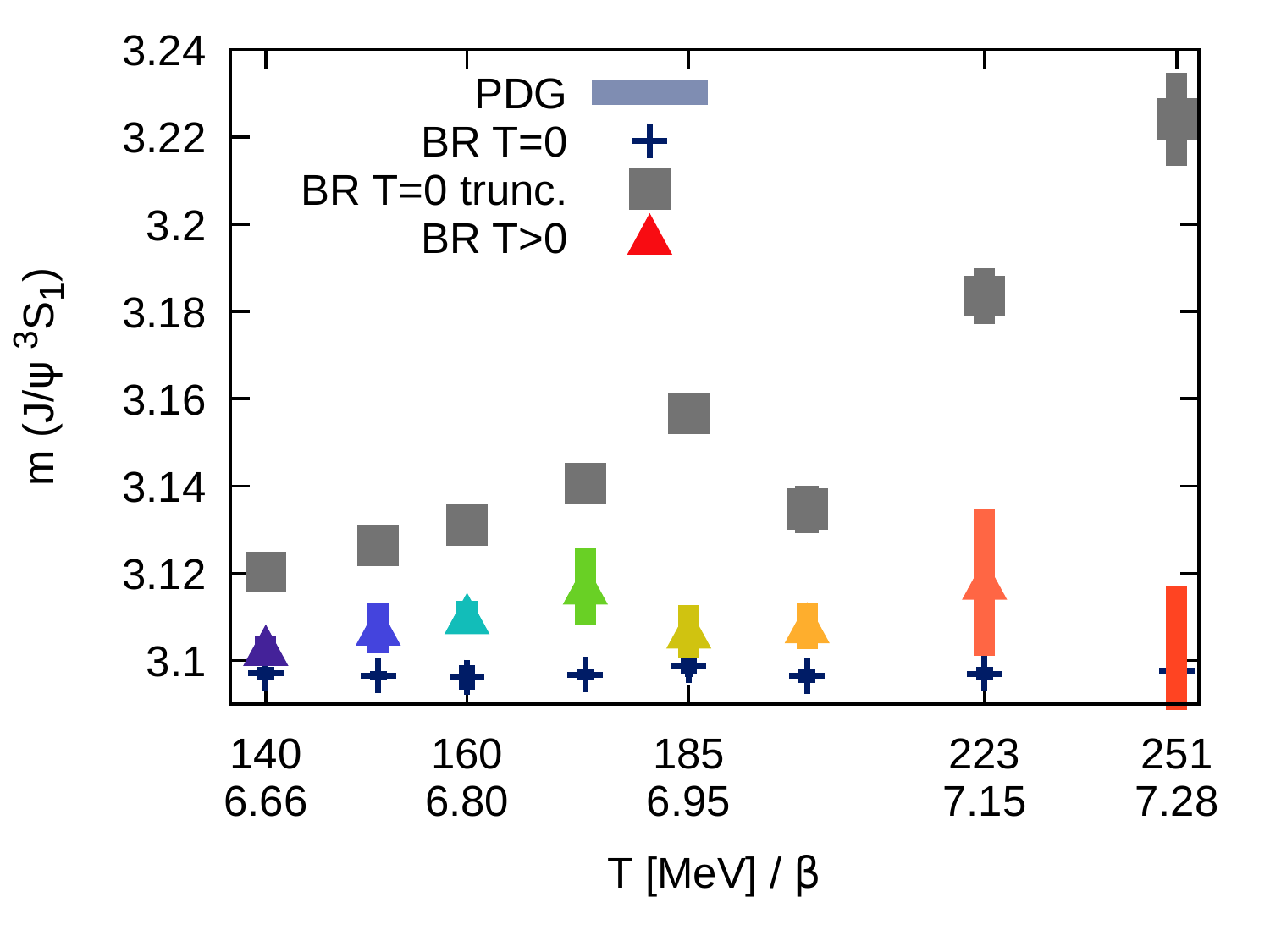}
\includegraphics[scale=0.5]{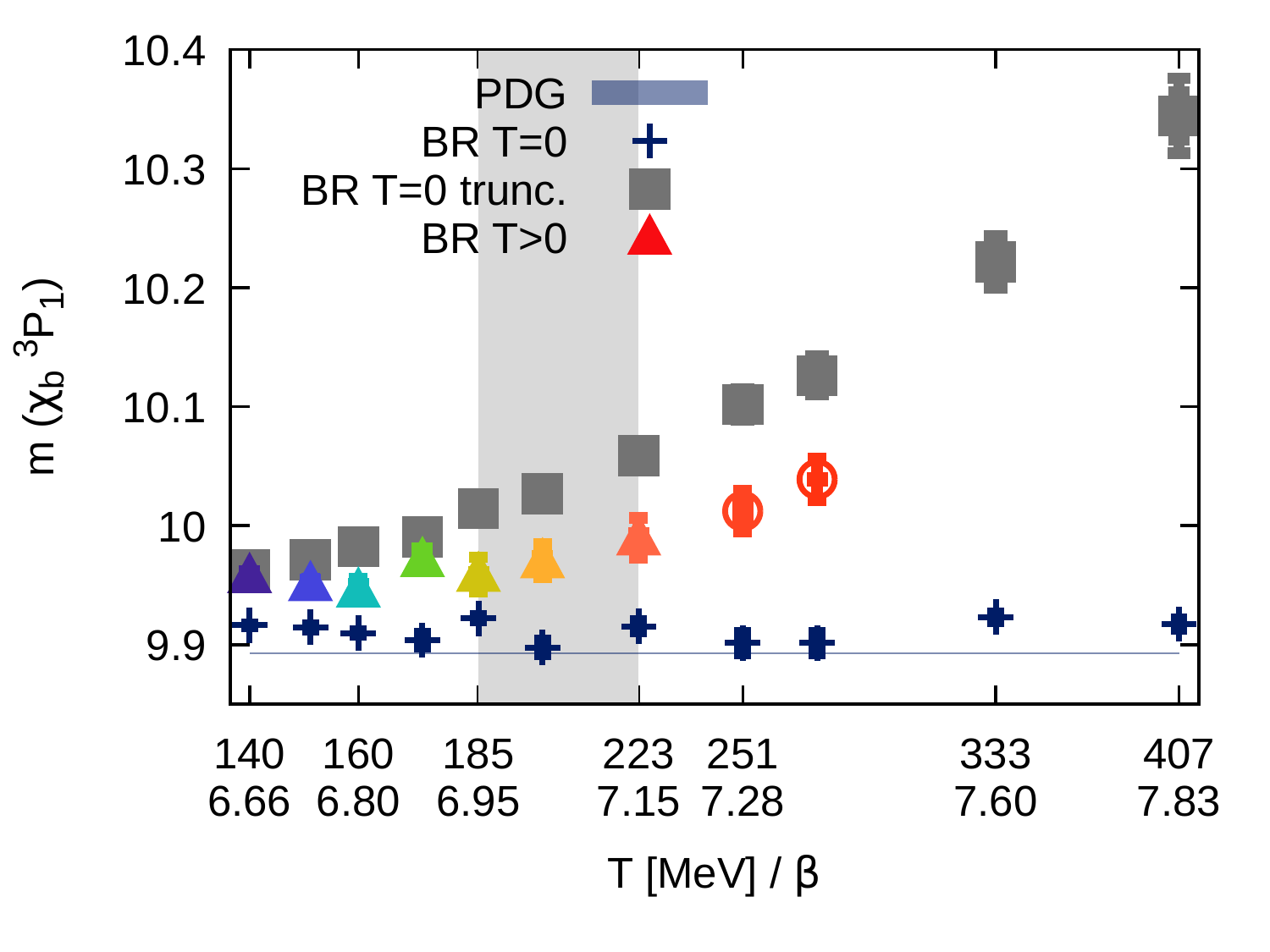}
\includegraphics[scale=0.5]{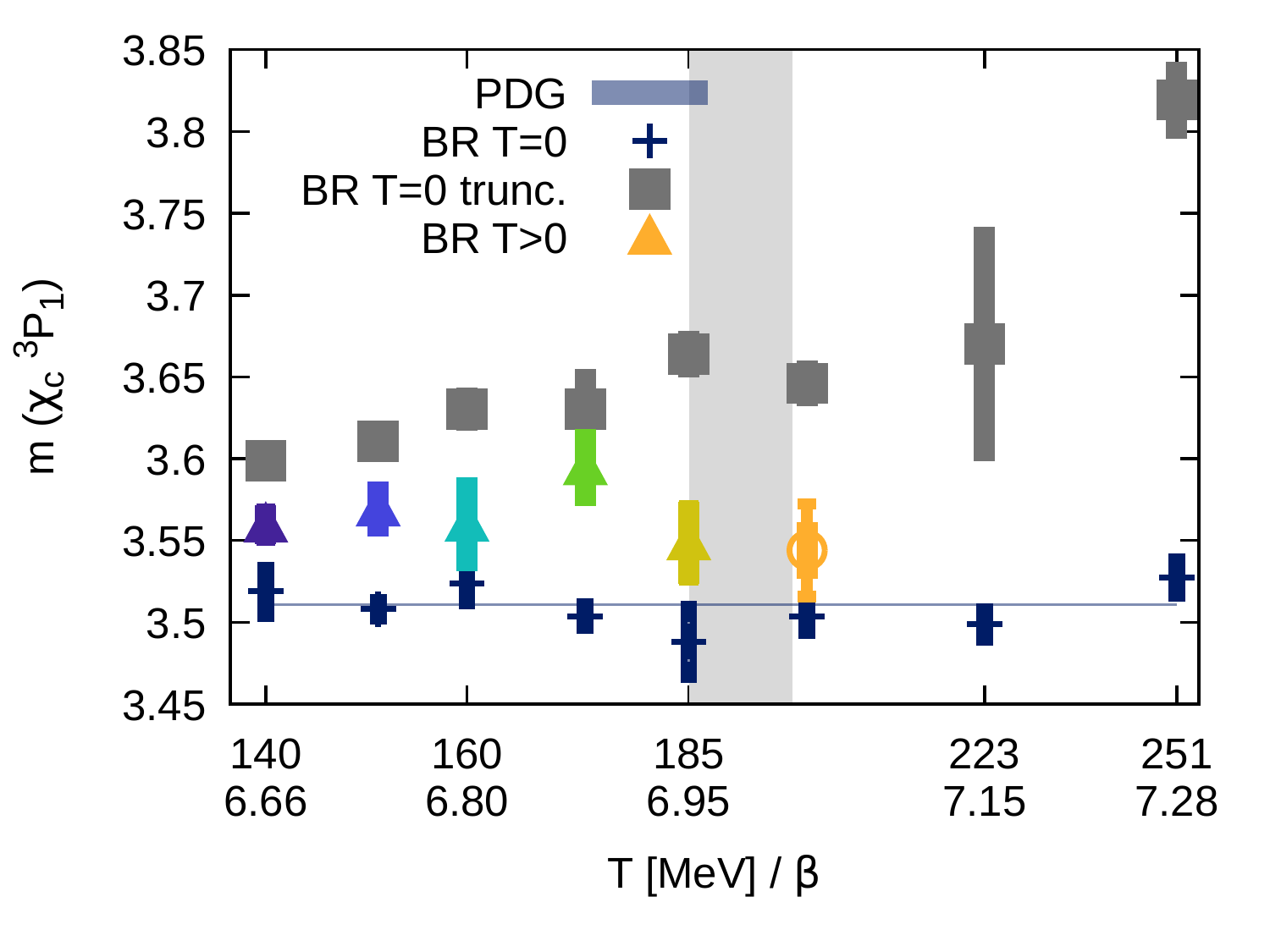}
\caption{The in-medium masses of quarkonium in lattice NRQCD together with the appropriate baseline from a truncated $T=0$ reconstruction. All data points are obtained with the standard BR method with bottomonium results on the left and charmonium on the right. We present the S-wave channels in the top row and the P-wave channels in the bottom row. Three sets of points are contained in each panel: Dark blue crosses denote the reconstructed peak position at zero temperatures, arising from lattices with the same lattice spacing used in the corresponding finite temperature mass estimate given as colored triangles. The x-axis therefore carries both labels for temperature in MeV and the lattice $\beta$ parameter. The PDG ground state mass is given as blue solid line. The gray filled squares on the other hand denote the masses extracted from the truncated $T=0$ input data sets and represent the correct baseline to which the in-medium masses need to be compared to. The area right of the gray separator refers to the temperature range in which the smooth BR method indicates that the standard BR method may not provide a robust determination of the ground state peak.}\label{Fig:FiniteTInMediumMasses}
\end{figure}

\subsection{Ground state mass shifts} 
\label{sec:FiniteTMassShifts}

The main quantitative result of this study is presented in this section, a determination of the in-medium mass shifts of S-wave and P-wave quarkonium states. This analysis combines several pieces of information, which we have gathered along the way of the preceding sections. 
All necessary ingredients are shown in Fig.\ref{Fig:FiniteTInMediumMasses} with the bottomonium results on the left and charmonium on the right. The top row showcases the S-wave channels, while the P-waves are found in the bottom row. In each of the four panels we put side by side three different data sets, all obtained with the standard BR method. 

The dark blue crosses denote the masses extracted from the full $T=0$ correlator data sets and we put the PDG value as reference as solid blue line. As we have discussed in sec.\ref{sec:prepFiniteT} these values, however, are not the correct baseline to which our in-medium results shall be compared, as they are obtained from a significantly different set of data points, most importantly with a much larger Euclidean physical extent. Instead we have argued that the masses obtained after truncating the $T=0$ correlators to $\tau_{\rm max}/a_\tau=12$ provide us with the same methods systematics, as the reconstructions actually carried out at $T>0$. The values of these truncated reconstructions are plotted as gray squares in Fig.\ref{Fig:FiniteTInMediumMasses}. The actual in-medium masses obtained from fitting the finite temperature spectral functions of Fig.\ref{Fig:FiniteTSpectra} with a Breit-Wigner type ansatz are given as colored triangles. We have inserted a gray separator in the P-wave plots indicating the temperature beyond which the smooth BR method indicates that the standard BR method may not provide a robust determination of the ground state peak.

Choosing the appropriate baseline is essential for understanding the physics of the in-medium mass shift. While at first sight a comparison of the full $T=0$ masses with their $T>0$ counterpart would have suggested that the in-medium mass shift actually moves the particle masses to larger values, we argue that the opposite is true. The reconstructed in-medium mass is actually smaller that the proper T=0 reference
obtained with twelve data points.

\begin{figure}[t]
\includegraphics[scale=0.5]{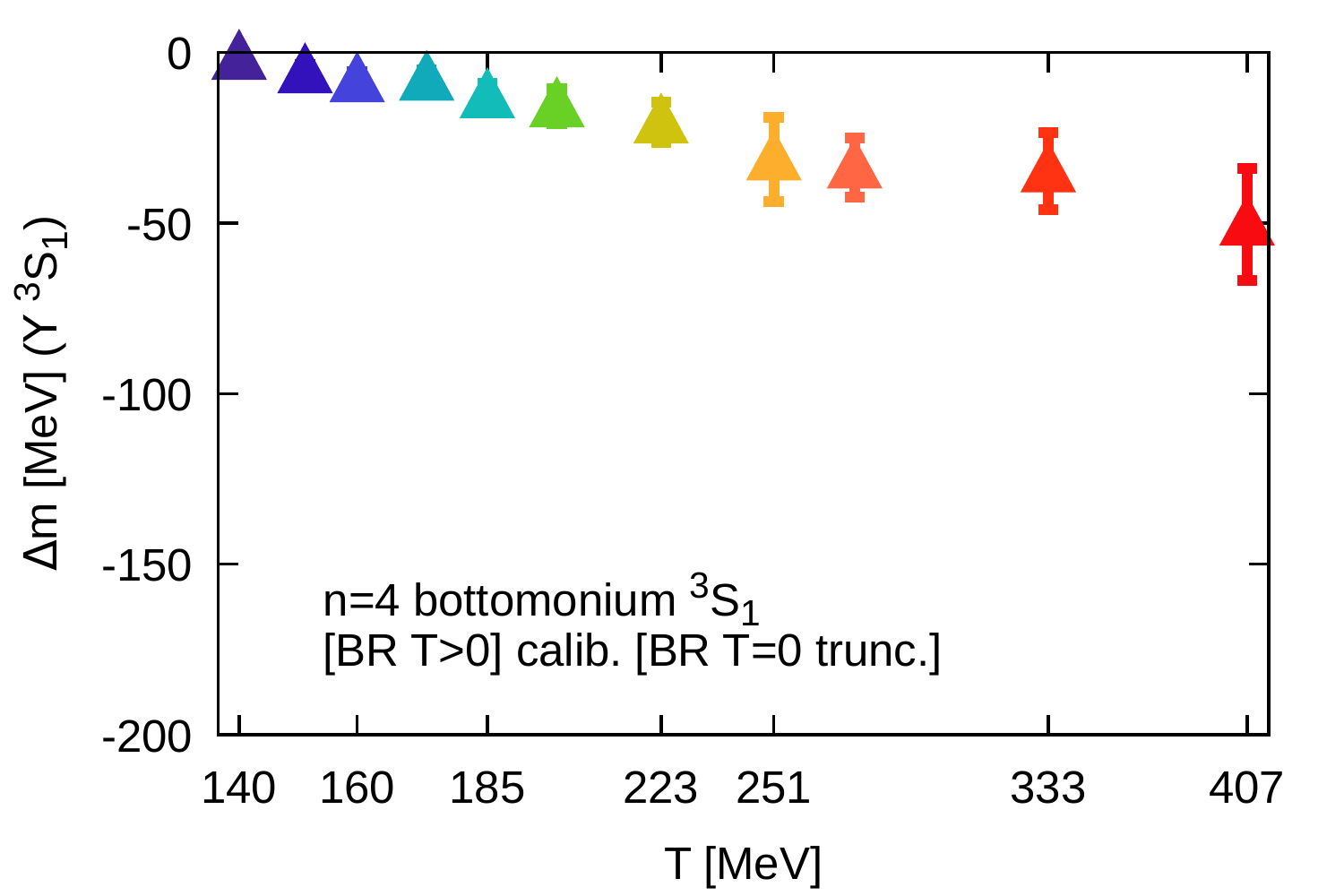}
\includegraphics[scale=0.5]{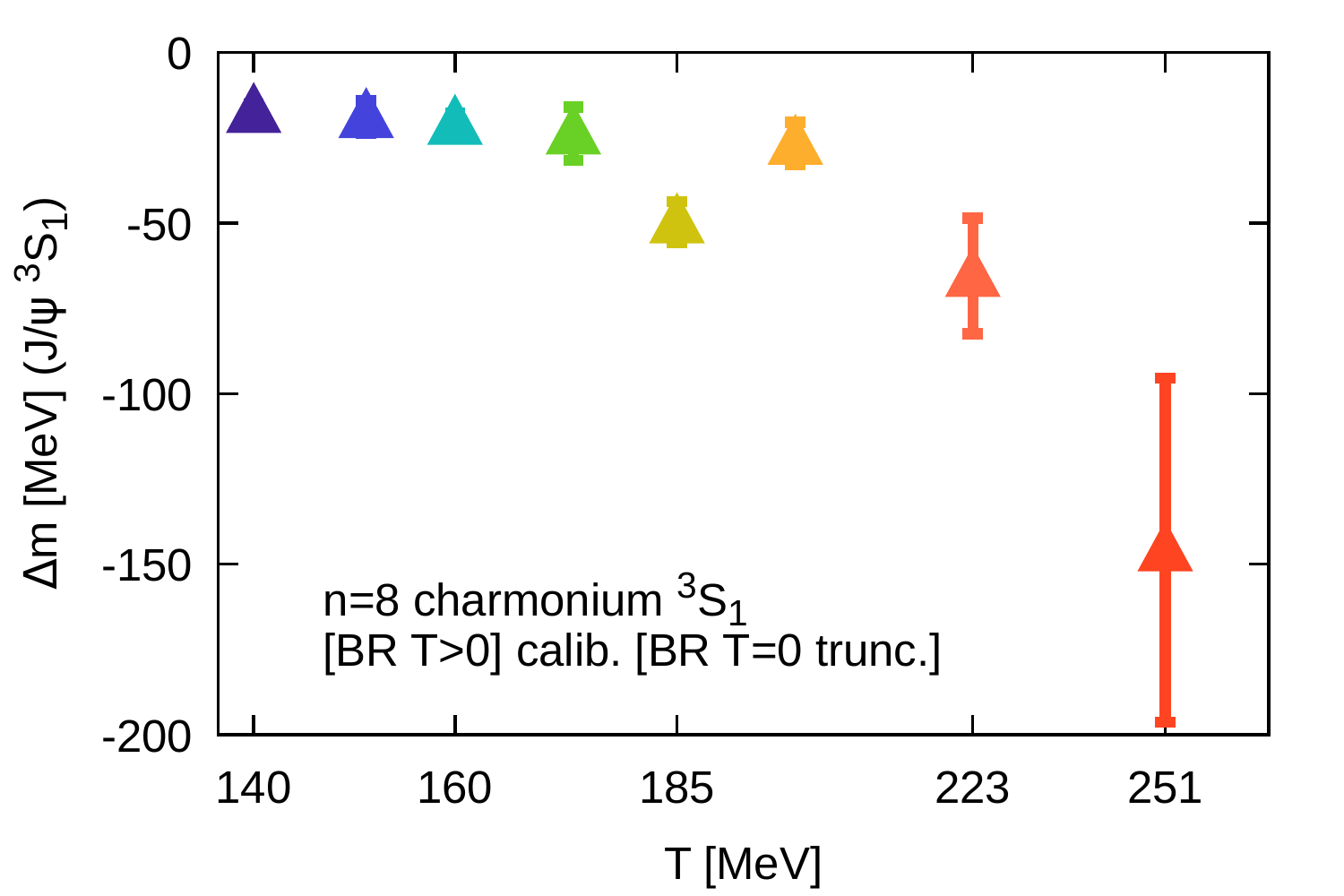}
\includegraphics[scale=0.5]{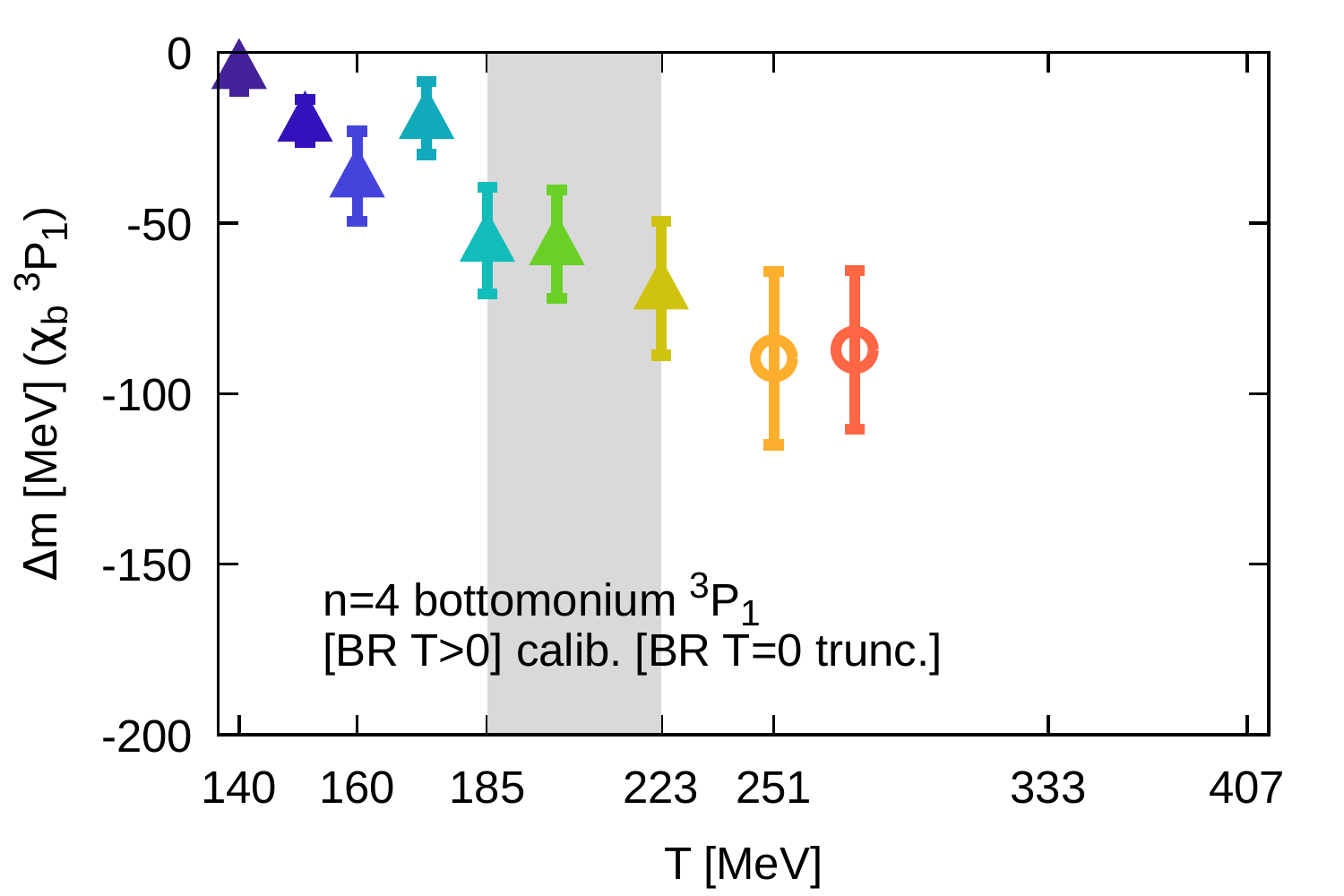}
\includegraphics[scale=0.5]{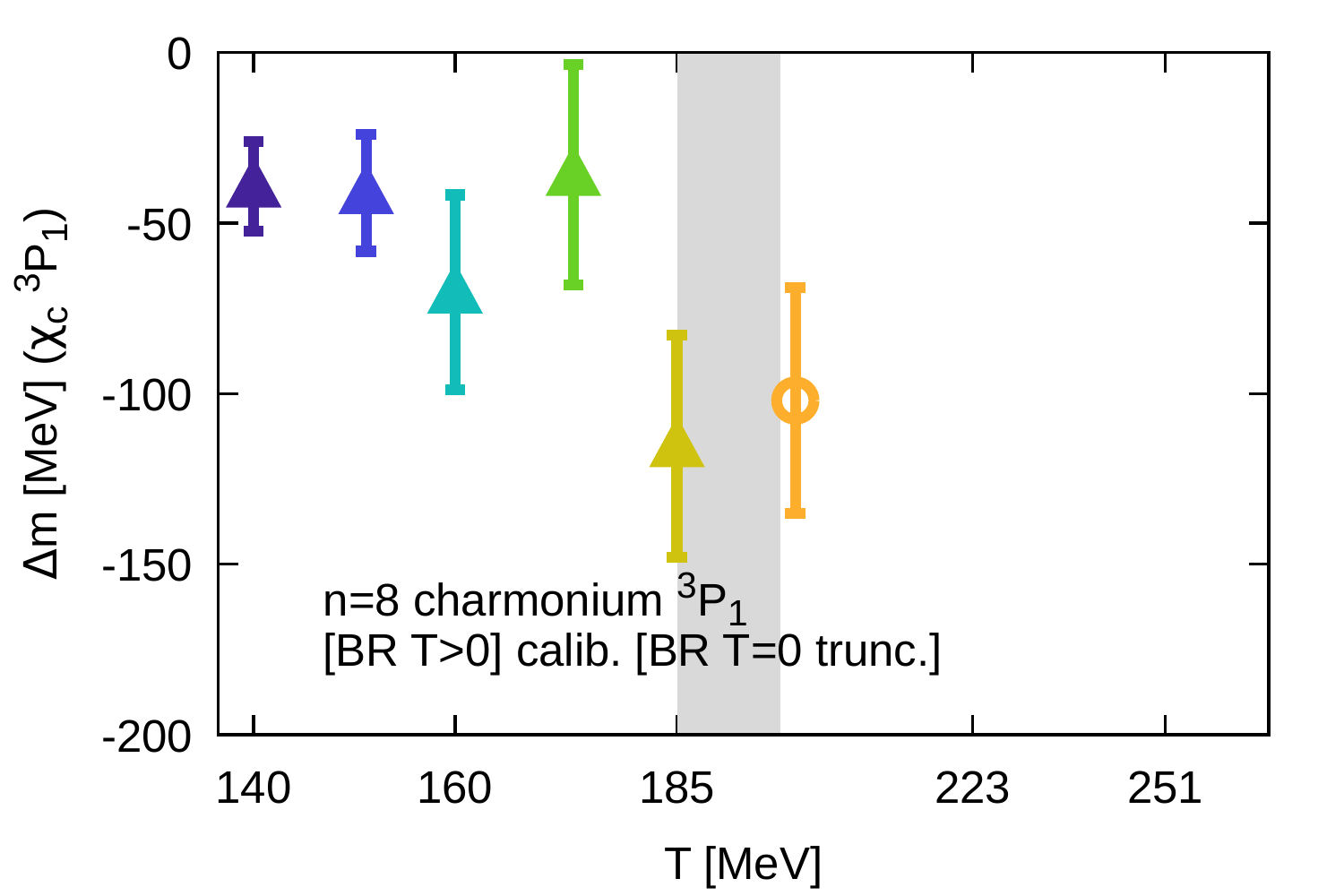}
\caption{In medium mass shifts for bottomonium (left) and charmonium (right). The S-wave channel is depicted in the top row, the P-wave channel in the bottom row. The area right of the gray separator refers to the temperature range in which the smooth BR method indicates that the standard BR method may not provide a robust determination of the ground state peak.}\label{Fig:FiniteTMassShifts}
\end{figure}

This behavior of a shift towards lighter masses is fully consistent with findings from non-perturbative tree-level pNRQCD computations of spectral functions based on a recently computed complex heavy quark potential from the lattice \cite{Burnier:2015tda}. When in the potential picture the real-part of the potential becomes more and more screened the effect it has on the in-medium states is that their masses move towards lower values before smoothly dissolving into the continuum. Note that weak-coupling pNRQCD computations that assume the scale 
hierarchy $mv \sim 1/r \gg T$ do lead to a positive mass shift instead \cite{Brambilla:2010vq}, while the weak coupling pNRQCD computation that assume $m v \sim g T$ lead to a negative mass shift \cite{Burnier:2007qm}.

Our study is the first and the only study in which a behavior consistent with the potential picture emerges. Note that in previous NRQCD studies by the FASTSUM collaboration \cite{Aarts:2014cda}, as well as in the works with the relativistic heavy quark formulations \cite{Ikeda:2016czj} a similar upward trend in the raw results (colored points in Fig.\ref{Fig:FiniteTMassShifts}) is observed. However, since these studies use the naive baseline from the untruncated $T=0$ correlators, they arrive at an opposite conclusion compared to us.

The behavior of the in-medium masses compared to the appropriate baseline also agrees with intuition. For the bottomonium states at $T=140$MeV the medium is still too weak to induce any significant changes in the ground state. For $\Upsilon$ essentially up to $T=173$MeV only a minute shift to lower values persists. Above $T=185$MeV on the other hand this shift becomes substantial and grows up to the highest temperature considered, $T=407$MeV. For $\chi_b$ the shift appears to go beyond the uncertainties already at $T=160$MeV.

In charmonium, as was expected already at $T=140$MeV both S-wave and P-wave states show a significant in-medium shift towards smaller values of the mass. While it may be difficult to judge from due to different resolution of the y-axis, the shift in the P-wave channel of charmonium starts off from a larger value than in the S-wave case.

We have collected the values of the in-medium mass shifts in Fig.\ref{Fig:FiniteTMassShifts} all plotted with the same scale for comparison purposes. The bottomonium mass shifts show a clear linear tendency, with only the data point at $T=173$ being slightly out of place. We have rechecked the analysis of this ensemble but could not find any indications for pathological behavior of the spectral reconstructions. 

The charmonium S-wave mass shifts seem to also follow a linear trend up to $T=173$MeV and then falling off more rapidly. The uncertainties for the P-wave mass shift are still too large to make a conclusive statement about a trend there. We should remark that due to the relatively small value of $M_c a$ on our lattices there are certainly significant radiative corrections expected, which we have not yet taken into account here.

We are therefore confident that the observed negative mass shifts are a manifestation of the physical in-medium modification of heavy quarkonium.

\section{Summary and Outlook}
\label{sec:disc}
\subsection{Summary of the study}

We have presented the final results of a multi-year investigation of in-medium quarkonium properties, based on a combination of state-of-the-art lattice QCD simulations of the QCD medium with $N_f=2+1$ light flavors by the HotQCD collaboration (sec. \ref{sec:latdetails}) and the effective field theory, NRQCD up to order ${\cal O}(v^4)$ (sec. \ref{Num:LatNRQCD}). Compared to our previous published study, we have enlarged the temperature range under consideration, have significantly increased the statistics in our simulations and have included charmonium in our investigation. 

We furthermore deploy improved Bayesian strategies for spectral function reconstruction (sec. \ref{sec:bayesrec}). On the one hand the standard BR method has been set up to take into account data not only along the Euclidean time domain but also in imaginary frequencies, which stabilizes the high frequency behavior of the reconstructions. On the other hand, to systematically address the issue of ringing in Bayesian reconstructions, we deploy a recently developed smooth variant of the BR method, in which an additional smoothing mechanism is incorporated in a genuinely Bayesian and self-consistent fashion. Using the known free spectral functions, we tune the smoothing such that no ringing artifacts survive in the reconstructions (Fig. \ref{Fig:TuningHyp}).

Considering several different splitting between the P-wave and S-wave states as a benchmark (Fig.\ref{Fig:T0Splittings}), we quantified the inherent systematics in the NRQCD simulation at zero temperature (sec.\ref{sec:T0groundstate}). We found that even though we did not tune the heavy quark masses it was possible to reproduce most splittings in a satisfactory manner, i.e. within at most $35$MeV. The spin-orbit coupling related splittings for charmonium and the S-wave hyperfine splitting for both bottomonium and charmonium were the most challenging. It is known that the latter requires both higher orders in the velocity expansion and one loop order corrected Wilson coefficients.

The higher statistics allowed us to extract the ground state masses consistently with both simple effective mass fits (Tab.\ref{Tab:zeroTmass}), as well as full Bayesian spectral reconstructions (Tab.\ref{Tab:zeroTmassBR} and Fig.\ref{Fig:T0BRMassesCmp}) . With the largest deviations of the the masses from their PDG values being around $35$MeV the outcome of the simulation proved satisfactory.  Interestingly several mass splitting for charmonium are even better reproduced than for bottomonium. I.e. for charmonium all computed ground states masses, except for the $^3P_2$ state, are within $2\sigma$ of their PDG values. Since in anticipation of the $T>0$ study we use simple point source correlators the estimates for the excited states were still rather imprecise.

In order to understand better the physics content of the correlators, we also carried out constrained model fits using a two peak and two-box spectrum (Fig.\ref{Fig:T0BRModelCmpBottom}), which allowed us to reproduce the correlator within its uncertainties. An inspection of the raw correlator after subtracting ground and first excited state contributions further revealed that only four to five $\tau$ points remain to encode all intricate structure around the threshold. This clearly showed that we cannot expect to recover more detailed information about the excited states content of the quarkonium channels with the current data quality.

Once the smooth BR method had been tuned to avoid ringing artifacts, we further checked with $T=0$ data that in case of a genuine peak signal being present in the correlator, the method remains capable of identifying it (Fig.\ref{Fig:T0SpectraCmpMethods}). Furthermore, in preparation for the $T>0$ study we investigated how limiting the accessible Euclidean time extent affects the reconstruction outcome. To this end we truncated the correlator at each lattice spacing to the same physical extent as is available at the corresponding finite temperature simulations and repeated the spectral reconstruction based on it(Fig. \ref{Fig:T0BRSpectraCmpTrunc}). This procedure allowed us to identify a shift to higher frequencies as well as a broadening as main artifacts (Fig.\ref{Fig:T0BRMaseesCmpTrunc}), similar to what we observed in our previous publication. 

Moving toward finite temperature we first investigated the ratio of in-medium correlators to their vacuum counterparts (Fig.\ref{Fig:FiniteTCorrelatorRatios}). Their behavior corroborated with improved precision the findings from our previous study. The overall in-medium modification is hierarchically ordered with the vacuum binding energy of the ground state. Using intuition from a pNRQCD computation with a complex potential (Fig.\ref{Fig:FiniteTCorrelatorRatioCmpModel}) we interpreted the temperature dependence of the ratios in more detail. We note that unprecedented
precision on the finite temperature correlators have been reached, the statistical errors on the correlators are at the level of one tenth of a percent on less. This implies that our lattice results can provide stringent checks on the validity of pNRQCD calculations of the spectral 
functions.

The in-medium spectral reconstructions we presented subsequently (Fig.\ref{Fig:FiniteTSpectra}) were used to study different properties of in-medium quarkonium. We started with the more challenging property of peak-area or peak height in order to determine a melting temperature for the ground state. Comparing the MEM, the standard BR, as well as the smooth BR method (Fig.\ref{Fig:FiniteTModelCmpBottom} and Fig.\ref{Fig:FiniteTModelCmpCharm}) we found that for the bottomonium S-wave all three methods shows a ground state remnant up to $T=407$MeV$=2.63T_c$. 
For the P-wave all methods indicate an in-medium bound state up to $T=185$MeV$=1.19T_c$. 
Investigating in detail the behavior of the standard and smooth BR method, 
we conclude that melting happens between $T=185-223$MeV$=1.19-1.44T_c$. 
A similar analysis for charmonium gives a melting range for the S-wave 
of $T=200-210$MeV$=1.29-1.35T_c$. The charmonium P-wave appears to melt just above $T=185$MeV$=1.19T_c$.

Let us stress that using the same Bayesian method than previous studies (MEM), we obtain very similar reconstructions. Our point is that using just a single method does not allow us to estimate the methods uncertainty and indeed a comparison with BR and smooth BR leads us to melting regions instead of a single temperature.

\begin{figure}[t]
\includegraphics[scale=0.5]{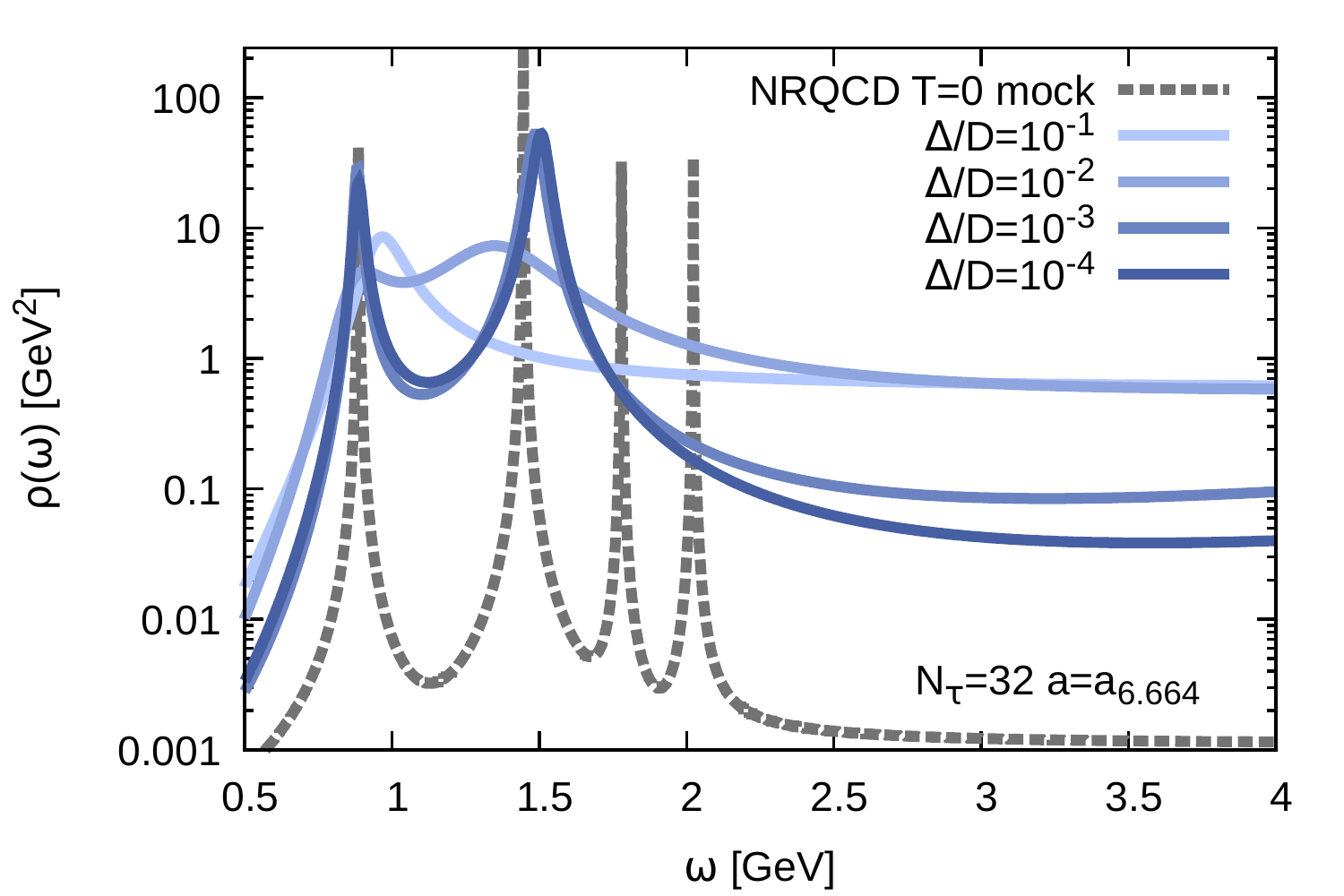}
\includegraphics[scale=0.5]{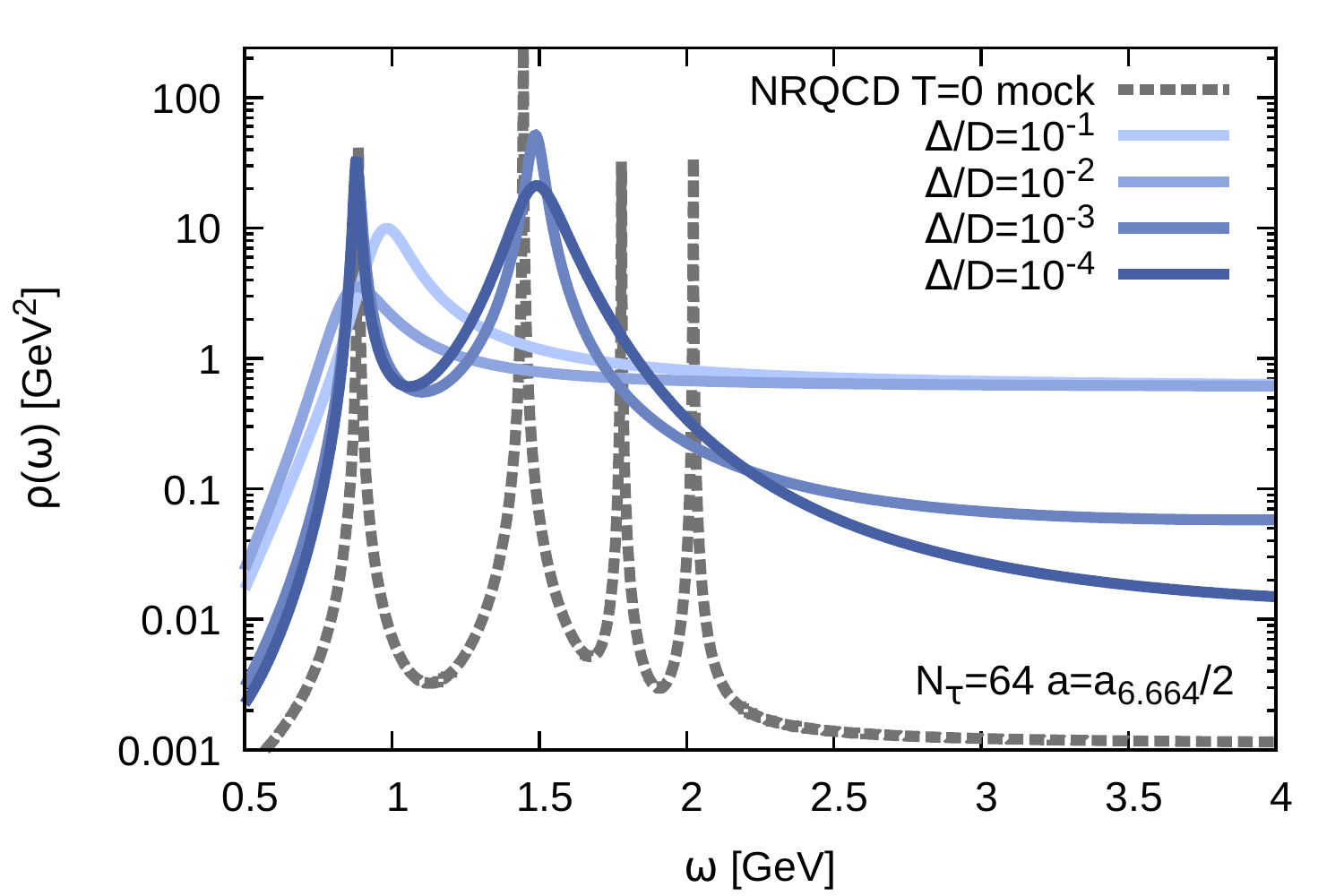}
\includegraphics[scale=0.5]{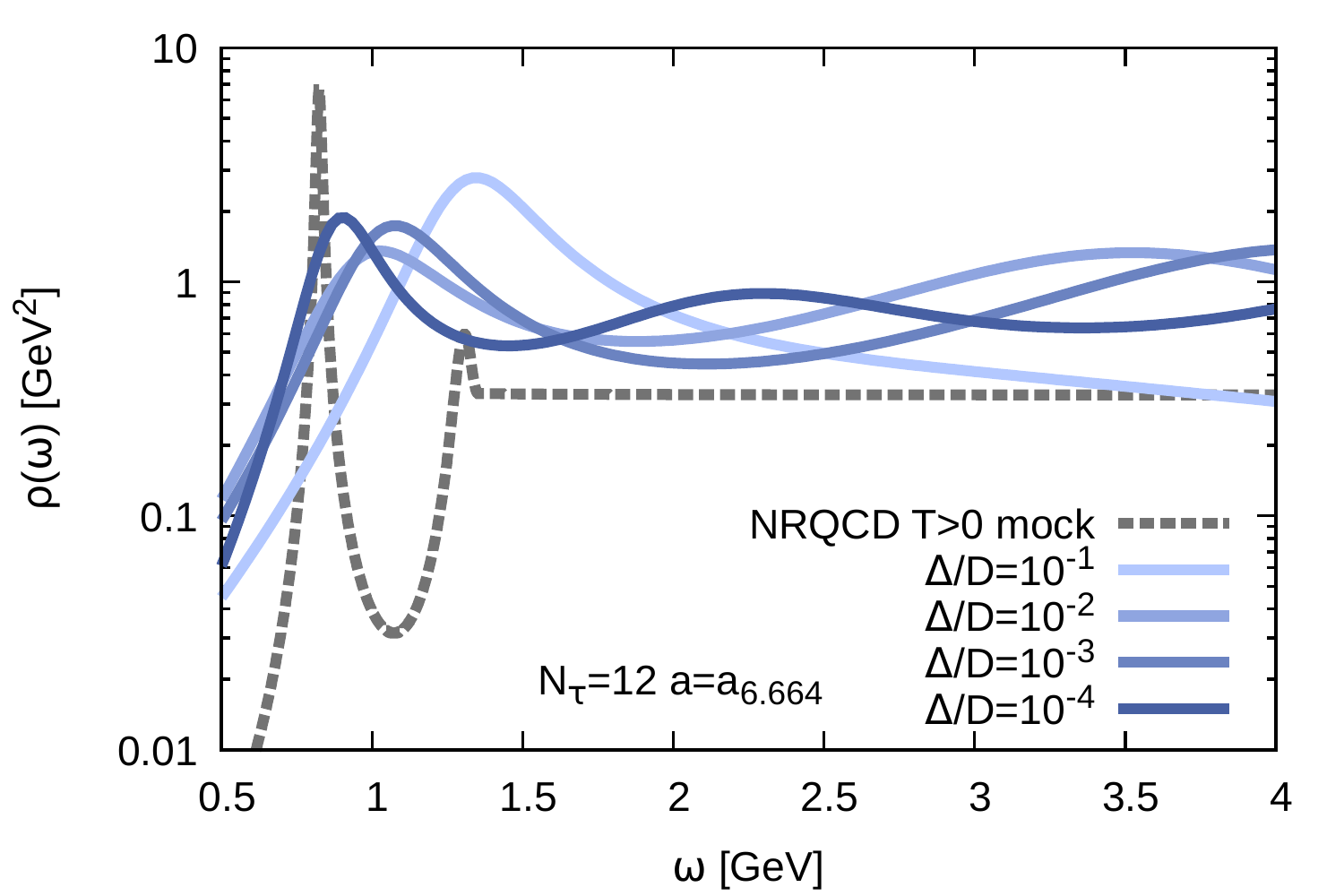}
\includegraphics[scale=0.5]{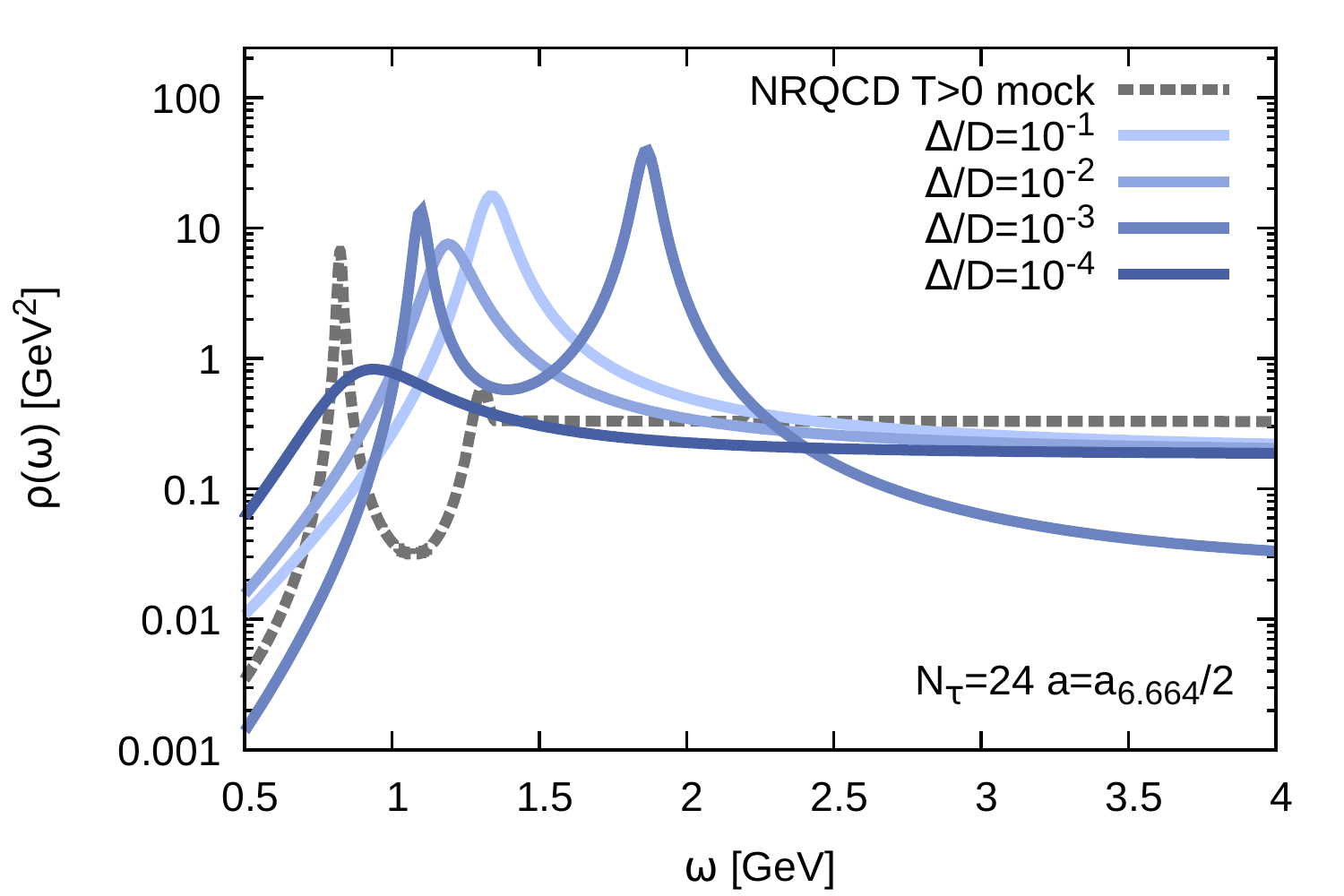}
\caption{Mock data tests on the prospects for improving spectral reconstruction by going closer to the continuum limit. In the top row we show a $T=0$ mock test with the mock spectrum (dashed gray) consisting of four sharp Breit-Wigner peaks ordered according to the PDG $\Upsilon$ channel masses. The lowest peak is located at the frequency found in the raw NRQCD correlator at $\beta=6.664$. in the top left panel the mock spectrum is encoded as $N_\tau=32$ correlator with the same temporal spacing as at $\beta=6.664$. Reconstructions based on input with different relative errors are given as blue solid lines. The top right panel shows the outcome if instead a twice as small lattice spacing and $N_\tau=64$ is used. In the bottom row a challenging $T>0$ scenario is considered with a widened Breit Wigner and a small threshold enhancement. Reconstructions in the bottom right panel arise from correlator data with $N_\tau=12$ and the same spacing as at $\beta=6.664$, while in the bottom right panel twice the resolution and $N_\tau=24$ is used.}\label{Fig:MockDataTests}
\end{figure}

The last part of this study also bears its main quantitative result, the in-medium mass shifts of the different quarkonium ground state particles. Since peak position is a much more robust feature than peak height or area, we were able to present values here with relatively high precision, indicating that the medium leads the former vacuum states to become lighter (Fig.\ref{Fig:FiniteTMassShifts}). An important part of understanding the physics of the in-medium mass shifts is related to using the correct baseline for comparison with the vacuum results, for which we argue one should not use the results from the full $T=0$ spectral reconstruction but instead that from a reconstruction based on the truncated correlator data sets (Fig.\ref{Fig:FiniteTInMediumMasses}). Only in this way the same systematics, related to the same available Euclidean time extent, are present in both reconstructions.

The in-medium mass shifts obtained this way are in qualitative agreement with those from a non perturbative tree-level pNRQCD based computation using the recently determined complex heavy quark potential from the lattice \cite{Burnier:2015tda}. The behavior is opposite to that obtained in weak-coupling pNRQCD which tells us that the underlying scale hierarchy $m v \gg T$ \cite{Brambilla:2010vq} is not applicable here.

Our results are the first ones that conclude on a consistent behavior among different non-relativistic non-perturbative descriptions, i.e. pNRQCD with a lattice QCD potential and direct lattice NRQCD. The reason lies in the careful construction of an appropriate $T=0$ baseline, the raw values for the in-medium masses are qualitatively consistent with those found in previous works.

On the one hand our results are encouraging: we are able to extract quarkonium spectral functions with a level of precision, sufficient that relevant quantitative information such as the in-medium mass shifts of the ground state can be robustly estimated. 
However, the present calculations do not constrain the shape of the spectral function very well.
Therefore we critically discuss in the next section  the prospects for improving on the results presented here.

\subsection{Future prospects}
\label{Sec:FutureProsp}

When contemplating how to improve spectral reconstructions there are two routes to consider. On the one hand we may attempt to further push the improvement of Bayesian techniques. We now understand that ringing affects both the MEM, as well as the BR method and constitutes an issue that needs to be tackled. In this study e.g. we have deployed a recently developed smooth version of the BR method, which however is more of a cure for symptoms than of the origin of the ringing deficiency it is supposed to eliminate. Therefore it can only constitute an intermediate step towards a novel Bayesian approach, with a regulator specifically designed to avoid such artifacts. 

Ringing in the spectral function of two-point correlation functions may be understood to arise from poles artificially populating regions of the complex plane inadmissible in physical field theory. Therefore we need to ask how our knowledge about this analytic structure of correlation functions can be encoded in the language of the prior probability, i.e. in the form of a regulator functional. Recent studies of the spectral function of gluons \cite{Cyrol:2018xeq} have e.g. proposed to introduce different types of functional search spaces, encoding the admissible structure of poles in the complex plane. At this stage this is an external constraint added to the reconstruction and further work is necessary to understand how it can be implemented in a genuine functional language.

The second route to consider is an improvement in the quality of the available simulation data. This in particular means how many points along Euclidean time are available and with what size of statistical errors are they known. 
In NRQCD the spatial lattice spacing, $a_s$  cannot be too small. To avoid large radiative corrections 
$a_s M$ has to be sufficiently large. Therefore, in order to have large $N_{\tau}$ one has to use anisotropic lattice
with temporal lattice spacing $a_{\tau} < a_s$. Efforts in this direction have been undertaken by FASTSUM collaboration
using lattice simulations with Wilson fermions at larger than the physical pion mass. A dedicated high statistics calculation
with physical quark masses and large anisotropies would be desirable. One should be aware, however, that
increasing $N_{\tau}$ alone cannot solve all the problems with reconstruction of the spectral functions.
Ground state properties, such as in-medium mass and width are encoded in the low $\omega$ region of the spectral
function, which in turn is sensitive to the large $\tau$ behavior of the Euclidean correlation functions. The Euclidean
temporal extent in physical units is limited by the inverse temperature and becomes small when the temperature is high.
Therefore, resolving bound state properties at high temperatures will be always more difficult compared to the case of
zero and low temperature, no matter how large $N_{\tau}$ is. Finally to benefit from the larger number of time slices 
available for Bayesian reconstruction eventually also the statistical error should be reduced. Adding additional data points on
the correlation function only helps when the data at two neighboring time-slices are different by few standard deviations.

To understand the various issues listed above we have carried out mock data tests shown in Fig.\ref{Fig:MockDataTests}. 
In the top row we have used a mock spectrum consisting of four Breit-Wigner peaks with the physical 
spacing between the Upsilon channel bound states. 
The position of the lowest peak has been chosen 
to agree with that found in NRQCD at $\beta=6.664$. 
We omit any continuum like structure in the UV in this first test.
In the top right panel we have encoded the mock spectrum in a correlator of $N_\tau=32$ points using the same lattice spacing as at $\beta=6.664$. We find, consistent with our study, that with a relative error of $\Delta D/D=10^{-3}$ in the input data, we can capture the ground state peak quite well and get a partial handle on the first excited state. Interestingly, even with  $\Delta D/D=10^{-4}$ the third peak 
cannot be resolved.
Now if we encode the same spectrum with a twice as fine resolution along Euclidean time 
and hence $N_\tau=64$ we obtain the results shown in the top right panel. We find that the reconstruction does not improve significantly,
the third and fourth peaks cannot be resolved. Thus to obtain information about second and higher excited states even larger
($N_{\tau}>64$) lattices should be used. The above mock data test as well as the calculation performed in  the present study are
based on point like meson sources. One way to get a better control over the excited states in such a vacuum-like scenario would be the use of extended sources.

Let us proceed with an even more challenging mock data test, shown in the lower two panels of Fig.\ref{Fig:MockDataTests}. Here we use a mock spectrum consisting of a widened ground state peak and a tiny threshold enhancement on top of a well pronounced continuum, extending into the UV. When reconstructed from Euclidean data with $N_\tau=12$ with the same resolution as for $\beta=6.664$ we find for realistic $\Delta D/D=10^{-3}$ the expected shift to higher frequencies, as well as a failure to capture the continuum. 
Now if in this case we increase the resolution, the reconstructions of the lower right panel ensue. 
We do not see significant improvement from using $N_{\tau}$ unless the relative errors are $10^{-4}$. This corroborates
our previous statement that increase in $N_{\tau}$ should be accompanied by decrease in statistical errors. When 
the relative error is at the level of $10^{-4}$ the Bayesian reconstruction correctly reproduces the continuum part of the spectral
function. Thus, increasing $N_\tau$ will eventually solve the ringing problem in the Bayesian analysis. On the other hand the description
of the spectral function in the peak region did not improve when increasing $N_{\tau}$ from $12$ to $24$. Thus, for any improved
description of the ground state peak even larger $N_{\tau}$ should be used.

In summary, dedicated NRQCD calculations using anisotropic lattices with large $N_{\tau}$ and physical pion
masses have the potential to improve the current situation. In particular the ringing problem can be significantly reduced by using larger $N_{\tau}$.
However, the determination of quarkonium bound state properties at high temperature will be still challenging. 

On the other hand, NRQCD calculations
can provide very precise information on the Euclidean correlation functions. This information can be used to constrain pNRQCD
calculations and a combination of NRQCD and pNRQCD analyses, as already hinted at in the discussion of Fig.\ref{Fig:FiniteTCorrelatorRatioCmpModel} in this manuscript, should lead to much improved understanding of in-medium quarkonium properties.

\begin{figure}[t]
\includegraphics[scale=0.5]{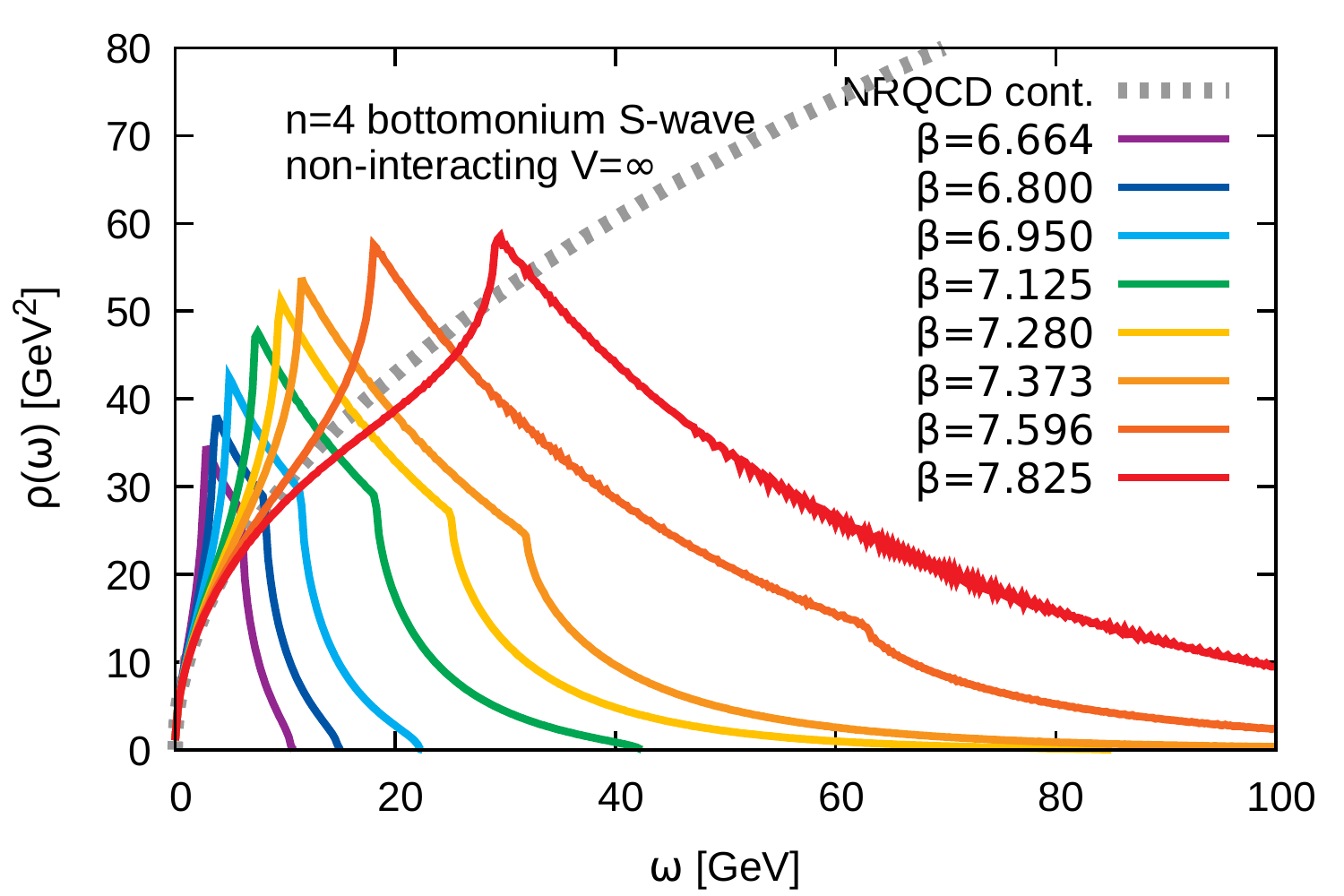}
\includegraphics[scale=0.5]{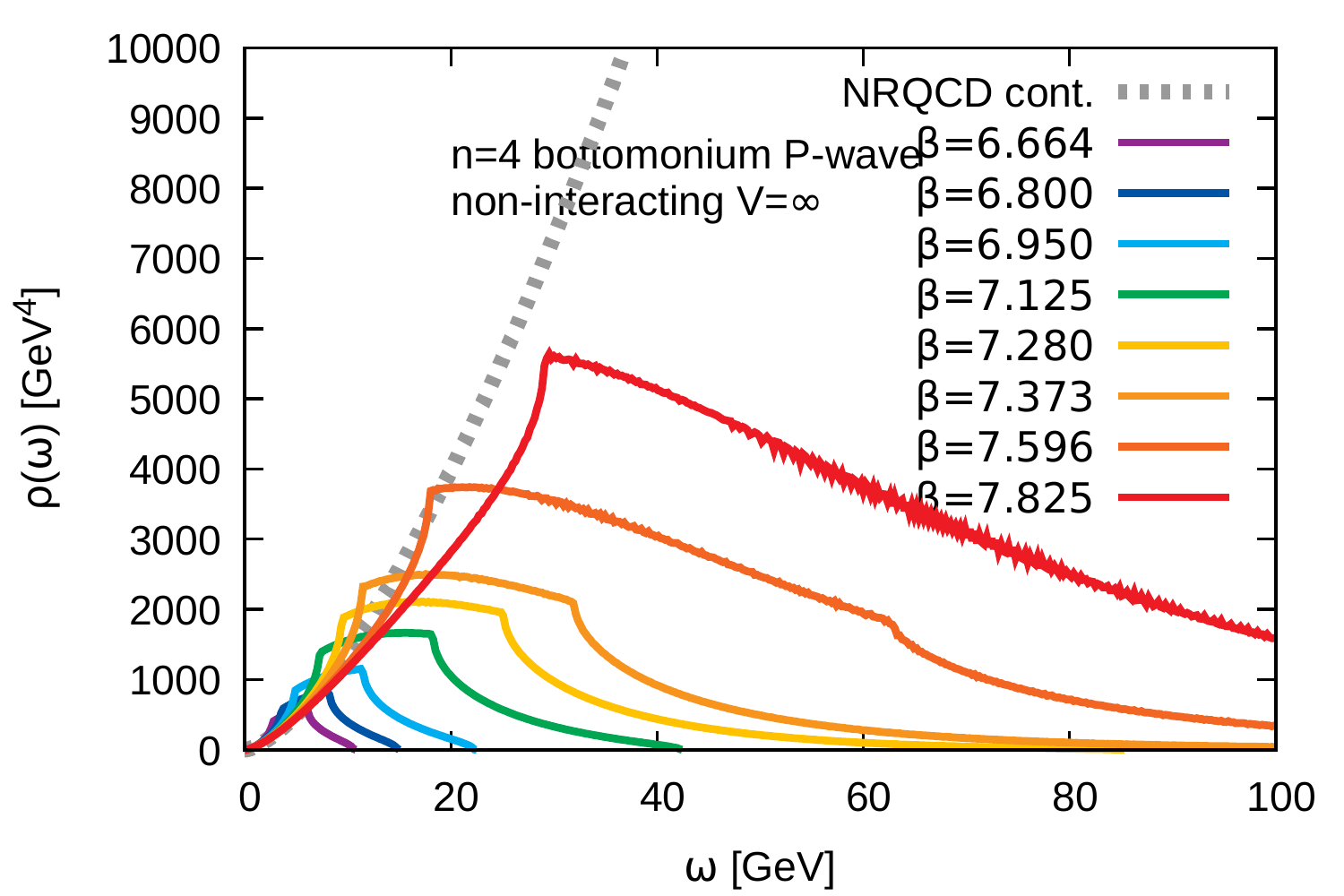}
\caption{ A selection of Lattice NRQCD bottomonium spectral function in the absence of 
interactions in the thermodynamic limit at Lepage parameter $n=4$.  
The S-wave channel is shown on the left,
the P-wave channel on the right.The curves correspond to a representative selection of
the values of $a_s M_b$ labeled by the corresponding $\beta$ values of the HotQCD ensembles.}\label{fig:FreeBottomInfVol}
\end{figure}

\acknowledgments

S.K. is supported by the NRF grant NRF-2018R1A2A2A05018231. PP is supported by the U.S. DOE under contract No.DE-SC001270 and A.R. acknowledges partial support by the DFG collaborative research center SFB1225 ISOQUANT. The authors furthermore acknowledge two USQCD computing time grants at the JLAB facility.

\appendix

\section{Free Theory Computations}
\label{app:freetheory}

Determining the quarkonium spectra in the absence of interactions allow us
to learn vital aspects of the lattice discretized theory of NRQCD, which also affect 
the interacting spectrum. In particular it elucidates the deformation of the UV regime. 
In a relativistic lattice theory the spectrum can maximally be populated up to 
to a limiting value \cite{Karsch:2003wy,Aarts:2005hg}, e.g. for Wilson fermions 
\begin{align}
\frac{\omega_{\rm max}^{\rm Wilson}}{T}\approx 2 n_\tau{\rm log}\Big[ 1+\frac{6+am}{\xi}\Big].
\end{align}
The behavior in a non-relativistic theory such as NRQCD is very different, in 
particular since no naive continuum limit exists.

The EFT approach instead relies on the separation of scales between
$\Lambda_{\rm QCD}/M_b\ll1$, $T/M_b\ll1$ and $\mathbf{p}^2/2M_b\ll1$.
For each ratio of these scales an individual effective theory is formulated,
which is distinguished by the actual values of the Wilson coefficients entering
its Lagrangian. If we wish to approach the continuum limit on the lattice 
the corresponding EFT would receive ever larger contribution to the Wilson
coefficients as higher and higher scales need to be integrated out.

\begin{figure}[t]
\includegraphics[scale=0.5]{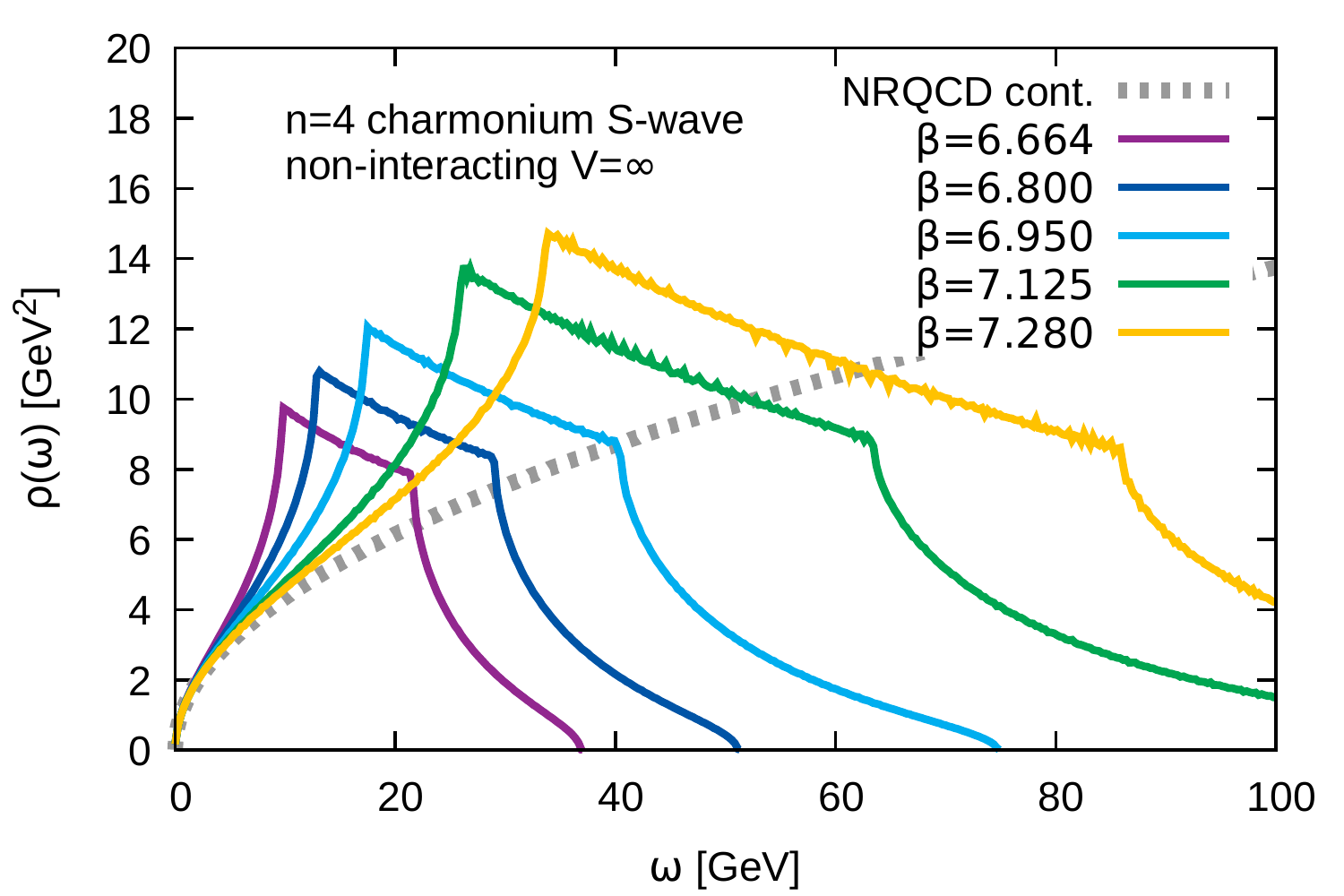}
\includegraphics[scale=0.5]{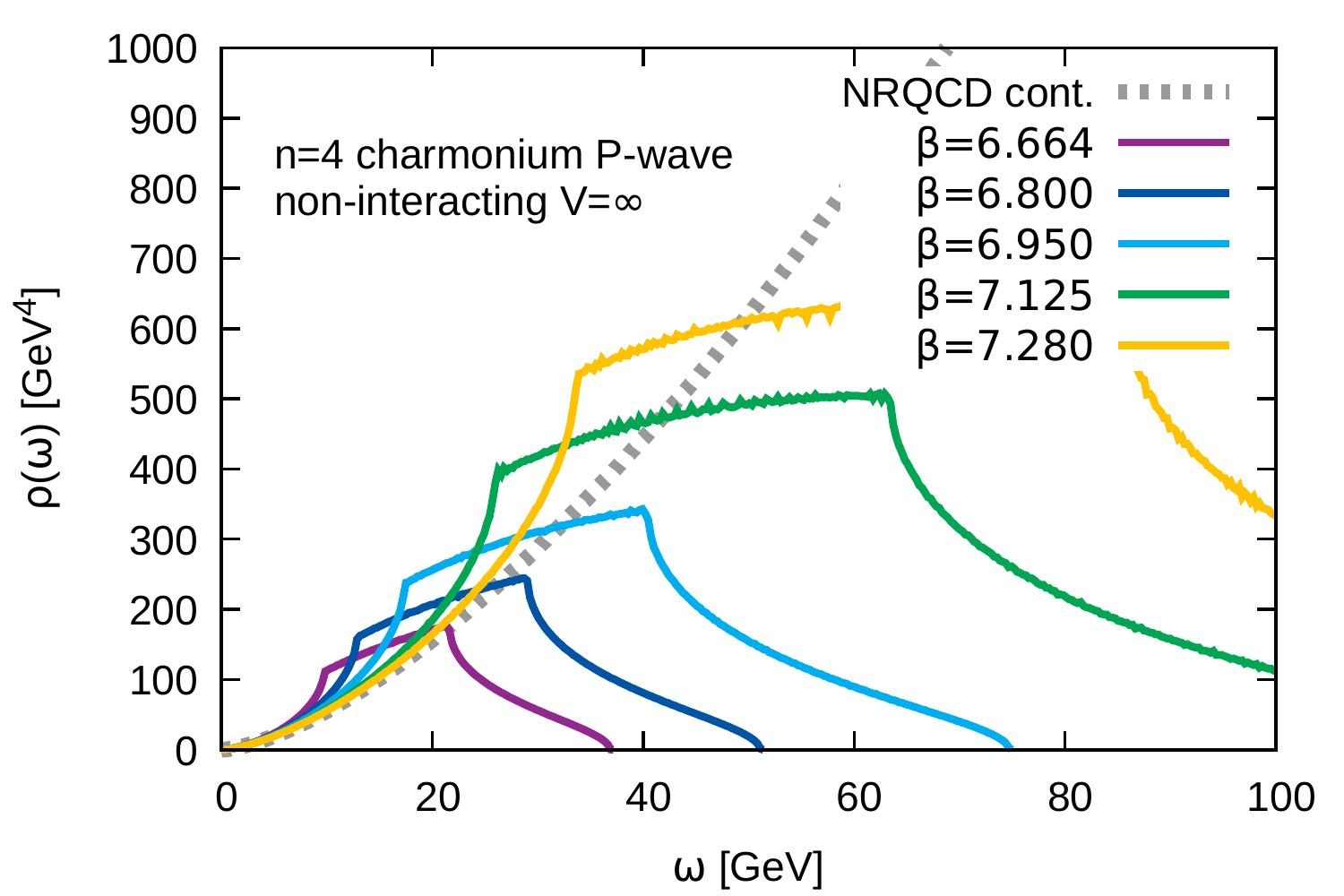}
\caption{ A selection of lattice NRQCD charmonium spectral function in the absence of 
interactions in the thermodynamic limit  at Lepage parameter $n=8$.  
The S-wave channel is shown on the left,
the P-wave channel on the right.}\label{fig:FreeCharmInfVol}
\end{figure}
\begin{figure}[t]
\includegraphics[scale=0.5]{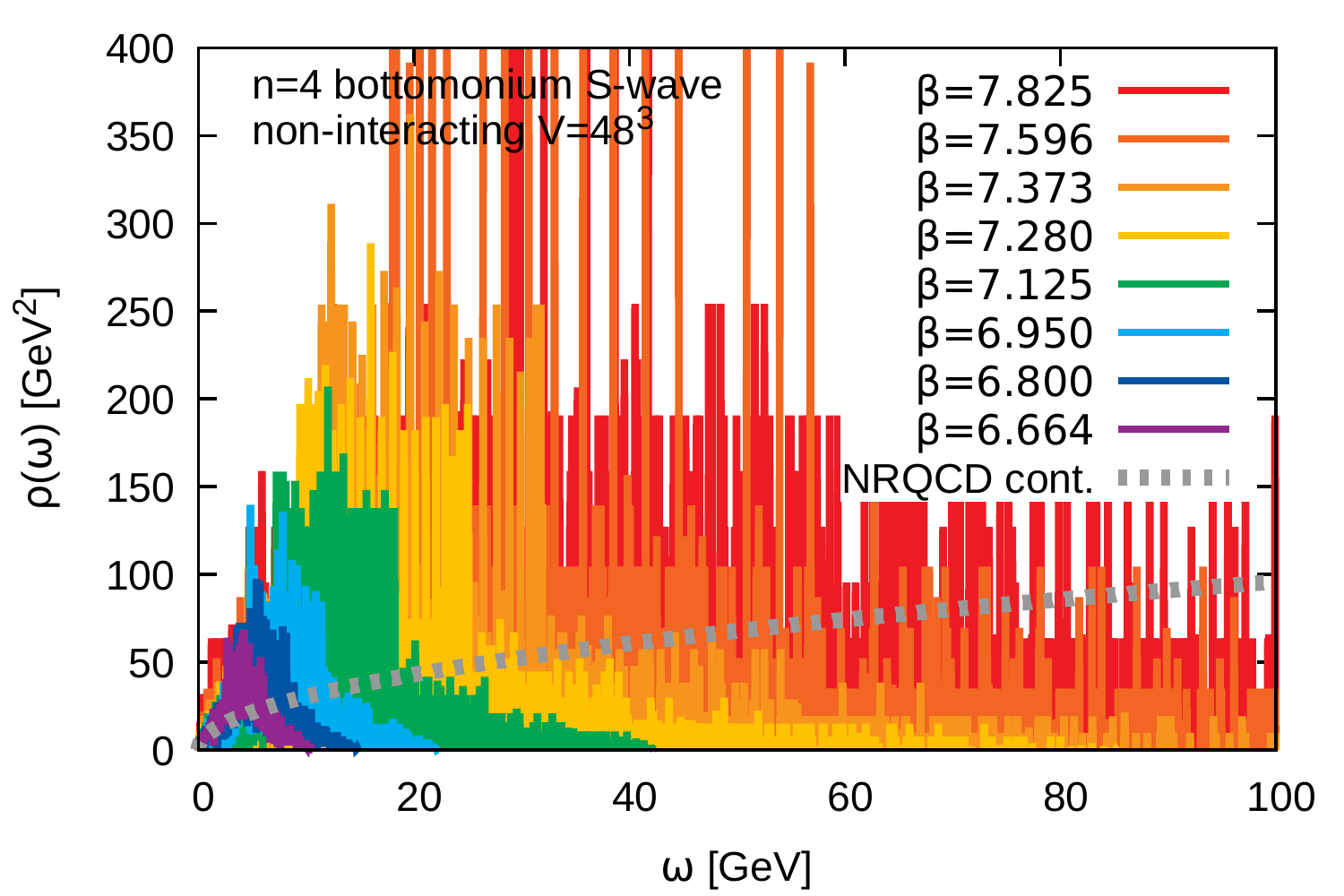}
\includegraphics[scale=0.5]{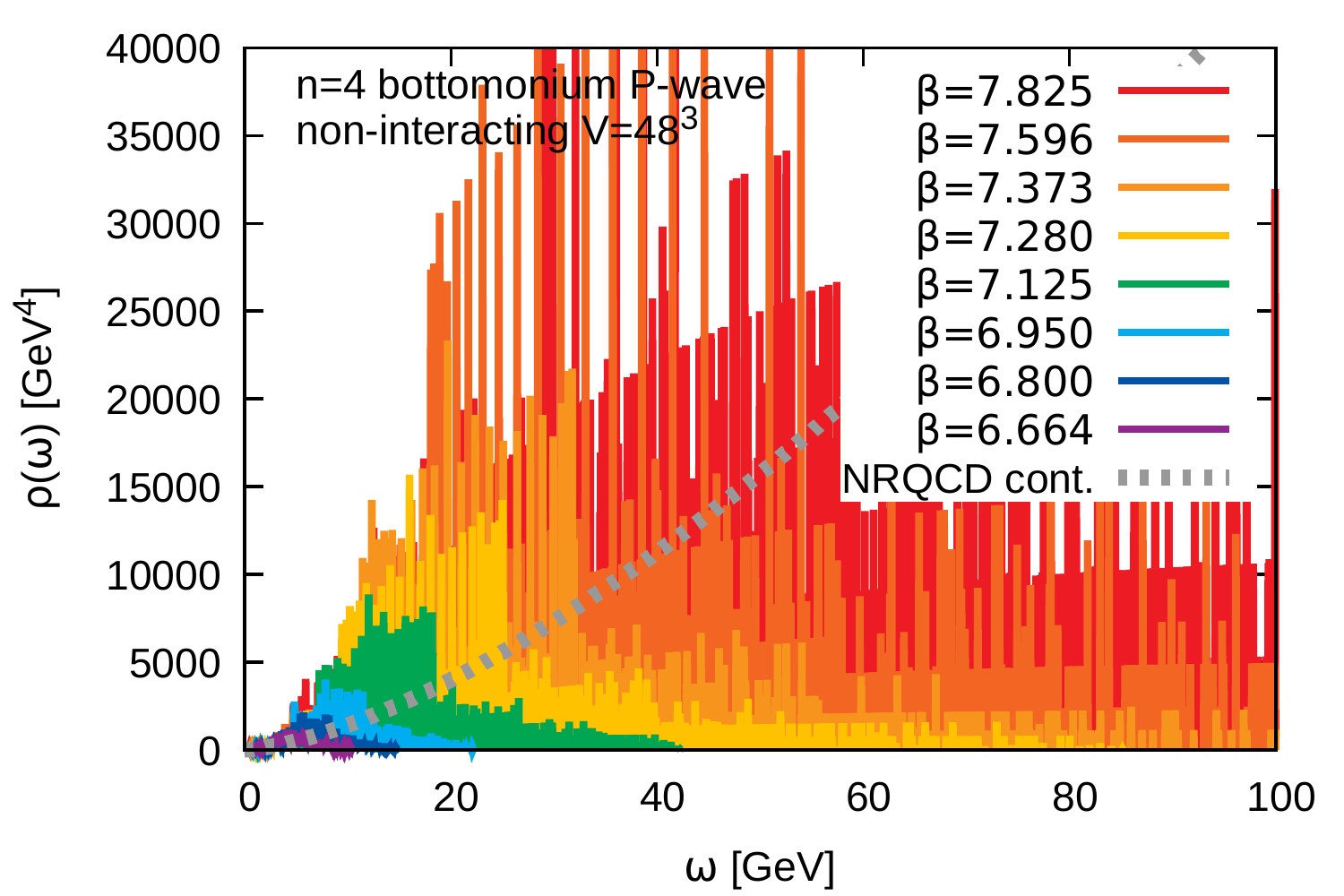}
\includegraphics[scale=0.5]{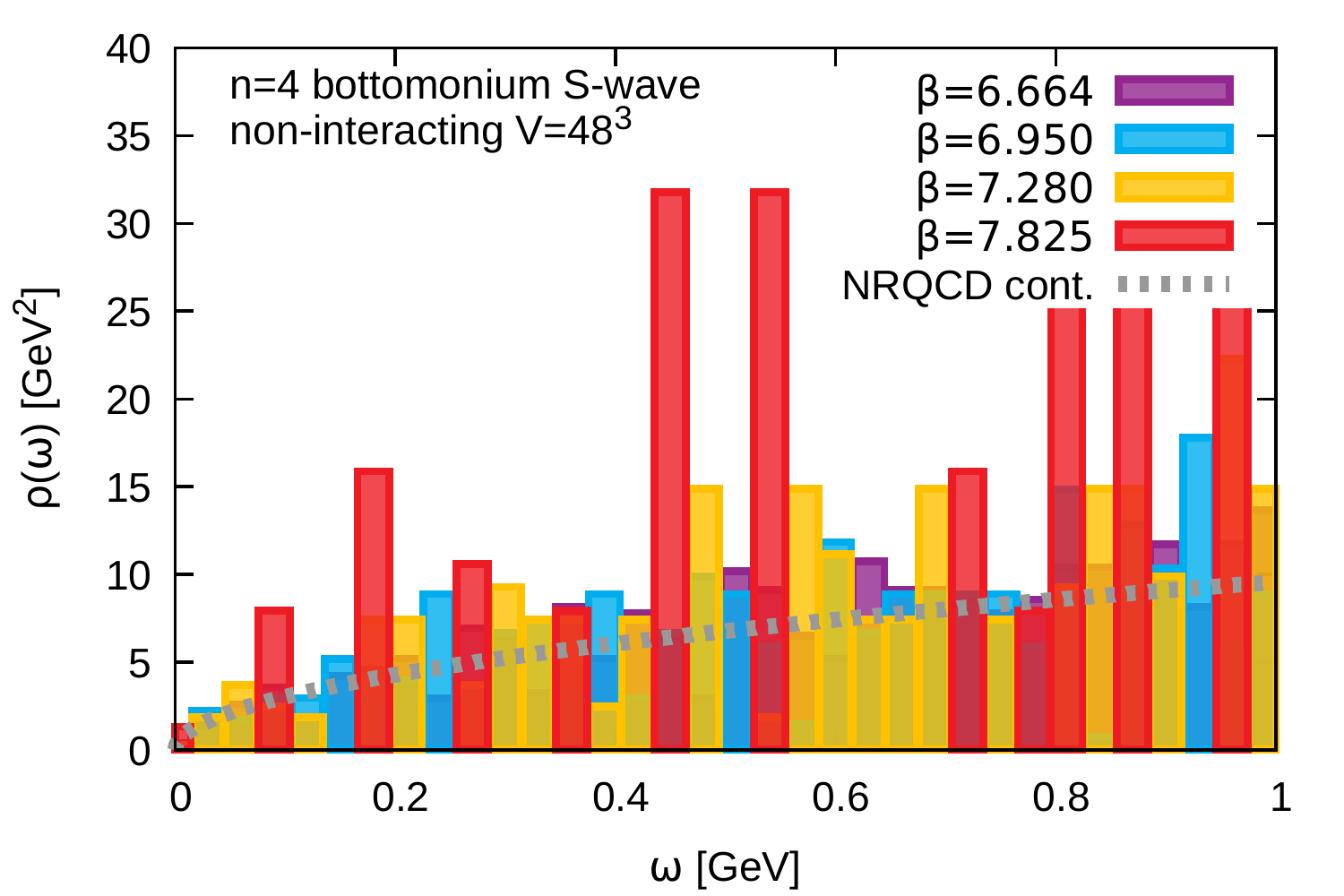}
\includegraphics[scale=0.5]{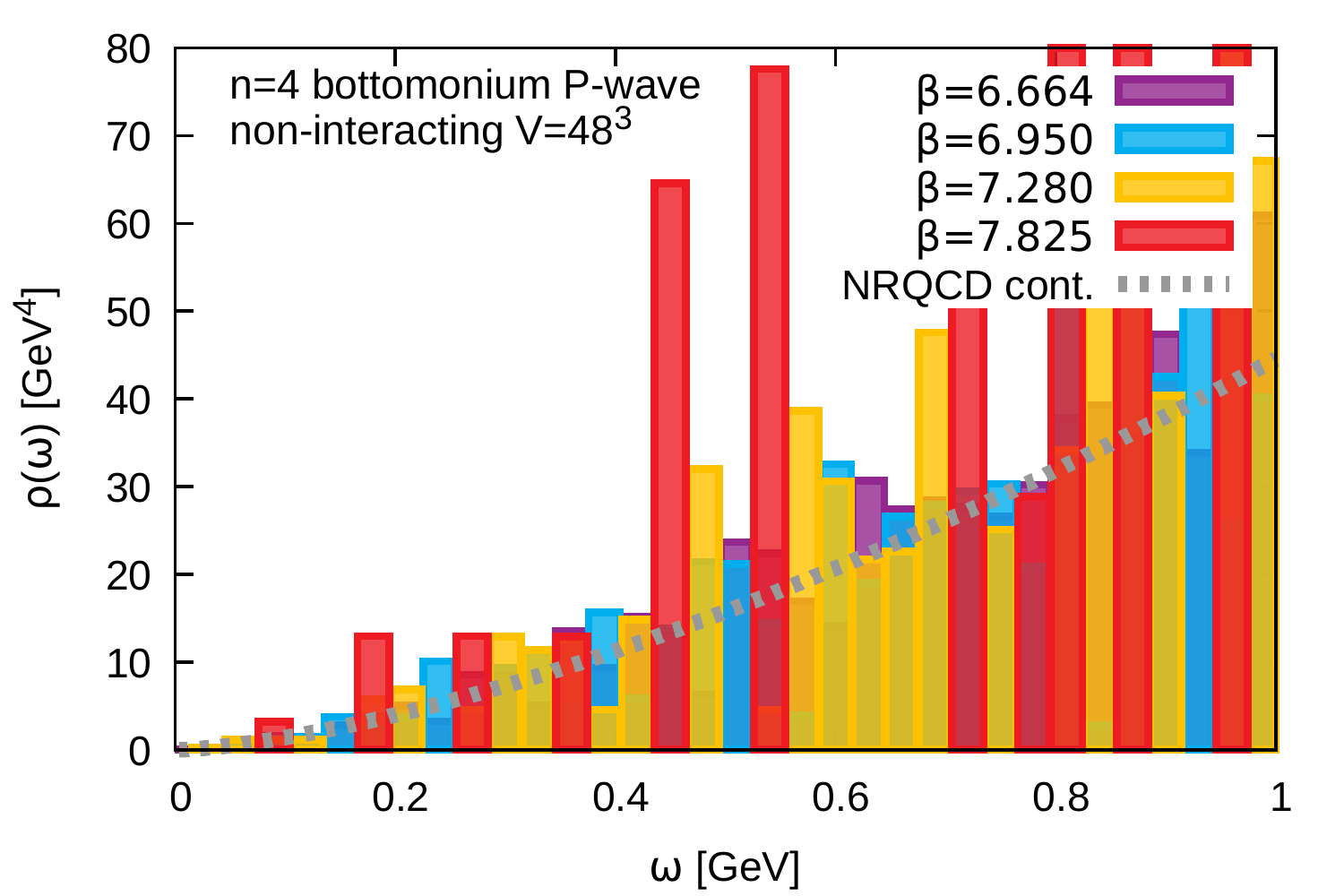}
\caption{ A selection of lattice NRQCD bottomonium spectral function in the absence of 
interactions on $48^3$ lattices at Lepage parameter $n=4$.  
The S-wave channel is shown on the left,
the P-wave channel on the right.
The different colors correspond to a representative selection of
the values of $a_s M_b$ labeled by the corresponding $\beta$ values of the HotQCD ensembles. (top) The full frequency range populated by the free spectral functions is shown, note its extrusion up to very high frequencies for fine lattices. (bottom) Zoom-in of the low frequency regime. Here we wish to emphasize that the finer the lattice is, the fewer modes are available as constituents of possible spectral features.}\label{fig:BottomActualVol}
\end{figure}

On the lattice we may formulate the scale separation for the spatial momenta
in terms of their lattice representation in the first Brillouin zone
\begin{align}
 \hat{\mathbf{p}}^2=4 \sum_{i=1}^{3}{\rm sin}^2\Big(\frac{\pi n_i}{N_s}\Big), \quad n_i=-\frac{N_s}{2}+1,\ldots,\frac{N_s}{2}
\end{align}
and the so called effective mass parameter $\hat{M}=2n \xi a_s M_b$.
$a_s$ denotes the spatial lattice spacing, $\xi$ the renormalized anisotropy
and $n$ the Lepage parameter used to discretize the time evolution of the
propagator in a simple forward scheme in \eqref{NRQCDEvolEq}.

In this Euler-like scheme it was shown \cite{Davies:1991py} that 
on isotropic lattices $(\xi=1)$ a value of $n=1$ leads to a well defined
UV behavior of the evolution if $a_s M_b>3$ or $a_s M_b>1.5$ if $n=2$.
The eventual breakdown of the NRQCD expansion is understood from 
considering the non-interacting dispersion relation discussed in \cite{Aarts:2011sm,Aarts:2014cda}.

Defining the non-interacting lattice Hamiltonian via \eqref{NRQCDEvolEq},
we obtain the free dispersion relation
\begin{align}
a_{\tau}&E_{\hat{\mathbf{p}}} = 2n {\rm Log}\big[1-\frac{1}{2}\frac{\hat{\mathbf{p}}^2}{ 2 n\xi a_s M_b }\big]\label{Eq:NRQCDDispRel}\\
\nonumber  +& {\rm Log}\big[1 + \frac{ (\hat{\mathbf{p}}^2)^2 }{ 16 n\xi (a_s M_b)^2 } +  \frac{ (\hat{\mathbf{p}}^2)^2 }{ 8 \xi (a_s M_b)^3 }  - \frac{ \hat{\mathbf{p}}^4 }{ 24 \xi (a_s M_b) } \big]
 \end{align}
 where the negative sign already in the first term signals that very large 
 energy eigenvalues may be encountered.

The corresponding spectral functions are obtained from \eqref{Eq:NRQCDDispRel}
and take the following form
\begin{align}
 \nonumber \rho_S(\omega)=\frac{4\pi N_c}{N_s^3}\sum_{\hat{\mathbf{p}}}\delta(\omega-2E_{\hat{\mathbf{p}}} ),\\
 \rho_P(\omega)=\frac{4\pi N_c}{N_s^3}\sum_{\hat{\mathbf{p}}} \hat{\mathbf{p}}^2\delta(\omega-2E_{\hat{\mathbf{p}}} )\label{Eq:AnalytFreeSpec},
\end{align}
which needs to be evaluated numerically. As expected we are facing a spectrum
composed out of a large number of delta peaks, which for any finite volume require binning.
Often the free spectral function in the thermodynamic limit is quoted in the literature,
which we plot in  Fig.\ref{fig:FreeBottomInfVol} for bottomonium and in  Fig.\ref{fig:FreeCharmInfVol}
for charmonium. The only parameter entering the computation is $a_s M_q$, for which
we take the values on the HotQCD lattices from Tab.\ref{tab:parameters} and label each curve with the
corresponding value of $\beta$. We would like to note that on the other hand on realistic 
finite lattices, when using a frequency resolution that is equal to that of the frequency bins 
selected in the spectral reconstruction approaches,
the actual spectrum is far from the smooth function at infinite volume. As an example we plot
the free bottomonium spectral functions with a frequency binning of $\Delta \omega=30$MeV
in Fig.\ref{fig:BottomActualVol}. In this case it is much more difficult to make out distinctive features,
such as the kink observed at $V=\infty$.

The normalization of this spectral function differs from that naively computed
on a lattice with unit links, due to the choice of initial sources in the propagator evolution.
Another reason is that the first step in \eqref{NRQCDEvolEq} only contains
the leading order terms of the Hamiltonian.

The choice of $n=4$ for bottomonium and $n=8$ for charmonium can also be
discussed in this context. For the highest temperature at which we consider 
bottomonium $T=2.56T_c$ the relevant product $a_s(\beta=7.825) M_b=0.95$,
which only allows a stable propagator evolution if $n>3$. One may be alerted
by the fact that the naive expansion parameter in this case is larger than unity,
however with the appropriate choice of $n$ the IR properties of the spectrum 
are only mildly affected. On the other hand the UV regime of the
spectrum now contains quite a lot of non-physical contributions and is only
cut off far beyond the original inverse lattice spacing. For charmonium the situation
is worse. To make sure that our choice of a large $n$ does not start to influence
the low momentum regime significantly, we decide to keep $n=8$ and limit the
temperature range considered to the maximum of $T=1.58T_c$ where
$a_s(\beta=7.280) M_c=0.42$ is around half of that of bottomonium at its highest temperature.
Note that if we had over-stepped the stability bound for NRQCD the spectrum
would become populated up to infinitely high frequencies.

For bottomonium we had previously studied the effect of varying $n$ on the interacting spectral 
function \cite{Kim:2014iga} and found that changing from $n=2$ to $n=4$ lead to changes in the spectrum
only above $\omega>1$GeV and even there only very mildly, so that a further increase to
$n=8$ appears unproblematic.

\section{Reconstruction methods survey}
\label{sec:methoddep}

While we have already discussed differences in the spectral reconstruction based on the standard BR method, its smoothed variant with $\kappa=1$ and the MEM, let us compare here to further reconstruction prescriptions discussed in the literature. This includes the MEM with extended search space \cite{Rothkopf:2011ef}, as well as as the Jakovac prescription \cite{Jakovac:2006sf} for the selection of an alternative search space.

Let us begin with the extended MEM. Instead of using only the first $N_\tau$ columns of the $N_\omega\times N_\omega$ sized matrix $U$ arising in the singular value decomposition of the transpose Kernel $K^t=U\Sigma V^t$ it uses a larger number of columns. In that way one may systematically enlarge Bryan's search space eventually recovering the full search space. It has been shown that in general the global extremum of the posterior is not located in Bryan's search space \cite{Rothkopf:2011ef}, while it has been proven to be unique in the full search space \cite{Asakawa:2000tr}. 

\begin{figure}[t]
\includegraphics[scale=0.5]{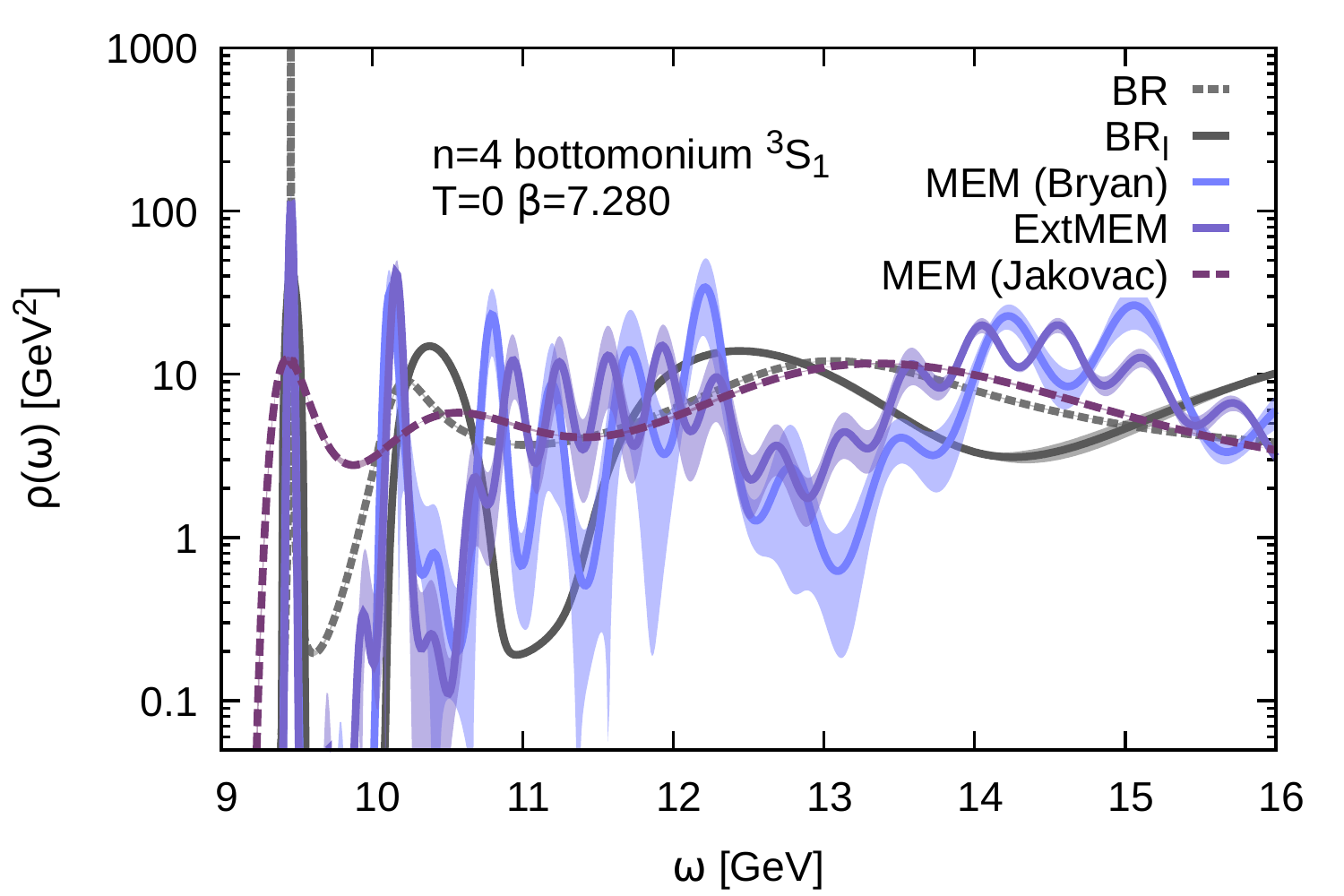}
\includegraphics[scale=0.5]{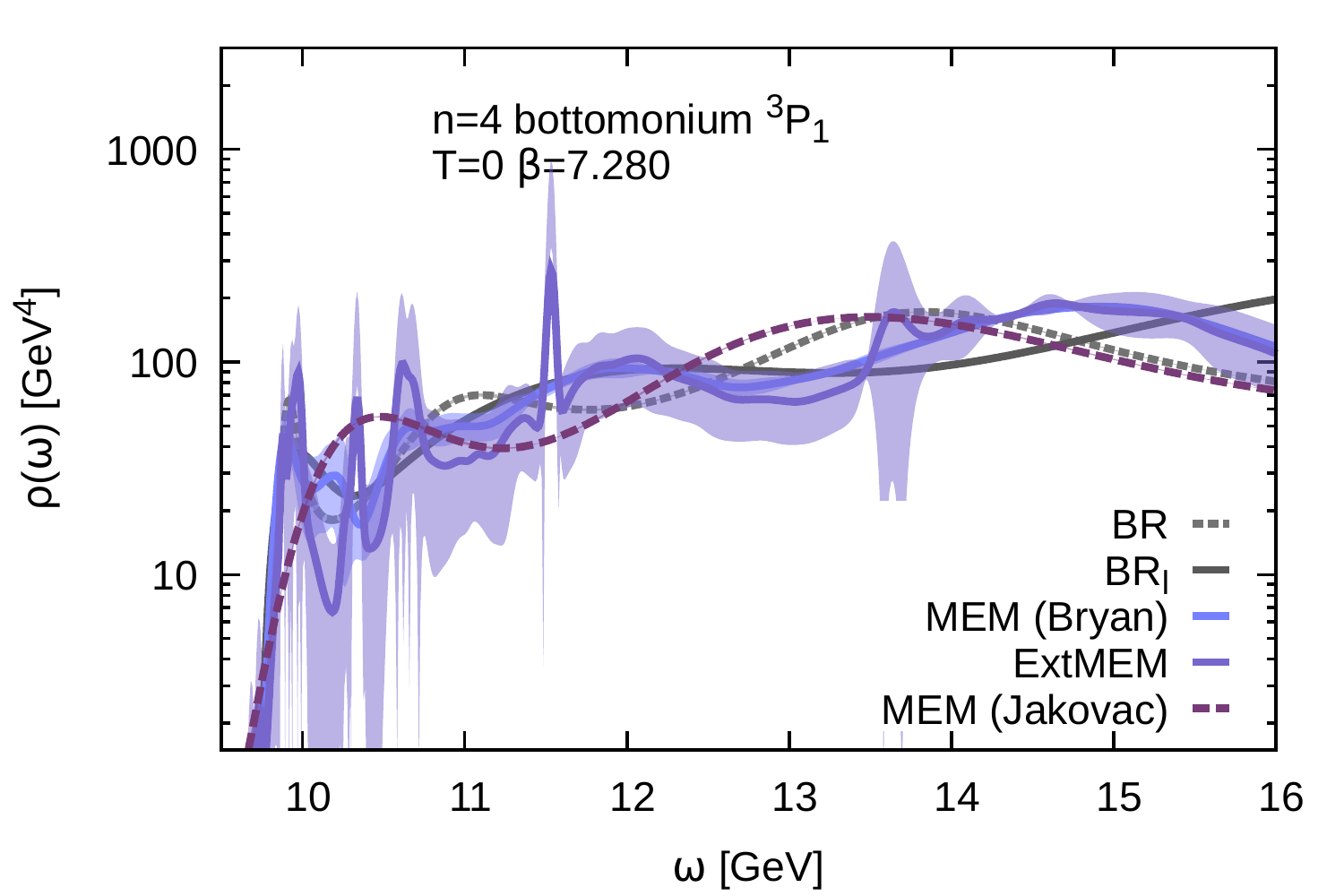}
\caption{Representative examples of the methods dependence of the bottomonium spectral reconstructions at zero temperatures ($\beta=7.280$). We depict the S-wave channel in the top row, the P-wave channel in the bottom row. There are five different methods compiled here: two versions of the BR method, the standard one as light gray dashed line and the smooth BR method as dark gray solid line. They are contrasted to three different MEM realizations, the Bryan implementation (light blue solid), the extended MEM with additional $32$ basis functions (darker blue solid) as well as the Jakovac MEM as dark violet dashed line.}\label{Fig:MethodDepT0}
\end{figure}

As we argued in sec.\ref{sec:RecBayesRes} the Shannon Jaynes entropy is not intrinsically safe from ringing artifacts, just as the BR regulator. It is instead the limited dimension of the search space that manually suppresses wiggly features. This issue is clear in the case of $T=0$ reconstruction compared in Fig.\ref{Fig:MethodDepT0}. We find that with $N_\tau=64$ and our choice of $\omega_{\rm min}^{\rm num}=-0.15$ the MEM manages to capture the ground state signal reasonable well in both the S-wave (top) and P-wave (bottom) channel. On the other hand it also produces highly oscillatory artifacts in the UV, which make the distinction between physical bound state and methods artifact rather difficult.

Now if we further enlarge the search space by e.g. $32$ additional columns of $U$ in the extended MEM, the situation worsens, since the additional basis functions are highly oscillatory and their contribution is not suppressed by the regulator. The relatively large error-bands from the Jackknife re-sampling are a manifestation of this issue. 

As further comparison we carry out the MEM based on the Jakovac prescription for the search space. By solving a self-consistent equation for the form of its basis functions it proposed to use the $N_\tau$ columns of the transpose Kernel themselves as basis functions. If the spectral reconstruction were a genuinely linear problem, this prescription, as well as Bryan's SVD search space should in principle be equivalent. What we find is that the Jakovac prescription produces very smooth reconstructions, shown as dark dashed lines in Fig.\ref{Fig:MethodDepT0}. Interestingly it shows a very similar behavior as the standard BR method in the UV (gray dashed line). On the other hand we find that its ability to reconstruct the ground state position is comparatively weak, leading to either values slightly below (S-wave) or slightly above (P-wave) the value determined consistently by the BR method and the effective mass fits. 

\begin{figure}[t]
\includegraphics[scale=0.5]{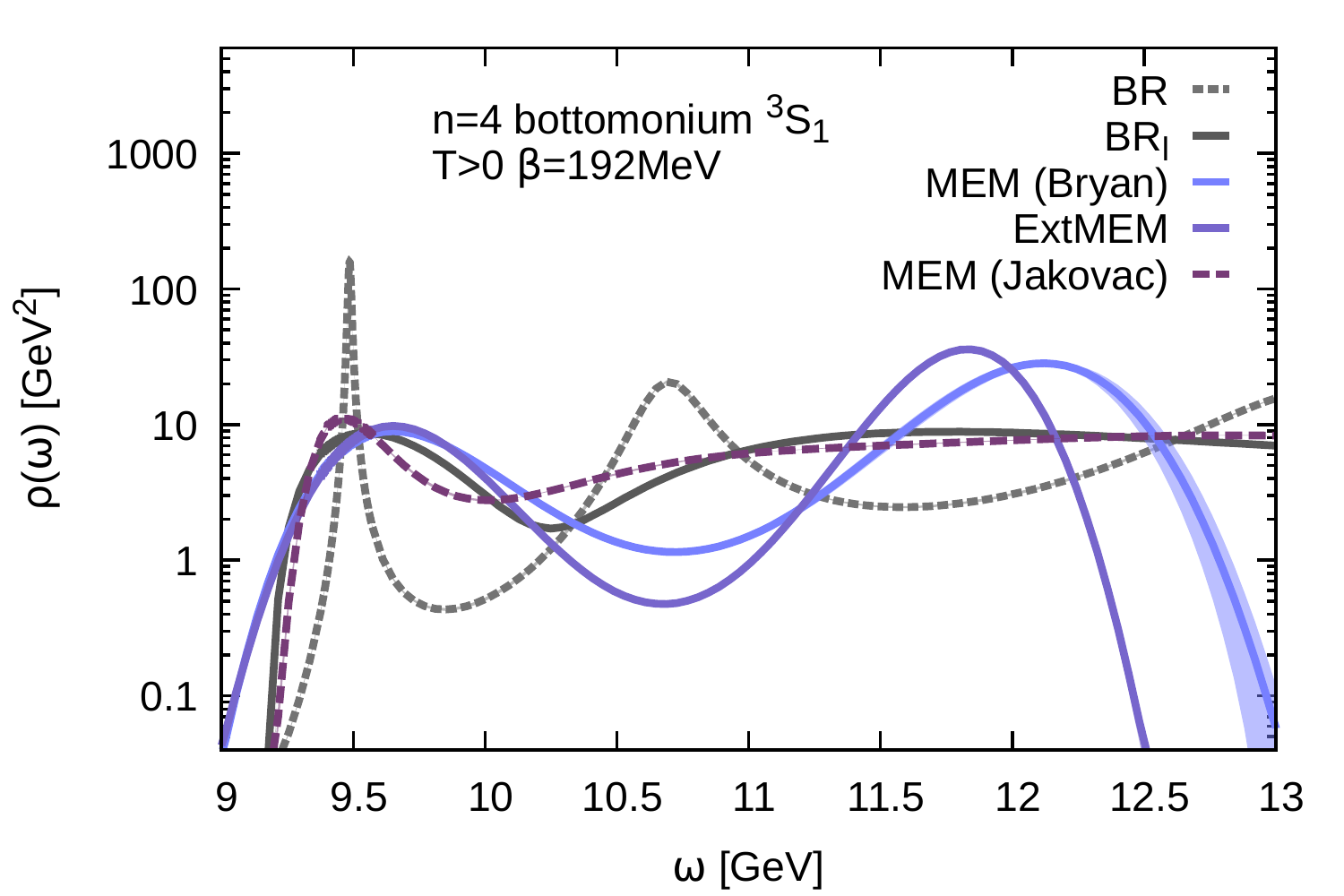}
\includegraphics[scale=0.5]{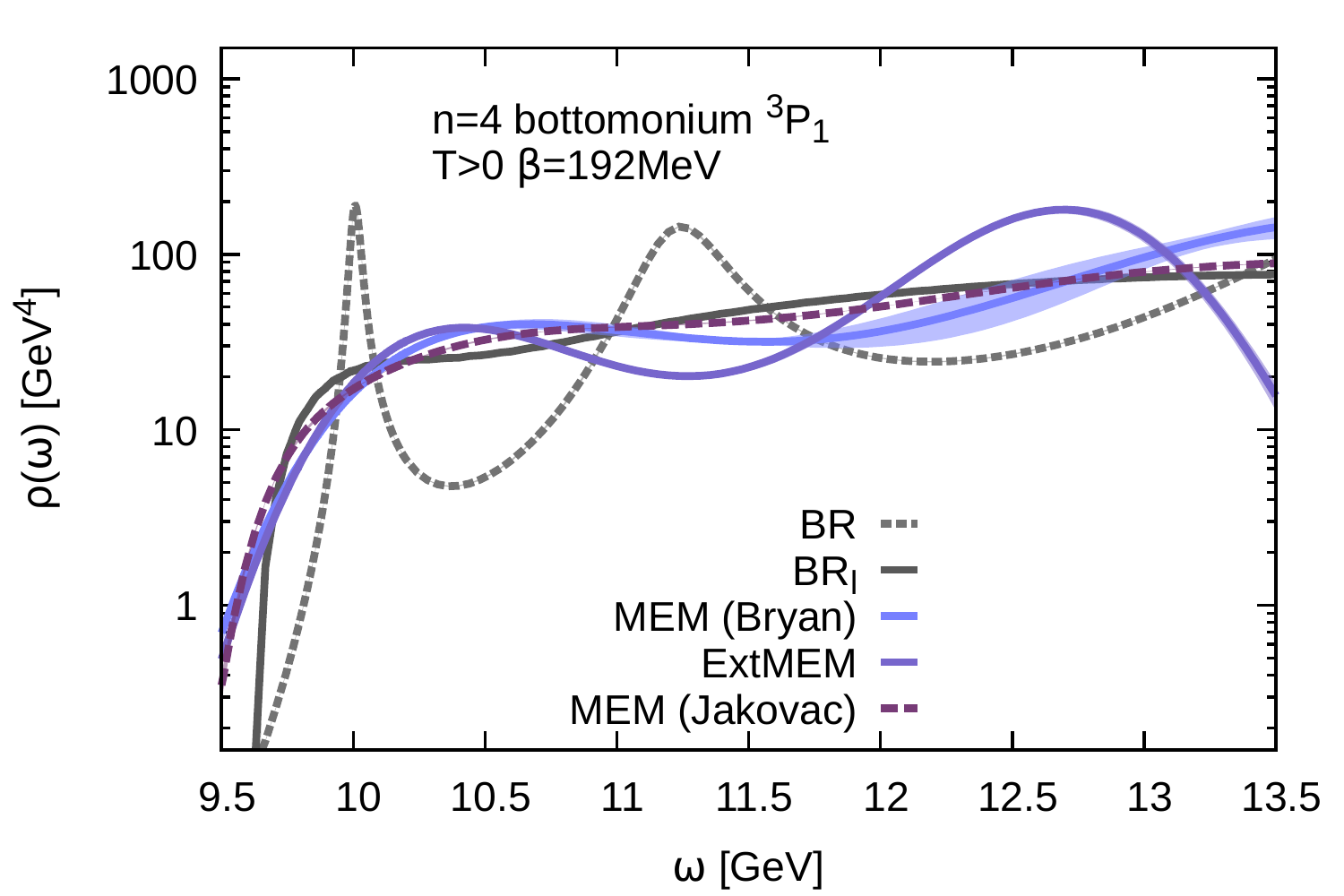}
\caption{Representative examples of the methods dependence of the bottomonium spectral reconstructions at $T=192$MeV. We depict the S-wave channel in the top panel, the P-wave channel in the bottom panel with the same color and line coding as in the $T=0$ comparison.}\label{Fig:MethodDepFiniteT}
\end{figure}

At finite temperature the number of basis functions is significantly reduced compared to the $T= 0 $ case. As is to be expected and visible in Fig.\ref{Fig:MethodDepFiniteT}, the MEM thus also produces smooth results. Adding additional columns of $U$ to the search space in the extended MEM does not yet lead to the appearance of significant wiggling in the range of frequencies considered here. In the S-wave case (top) the Bryan and extended MEM show a similar behavior as the smooth BR method at small frequencies but start to show wiggles at higher frequencies. The Jakovac method on the other hand shows a very similar UV behavior as the smooth BR method, however its ground state peak sits at a position even below the agreed upon $T=0$ position of the peak. 

In the P-wave channel at high temperatures all MEM implementations show a similar behavior at low frequencies, again with a systematic pull of the lowest lying structure to higher frequencies than what is picked up by the smooth BR method. Adding more basis functions the Bryan's search space in this case leads to the appearance of stronger wiggly structures in the UV.

The results of this section tell us that on the one hand extending the MEM search space beyond Bryan's prescription will lead to an increased presence of ringing artifacts, as the Shannon Jaynes prior is unable to suppress unphysical wiggles by itself. Secondly the Jakovac implementation of the MEM seems to provide a well controlled reconstruction of the UV features of the spectra, however it comes at the cost of a distorted reconstruction at small frequencies, which prevents its use as precision tool for the determination of ground state masses. As a tool for the qualitative elucidation of spectral features it still remains a valid option, as it does not seem to suffer from significant ringing artifacts.

For a quantitative investigation of spectral features however the combination of smooth and standard BR method seems to be the current best option. The former prevents ringing and thus makes possible a robust qualitative assessment of whether genuine bound state features are present, while the latter then permits the most accurate determination of their properties.

\subsection*{Hyperparameter marginalization in the MEM}
The MEM based reconstruction, at low and finite temperature, shown in the main part of this text use the standard Bryan approach, in which the most probable spectrum is obtained from a scan through the hyperparameter $\alpha$. For each spectrum reconstructed for a specific value of $\alpha$, a probability is computed over which the results are finally averaged. In this appendix we collect for completeness representative examples of these probability distributions.

\begin{figure}[t]
\includegraphics[scale=0.5]{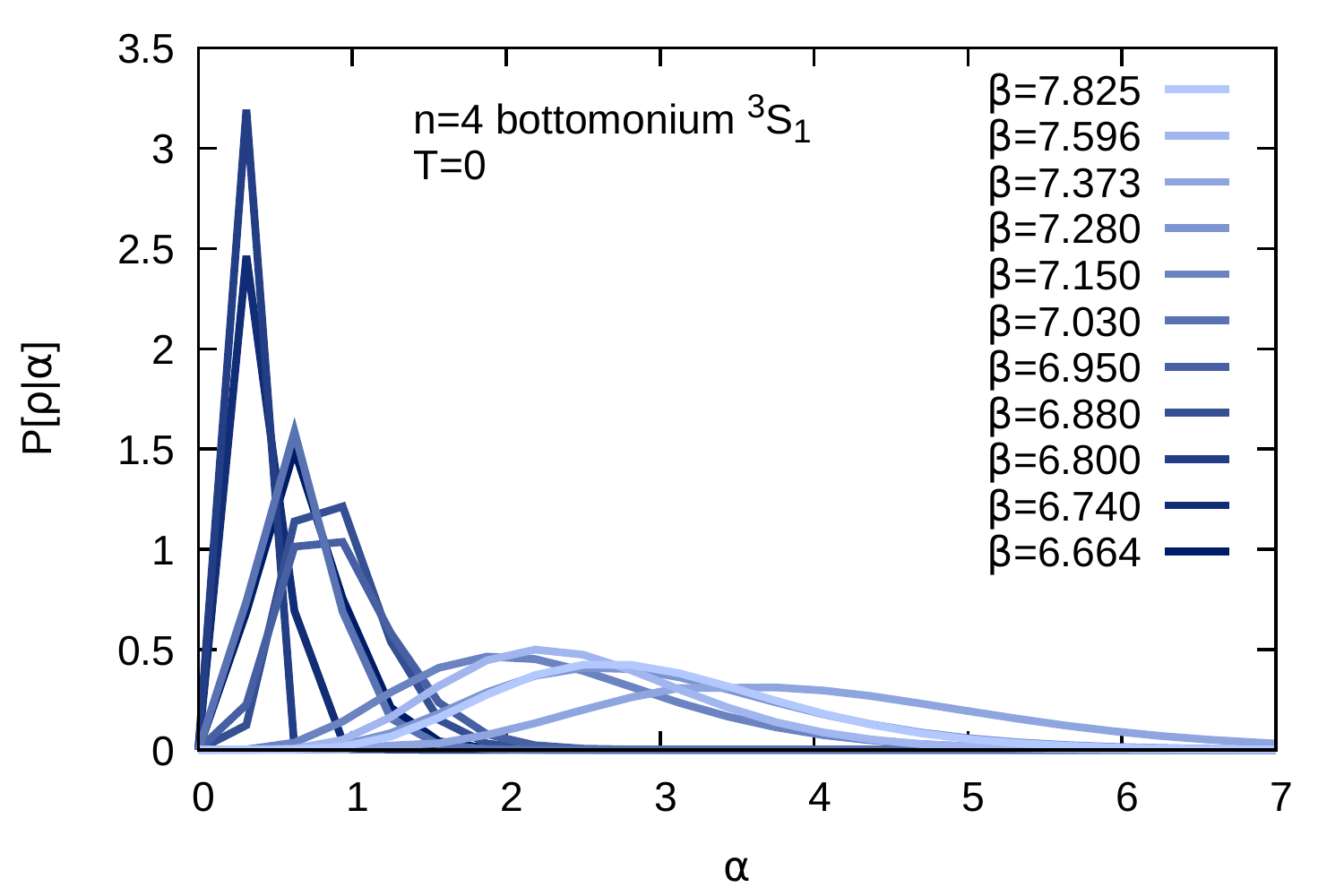}
\includegraphics[scale=0.5]{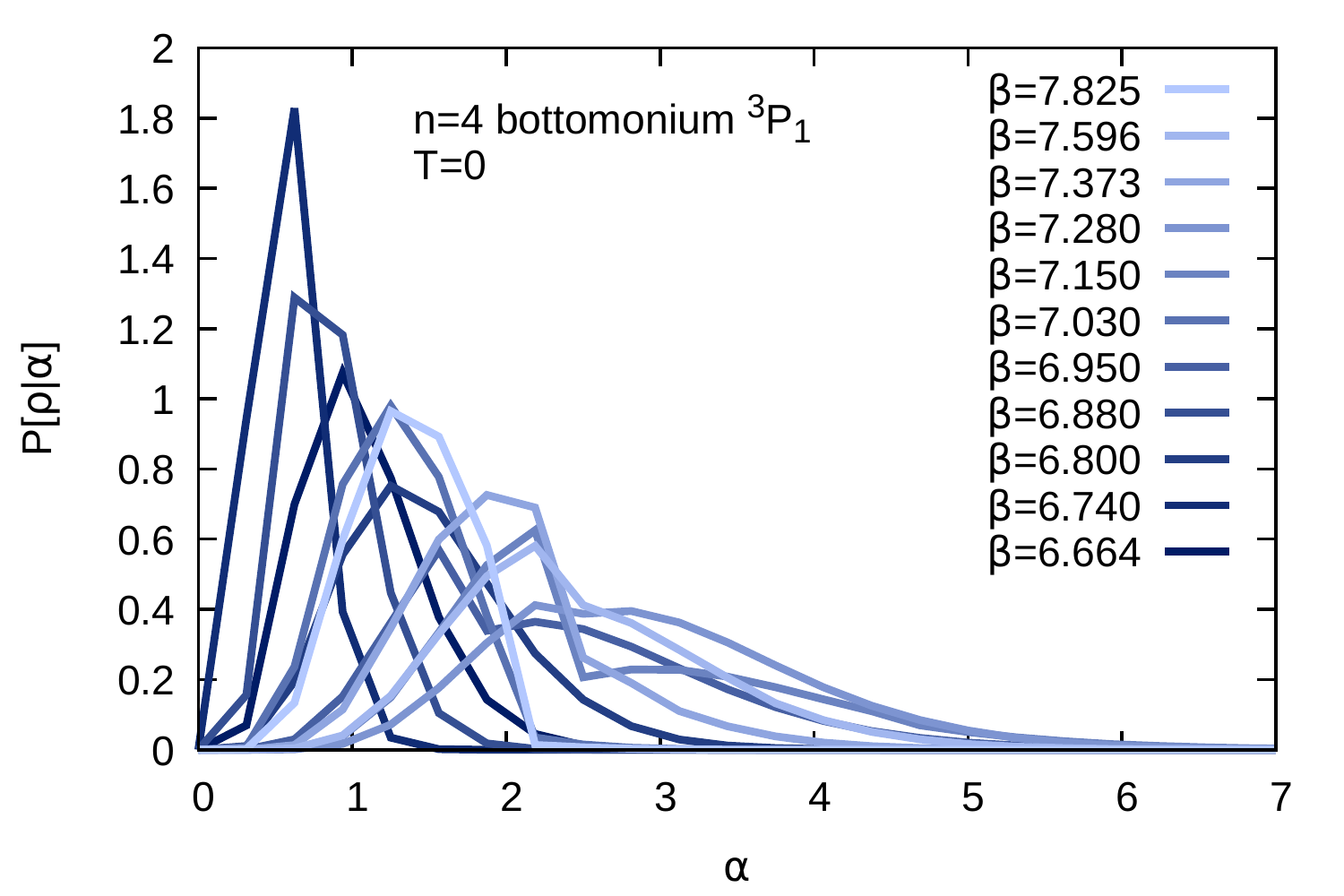}
\caption{Representative examples of the probability distribution over the hyperparameter $\alpha$ used in the self consistent determination of the most probable spectrum at zero temperatures. We show here bottomonium results, the top row contains the S-wave results, the bottom row the P-wave channel.}\label{Fig:ZeroTMEMAlphasBottom}
\end{figure}

\begin{figure}[t]
\includegraphics[scale=0.5]{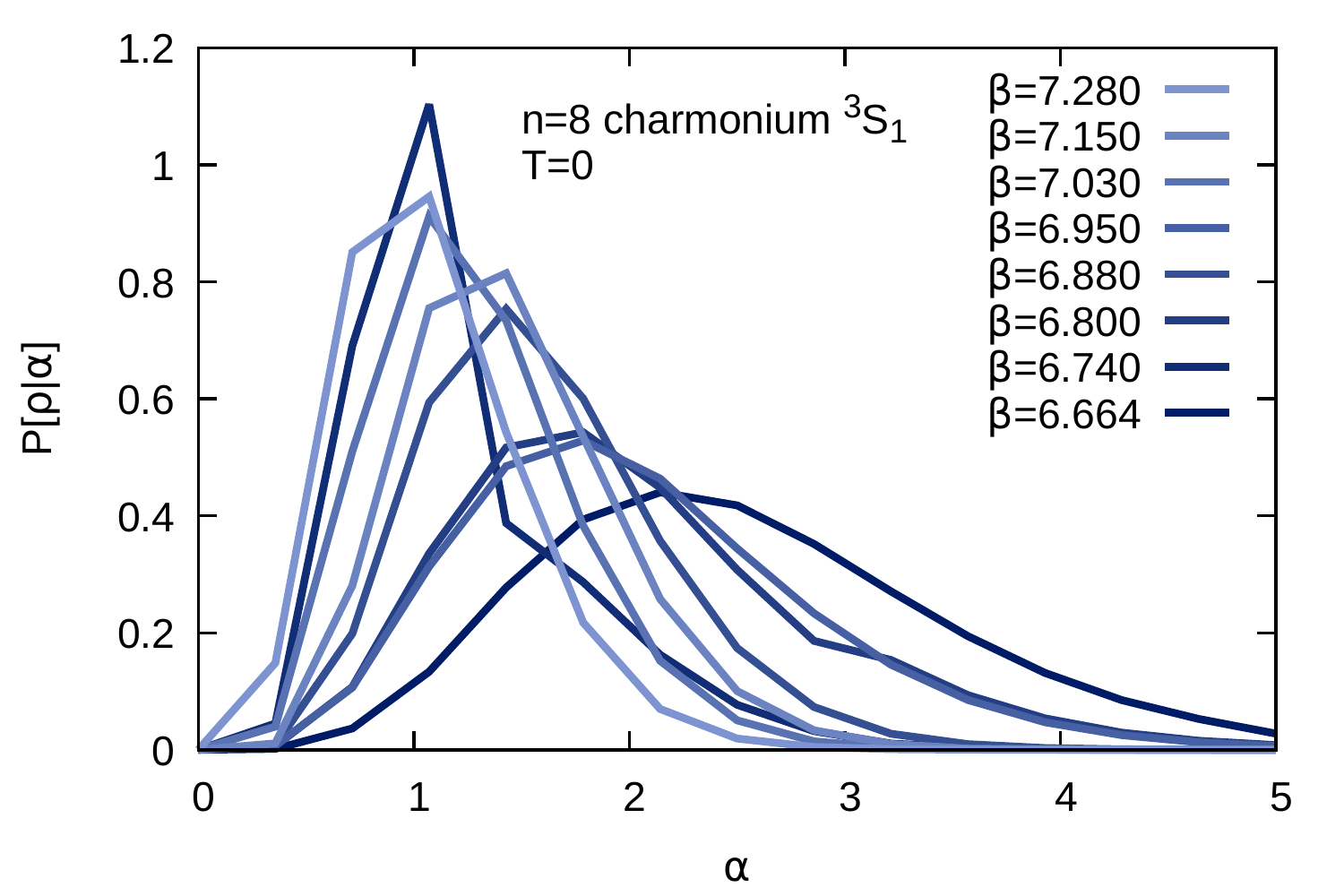}
\includegraphics[scale=0.5]{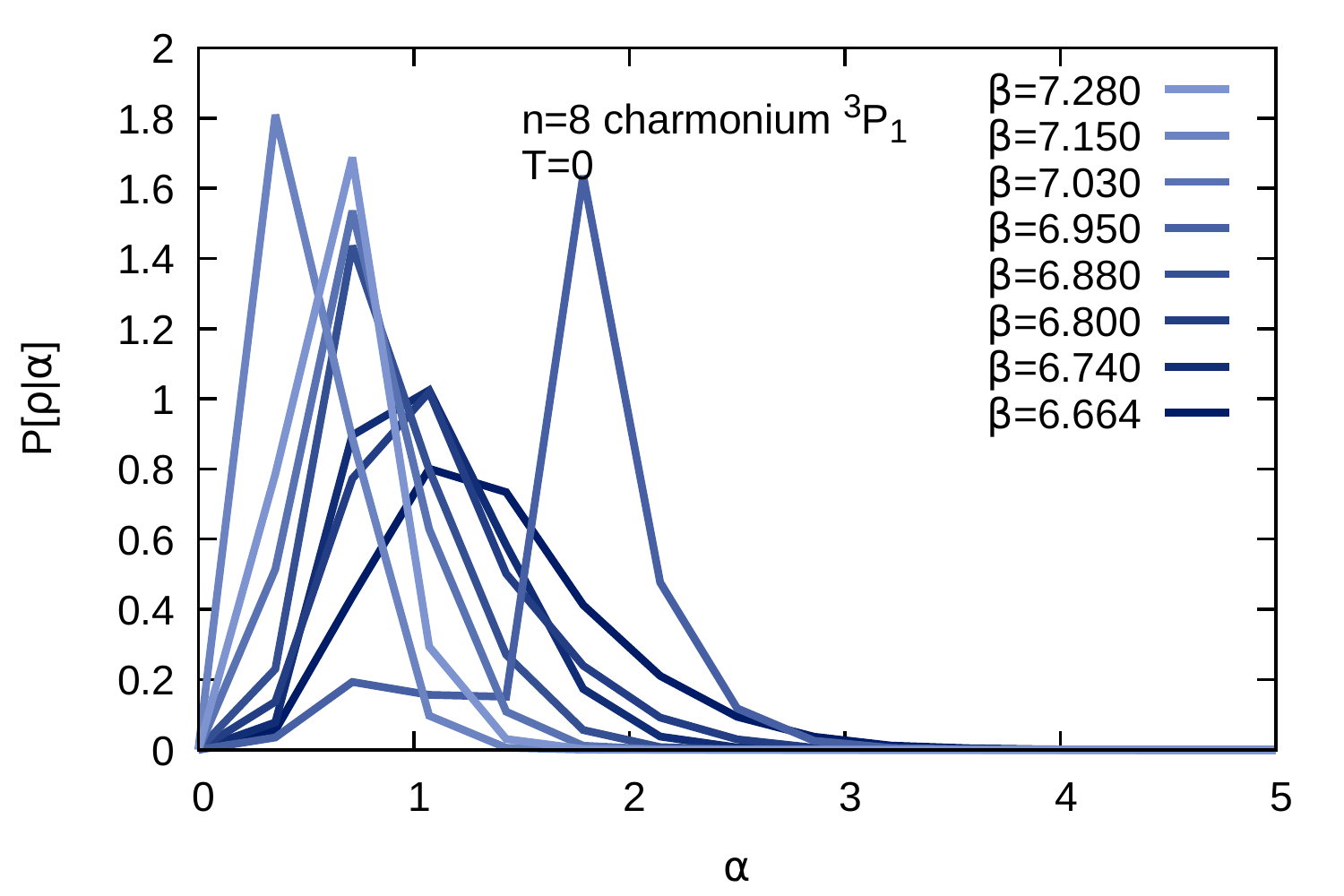}
\caption{Representative examples of the probability distribution over the hyperparameter $\alpha$ used in the self consistent determination of the most probable spectrum at zero temperatures. We show here charmonium results, the top row contains the S-wave results, the bottom row the P-wave channel.}\label{Fig:ZeroTMEMAlphasCharm}
\end{figure}

\begin{figure}[t]
\includegraphics[scale=0.5]{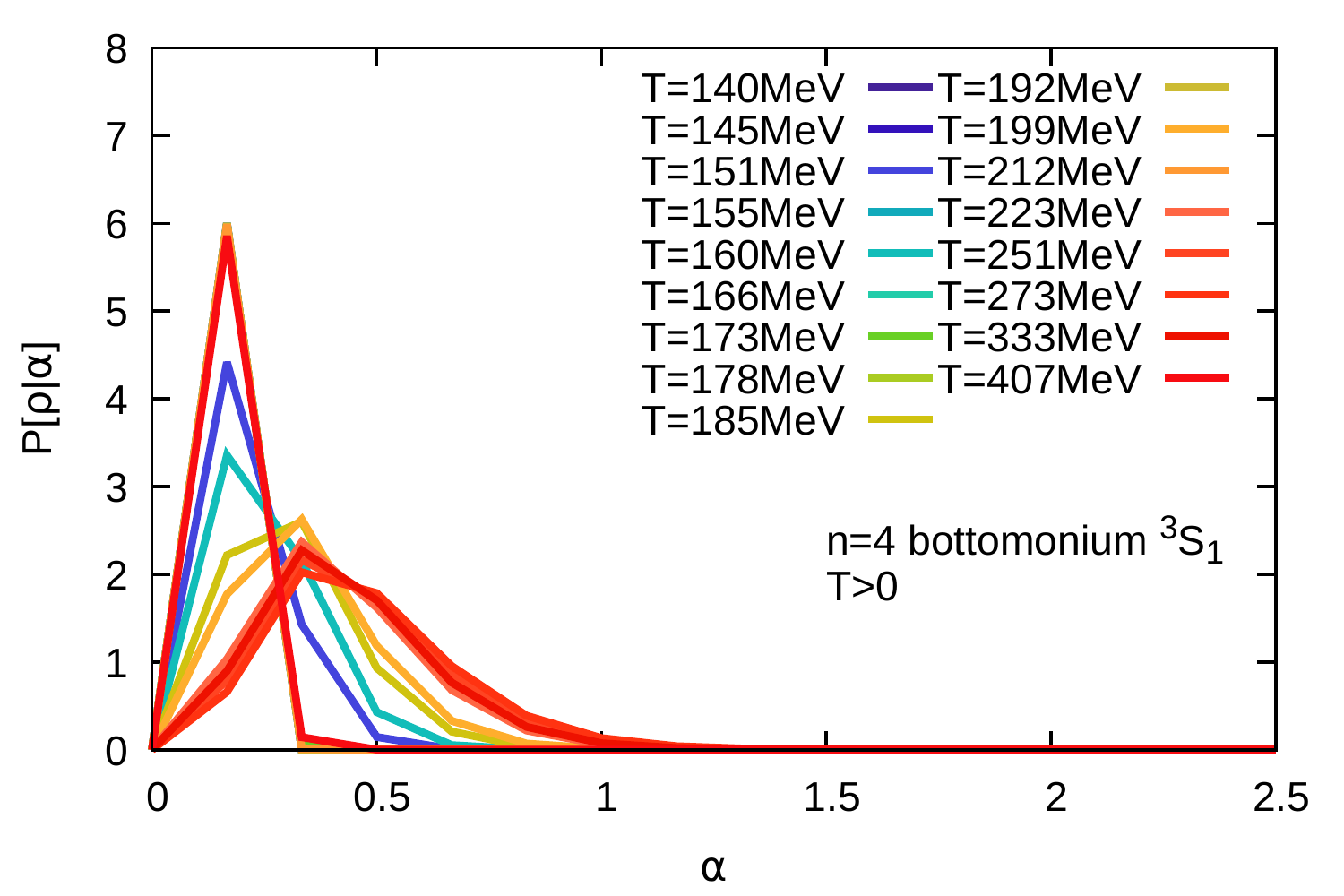}
\includegraphics[scale=0.5]{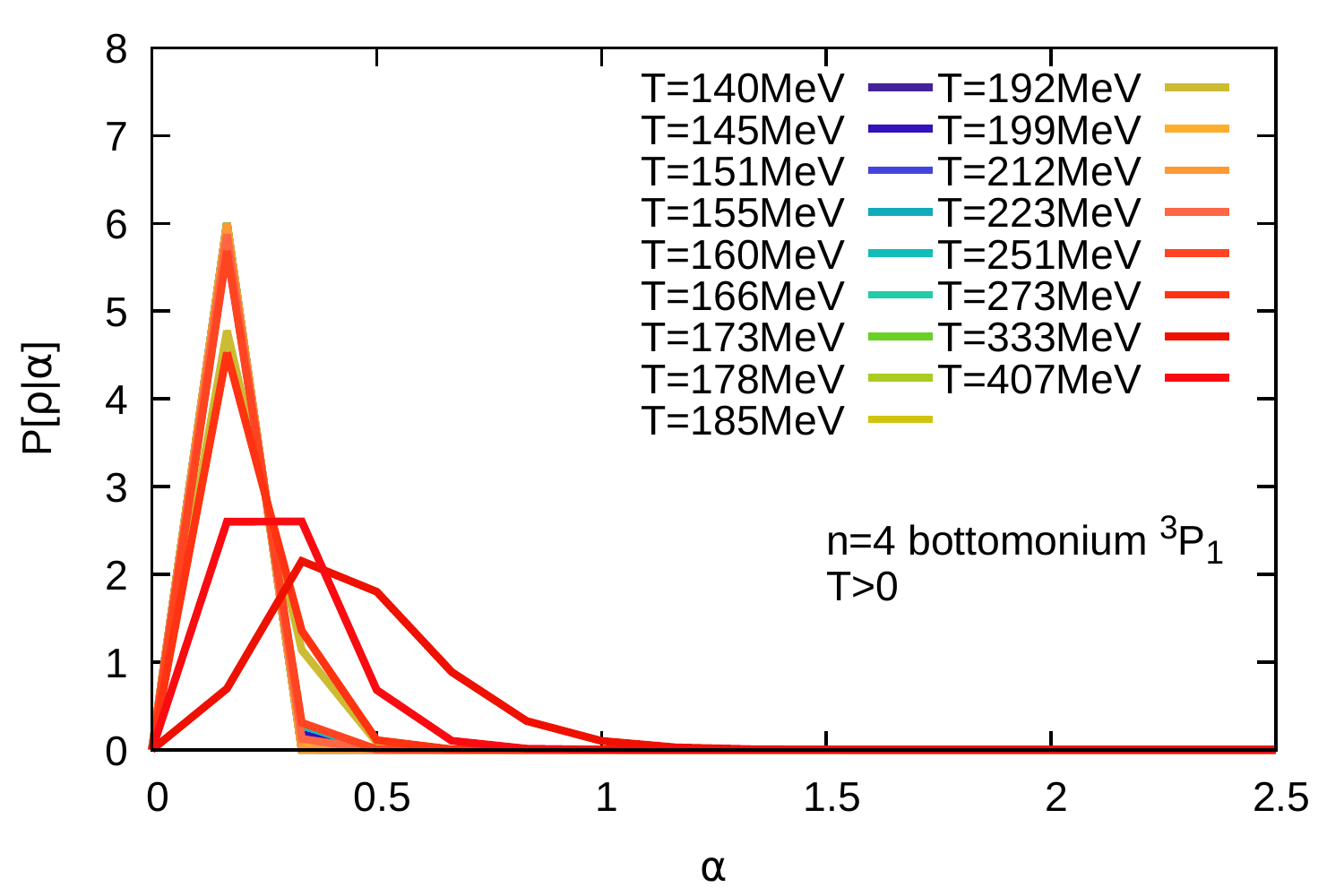}
\caption{Representative examples of the probability distribution over the hyperparameter $\alpha$ used in the self consistent determination of the most probable spectrum at finite temperatures. We show here bottomonium results, the top row contains the S-wave results, the bottom row the P-wave channel.}\label{Fig:FiniteTTMEMAlphasBottom}
\end{figure}

\begin{figure}[t]
\includegraphics[scale=0.5]{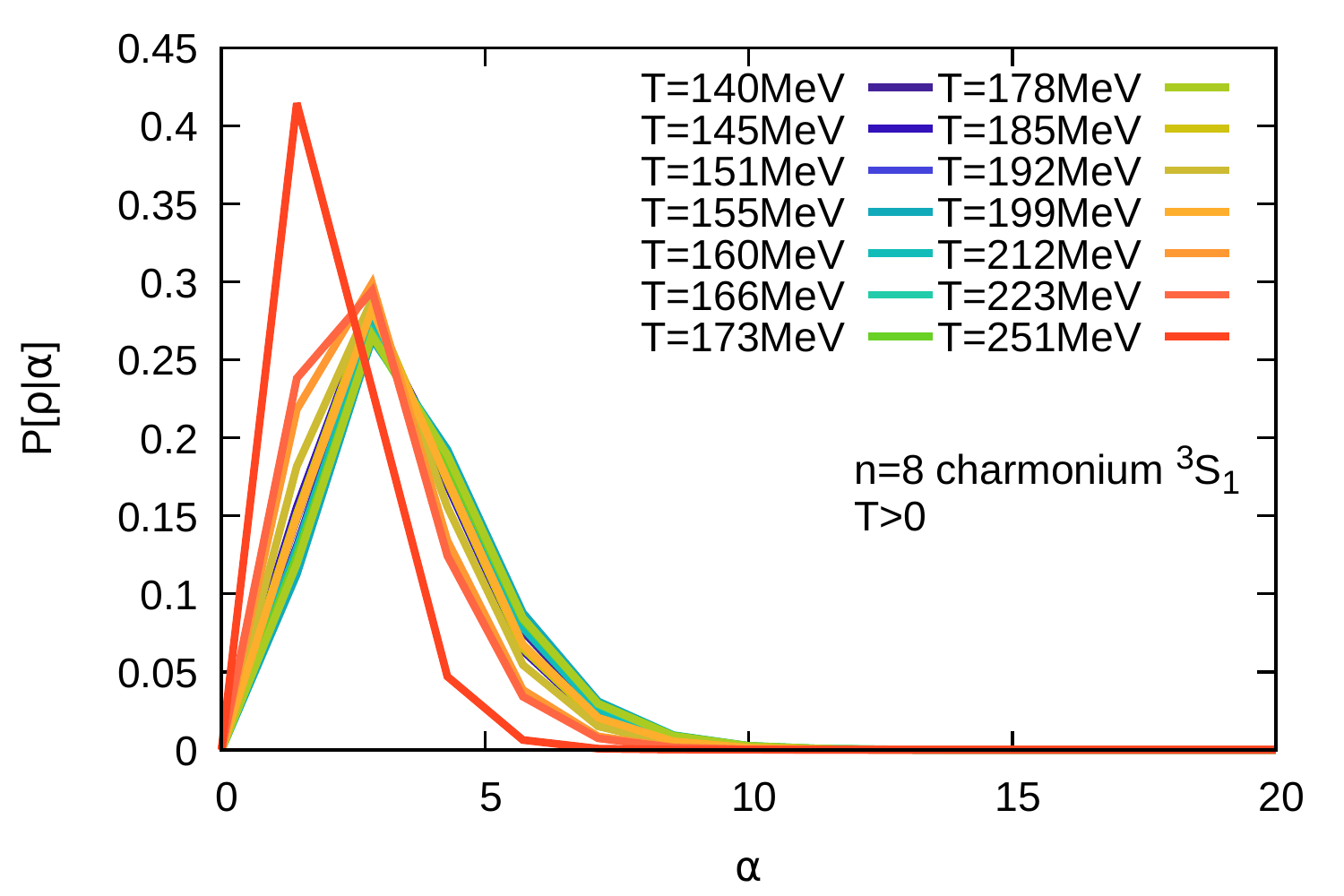}
\includegraphics[scale=0.5]{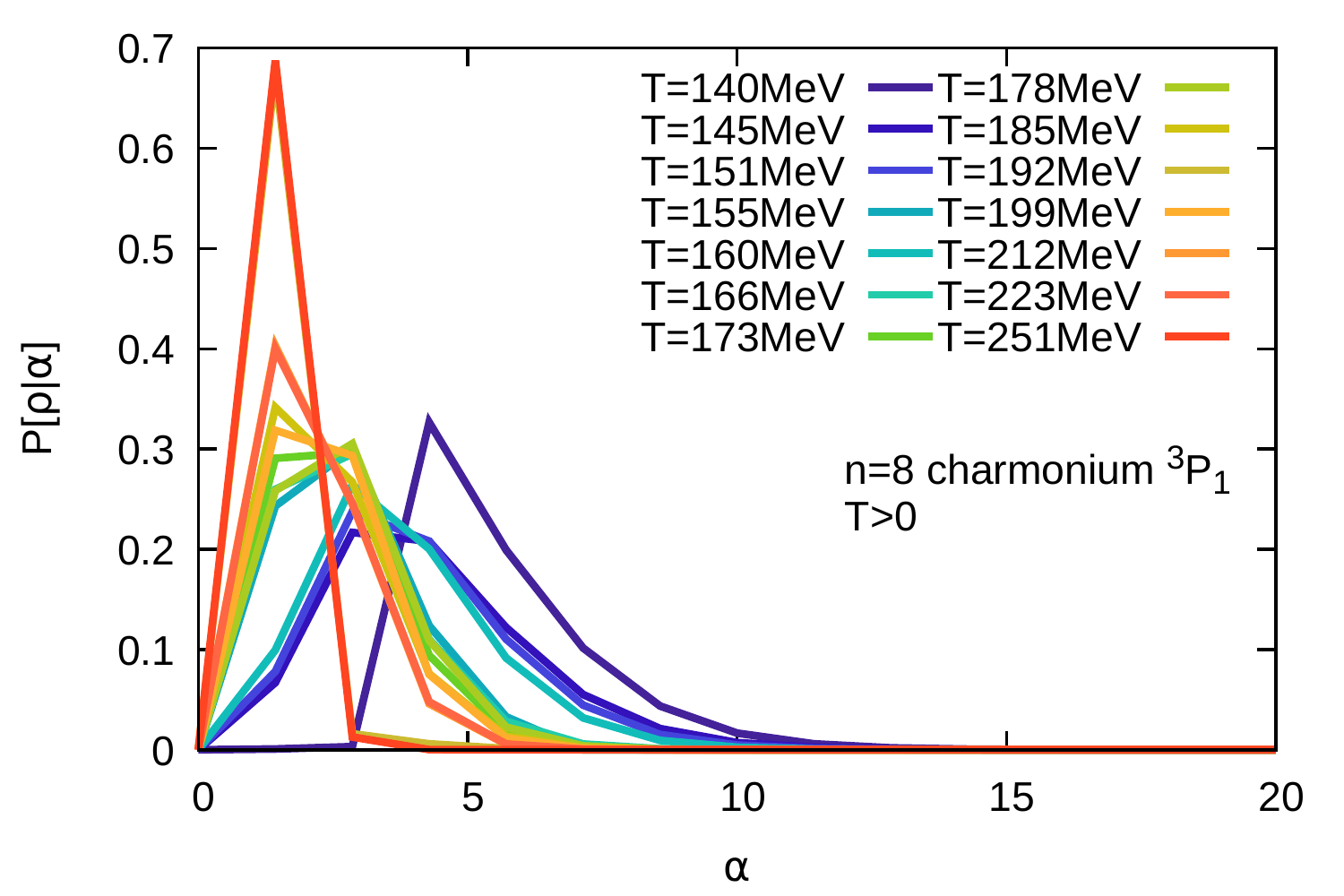}
\caption{Representative examples of the probability distribution over the hyperparameter $\alpha$ used in the self consistent determination of the most probable spectrum at finite temperatures. We show here charmonium results, the top row contains the S-wave results, the bottom row the P-wave channel.}\label{Fig:FiniteTTMEMAlphasCharm}
\end{figure}

\section{Additional reconstruction comparison and robustness tests }

\subsection*{Charmonium constrained model fits}

\label{app:ConstrModT0}

\begin{figure}[t]
\includegraphics[scale=0.5]{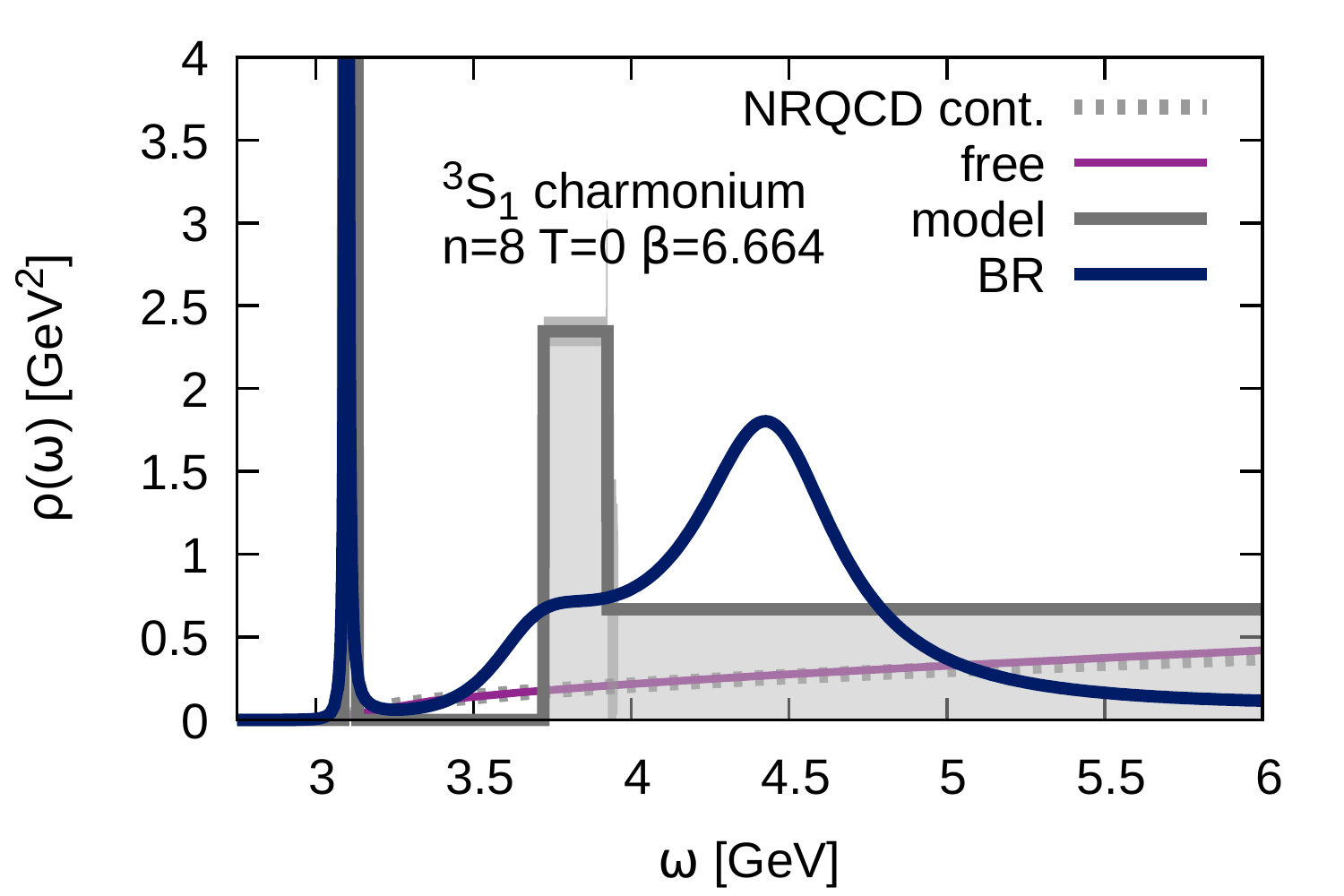}
\includegraphics[scale=0.5]{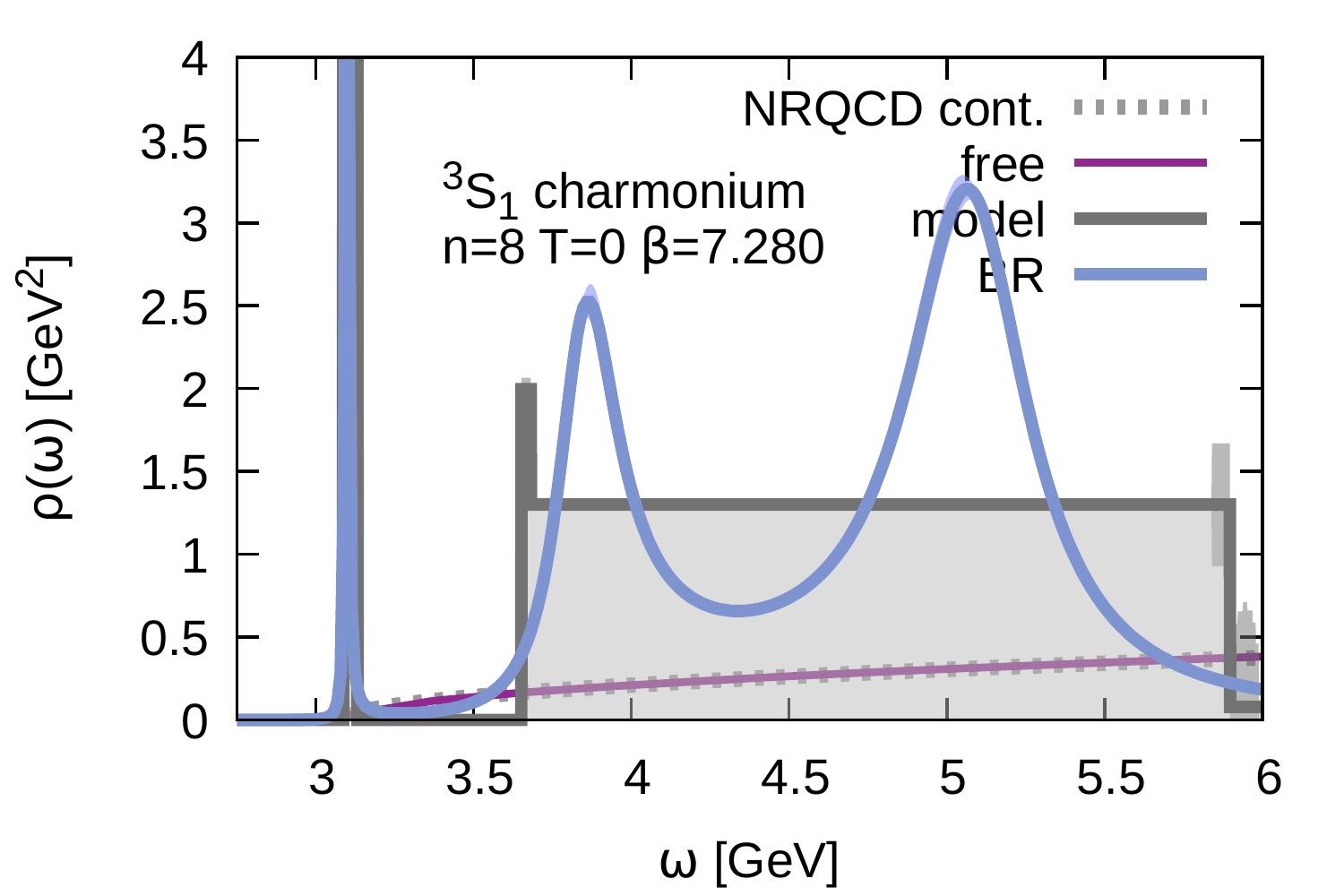}
\includegraphics[scale=0.5]{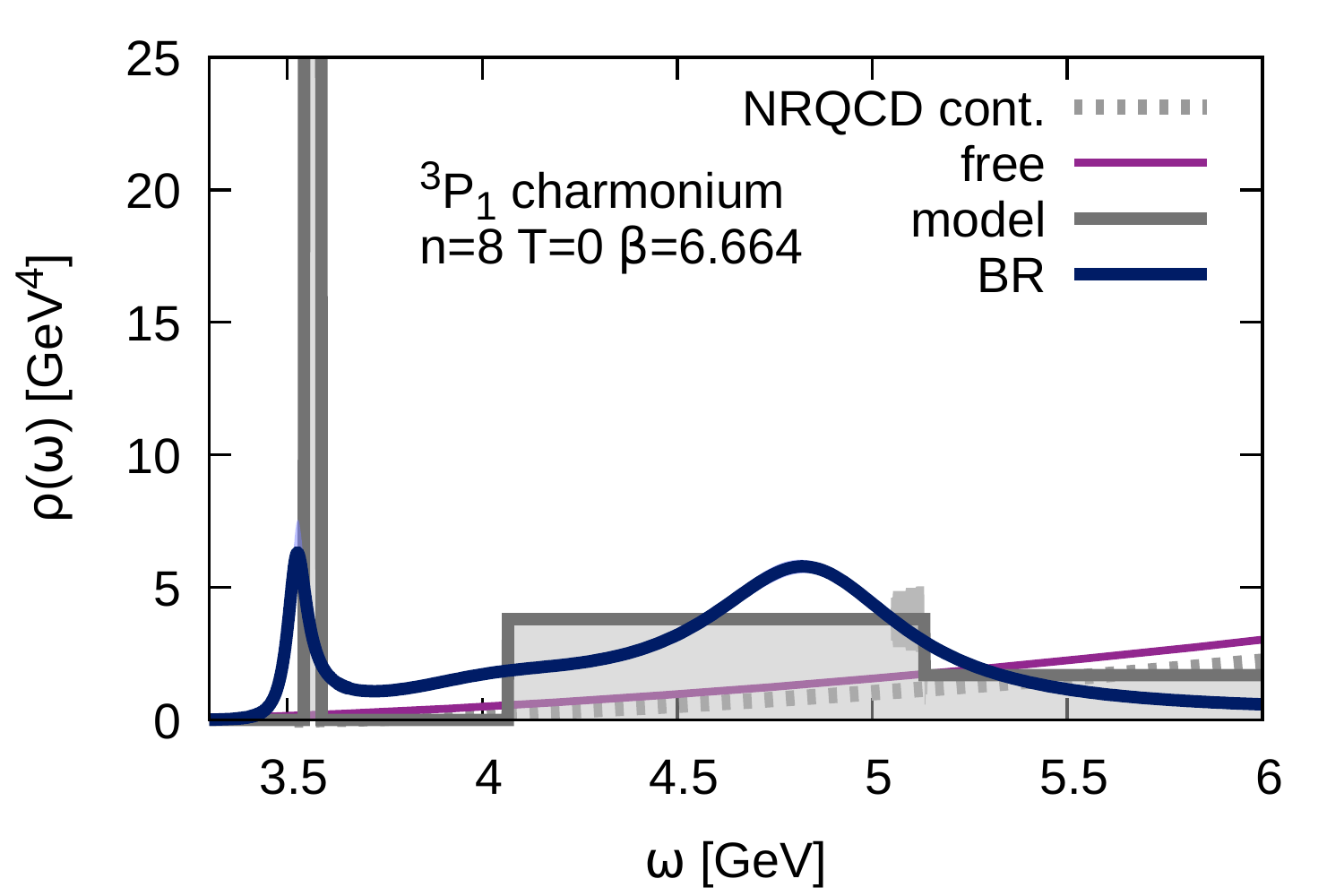}
\includegraphics[scale=0.5]{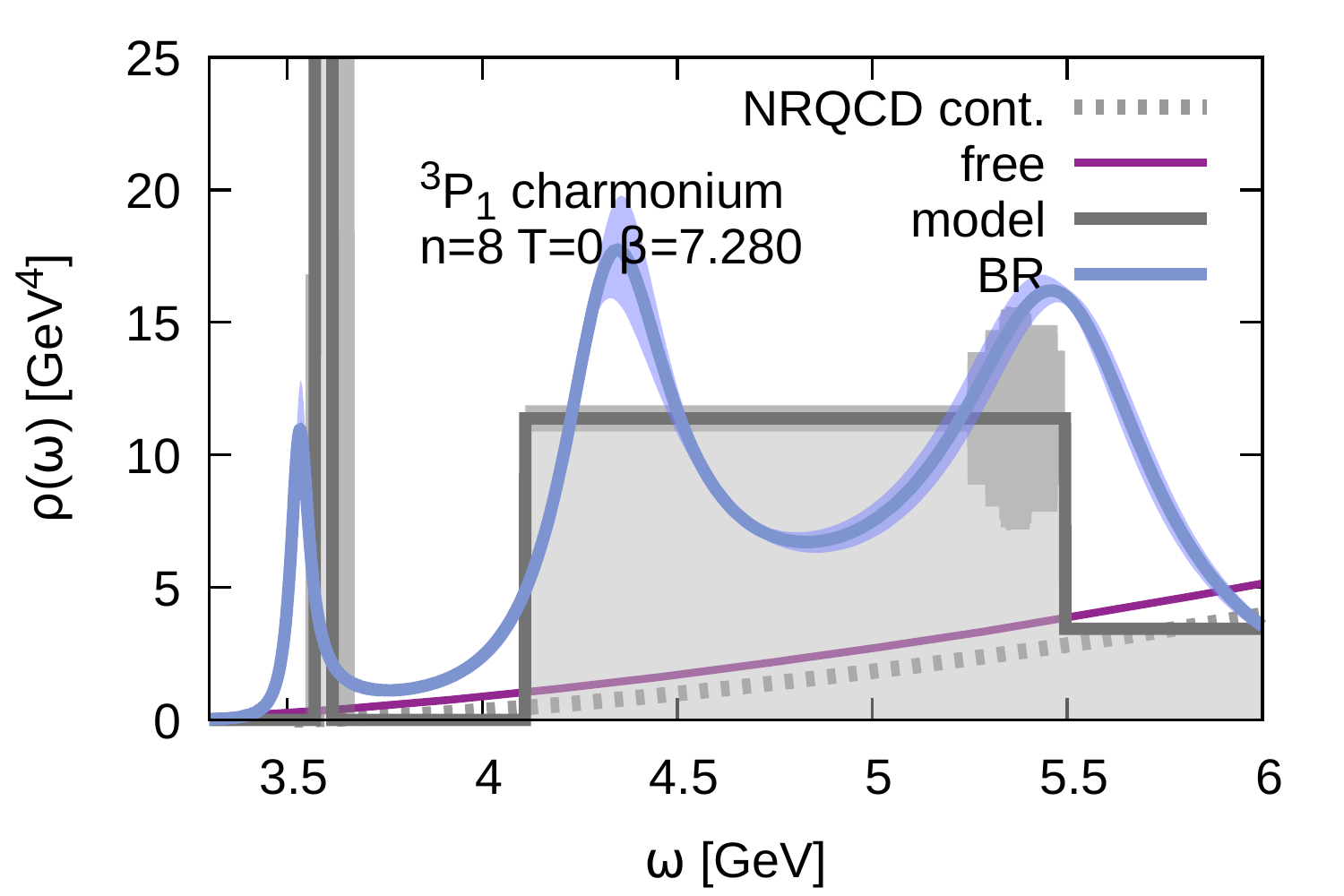}
\caption{Constrained model fits (gray) compared to the spectral reconstruction based on the standard BR method at $T= 0$.The top two panels show S-wave charmonium on the coarsest $\beta=6.664$ (left) and finest $\beta=7.825$ (right) lattices. The lower two panels on the other hand contain the results for P-wave charmonium again for both the coarsest $\beta=6.664$ (left) and finest $\beta=7.825$ (right) lattices.}\label{Fig:T0BRModelCmpCharm}
\end{figure}

In Fig.\ref{Fig:T0BRModelCmpCharm} we present for completeness the results for the zero temperature constrained model fits for charmonium, which was not discussed explicitly in the main text. The conclusion remain the same. One may note the slight discrepancy between the constrained model fit and the BR reconstruction for the P-wave ground state peak, which we understand to arise from the difficulty of choosing an appropriate fit range for the former by eye.

\subsection*{Truncation tests at T=0}

\label{app:truncTstT0}

\begin{figure}
\includegraphics[scale=0.5]{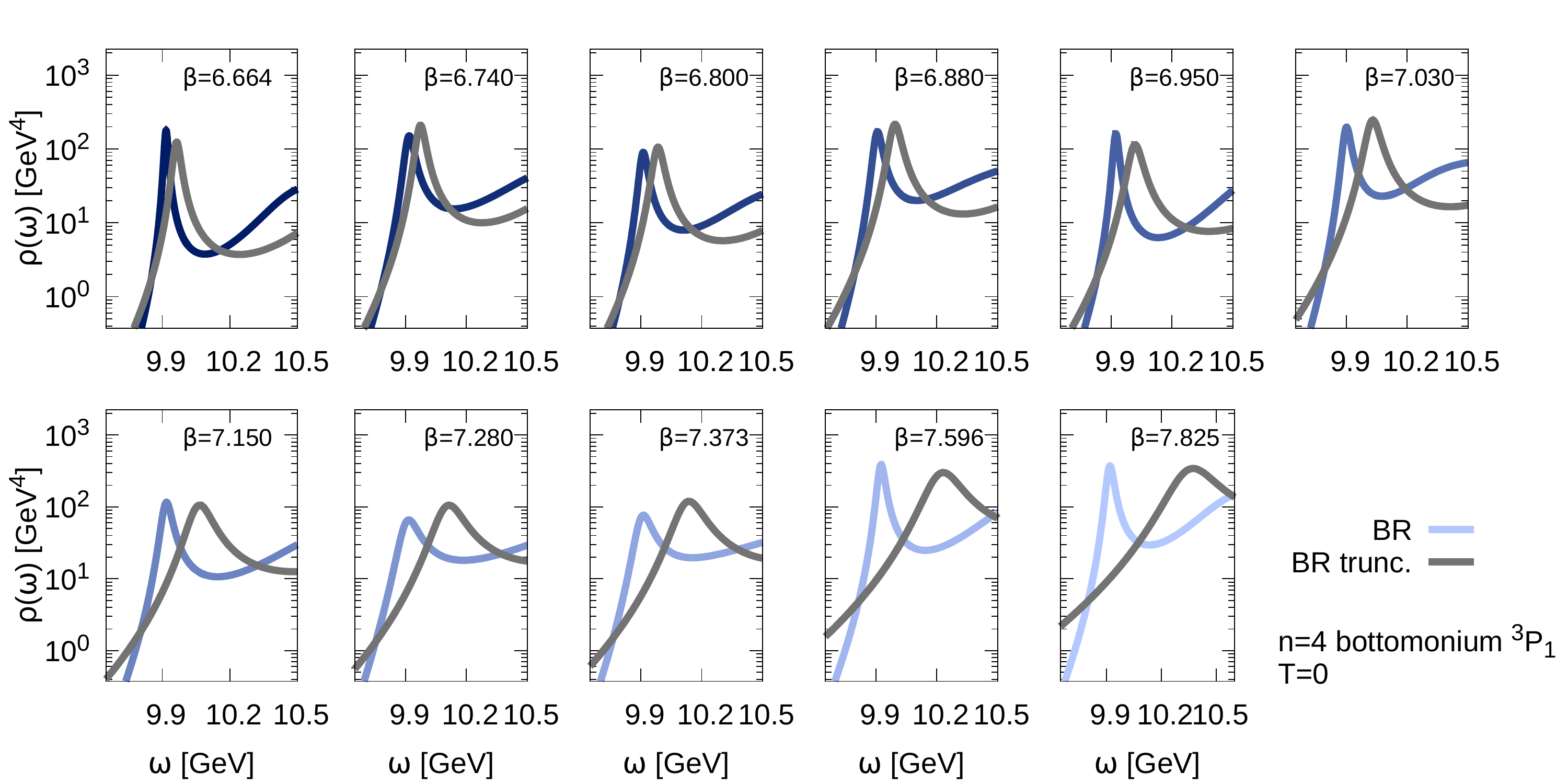}
\caption{Influence of truncating the low-temperature simulation data to the same $\tau_{\rm max}/a_\tau=12$ Euclidean time extent available at finite temperature for bottomonium $^3P_1$ channel. The full reconstruction given by colored solid lines, while the result after truncation is given by the gray solid curves.}\label{Fig:T0BRSpectraCmpTruncbb3P1}
\end{figure}

\begin{figure}
\includegraphics[scale=0.5]{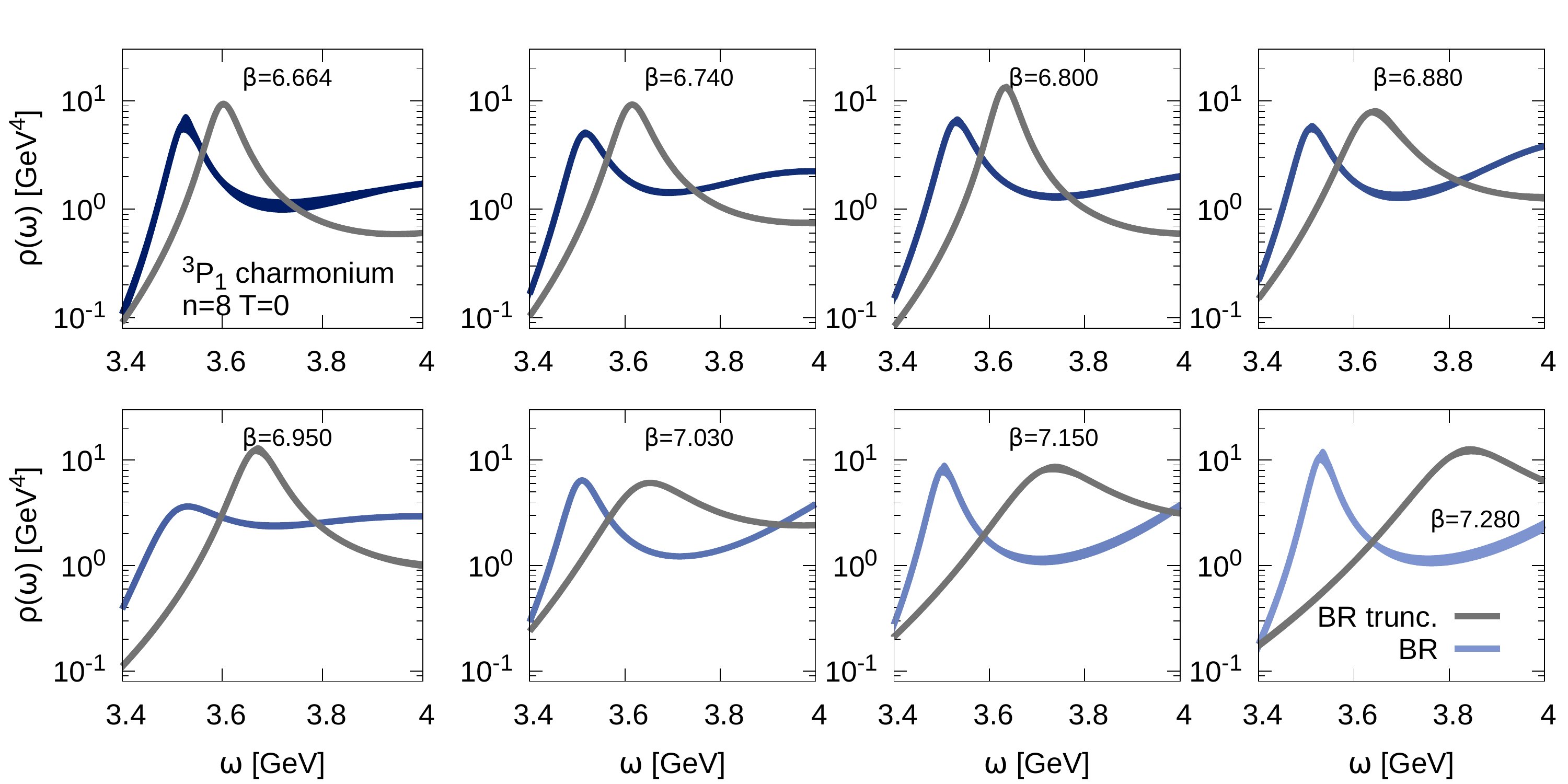}
\caption{Influence of truncating the low-temperature simulation data to the same $\tau_{\rm max}/a_\tau=12$ Euclidean time extent available at finite temperature for charmonium $^3P_1$ channel. The full reconstruction given by colored solid lines, while the result after truncation is given by the gray solid curves.}\label{Fig:T0BRSpectraCmpTrunccc3P1}
\end{figure}

The results of the truncation tests at $T=0$ of the bottomonium $^3P_1$, the charmonium $^3S_1$, as well as $^3P_1$ channels not discussed explicitly in the main text can be found in Fig.\ref{Fig:T0BRSpectraCmpTruncbb3P1}, Fig.\ref{Fig:T0BRSpectraCmpTrunccc3P1} and Fig.\ref{Fig:T0BRSpectraCmpTrunccc3P1} respectively. In all cases the same type artifacts appear, which include a shift to higher masses of the ground state peak and a significant broadening of its width. The values for the mass shifts are depicted in Fig.\ref{Fig:FiniteTInMediumMasses}

\begin{figure}
\includegraphics[scale=0.5]{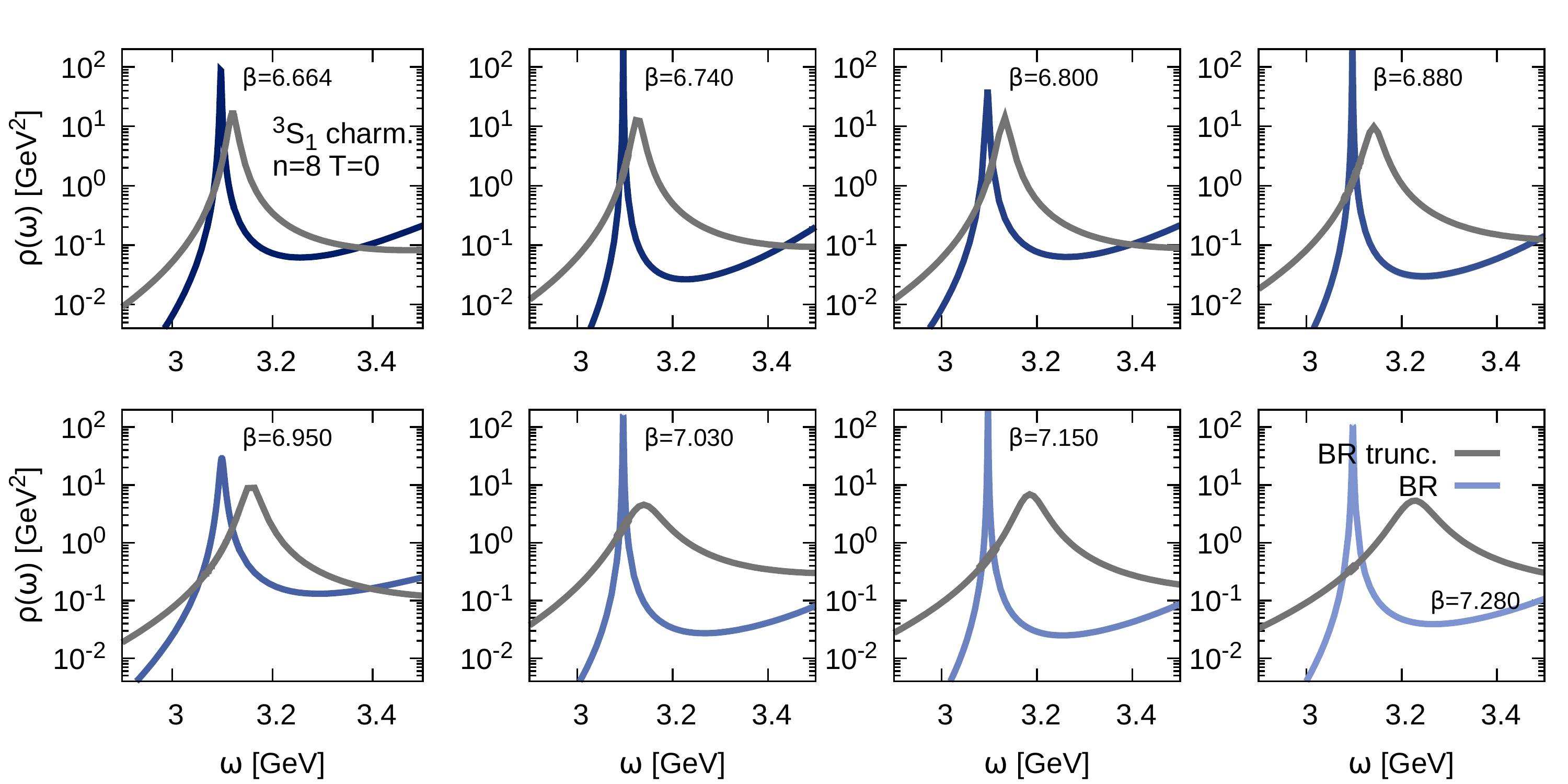}
\caption{Influence of truncating the low-temperature simulation data to the same $\tau_{\rm max}/a_\tau=12$ Euclidean time extent available at finite temperature for charmonium $^3S_1$ channel. The full reconstruction given by colored solid lines, while the result after truncation is given by the gray solid curves.}\label{Fig:T0BRSpectraCmpTrunccc3S1}
\end{figure}

\begin{figure}[t]
\includegraphics[scale=0.5]{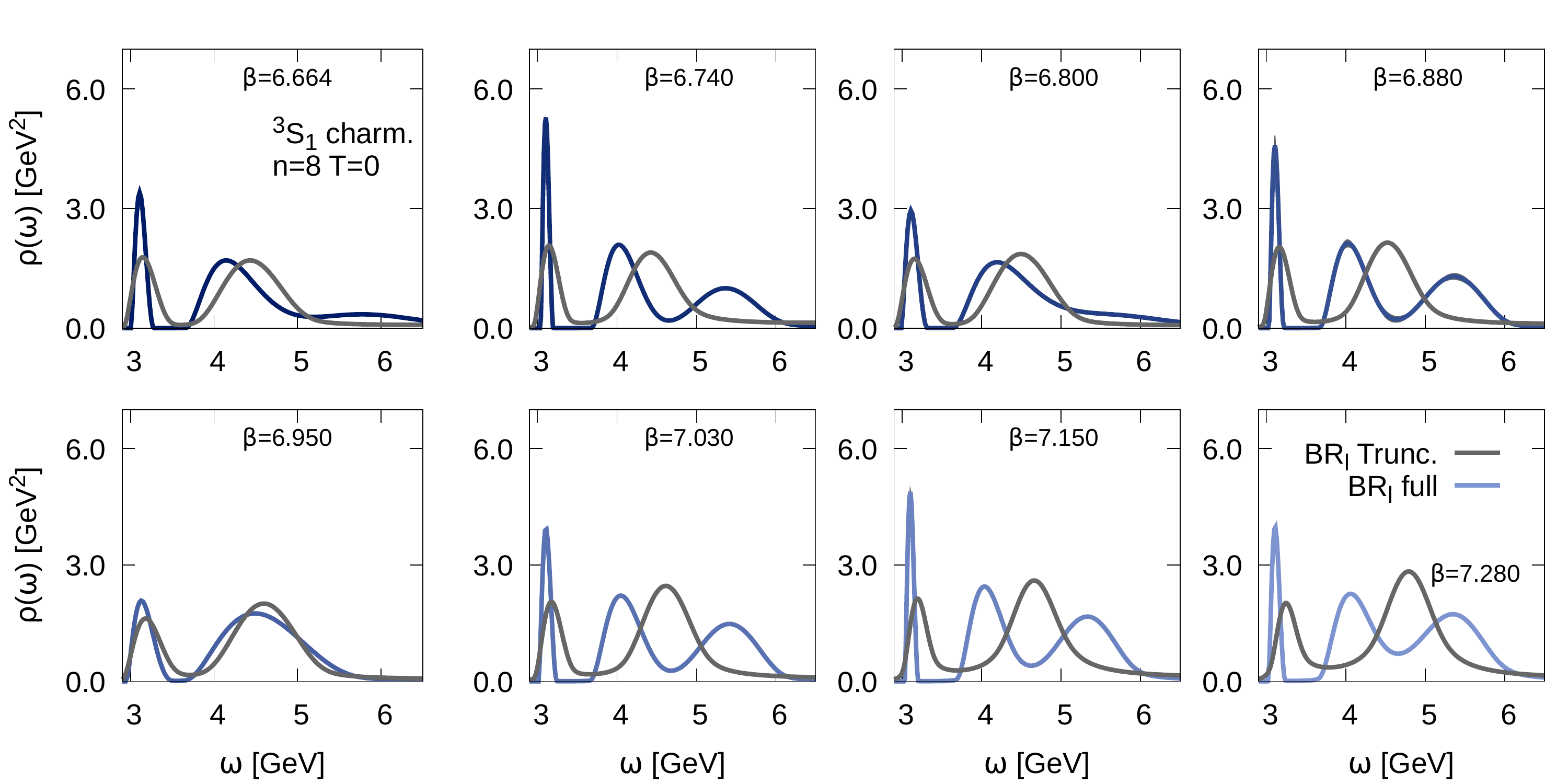}
\includegraphics[scale=0.5]{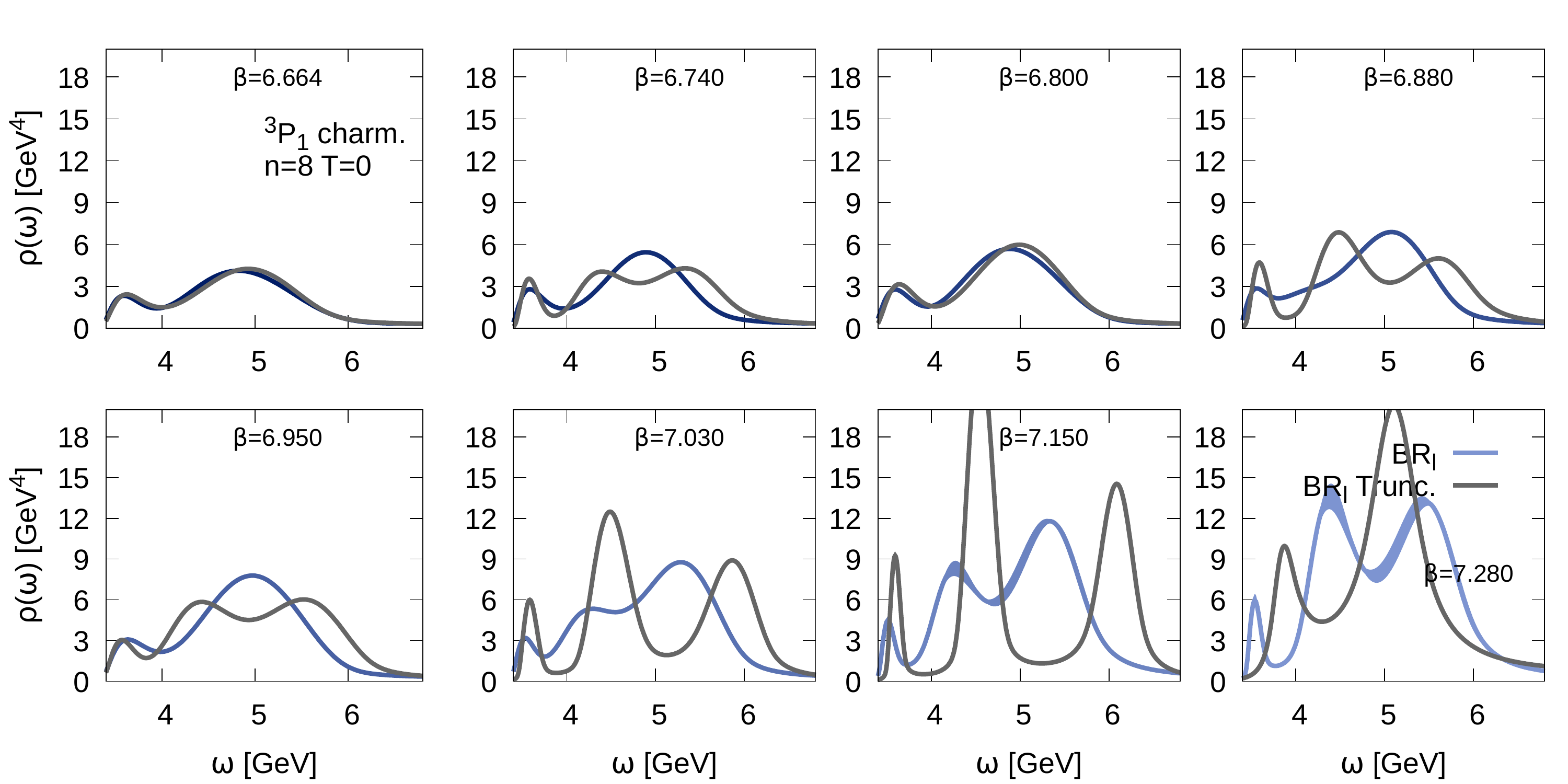}
\caption{Influence on the smooth BR reconstruction from a truncation of the low-temperature simulation data to the same $\tau_{\rm max}/a_\tau=12$ Euclidean time extent available at finite temperature. The spectra shown here correspond to the $^3S_1$ (top) and $^3P_1$ channel of charmonium with the full reconstruction given by colored solid lines, while the result after truncation is given by the gray solid curves.}\label{Fig:T0BRSpectraCmpTrunc}
\end{figure}

\subsection*{Comparison of truncation vs. finite temperature effects}

\label{app:truncTstT0}

\begin{figure}
\includegraphics[scale=0.5]{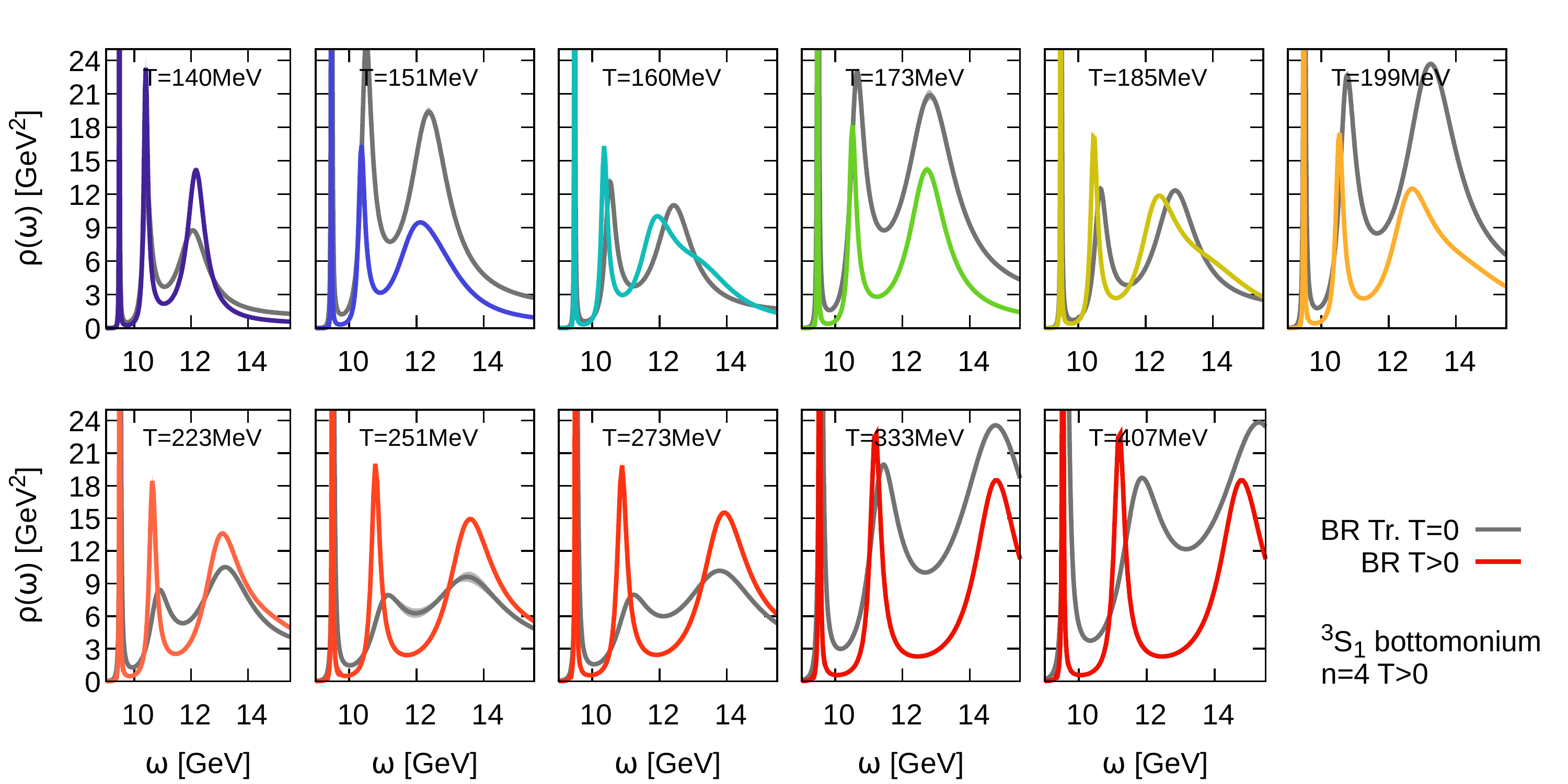}
\includegraphics[scale=0.5]{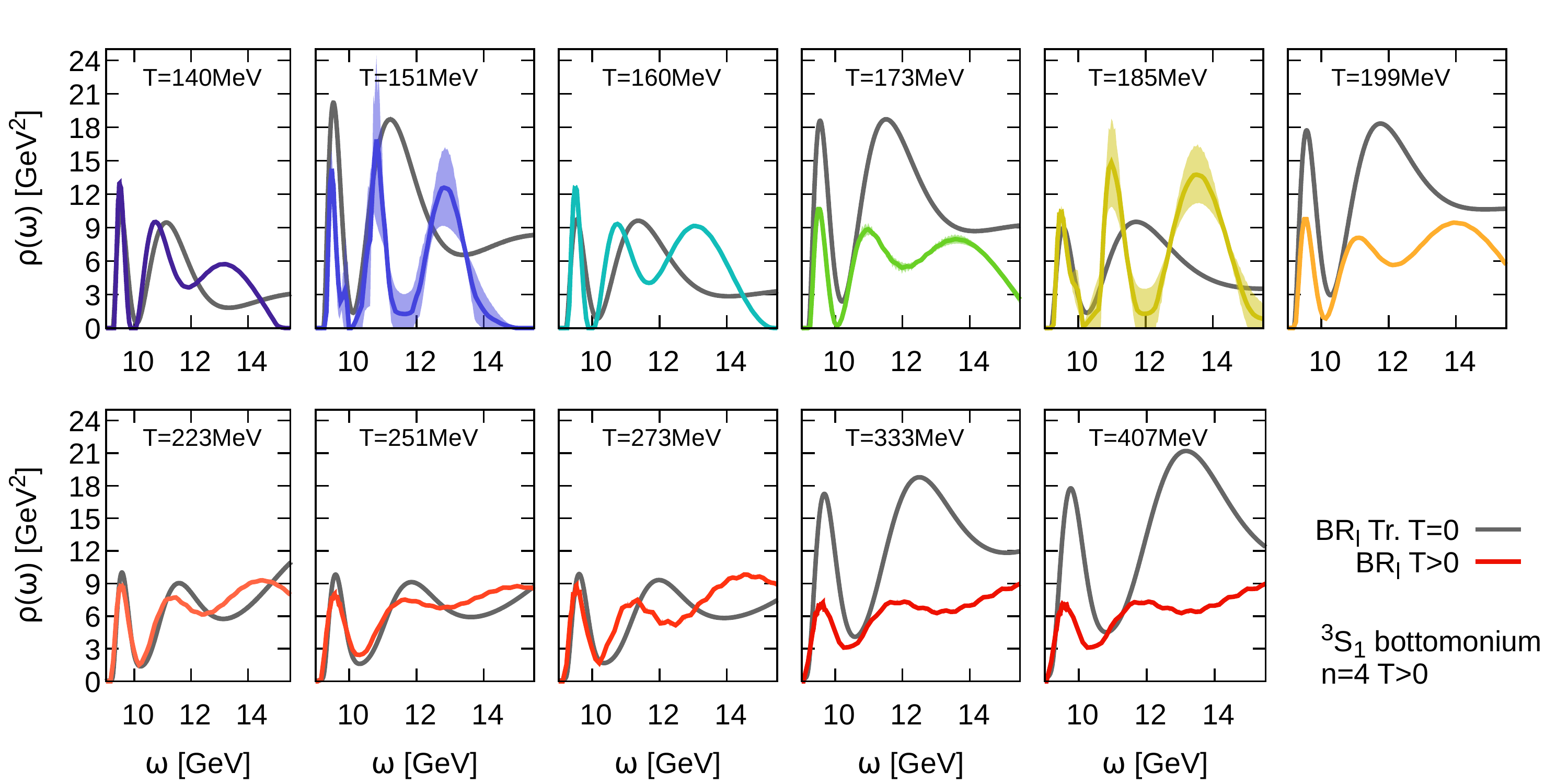}
\caption{Comparison of the (top) BR truncated at $T=0$ and at $T>0$ (bottom) the same comparison for the smooth BR method in the $^3S_1$ bottomonium channel. Note the difference in statistics available between $T=0$ and $T>0$ here. For Charmonium the statistics in the two cases are the same.}\label{Fig:CmpBottom3S1TruncVsFiniteT}
\end{figure}

\begin{figure}
\includegraphics[scale=0.5]{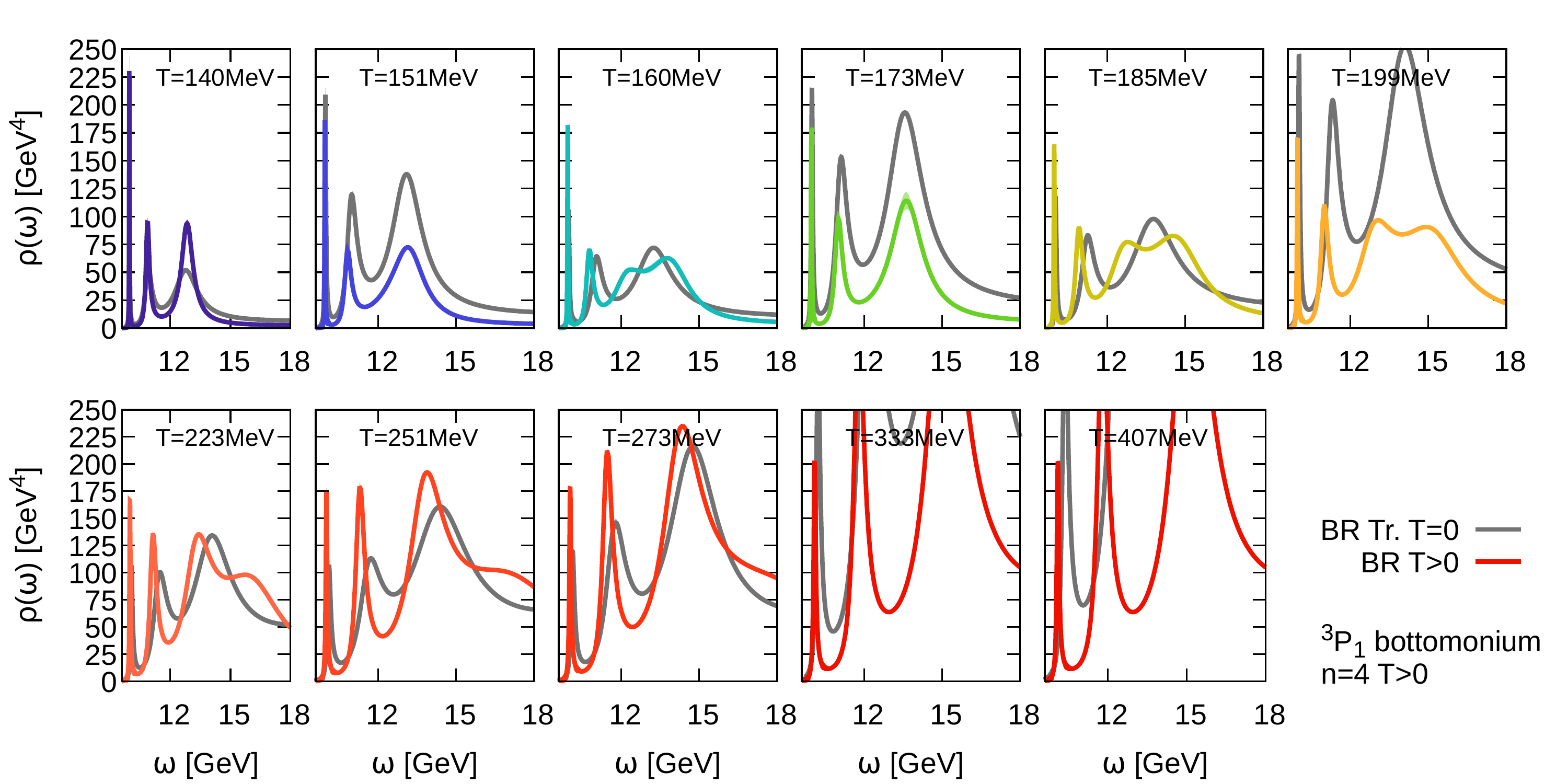}
\includegraphics[scale=0.5]{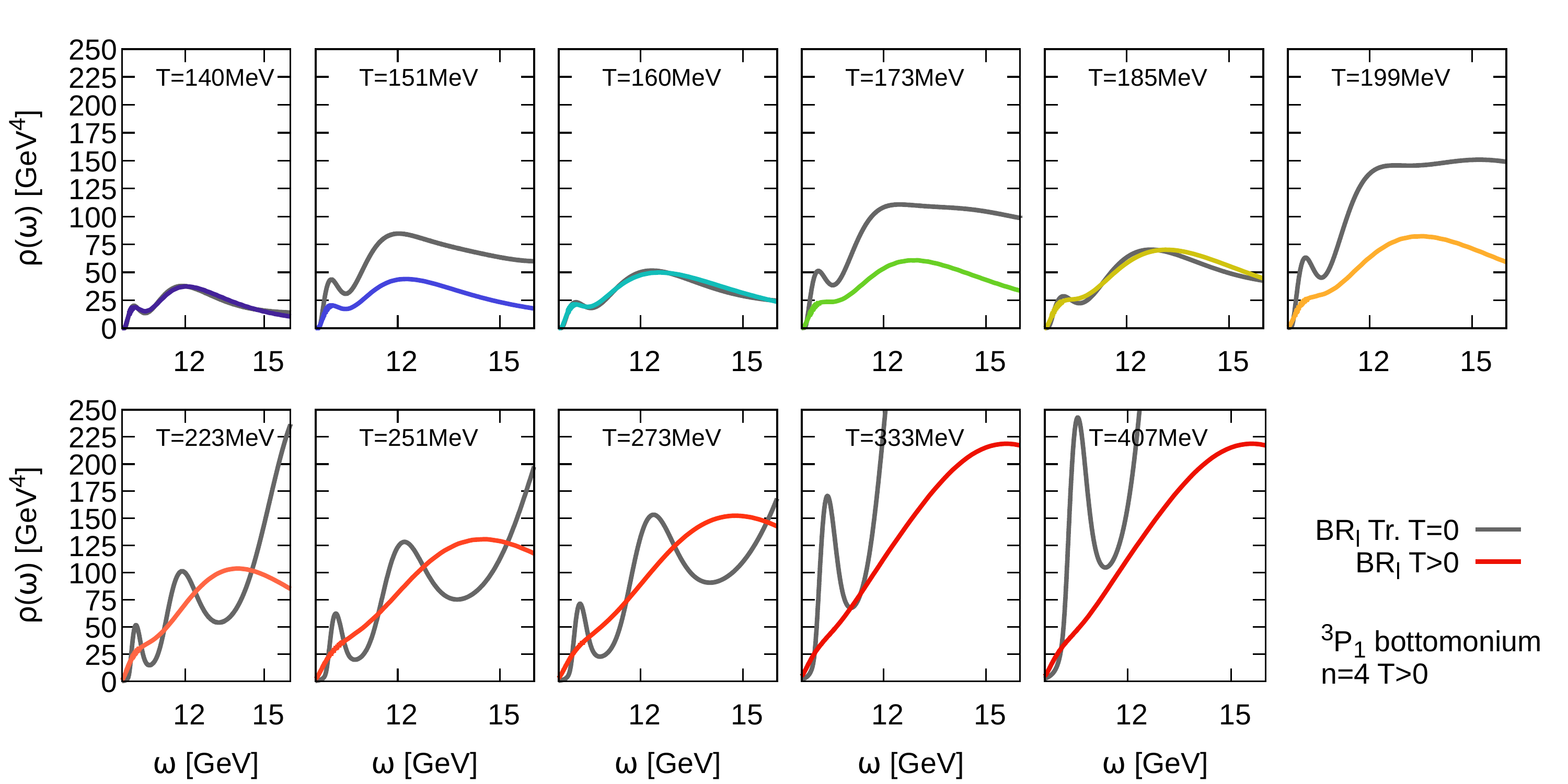}
\caption{Comparison of the (top) BR truncated at $T=0$ and at $T>0$ (bottom) the same comparison for the smooth BR method in the $^3P_1$ bottomonium channel. Note the difference in statistics available between $T=0$ and $T>0$ here. For Charmonium the statistics in the two cases are the same.}\label{Fig:CmpBottom3P1TruncVsFiniteT}
\end{figure}

\begin{figure}
\includegraphics[scale=0.5]{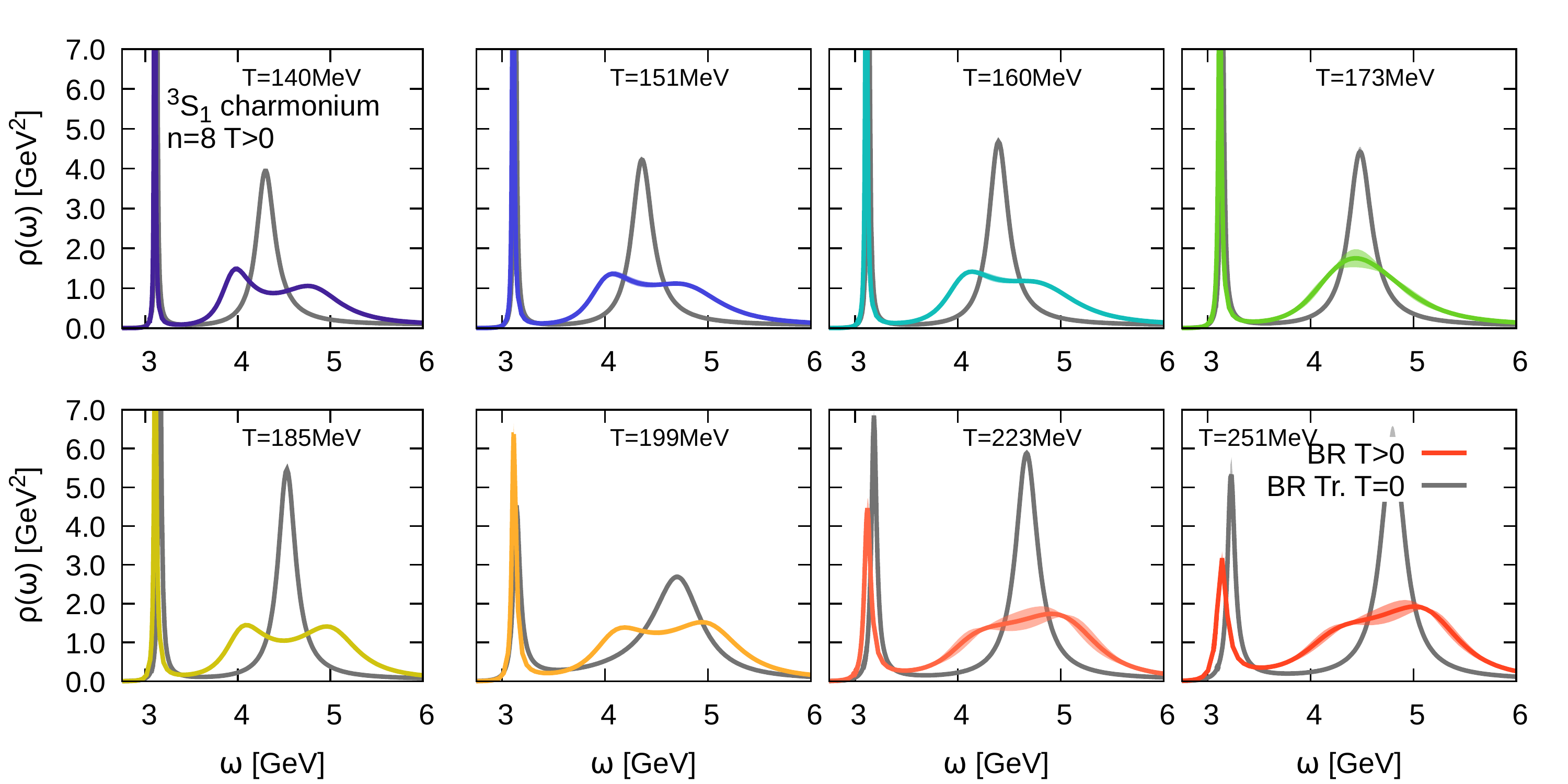}
\includegraphics[scale=0.5]{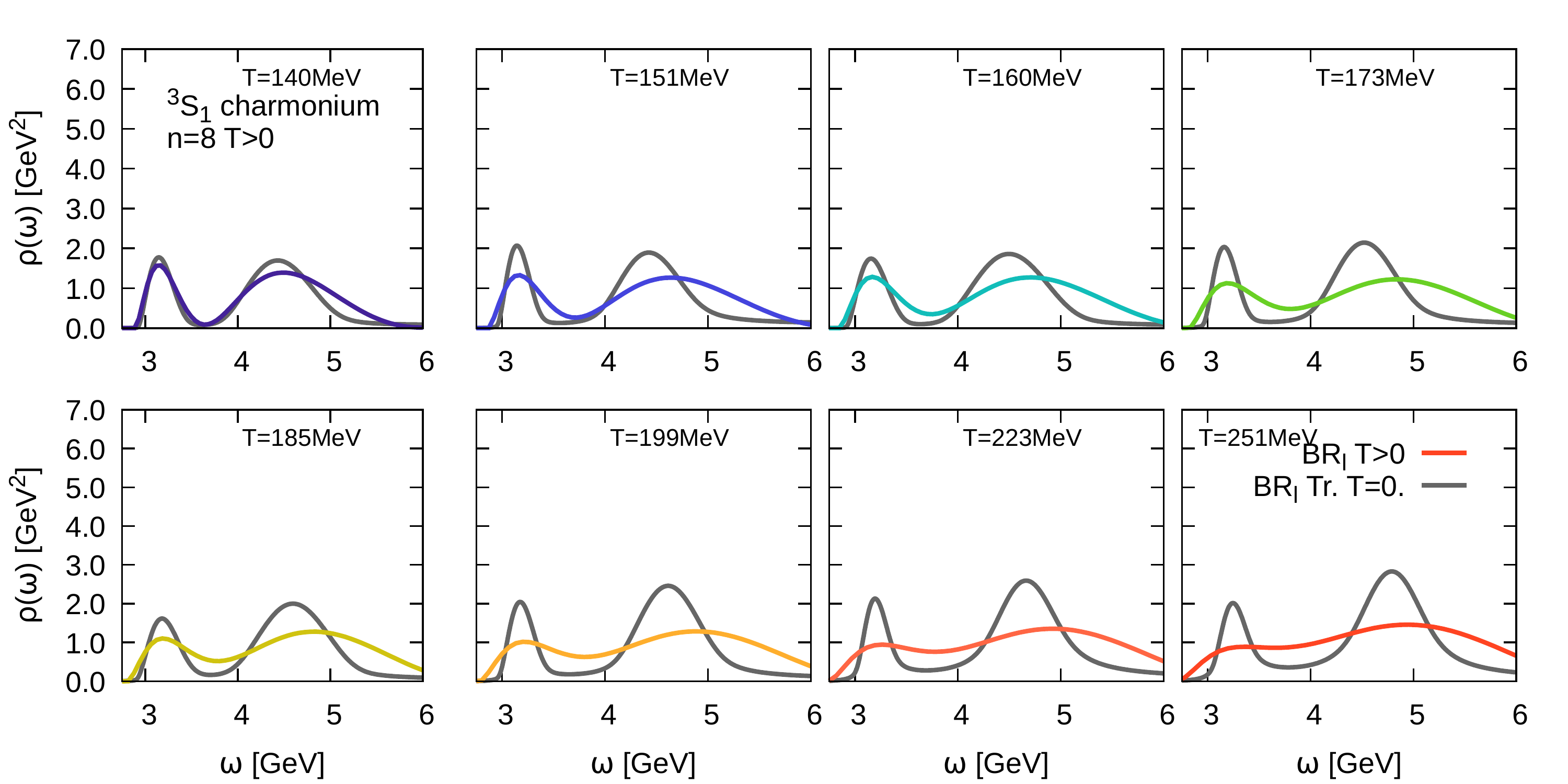}
\caption{Comparison of the (top) BR truncated at $T=0$ and at $T>0$ (bottom) the same comparison for the smooth BR method in the $^3S_1$ charmonium channel. Here we have the same statistics at $T=0$ and at $T>0$. }\label{Fig:CmpCharm3S1TruncVsFiniteT}
\end{figure}

\begin{figure}
\includegraphics[scale=0.5]{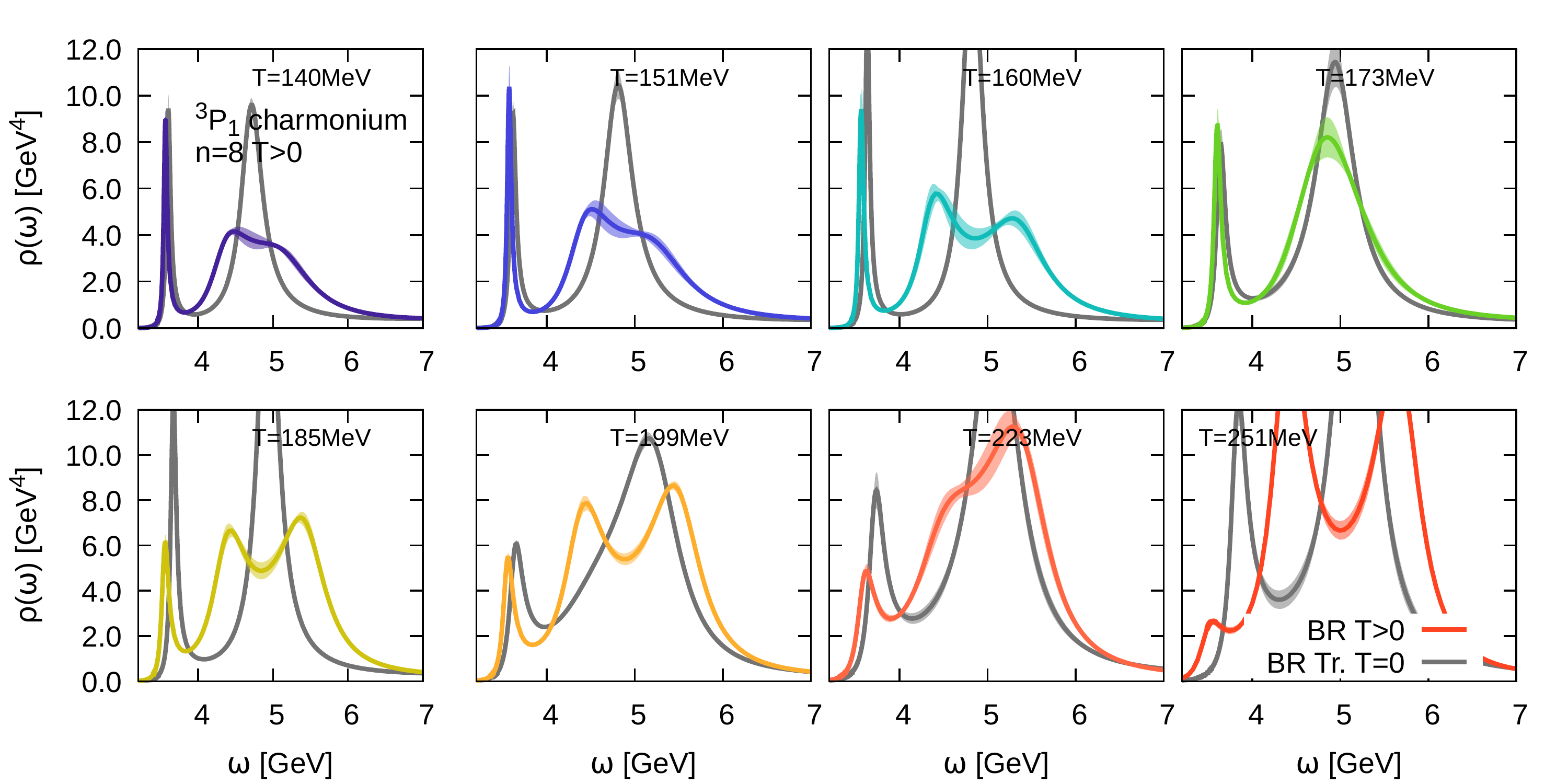}
\includegraphics[scale=0.5]{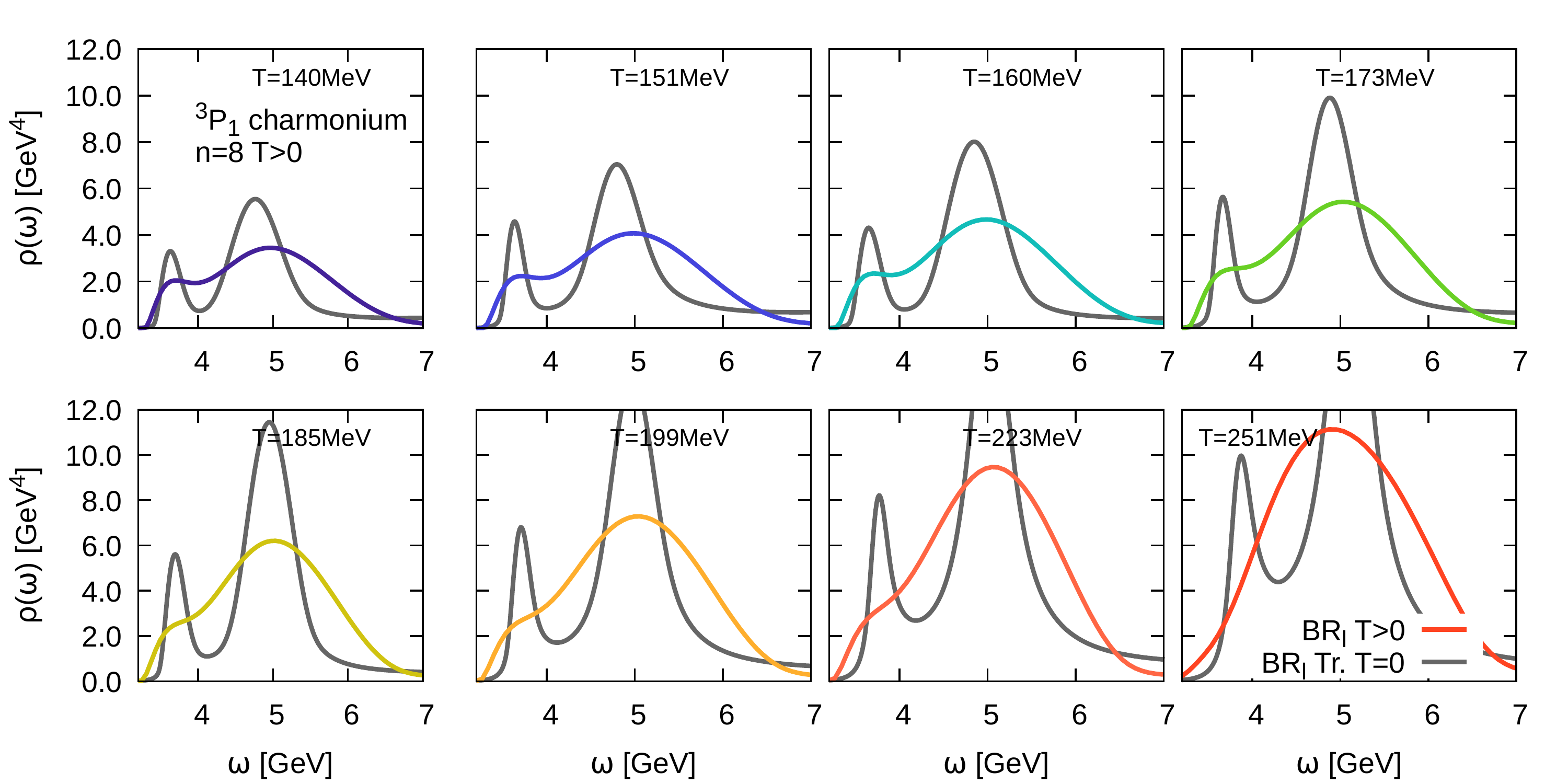}
\caption{Comparison of the (top) BR truncated at $T=0$ and at $T>0$ (bottom) the same comparison for the smooth BR method in the $^3P_1$ charmonium channel. Here we have the same statistics at $T=0$ and at $T>0$.}\label{Fig:CmpCharm3P1TruncVsFiniteT}
\end{figure}

Further illuminating comparisons supporting our discussion of the melting pattern of the quarkonium states concern spectral reconstructions obtained at $T=0$ from truncated correlator sets with those obtained directly at finite temperature. Starting with the bottomonium $^3S_1$ channel in Fig.\ref{Fig:CmpBottom3S1TruncVsFiniteT}, we show in the top two rows the comparison for the standard BR method, while in the bottom two rows that for the smooth BR method. The bottomonium $^3P_1$, as well as charmonium $^3S_1$ and $^3P_1$ channel can be found in Fig.\ref{Fig:CmpBottom3P1TruncVsFiniteT}, Fig.\ref{Fig:CmpCharm3S1TruncVsFiniteT} and Fig.\ref{Fig:CmpCharm3P1TruncVsFiniteT} respectively.

\subsection*{Explicit default model dependencies}
\label{sec:dmdep}

All of our quantitative determinations of zero temperature and in-medium masses include an estimation of statistical and systematic uncertainties. For the former we deploy a ten-bin Jackknife, while for the latter we repeated the spectral reconstructions with different default models, varying both the amplitude and functional form of 
$m(\omega)=m_0 (\omega-\omega_{\rm min}+1)^\gamma$ between $m_0\in[0.1,10]$ as well as $\gamma=\{0,1,2,-1,-2\}$. Here we provide explicit illustrations of the effect this change in default model has on the reconstruction of the full spectral function. 

\begin{figure}[t]
\includegraphics[scale=0.5]{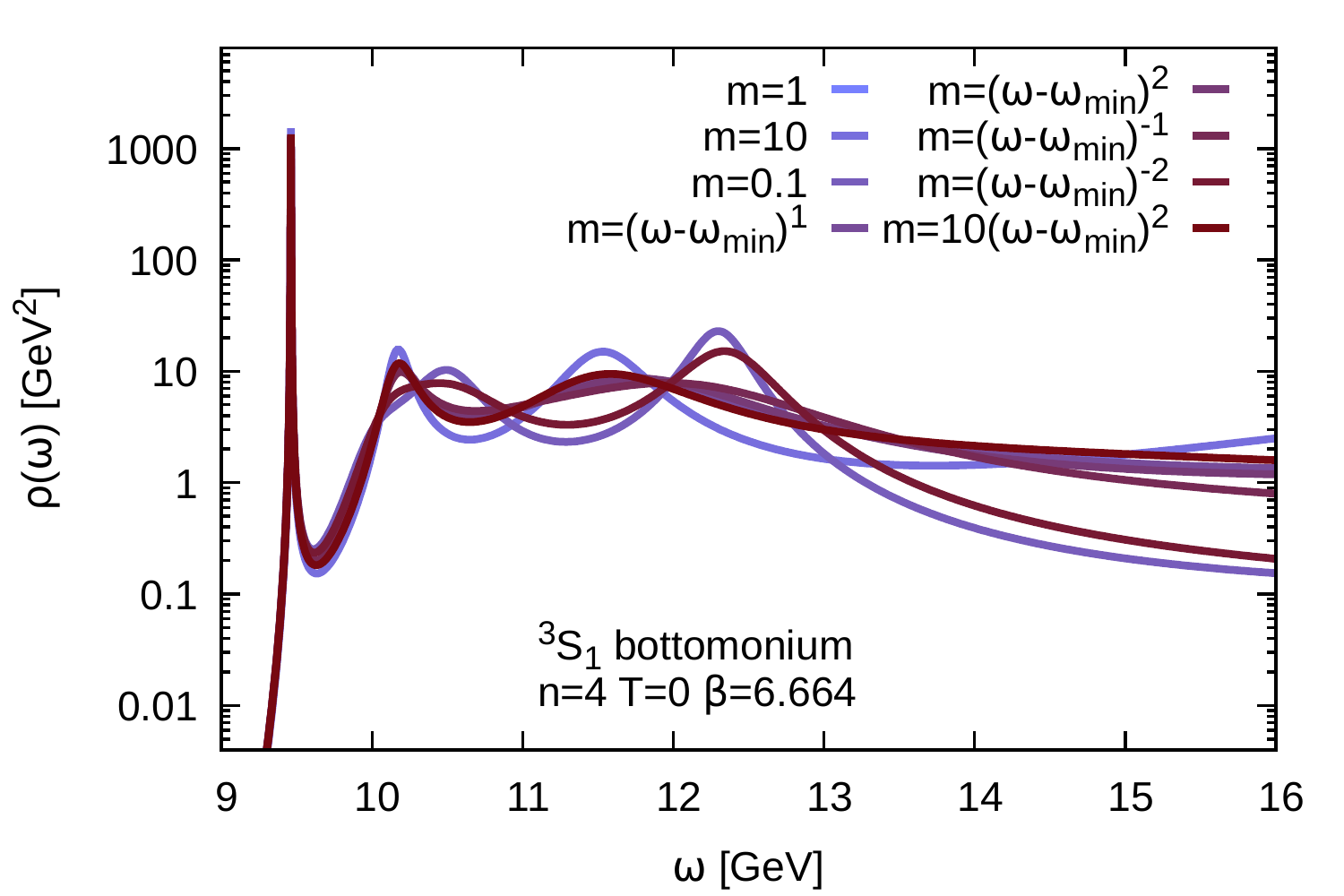}
\includegraphics[scale=0.5]{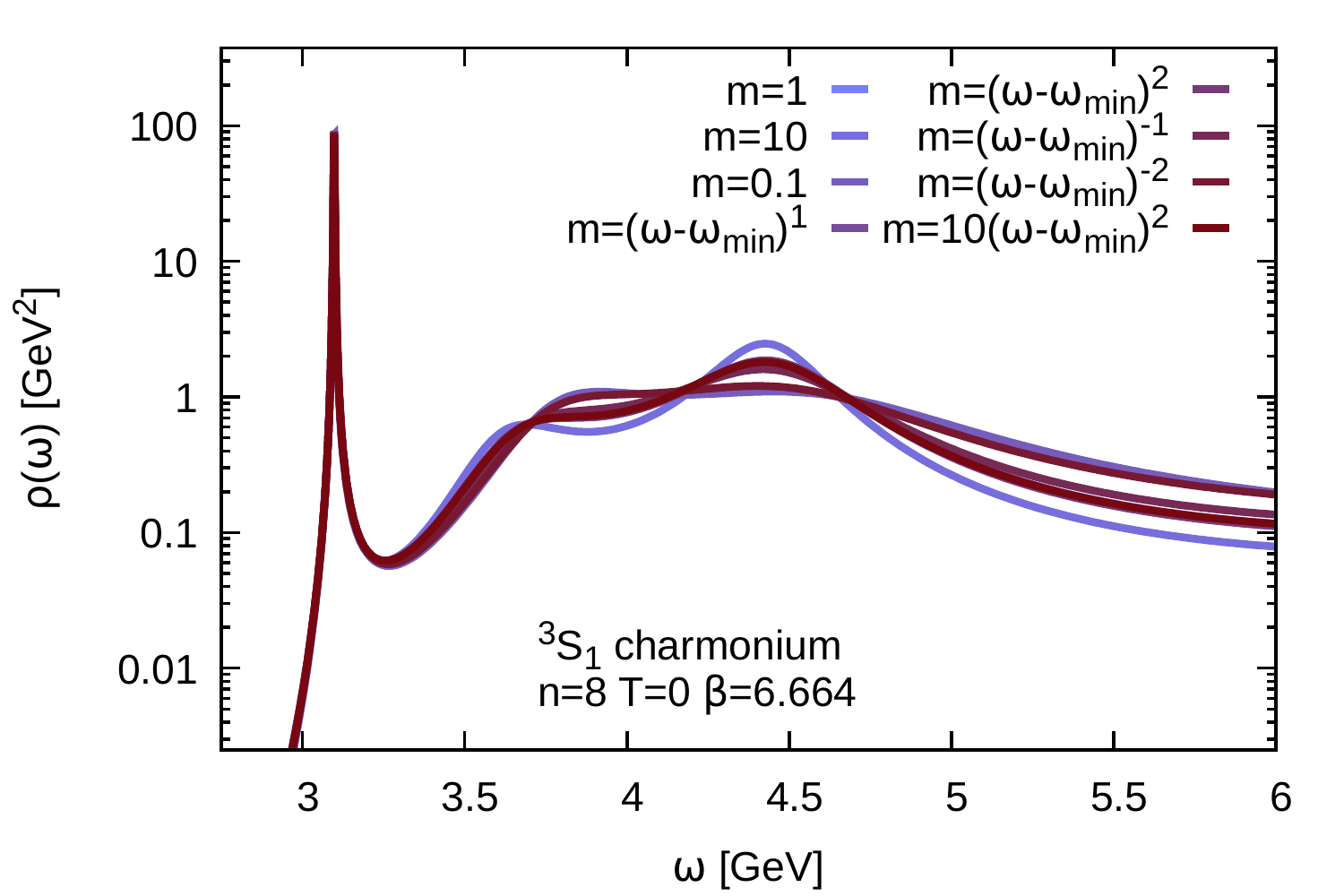}
\includegraphics[scale=0.5]{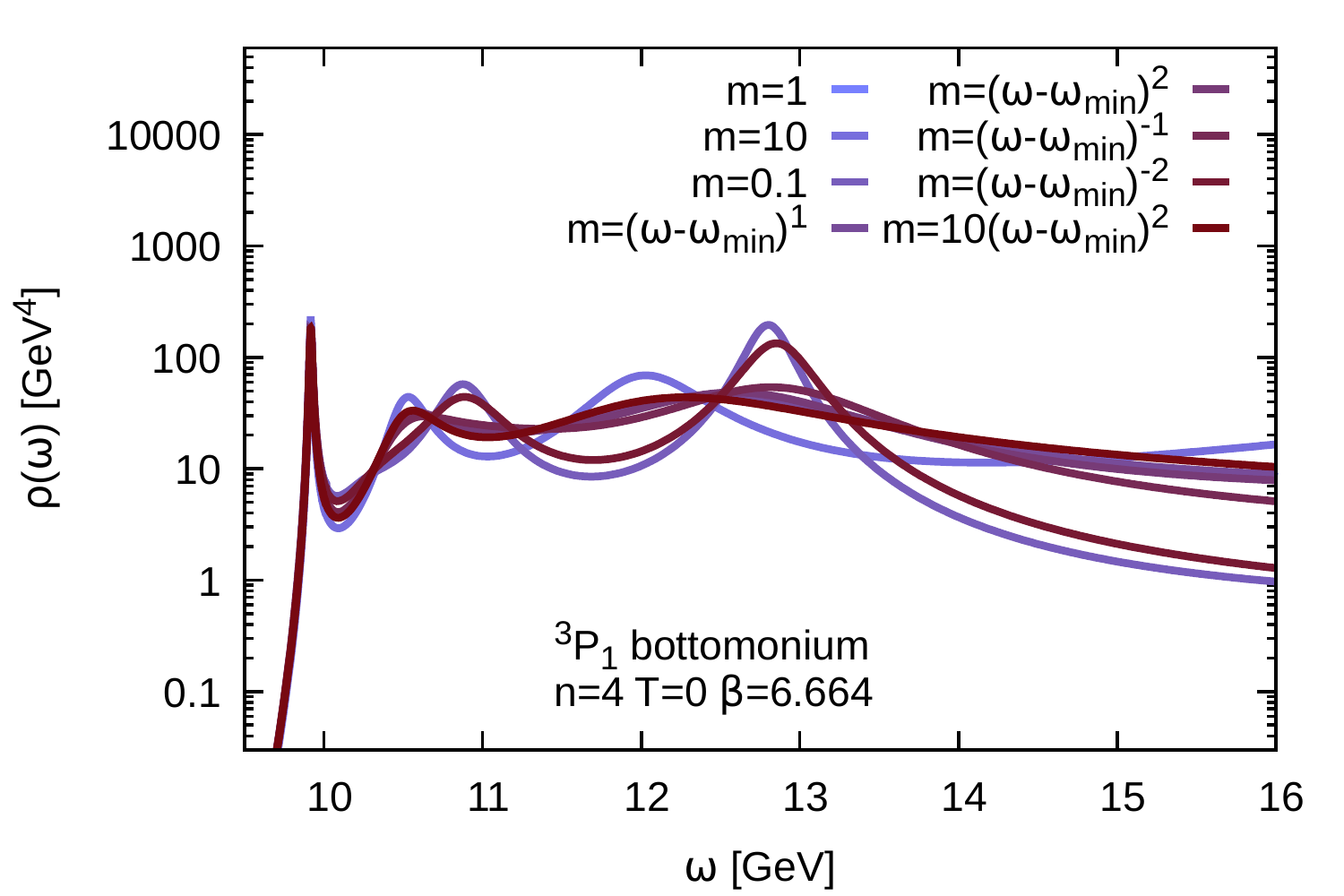}
\includegraphics[scale=0.5]{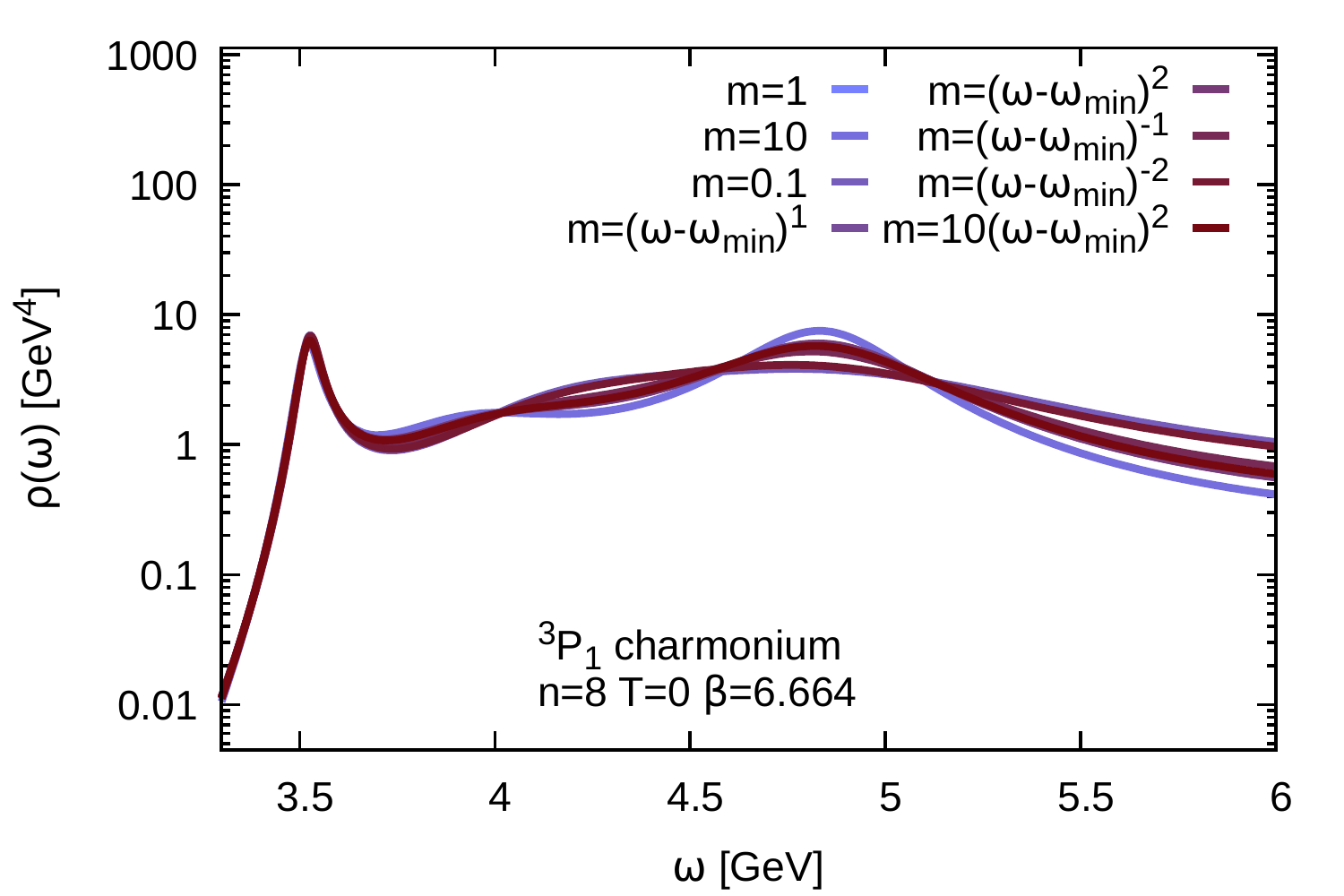}
\caption{Representative examples of the default model dependence of the spectral reconstructions at zero temperatures ($\beta=6.664$). Bottomonium results given on the left, charmonium on the right. We depict the S-wave channel in the top row, the P-wave channel in the bottom row.}\label{Fig:DefModeDepT0}
\end{figure}

We see in Fig.\ref{Fig:DefModeDepT0} that at zero temperatures the ground state peaks in both the S-wave and P-wave channel remain virtually unaffected, the availability of a relatively large number of data points $\tau_{\rm max}/a \geq 32$ as well as correspondingly large Euclidean extent are the reason. On the other hand the higher lying structures, either a manifestation of excited states or the continuum do show a significant dependence on the choice of $m$. This is reflected also in the large systematic errorbars on the excited states masses in Fig.\ref{Fig:T0BRMassesCmp}. 

\begin{figure}[t]
\includegraphics[scale=0.5]{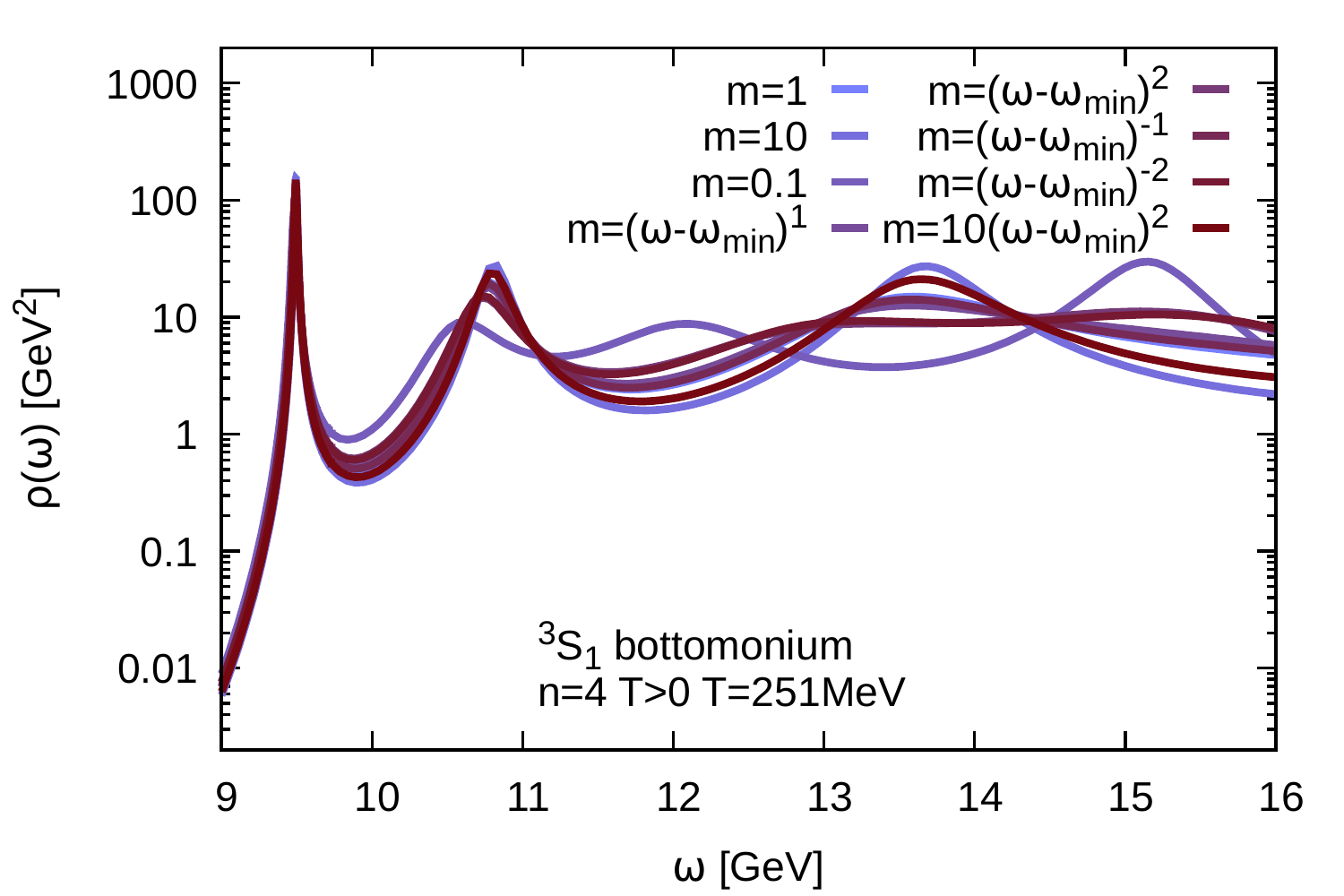}
\includegraphics[scale=0.5]{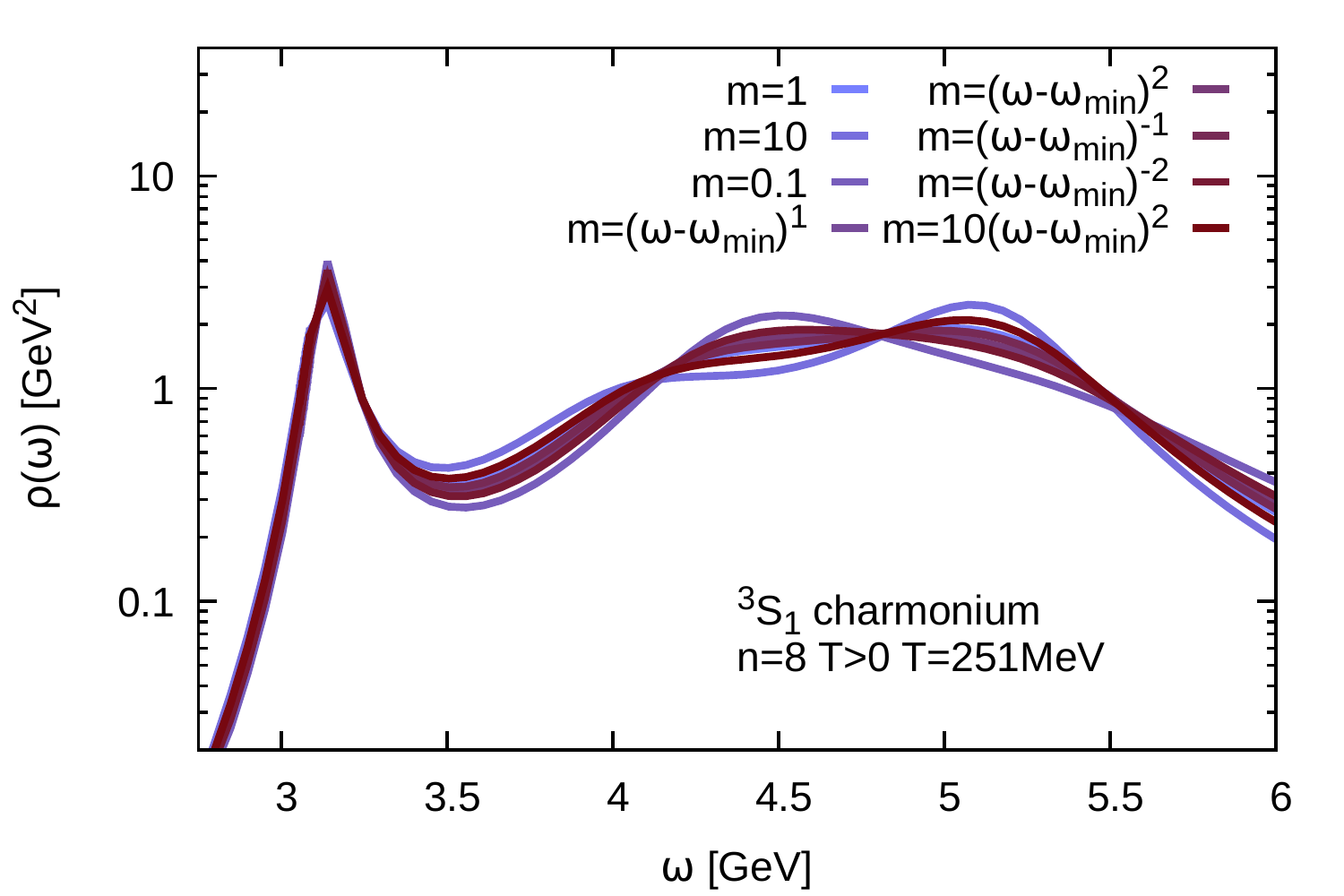}
\includegraphics[scale=0.5]{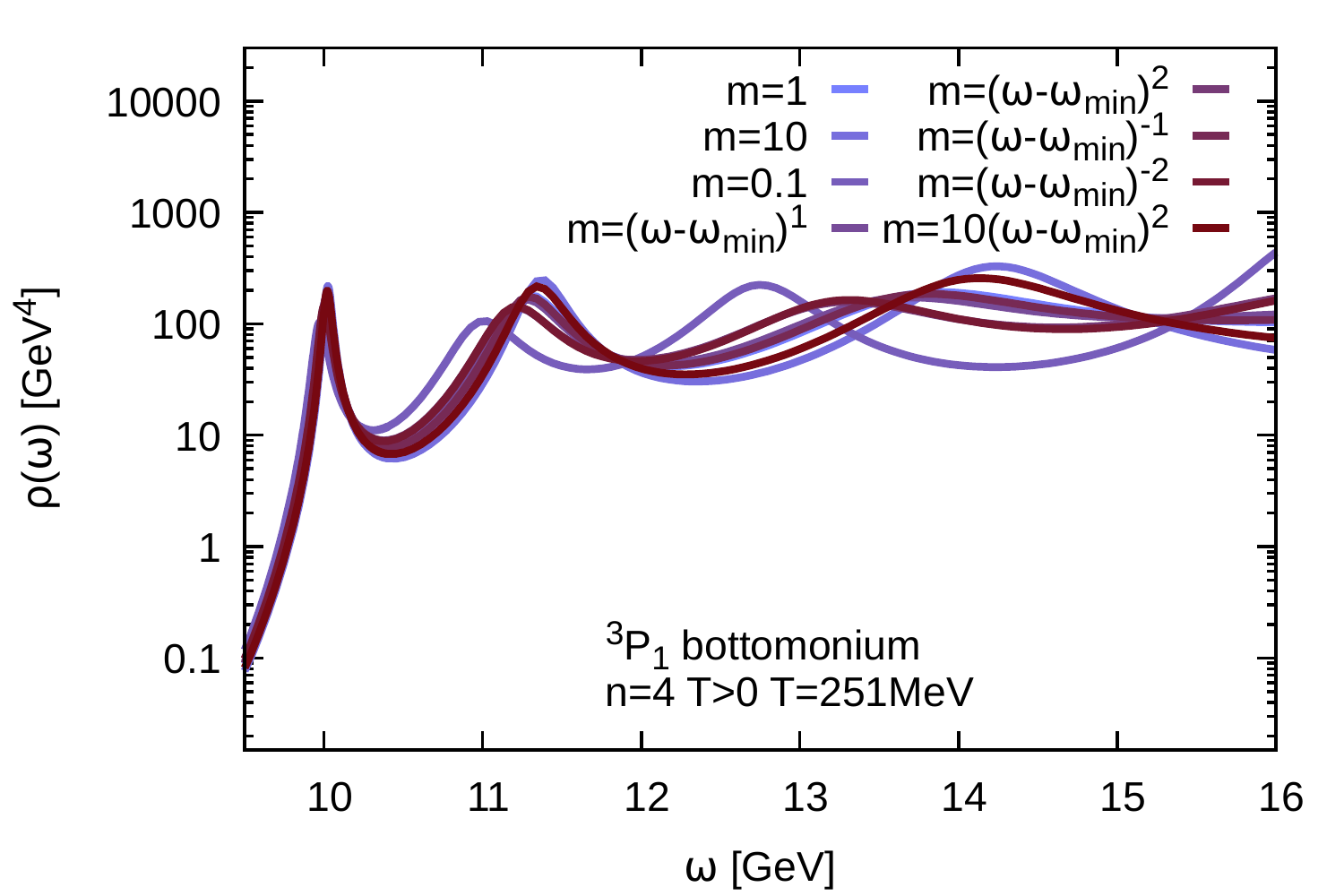}
\includegraphics[scale=0.5]{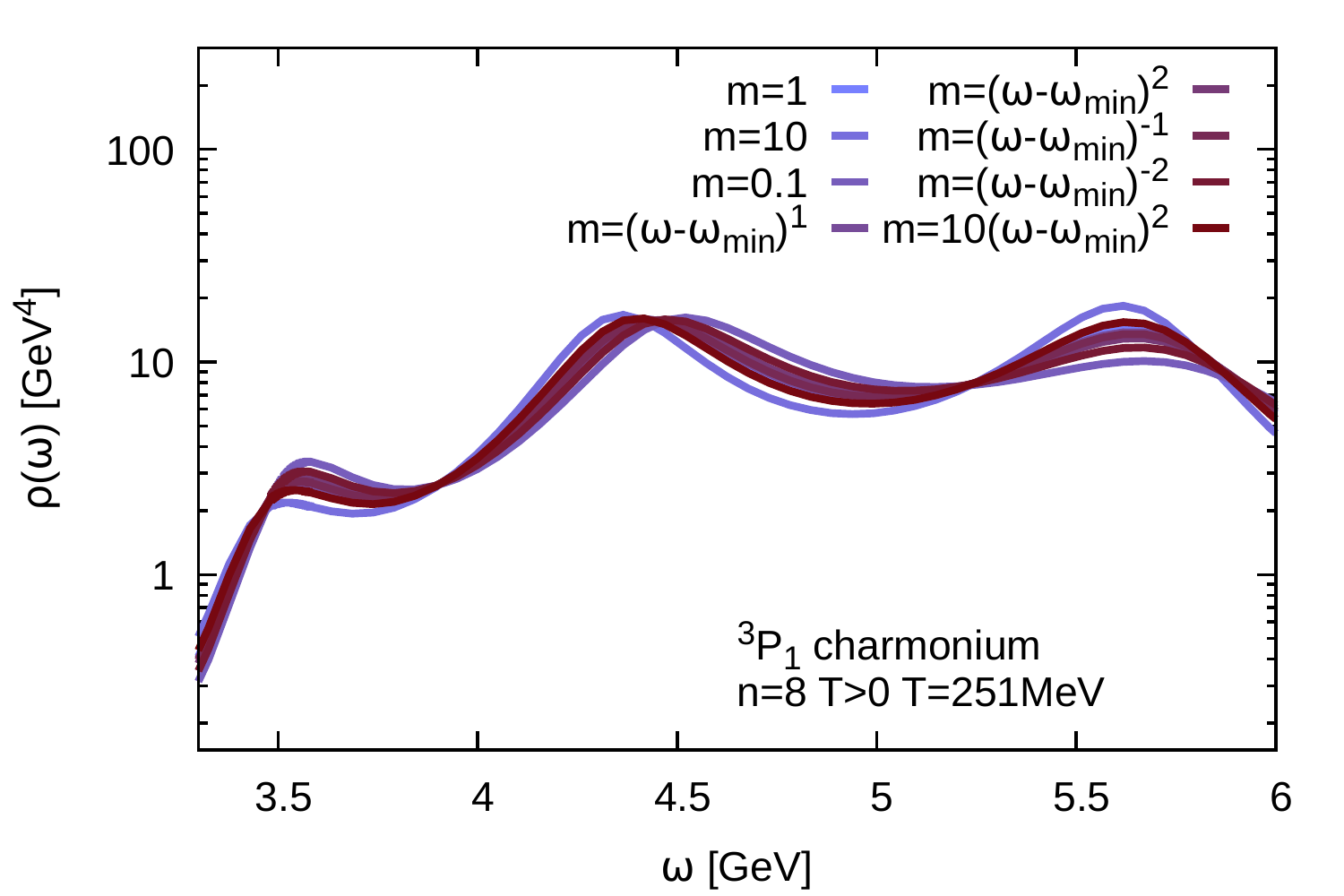}
\caption{Representative examples of the default model dependence of the spectral reconstructions at $T=251$MeV. Bottomonium results given on the left, charmonium on the right. We depict the S-wave channel in the top row, the P-wave channel in the bottom row.}\label{Fig:DefModeDepFiniteT}
\end{figure}

At finite temperature in Fig.\ref{Fig:DefModeDepFiniteT} the robustness of the ground state extraction is weaker, due to the smaller Euclidean extent, as well as the relatively small $\tau_{\rm max}/a=12$. Nevertheless due to the comparatively high statistics present in our study the ground state features appear well under control. The corresponding uncertainty is reflected in the errorbars of Fig.\ref{Fig:FiniteTInMediumMasses} and Fig.\ref{Fig:FiniteTMassShifts}.

\bibliographystyle{JHEP}
\bibliography{NRQCD}

\end{document}